\pdfoutput=1
\documentclass[%
 superscriptaddress,
preprint,
showpacs,preprintnumbers,
nofootinbib,
 amsmath,amssymb,
 aps,
prd,
floatfix,
]{revtex4-1}

\usepackage{graphicx}
\usepackage{dcolumn}
\usepackage{bm}
\usepackage{color}
\usepackage{subfig}
\usepackage{xspace}
\usepackage{hyperref}
\usepackage{natbib}
\usepackage{eqnarray,amsmath}
\usepackage{epstopdf}
\usepackage{multirow}
\input symbols

\def\bw{bw}
\def\figstyle{bw}

\begin{document}
\title{Measurements of neutrino oscillation in appearance and disappearance
channels by the T2K experiment \\ with $6.6\times10^{20}$ protons on target}


\newcommand{\INSTC}{\affiliation{University of Alberta, Centre for Particle Physics, Department of Physics, Edmonton, Alberta, Canada}}
\newcommand{\INSTEE}{\affiliation{University of Bern, Albert Einstein Center for Fundamental Physics, Laboratory for High Energy Physics (LHEP), Bern, Switzerland}}
\newcommand{\INSTFE}{\affiliation{Boston University, Department of Physics, Boston, Massachusetts, U.S.A.}}
\newcommand{\INSTD}{\affiliation{University of British Columbia, Department of Physics and Astronomy, Vancouver, British Columbia, Canada}}
\newcommand{\INSTGA}{\affiliation{University of California, Irvine, Department of Physics and Astronomy, Irvine, California, U.S.A.}}
\newcommand{\INSTI}{\affiliation{IRFU, CEA Saclay, Gif-sur-Yvette, France}}
\newcommand{\INSTGB}{\affiliation{University of Colorado at Boulder, Department of Physics, Boulder, Colorado, U.S.A.}}
\newcommand{\INSTFG}{\affiliation{Colorado State University, Department of Physics, Fort Collins, Colorado, U.S.A.}}
\newcommand{\INSTFH}{\affiliation{Duke University, Department of Physics, Durham, North Carolina, U.S.A.}}
\newcommand{\INSTBA}{\affiliation{Ecole Polytechnique, IN2P3-CNRS, Laboratoire Leprince-Ringuet, Palaiseau, France }}
\newcommand{\INSTEF}{\affiliation{ETH Zurich, Institute for Particle Physics, Zurich, Switzerland}}
\newcommand{\INSTEG}{\affiliation{University of Geneva, Section de Physique, DPNC, Geneva, Switzerland}}
\newcommand{\INSTDG}{\affiliation{H. Niewodniczanski Institute of Nuclear Physics PAN, Cracow, Poland}}
\newcommand{\INSTCB}{\affiliation{High Energy Accelerator Research Organization (KEK), Tsukuba, Ibaraki, Japan}}
\newcommand{\INSTED}{\affiliation{Institut de Fisica d'Altes Energies (IFAE), Bellaterra (Barcelona), Spain}}
\newcommand{\INSTEC}{\affiliation{IFIC (CSIC \& University of Valencia), Valencia, Spain}}
\newcommand{\INSTEI}{\affiliation{Imperial College London, Department of Physics, London, United Kingdom}}
\newcommand{\INSTGF}{\affiliation{INFN Sezione di Bari and Universit\`a e Politecnico di Bari, Dipartimento Interuniversitario di Fisica, Bari, Italy}}
\newcommand{\INSTBE}{\affiliation{INFN Sezione di Napoli and Universit\`a di Napoli, Dipartimento di Fisica, Napoli, Italy}}
\newcommand{\INSTBF}{\affiliation{INFN Sezione di Padova and Universit\`a di Padova, Dipartimento di Fisica, Padova, Italy}}
\newcommand{\INSTBD}{\affiliation{INFN Sezione di Roma and Universit\`a di Roma ``La Sapienza'', Roma, Italy}}
\newcommand{\INSTEB}{\affiliation{Institute for Nuclear Research of the Russian Academy of Sciences, Moscow, Russia}}
\newcommand{\INSTHA}{\affiliation{Kavli Institute for the Physics and Mathematics of the Universe (WPI), Todai Institutes for Advanced Study, University of Tokyo, Kashiwa, Chiba, Japan}}
\newcommand{\INSTCC}{\affiliation{Kobe University, Kobe, Japan}}
\newcommand{\INSTCD}{\affiliation{Kyoto University, Department of Physics, Kyoto, Japan}}
\newcommand{\INSTEJ}{\affiliation{Lancaster University, Physics Department, Lancaster, United Kingdom}}
\newcommand{\INSTFC}{\affiliation{University of Liverpool, Department of Physics, Liverpool, United Kingdom}}
\newcommand{\INSTFI}{\affiliation{Louisiana State University, Department of Physics and Astronomy, Baton Rouge, Louisiana, U.S.A.}}
\newcommand{\INSTJ}{\affiliation{Universit\'e de Lyon, Universit\'e Claude Bernard Lyon 1, IPN Lyon (IN2P3), Villeurbanne, France}}
\newcommand{\INSTHB}{\affiliation{Michigan State University, Department of Physics and Astronomy,  East Lansing, Michigan, U.S.A.}}
\newcommand{\INSTCE}{\affiliation{Miyagi University of Education, Department of Physics, Sendai, Japan}}
\newcommand{\INSTDF}{\affiliation{National Centre for Nuclear Research, Warsaw, Poland}}
\newcommand{\INSTFJ}{\affiliation{State University of New York at Stony Brook, Department of Physics and Astronomy, Stony Brook, New York, U.S.A.}}
\newcommand{\INSTGJ}{\affiliation{Okayama University, Department of Physics, Okayama, Japan}}
\newcommand{\INSTCF}{\affiliation{Osaka City University, Department of Physics, Osaka, Japan}}
\newcommand{\INSTGG}{\affiliation{Oxford University, Department of Physics, Oxford, United Kingdom}}
\newcommand{\INSTBB}{\affiliation{UPMC, Universit\'e Paris Diderot, CNRS/IN2P3, Laboratoire de Physique Nucl\'eaire et de Hautes Energies (LPNHE), Paris, France}}
\newcommand{\INSTGC}{\affiliation{University of Pittsburgh, Department of Physics and Astronomy, Pittsburgh, Pennsylvania, U.S.A.}}
\newcommand{\INSTFA}{\affiliation{Queen Mary University of London, School of Physics and Astronomy, London, United Kingdom}}
\newcommand{\INSTE}{\affiliation{University of Regina, Department of Physics, Regina, Saskatchewan, Canada}}
\newcommand{\INSTGD}{\affiliation{University of Rochester, Department of Physics and Astronomy, Rochester, New York, U.S.A.}}
\newcommand{\INSTBC}{\affiliation{RWTH Aachen University, III. Physikalisches Institut, Aachen, Germany}}
\newcommand{\INSTFB}{\affiliation{University of Sheffield, Department of Physics and Astronomy, Sheffield, United Kingdom}}
\newcommand{\INSTDI}{\affiliation{University of Silesia, Institute of Physics, Katowice, Poland}}
\newcommand{\INSTEH}{\affiliation{STFC, Rutherford Appleton Laboratory, Harwell Oxford,  and  Daresbury Laboratory, Warrington, United Kingdom}}
\newcommand{\INSTCH}{\affiliation{University of Tokyo, Department of Physics, Tokyo, Japan}}
\newcommand{\INSTBJ}{\affiliation{University of Tokyo, Institute for Cosmic Ray Research, Kamioka Observatory, Kamioka, Japan}}
\newcommand{\INSTCG}{\affiliation{University of Tokyo, Institute for Cosmic Ray Research, Research Center for Cosmic Neutrinos, Kashiwa, Japan}}
\newcommand{\INSTGI}{\affiliation{Tokyo Metropolitan University, Department of Physics, Tokyo, Japan}}
\newcommand{\INSTF}{\affiliation{University of Toronto, Department of Physics, Toronto, Ontario, Canada}}
\newcommand{\INSTB}{\affiliation{TRIUMF, Vancouver, British Columbia, Canada}}
\newcommand{\INSTG}{\affiliation{University of Victoria, Department of Physics and Astronomy, Victoria, British Columbia, Canada}}
\newcommand{\INSTDJ}{\affiliation{University of Warsaw, Faculty of Physics, Warsaw, Poland}}
\newcommand{\INSTDH}{\affiliation{Warsaw University of Technology, Institute of Radioelectronics, Warsaw, Poland}}
\newcommand{\INSTFD}{\affiliation{University of Warwick, Department of Physics, Coventry, United Kingdom}}
\newcommand{\INSTGE}{\affiliation{University of Washington, Department of Physics, Seattle, Washington, U.S.A.}}
\newcommand{\INSTGH}{\affiliation{University of Winnipeg, Department of Physics, Winnipeg, Manitoba, Canada}}
\newcommand{\INSTEA}{\affiliation{Wroclaw University, Faculty of Physics and Astronomy, Wroclaw, Poland}}
\newcommand{\INSTH}{\affiliation{York University, Department of Physics and Astronomy, Toronto, Ontario, Canada}}

\INSTC
\INSTEE
\INSTFE
\INSTD
\INSTGA
\INSTI
\INSTGB
\INSTFG
\INSTFH
\INSTBA
\INSTEF
\INSTEG
\INSTDG
\INSTCB
\INSTED
\INSTEC
\INSTEI
\INSTGF
\INSTBE
\INSTBF
\INSTBD
\INSTEB
\INSTHA
\INSTCC
\INSTCD
\INSTEJ
\INSTFC
\INSTFI
\INSTJ
\INSTHB
\INSTCE
\INSTDF
\INSTFJ
\INSTGJ
\INSTCF
\INSTGG
\INSTBB
\INSTGC
\INSTFA
\INSTE
\INSTGD
\INSTBC
\INSTFB
\INSTDI
\INSTEH
\INSTCH
\INSTBJ
\INSTCG
\INSTGI
\INSTF
\INSTB
\INSTG
\INSTDJ
\INSTDH
\INSTFD
\INSTGE
\INSTGH
\INSTEA
\INSTH

\author{K.\,Abe}\INSTBJ
\author{J.\,Adam}\INSTFJ
\author{H.\,Aihara}\INSTCH\INSTHA
\author{T.\,Akiri}\INSTFH
\author{C.\,Andreopoulos}\INSTEH\INSTFC
\author{S.\,Aoki}\INSTCC
\author{A.\,Ariga}\INSTEE
\author{S.\,Assylbekov}\INSTFG
\author{D.\,Autiero}\INSTJ
\author{M.\,Barbi}\INSTE
% ...

\collaboration{The T2K Collaboration}\noaffiliation

\date{\today}

\begin{abstract}
We report on measurements of neutrino oscillation using data from the T2K long-baseline neutrino experiment collected between 2010 and 2013. In an analysis of muon neutrino disappearance alone, we find the following estimates and 68\% confidence intervals for the two possible mass hierarchies: 
\begin{eqnarray*}
{\textrm{Normal Hierarchy:}}& \sin^2\theta_{23}=0.514^{+0.055}_{-0.056} \ \  {\textrm{~and~}} \ \ \Delta m_{32}^{2}=(2.51\pm0.10)\times10^{-3}\evvcccc\\
{\textrm{Inverted Hierarchy:}}& \sin^2\theta_{23}=0.511\pm0.055 \ \ {\textrm{~and~}} \ \  \Delta m_{13}^{2}=(2.48\pm0.10)\times10^{-3}\evvcccc
\end{eqnarray*}
The analysis accounts for multi-nucleon mechanisms in neutrino interactions which were found to introduce negligible bias.

We describe our first analyses that combine measurements of muon neutrino disappearance 
and electron neutrino appearance to estimate four oscillation parameters, \Dmsq, \stt, \sot, \dcp, and the mass hierarchy. 
Frequentist and Bayesian intervals are presented for combinations of these parameters, with and without including recent
reactor measurements.
At 90\% confidence level and including reactor measurements, we exclude the region
$\delta_{CP} = [0.15,0.83]\pi$ for normal hierarchy and
$\delta_{CP} = [-0.08,1.09]\pi$ for inverted hierarchy.
The T2K and reactor data weakly favor the normal hierarchy with a Bayes Factor of 2.2.
The most probable values and 68\% 1D credible intervals for the other oscillation parameters,
when reactor data are included, are:
\begin{equation*}
\sin^2\theta_{23}=0.528^{+0.055}_{-0.038} \ \ {\textrm{~and~}} \ \  |\dmsq|=(2.51\pm0.11)\times10^{-3}\evvcccc \ \ .
\end{equation*}
\end{abstract}
\pacs{14.60.Pq}

\maketitle

\section{\label{sec:intro} Introduction}
Neutrino oscillation was firmly established in the late 1990's with the observation by the 
Super-Kamiokande (SK) experiment that muon neutrinos produced by cosmic ray interactions 
in our atmosphere changed their flavor~\cite{Ashie:2005ik}. 
Measurements from the Sudbury Neutrino Observatory a few years later,
in combination with SK data,
revealed that neutrino oscillation was responsible for the 
apparent deficit of electron neutrinos produced in the Sun~\cite{PhysRevLett.89.011301}.
In the most recent major advance, the T2K experiment~\cite{Abe:2013xua,Abe:2013hdq} and 
reactor experiments~\cite{An:2012eh,An:2013zwz,Ahn:2012nd,PhysRevLett.108.131801} 
have established that all three neutrino mass states are mixtures of all three flavor states,
which allows the possibility of CP violation in neutrino oscillation.
This paper describes our most recent measurements of neutrino oscillation including
our first results from analyses that combine measurements of muon neutrino disappearance 
and electron neutrino appearance.

The Tokai to Kamioka (T2K) experiment~\cite{Abe:2011ks} was made possible 
by the construction of the J-PARC high-intensity proton
accelerator at a site that is an appropriate distance from the SK 
detector for precision measurements of neutrino oscillation.
Protons, extracted from the J-PARC main ring, strike a target to produce secondary hadrons, 
which are focused and subsequently decay in-flight to produce an intense neutrino beam, 
consisting mostly of muon neutrinos.
The neutrino beam axis is directed 2.5~degrees away from the SK detector, in order 
to produce a narrow-band 600~MeV flux at the detector,
the energy that maximizes muon neutrino oscillation at the 295~km baseline.
Detectors located 280~m downstream of the production target measure the properties of the neutrino
beam, both on-axis (INGRID detector) and off-axis in the direction of SK (ND280 detector).

T2K began operation in 2010 and was interrupted for one year by the Great East Japan Earthquake in 2011.
The results reported in this paper use data collected through 2013, as summarized in Tab.~\ref{tbl:run1to4_pot}. 
With these data, almost 10\% of the total proposed for the experiment,
T2K enters the era of precision neutrino oscillation measurements.
In 2014, we began to collect our first data in which the current in the magnetic focusing horns is reversed, so
as to produce a beam primarily of muon anti-neutrinos.
Future publications will report on measurements using that beam configuration.

We begin this paper by describing the neutrino beamline and how we model neutrino production and
interactions.
We then summarize the near detectors
and explain how we use their data to improve model predictions of neutrino interactions
at the far detector.
This is followed by an overview of the far detector, how neutrino candidate events are selected, and how
we model the detector response.
Next, we describe the neutrino oscillation model, 
list the external inputs for the oscillation parameters,
summarize the approaches used in the oscillation analyses,
and characterize our main sources of systematic uncertainty.
The final sections give detailed descriptions and results for the analysis of \num\ disappearance 
alone~\cite{Abe:2014ugx} 
and for the joint analyses of \num\ disappearance and \nue\ appearance.

\begin{table}[h]
\caption{
T2K data-taking periods and the protons on target (POT) used in the analyses presented in this paper.
The maximum stable proton beam power achieved was 230~kW.
}
\label{tbl:run1to4_pot}
\begin{tabular}{ l c c }
\hline\hline
Run Period & Dates & POT \\ \hline
Run 1 & Jan. 2010-Jun. 2010 & \(0.32\times10^{20}\) \\
Run 2 & Nov. 2010-Mar. 2011 & \(1.11\times10^{20}\) \\
Run 3 & Mar. 2012-Jun. 2012 & \(1.58\times10^{20}\) \\
Run 4 & Oct. 2012-May 2013& \(3.56\times10^{20}\) \\
\hline
Total & Jan. 2010-May 2013 & \(6.57\times10^{20}\) \\
\hline \hline
\end{tabular}
\end{table}

\clearpage
\section{\label{sec:beam} Neutrino Beamline}
The T2K primary beamline transports and focuses the 30\,\gev proton beam extracted from the J-PARC 
Main Ring onto a 91.4-cm long graphite target. 
The secondary beamline consists of the target station, decay volume, and beam dump.
The apparatus has been described in detail elsewhere~\cite{Abe:2011ks}.

The upstream end of the target station contains a collimator to protect 
the three downstream focusing horns.  The graphite target sits inside the first horn, and pions 
and other particles exiting the target are focused by these 
magnetic horns and are allowed to decay in the 96-m-long decay volume.
Following the decay volume, protons and other particles that have not decayed
are stopped in a beam dump consisting of 3.2~m of graphite and 2.4~m of iron, while muons above 5\,\gev pass through and are
detected in a Muon Monitor, designed to monitor the beam stability.  
With further absorption by earth, a beam of only neutrinos (primarily \num) continues to the near and far detectors.

\subsection{Neutrino flux simulation}
\label{sec:beam:fluxmc}
The secondary beamline is simulated in order to 
estimate the nominal neutrino flux (in absence of neutrino oscillations) at
the near and far detectors and the
covariance arising from uncertainties in hadron production and the beamline
configuration~\cite{PhysRevD.87.012001}.
We use the FLUKA 2008 package~\cite{Ferrari:2005zk,Battistoni:2007zzb} to model the
interactions of the primary beam protons and the subsequently-produced pions and kaons in
the graphite target.  
As described below, we tune this simulation using external hadron production data.
Particles exiting the target are tracked through the magnetic
horns and decay volume in a GEANT3~\cite{GEANT3} simulation using the GCALOR~\cite{GCALOR} package
to model the subsequent hadron decays.  

In order to precisely predict the neutrino flux,
each beam pulse is measured in the primary neutrino beamline. 
The suite of proton beam monitors consists of five current transformers which measure 
the proton beam intensity, 21 electrostatic monitors which measure the 
proton beam position, and 19 segmented secondary emission monitors and an optical transition
radiation monitor~\cite{Bhadra:2012st} which
measure the proton beam profile.  The proton beam properties have been stable
throughout T2K operation, and their values and uncertainties for the most
recent T2K run period, Run 4, are given in Tab.~\ref{tbl:pbeam_run4}.  
The values for other run periods have been published previously~\cite{PhysRevD.87.012001}.
The neutrino beam position and width stability is also monitored by the
INGRID detector, and the results are given
in Sec.~\ref{sec:INGRID}.

\begin{table}[tbp]
\caption{Summary of the estimated proton beam properties and their systematic errors
at the collimator for the T2K Run 4 period.  Shown are the mean position (\(X, Y\)), angle
(\(X', Y'\)), width (\(\sigma\)), emittance (\(\epsilon\)), and Twiss parameter (\(\alpha\))~\cite{McDonald:1989}.}
\label{tbl:pbeam_run4}
\begin{tabular}{ l c c c c }
\hline\hline
& \multicolumn{2}{c}{X Profile} & \multicolumn{2}{c}{Y Profile} \\
Parameter & Mean & Error & Mean & Error \\ \hline
\(X,Y\) (mm) & 0.03 &  0.34 & -0.87 & 0.58 \\
\(X',Y'\) (mrad) & 0.04 &  0.07 & 0.18  & 0.28 \\
\(\sigma\) (mm) & 3.76 & 0.13 & 4.15 & 0.15 \\
\(\epsilon\) (\(\pi\) mm mrad) & 5.00 & 0.49 & 6.14 & 2.88 \\
\(\alpha\) & 0.15 & 0.10 & 0.19 & 0.35 \\
\hline \hline
\end{tabular}
\end{table}

To improve the modeling of
hadron interactions inside and outside the target, we use data from
the NA61/SHINE experiment~\cite{Abgrall:2011ae,Abgrall:2011ts} collected at 31\,\gevc
and several other experiments~\cite{eichten,allaby,e910}.  
The hadron production data used for the oscillation analyses described
here are equivalent to those used in our previous publications~\cite{PhysRevD.87.012001,Abe:2013xua}, 
including the statistics-limited NA61/SHINE dataset taken in 2007 on a thin carbon target. 
The NA61/SHINE data analyses of the 2009 thin-target and T2K-replica-target data are ongoing, 
and these additional data will be used in future T2K analyses.
We incorporate the external hadron production data by
weighting each simulated hadron interaction
according to the measured multiplicities and particle production cross sections,
using the true initial and final state hadron kinematics, as well
as the material in which the interaction took place. 
The predicted flux at SK from the T2K beam is shown in Fig.~\ref{fig:flux_at_sk}.

\subsection{Neutrino flux uncertainties}
\label{sec:beam:fluxerrs}
Uncertainty in the neutrino flux prediction arises from the hadron production model,
proton beam profile, horn current, horn alignment, and other factors.
For each source of uncertainty, we vary the underlying parameters to evaluate the
effect on the flux prediction in bins of neutrino energy for each neutrino 
flavor~\cite{PhysRevD.87.012001}.
Table~\ref{tbl:beam_errs} shows the breakdown for the \num\ and \nue\ flux uncertainties for
energy bins near the peak energy.

\begin{table}[tbp]
\caption{Contributions to the systematic uncertainties for the unoscillated \num\ and \nue\ 
flux prediction at SK, near the peak energy and without the use of near detector data.
The values are shown for the \num\ (\nue) energy bin 0.6~GeV $<E_\nu<$ 0.7~GeV (0.5~GeV $<E_\nu<$ 0.7~GeV).}

\label{tbl:beam_errs}
\begin{tabular}{ l c c }
\hline\hline
Error source & \multicolumn{2}{c}{Uncertainty in SK flux near peak (\%)} \\
 & \(\nu_\mu\) & \(\nu_e\) \\
 \hline
 Beam current normalization & 2.6  & 2.6 \\
 Proton beam properties & 0.3 & 0.2 \\
 Off axis angle & 1.0 & 0.2 \\
 Horn current & 1.0 & 0.1 \\
 Horn field & 0.2 & 0.8 \\
 Horn misalignment & 0.4 & 2.5 \\
 Target misalignment & 0.0 & 2.0 \\
 MC statistics & 0.1 & 0.5 \\
 \hline
 Hadron production &  & \\
 \qquad Pion multiplicities & 5.5 & 4.7 \\
 \qquad Kaon multiplicities & 0.5 & 3.2 \\
 \qquad Secondary nucleon multiplicities & 6.9 & 7.6 \\
 \qquad Hadronic interaction lengths & 6.7 & 6.9 \\
 Total hadron production & 11.1 & 11.7 \\
 \hline
 Total & 11.5 & 12.4 \\
 \hline \hline
 \end{tabular}
\end{table}

The largest uncertainty from beam monitor calibrations arises in the beam current measurement 
using a current transformer, but its effect on the oscillation analyses is reduced through the 
use of near detector data.  The remaining uncertainties due to the uncertain position and 
calibration of the other beam monitors are significantly smaller.
As described in Sec.~\ref{sec:INGRID}, 
the neutrino beam direction is determined with the INGRID detector, 
and therefore the assigned uncertainty on the off-axis angle comes directly from
the INGRID beam profile measurement.
To account for the horn current measurement that drifts over time and a possible scale uncertainty,
5~kA is assigned as a conservative estimate of the horn current error.
In the flux simulation, the horn magnetic field is assumed to have a \(1/r\) dependence.
Deviations from this field, measured using a Hall probe, are used to define the
uncertainty of the horn field.
Horn and target alignment uncertainties come from survey measurements. 

Systematic uncertainties in modeling particle multiplicities from hadronic interactions come
from several sources:
experimental uncertainties in the external data,
the uncertain scaling to different
incident particle momenta and target materials,
and extrapolation to
regions of particle production phase space not covered by external 
data~\cite{PhysRevD.87.012001}.
The overall uncertainty is described by calculating
the covariance of the pion, kaon, and 
secondary nucleon 
multiplicities 
and their interaction lengths.

\begin{figure}[tbp]
\begin{center}
\includegraphics[width=0.8\textwidth]{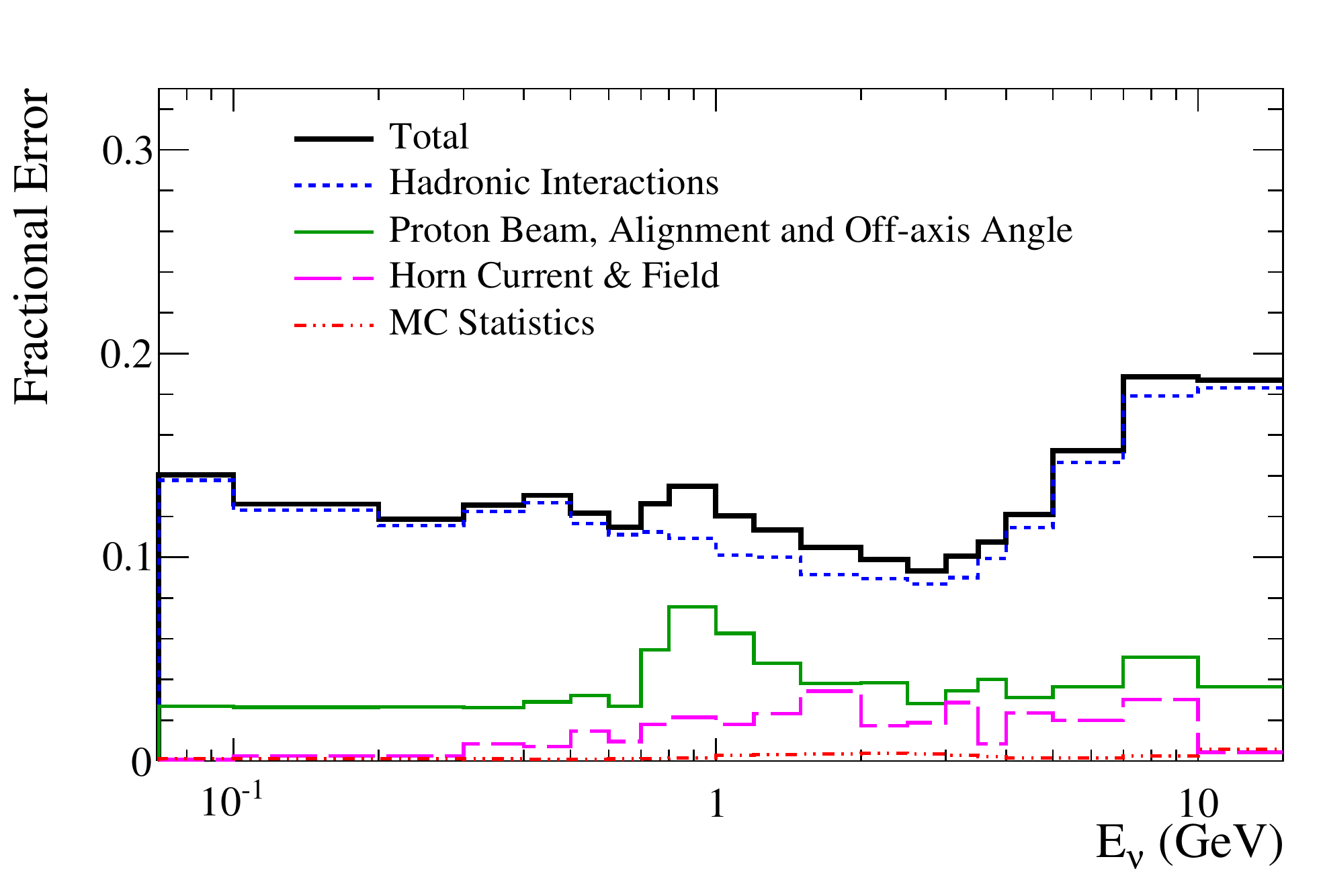}
\caption{Fractional systematic error on the \(\nu_\mu\) 
flux at SK arising from the beamline configuration and hadron production, prior to applying
near detector data constraints.}
\label{fig:beam_errs_breakdown}
\end{center}
\end{figure}

The systematic errors on the \(\nu_\mu\) flux at SK, without applying near detector data, 
are shown in bins of neutrino energy in Fig.~\ref{fig:beam_errs_breakdown}.
The dominant source of uncertainty is from hadron production.

\begin{figure}[tbp]
\begin{center}
\includegraphics[width=100mm]{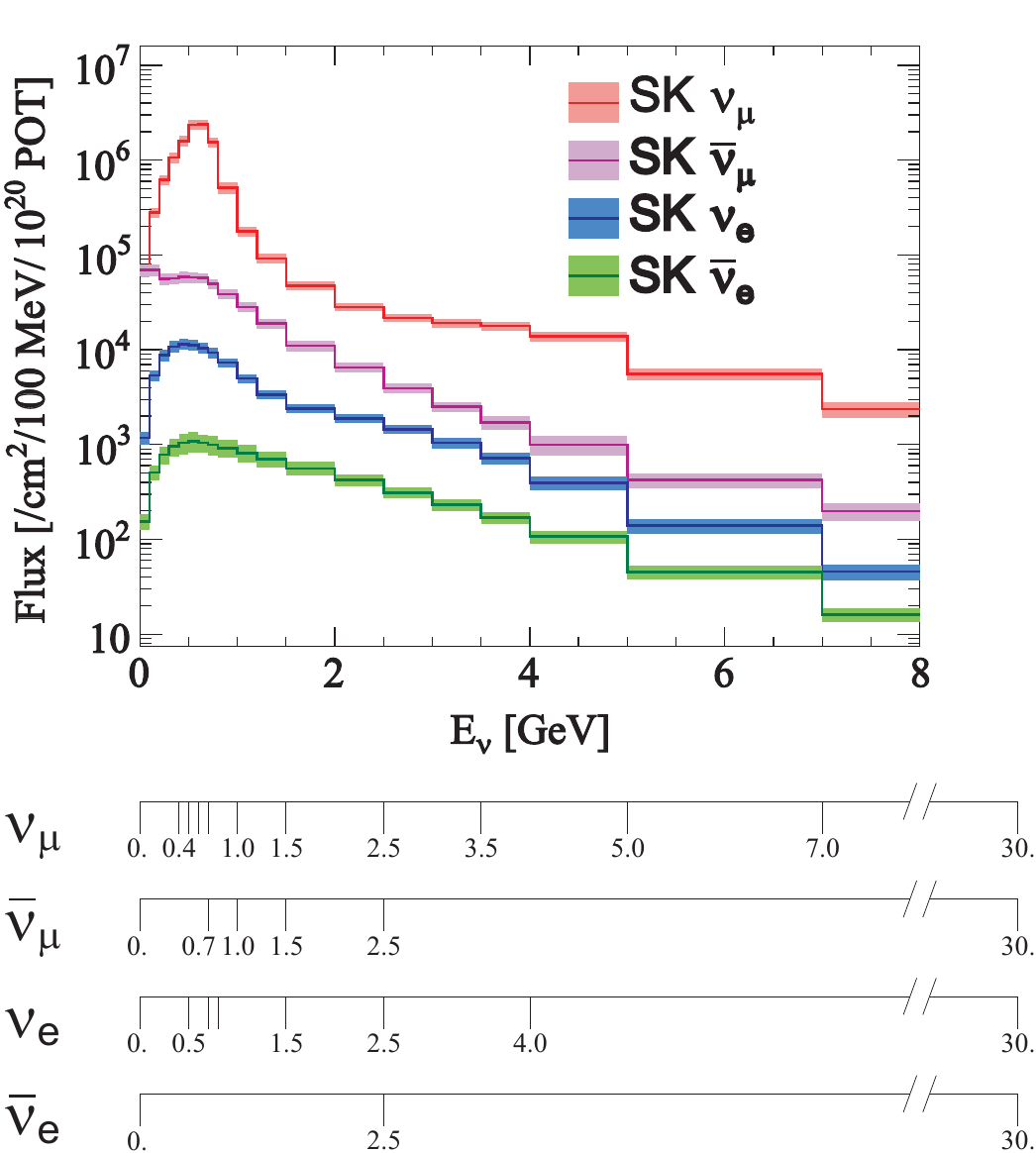}
\caption{The T2K unoscillated neutrino flux prediction at SK is shown with bands indicating 
the systematic uncertainty prior to applying near detector data. The
flux in the range 8~GeV $< E_\nu <$ 30~GeV is simulated but not shown.
The binning for the vector of systematic parameters, $\vec{b}$, for each neutrino component is shown by
the four scales. The same binning is used for the ND280 and SK flux systematic parameters, $\vec{b}_{n}$ and
$\vec{b}_{s}$.
}
\label{fig:flux_at_sk}
\end{center}
\end{figure}

For analyses of near and far detector data,
the uncertainties arising from the beamline configuration and hadron production are
propagated using a vector of systematic parameters, \(\vec{b}\), which scale the nominal
flux in bins of neutrino energy, for each neutrino type (\(\nu_e\), \(\nu_\mu\),
\(\bar{\nu}_e\), \(\bar{\nu}_\mu\)) at each detector (ND280 and SK).
The energy binning for each neutrino type is shown in Fig.~\ref{fig:flux_at_sk}.
The covariance for these parameters is calculated separately for each T2K run
period given in Tab.\ \ref{tbl:run1to4_pot}, and the POT-weighted average is the
flux covariance, \(V_b\), used by the near detector and oscillation analyses.
We define $\vec{b}_{n}$ and $\vec{b}_{s}$ as the sub-vector elements of $\vec{b}$ for
ND280 and SK.
It is through the covariance between $\vec{b}_{n}$ and $\vec{b}_{s}$ that the near detector measurements
of \num\ events constrain the expected unoscillated far detector \num\ and \nue\
event rates in the oscillation analyses.

\clearpage
\section{\label{sec:nuint} Neutrino Interaction Model}
Precision neutrino oscillation measurements rely on having an accurate neutrino interaction model.  The model is used to evaluate the selection efficiencies of the different signal and background interactions as well as the estimate of the neutrino energy from the detected final state particles.
Finally, the model forms the basis to account for differences in the predicted neutrino cross sections between different T2K detectors due to their different target nuclei compositions. All of these factors and their uncertainties are incorporated into the model for the T2K experiment through a set of systematic parameters $\vec{x}$ listed in Tab.~\ref{tbl:xsecpar}, and their covariance $V_x$.

This section describes the interaction model in \neut, the primary neutrino interaction generator used by T2K, explains how we use data from external experiments to provide initial constraints on the model before fitting to T2K data, discusses remaining uncertainties not constrained by external data sources, and discusses uncertainties based on differences between the \neut model and those found in other interaction generators.

\subsection{Neutrino Interaction Model}
\label{subsec:nuintmodel}
The interaction model used in this analysis is \neut~\cite{Hayato:2009} version 5.1.4.2, which models neutrino interactions on various nuclear targets over a range of energies from $\sim$100\,MeV to $\sim$100\,TeV. \neut simulates seven types of charged current (CC) and neutral current (\nc) interactions: (quasi-)elastic scattering, single pion production, single photon production, single kaon production, single eta production, deep inelastic scattering (DIS), and coherent pion production. Interactions not modeled in this version of \neut include, but are not limited to, multi-nucleon interactions in the nucleus~\cite{Nieves:2012,Martini:2010}, and neutrino-electron scattering processes.

The Llewellyn Smith model~\cite{LlewellynSmith:1972} is used as the basis to describe
charged current quasi-elastic (CCQE) and neutral current elastic scattering (NCEL) interactions.  
In order to take into account the fact that the target nucleon is in a nucleus, the Relativistic Fermi Gas (RFG) model by Smith and Moniz~\cite{SmithMoniz:1972,SmithMonizErratum} is used.  The model uses dipole axial form factors and the vector form factors derived from electron scattering experiments~\cite{Bradford:2006yz}.  The default quasi-elastic axial mass, \maqe, is 1.21\,\gevcc and the default Fermi momenta for the two dominant target nuclei carbon and oxygen are 217\,\mevc and 225\,\mevc, respectively.
Appropriate Fermi momenta, $p_F$, and binding energies, $E_B$, are assigned to the other
target nuclei.

The Rein and Sehgal model~\cite{ReinSehgal:1981} is used to simulate neutrino-induced single pion production.  The model assumes the interaction is split into two steps as follows: $\nu + N \to \ell + N^{\star}$, $N^{\star} \to \pi + N'$, where $N$ and $N'$ are nucleons, $\ell$ is an outgoing neutrino or charged lepton, and $N^{\star}$ is the resonance.  For the initial cross section calculation, the amplitude of each resonance production is multiplied by the branching fraction of the resonance into a pion and nucleon.  Interference between 18 resonances with masses below 2\,\gevcc are included in the calculation.  To avoid double counting processes that produce a single pion through either resonance or DIS in calculating the total cross section, the invariant hadronic mass $W$ is restricted to be less than 2\,\gevcc.  
The model assigns a 20\% branching fraction for the additional delta decay channel that can occur in the nuclear medium, $\Delta+ N\rightarrow N + N$, which we refer to as pion-less delta decay (PDD).
Since the Rein and Sehgal model provides the amplitudes of the neutrino resonance production, we adjust the \neut predictions for  the cross sections of single photon, kaon, and eta production by changing the branching fractions of the various resonances.

The coherent pion production model is described in~\cite{Rein:1982pf}.  The interaction is described as $\nu + A \to \ell + \pi + X$, where $A$ is the target nucleus, $\ell$ is the outgoing lepton, $\pi$ is the outgoing pion, and $X$ is the remaining nucleus.  The CC component of the model takes into account the lepton mass correction provided by the same authors~\cite{Rein:2006di}.

The DIS cross section is calculated over the range of $W>1.3$\,\gevcc.  The structure functions are taken from the GRV98 parton distribution function~\cite{Gluck:1998xa} with corrections proposed by Bodek and Yang~\cite{Bodek:2003wd} to improve agreement with experiments in the low-$Q^{2}$ region.  To avoid double counting single pion production with the resonance production described above, in the region $W\le2$\,\gevcc the model includes the probability to produce more than one pion only. For $W>2$\,\gevcc, \neut uses PYTHIA/JetSet~\cite{Sjostrand:1993yb} for hadronization while for $W\le2$\,\gevcc it uses its own model.

Hadrons that are generated in a neutrino-nucleus interaction can interact with the nucleus and these final state interactions (FSI) can affect both the total number of particles observed in a detector and their kinematics.  \neut uses a cascade model for pions, kaons, etas, and nucleons.  Though details are slightly different between hadrons, the basic procedure is as follows. 
The starting point for the cascade model is the neutrino interaction point in the nucleus based on a Woods-Saxon density distribution~\cite{Woods:1954zz} except in DIS, where a formation zone is taken into account.
The hadron is moved a small distance and interaction probabilities for that step are calculated.  The interaction types include charge exchange, inelastic scattering, particle production, and absorption.  If an interaction has occurred, then the kinematics of the particle are changed as well as the particle type if needed. The process is repeated until all particles are either absorbed or escape the nucleus.  

\subsection{Constraints From External Experiments}
\label{subsec:externalconstraints}
To establish prior values and errors for neutrino-interaction systematic parameters $\vec{x}$ and constrain a subset for which ND280 observables are insensitive, neutrino-nucleus scattering data from external experiments are used.

The datasets external to T2K come from two basic sources: pion-nucleus and neutrino-nucleus scattering experiments.  To constrain pion-nucleus cross section parameters in the \neut FSI model, pion-nucleus scattering data on a range of nuclear targets are used.  The most important external source of neutrino data for our interaction model parameter constraints is the \mb experiment~\cite{mb-nim}.
The \mb flux~\cite{mb-flux} covers an energy range similar to that of T2K and as a 4$\pi$ detector like SK has a similar phase space acceptance, meaning \neut is tested over a broader range of $Q^{2}$ than current ND280 analyses.  

\subsubsection{Constraints From Pion-Nucleus Scattering Experiments}
\label{sub:piA}
To evaluate the uncertainty in the pion transport model in the nucleus, we consider the effects of varying the pion-nucleus interaction probabilities via six scale factors.  These scale factors affect the following processes in the cascade model: absorption ($x^{FSABS}$), low energy QE scattering including single charge exchange ($x^{FSQE}$) and low energy single charge exchange (SCX) ($x^{FSCX}$) in a nucleus, high energy QE scattering ($x^{FSQEH}$), high energy SCX ($x^{FSCXH}$), and pion production ($x^{FSINEL}$).  The low (high) energy parameters are used for events with pion momenta below (above) 500\,\mevc with the high energy parameters explicitly given and the remaining parameters all low energy. 
The simulation used to perform this study is similar to the one in~\cite{Salcedo:1988}.  The model is fit to a range of energy-dependent cross sections comprising nuclear targets from carbon to 
lead~\cite{ashery:piscat,levenson:piscat,ingram:piscat,jones:piscat,giannelli:piscat,ransome:piscat,miller:piscat,nakai:piscat,navon:piscat,ashery:pioncx,rowntree:piscat,fujii:piscat,saunders:piscat,allardyce:piscat,cronin:piscat,crozon:piscat,binon:piscat,wilkin:piscat,clough:piscat,carroll:piscat,bowles:piscat,wood:piondcx,takahashi:piscat,gelderloos:piscat,grion:piscat,rahav:piscat,aoki:piscat}.  
The best-fit scale factors for these parameters are shown in Tab.~\ref{tab:fsi_parsets} as well as the maximum and minimum values for each parameter taken from 16 points on the 1$\sigma$ surface of the 6-dimensional parameter space.  The parameter sets are used for assessing systematic uncertainty in secondary hadronic interactions in the near and far detectors, as discussed in Secs.~\ref{sec:BANFF}B and~\ref{sec:SK}C, respectively.

\begin{table}[tbp]
\small
\caption[\neut FSI Parameter Sets]{\neut FSI parameters, $\vec{x}^{FSI}$, that scale each interaction cross section.  Shown are the best-fit and the maximum and minimum scaling values from the 16 parameter sets taken from the 6-dimensional 1$\sigma$ surface.}
\begin{tabular}{c c c c c c c}
\hline\hline
  & \ \ $x^{FSQE}$ \ \ & \ \ $x^{FSQEH}$ \ \  & \ \ $x^{FSINEL}$ \ \ & \ \ $x^{FSABS}$ \ \ & \ \ $x^{FSCX}$ \ \ & $x^{FSCXH}$ \ \ \\ \hline 
Best Fit  & 1.0  & 1.8  & 1.0    & 1.1   & 1.0  & 1.8    \\ \hline 
Maximum   & 1.6  & 2.3  & 1.5    & 1.6   & 1.6  & 2.3 \\
Minimum   & 0.6  & 1.1  & 0.5    & 0.6   & 0.4  & 1.3 \\ \hline \hline
\end{tabular}  
\label{tab:fsi_parsets}
\end{table}

\subsubsection{Constraints From \mb \ccqe Measurements}
To constrain parameters related to the CCQE model and its overall normalization, we fit the 2D cross-section data from \mb~\cite{mb-ccqe}, binned in the outgoing muon kinetic energy, $T_{\mu}$, and angle with respect to the neutrino beam direction, $\theta_{\mu}$.  The \neut interactions selected for the fit are all true CCQE interactions. Our fit procedure follows that described by Juszczak {\it et al.}~\cite{mb-ccqe-wroclaw}, with the \chisq defined as
\begin{equation}
\chi^{2}(\maqe,\lambda) = \sum\limits_{i=0}^n \Bigg\{\frac{p^{\textrm{d}}_{i} - p^{\textrm{p}}_{i}(\maqe,\lambda)}{\Delta p_{i}} \Bigg\}^{2}+\bigg( \frac{\lambda^{-1}-1}{\Delta\lambda}\bigg)^{2}
  \label{eq:ccqefit}
\end{equation}
where the index $i$ runs over the bins of the ($T_{\mu},\cos{\theta_{\mu}}$) distribution, $p^{\textrm{d(p)}}_{i}$
 is the measured (predicted) differential cross section,
$\Delta p_{i}$ is its 
uncertainty,
$\lambda$ is the CCQE 
normalization, and $\Delta\lambda$ is the normalization uncertainty, set at
10.7\% by MiniBooNE measurements.
The main difference from the procedure in~\cite{mb-ccqe-wroclaw} is that we include ($T_{\mu},\cos{\theta_{\mu}}$) bins where a large percentage of the events have 4-momentum transfers that are not allowed in the RFG model. We find $\maqe = 1.64 \pm 0.03$\,\gevcc and $\lambda$ = 0.88$\pm$0.02 with $\chi^{2}_{min}$/DOF = 26.9/135.  It should be noted that \mb does not report correlations, and without this information assessing the goodness-of-fit is not possible.
To take this into account, we assign the uncertainty to be the difference between the fit result and nominal plus the uncertainty on the fit result. 
The \maqe fit uncertainty is set to 0.45\,\gevcc, which covers (at 1 standard deviation) the point estimates from our fit to the MiniBooNE data, the K2K result~\cite{Gran:2006jn} and a world deuterium average, 1.03\,\gevcc~\cite{Bernard:2001rs}.
The normalization uncertainty for neutrinos with $E_{\nu}<1.5$\,\gev, $x_{1}^{QE}$, is set to 11\%, the \mb flux normalization uncertainty, since most of the neutrinos from \mb are created in this energy range.

\subsubsection{Constraints From \mb Inclusive $\pi$ Measurements}
To constrain single pion production parameter errors, we use published \mb differential cross-section datasets for CC single \piz production (\ccpi)~\cite{mb-cc1pi0}, CC single \pip production (\ccpip)~\cite{mb-cc1pip}, and NC single \piz production (\ncpi)~\cite{mb-nc1pi0}. Because the modes are described by a set of common parameters in \neut, we perform a joint fit to all three data sets.

The selection of \neut simulated events follows the signal definition in each of the \mb measurements.  
For the (\ccpi, \ccpip, \ncpi) selections, the signals are defined as
(\num, \num, $\nu$) interactions with (1,1,0) \mun and exactly one (\piz,\pip,\piz) exiting the target nucleus, with no additional
leptons or mesons exiting.
In all cases, there is no constraint on the number of nucleons or photons exiting the nucleus.

We consider a range of models by adjusting 9 parameters shown in 
Tab.~\ref{tab:singlepi-fitparams}. \mares is the axial vector mass
for resonant interactions, which affects both the rate and $Q^2$ shape
of interactions. The ``$W$ shape'' parameter is an empirical parameter
that we introduce in order to improve agreement with \ncpi \ppi data.
The weighting function used is a Breit-Wigner function with a phase space term:
\begin{equation}
  r(W; S) = \alpha \cdot \frac{S}{(W-W_0)^2 + S^2/4} \cdot P(W;m_\pi, m_N)
\label{eq:wshape}
\end{equation}
where $S$ is the ``$W$ shape'' parameter, $W_0=1218$\,\mevc, $P(W; m_\pi, m_N)$ is the phase
space for a two-body decay of a particle with mass $W$ into particles
with masses $m_\pi$ and $m_N$, and $\alpha$ is a normalization factor
calculated to leave the total nucleon-level cross section unchanged as
$S$ is varied.
The nominal values of $S$ and $W_{0}$ come from averages of fits to two $W$ distributions of \neut interactions, one with a resonance decaying to a neutron and $\pi^{+}$ and the other with it decaying to a proton and $\pi^{0}$.
The ``CCOther shape'' parameter, $x^{CCOth}$, modifies the neutrino energy dependence of the cross section for a combination of CC modes, as
described in Sec.~\ref{subsec:othermodelerrors}, along with the remaining parameters that
are normalizations applied to the \neut interaction modes.
Simulated events modified by $x^{CCOth}$ constitute a small fraction of the selected samples. 
As a result, the data have minimal power to constrain this parameter and likewise for the NC1$\pi^{+}$, NC coherent pion, and NCOther normalization parameters, $x^{NC1\pi^{\pm}}$, $x^{NCcoh\pi}$, and $x^{NCOth}$, respectively.
The T2K oscillation analyses are insensitive to these poorly determined parameters, and an arbitrary constraint is applied to stabilize the fits.
In our external data analysis the NC coherent normalization cannot be constrained independently of the \ncpi normalization, $x^{NC1\pi^{0}}$, because there is no difference in the \ppi spectrum between the two components.  The errors given in Tab.~\ref{tab:singlepi-fitparams} also include 
the variance observed when refitting using the 16 FSI 1$\sigma$ parameter sets
and scaling the errors when fitting multiple datasets following the approach of Maltoni and Schwetz~\cite{maltonischwetz}.  
The ``$W$ shape'' nominal prior is kept at the default of 87.7\,\mevcc and in the absence of reported correlations from \mb, the uncertainty is estimated as the difference between the best fit and default values.  The correlations between \mares, $x_{1}^{CC1\pi}$, and $x^{NC1\pi^{0}}$ are given in Table~\ref{tab:1pi_cov}.

\begin{table}[tp]
  \centering
  \caption{Parameters used in the single pion fits and their results from fitting the \mb data. Those with an
    arbitrary constraint applied have their $1\sigma$ penalty term shown.
    \mares, $x_{1}^{CC1\pi}$, and $x^{NC1\pi^{0}}$ fit results and their covariance are used in subsequent analyses.}
  \begin{tabular}{cccccc}
    \hline\hline
             & \ \ units  \ \   & \ \ Nominal value \ \ & \ \ Penalty  \ \ & \ \ Best fit \ \ & \ \ Error \ \  \\
    \hline
    \mares & \ \ \gevcc \ \              & 1.21 &        & 1.41   & 0.22  \\
    $W$ shape & \ \ \mevcc \ \              & 87.7 &        & 42.4 & 12 \\
    $x^{CCcoh\pi}$  &     & 1                &        & 1.423 & 0.462 \\
    $x_{1}^{CC1\pi}$  &      & 1                &        & 1.15  & 0.32 \\
    $x^{CCOth}$    &     & 0                & 0.4    & 0.360 & 0.386 \\
    $x^{NCcoh\pi}$  &    & 1                & 0.3    & 0.994 & 0.293 \\
    $x^{NC1\pi^{0}}$  &    & 1                &        & 0.963  & 0.330 \\
    $x^{NC1\pi^{\pm}}$ &   & 1                & 0.3    & 0.965  & 0.297 \\
    $x^{NCOth}$     &    & 1              & 0.3    & 0.987  & 0.297 \\
    \hline\hline
  \end{tabular}
  \label{tab:singlepi-fitparams}
\end{table}

\begin{table}[tp]
  \centering
  \caption{Correlation between \mares, $x_{1}^{CC1\pi}$, and $x^{NC1\pi^{0}}$.}
  \begin{tabular}{lccc}
   \hline\hline
                 & \ \  \mares \ \ &  \ \ $x_{1}^{CC1\pi}$ \ \ & \ \ $x^{NC1\pi^{0}}$ \ \ \\
    \hline
   \mares\ \ \ \  &     1  &  $-$0.26  &  $-$0.30 \\
   $x_{1}^{CC1\pi}$ &   $-$0.26  &   1  &   0.74 \\
   $x^{NC1\pi^{0}}$  &   $-$0.30  &   0.74  &   1 \\
    \hline\hline
  \end{tabular}
  \label{tab:1pi_cov}
\end{table}

\subsection{Other \neut Model Parameters}
\label{subsec:othermodelerrors}
The remaining uncertainties are in the modeling of the CC resonant, CCDIS, NC resonant charged pion, CC and NC coherent pion, anti-neutrino, as well as \nue CCQE interactions.
 An additional set of energy-dependent normalization parameters is added for CCQE and CC1$\pi$ interactions. Finally, a normalization parameter for the remaining NC interactions is included.

The CCOther shape parameter, $x^{CCOth}$, accounts for model uncertainties for CCDIS and resonant interactions where the resonance decays to a nucleon and photon, kaon, or eta. The nominal interaction model for these interactions is not modified.  From MINOS~\cite{minostotalxsec}, the uncertainty of their cross section measurement at 4\gev, which is dominated by CCDIS, is approximately 10\%.  Using this as a reference point, the cross section is scaled by the factor
$(1+x^{CCOth}/E_{\nu})$
where $E_{\nu}$ is the neutrino energy in \gev. The nominal value for $x^{CCOth}$ is 0 and has a 1$\sigma$ constraint of 0.4.

Normalization parameters are included for both CC and NC coherent pion interactions, $x^{CCcoh\pi}$ and $x^{NCcoh\pi}$, respectively. The CC coherent pion cross section is assigned an error of 100\% due to the fact that the CC coherent pion cross section had only 90\% confidence upper limits for sub-GeV neutrino energies at the time of this analysis.  In addition, when included in the \mb pion production fits, 
the data are consistent with the nominal \neut model at 1$\sigma$ and with zero cross section at 2$\sigma$.  The NC coherent pion production data~\cite{PhysRevD.81.111102} differ from \neut by 15\%, within the measurement uncertainty of 20\%.  
To account for the difference and the uncertainty, we 
conservatively assign a 30\% overall uncertainty to $x^{NCcoh\pi}$.

The anti-neutrino/neutrino cross section ratios are assigned an uncertainty of 40\%.  This is a conservative estimate derived from doubling the maximum deviation between the energy-dependent \mb CCQE neutrino cross section and the RFG model assuming an axial mass of $\maqe = 1.03$\,\gevcc, which was 20\%.

For \nue CCQE interactions, there may be some effects that are not accounted for in the \neut model, such as the existence of second class currents, as motivated in Ref.~\cite{Day:2012gb}. 
The dominant source of uncertainty is the vector component, which may be as large as 3\% at the T2K beam peak, and thus is assigned as an additional error on \nue CCQE interactions relative to \num CCQE interactions.

Table~\ref{tbl:xsecpar} shows
energy-dependent normalization parameters for CCQE and CC1$\pi$ interactions which are included to account for possible discrepancies in the model as suggested, for example, by the difference between the \mb and NOMAD~\cite{Lyubushkin:2008pe} results.  
As mentioned above, the uncertainties for $x_{1}^{QE}$ and $x_{1}^{CC1\pi}$ are assigned from our study of \mb data. 
The remaining CCQE energy regions are assigned a 30\% uncertainty to account for the aforementioned discrepancy while $x_{2}^{CC1\pi}$ has a 40\% uncertainty assigned since it is necessary to extrapolate from the \mb \ccpip inclusive measurement at 2\gev.

The NCOther category consists of NCEL, NC resonant production where the resonance decays to a nucleon and kaon, eta, or photon, and NCDIS interactions.   For fits to the ND280 data and \nue analyses at SK, resonant production that produces a nucleon and charged pion is also included in the NCOther definition, though kept separate in other analyses.  NCOther interactions have a 30\% normalization error assigned to them, which is given to the parameters $x^{NCOth}$ and $x^{NC1\pi^{\pm}}$.

\subsection{Alternative Models}
\label{subsec:othererrors}
As mentioned above, \neut's default model for CCQE assumes an RFG for the nuclear potential and momentum distribution of the nucleons.  An alternative model, referred to as the ``spectral function'' (SF)~\cite{Benhar:1994af}, appears to be a better model when compared to electron scattering data.  SF is a generic term for a function that describes the momentum and energy distributions of nucleons in a nucleus.  In the model employed in \cite{Benhar:1994af}, the SF consists of a mean-field term for single particles and a term for correlated pairs of nucleons, which leads to a long tail in the momentum and binding energy.  It also includes the nuclear shell structure of oxygen, the main target nucleus in the T2K far detector. The difference between the RFG and SF models is treated with an additional systematic parameter.

At the time of this analysis, the SF model had not been implemented in \neut, so the NuWro generator~\cite{Juszczak:2009qa} was used for generating SF interactions with the assumption that a \neut implementation of SF would produce similar results. The SF and RFG distributions were produced by NuWro and \neut, respectively, for \num and \nue interactions on both carbon and oxygen, while using the same vector and axial form factors.

The ratio of the SF and RFG cross sections in NuWro is the weight applied to each \neut CCQE event, according to the true lepton momentum, angle, and neutrino energy of the interaction. Overall, this weighting would change the predicted total cross section by 10\%. Since we already include in the oscillation analysis an uncertainty on the total CCQE cross section, the NuWro cross section is scaled so that at $E_{\nu}$=1\,\gev it agrees with the \neut CCQE cross section. 

A parameter $x_{SF}$ is included to allow the cross section model to be linearly adjusted between the extremes of the RFG ($x_{SF}$=0) and SF ($x_{SF}$=1) models. The nominal value for $x_{SF}$ is taken to be zero, and the prior distribution for $x_{SF}$ is assumed to be a standard Gaussian (mean zero and standard deviation one) but truncated outside the range [0,1].

\subsection{Summary of cross section systematic parameters}
\label{sec:xsecpriors}
All the cross section parameters, $\vec{x}$, are summarized in Tab.~\ref{tbl:xsecpar}, including the errors prior to the analysis of near detector data. They are categorized as follows:
\begin{enumerate}
\item Common between ND280 and SK; constrained by ND280 data. The parameters which are common with SK and well measured by ND280 are $M_A^{QE}$, $M_A^{RES}$ and some normalization parameters.
\item Independent between ND280 and SK, therefore unconstrained by ND280 data.  The 
parameters $p_F$, $E_B$ and SF are target nuclei dependent 
and so are independent between ND280 ($^{12}$C) and SK ($^{16}$O).
\item Common between ND280 and SK, but for which ND280 data have negligible sensitivity, so
no constraint is taken from ND280 data.
The remaining parameters in Tab.~\ref{tbl:xsecpar} are not expected 
to be measured well by ND280 and therefore are treated like independent 
parameters.
\end{enumerate}
We define $\vec{x}_n$ to be the set of cross section systematic parameters which are constrained by ND280 data (category 1), to
distinguish them from the remaining parameters $\vec{x}_s$ (categories 2 and 3).
 
\begin{table}[tbp] \centering
\caption[\small]{Cross section parameters $\vec{x}$ for the ND280 constraint and for the SK oscillation fits, showing 
the applicable range of neutrino energy, nominal value, and prior error. The 
category of each parameter describes the relation between ND280 and SK and is defined in Sec.~\ref{sec:xsecpriors}. Parameters marked with an asterisk
are not included in the parametrization for the appearance analysis.}
\begin{tabular}{cccccc}\hline\hline
\ \ Parameter \ \ & \ \ $E_{\nu}$/\gev Range \ \  & \ \ units \ \ & \ \ Nominal \ \  & \ \ Error \ \  & \ \ Category\ \ \\
\hline
\maqe   & all  & \gevcc            & 1.21   & 0.45  & 1 \\
$x_{1}^{QE}$     & $0<E_{\nu}<1.5$  & & 1.0              & 0.11 & 1 \\ 
$x_{2}^{QE}$     & $1.5<E_{\nu}<3.5$ & & 1.0              & 0.30  & 1 \\
$x_{3}^{QE}$     & $E_{\nu}>3.5$   &  & 1.0              & 0.30  & 1 \\
$p_F$ $^{12}$C & all          & \mevc  & 217      & 30      & 2\\
$E_B$ $^{12}$C *& all         & \mevc     & 25  & 9    & 2\\
$p_F\ ^{16}$O & all           & \mevc  & 225 & 30   & 2\\
$E_B\ ^{16}$O *& all          & \mev   & 27      & 9  & 2\\
$x_{SF}$ for C & all          &     & 0 (off) & 1 (on)  & 2\\
$x_{SF}$ for O & all          &     & 0 (off) & 1 (on)  & 2\\
\mares  & all              & \gevcc & 1.41   & 0.22  & 1\\
$x_{1}^{CC1\pi}$ & $0<E_{\nu}<2.5$ &  & 1.15             & 0.32 & 1\\ 
$x_{2}^{CC1\pi}$ & $E_{\nu}>2.5$   &  & 1.0       & 0.40   & 1 \\
$x^{NC1\pi^{0}}$  & all           &    & 0.96             & 0.33 & 1\\
$x^{CCcoh\pi}$ & all             &  & 1.0       & 1.0  & 3\\
$x^{CCOth}$  & all              & & 0.0       & 0.40  & 3\\
$x^{NC1\pi^{\pm}}$    & all       &        & 1.0       & 0.30  & 3\\
$x^{NCcoh\pi}$ & all            &   & 1.0       & 0.30  & 3\\
$x^{NCOth}$    & all           &    & 1.0       & 0.30  & 3\\
$W$ Shape   & all            & \mevcc   & 87.7   & 45.3  & 3\\
$x^{PDD}$  & all              & &  1.0     & 1.0     & 3\\
CC $\nu_e$  & all            &  & 1.0     & 0.03     & 3\\
$\nu$/\nub & all & & 1.0 & 0.40 & 3\\
$\vec{x}^{FSI}$ & all & & \multicolumn{2}{c}{Section~\ref{sub:piA}} & 3 \\
\hline\hline
\end{tabular}
\label{tbl:xsecpar}
\end{table}

\clearpage
\section{\label{sec:ND} Near Detectors}
Precision neutrino oscillation measurements require good understanding of
the neutrino beam properties and of neutrino interactions.
The two previous sections describe how we model these aspects 
for the T2K experiment and how we use external data to
reduce model uncertainty.
However, if only external data were used, the resulting
systematic uncertainty would limit the precision for oscillation analyses.

In order to reduce systematic 
uncertainty below the statistical uncertainty for the experiment,
an underground hall was constructed 280~m downstream of the production target
for near detectors to directly measure the neutrino
beam properties and neutrino interactions.
The hall contains the on-axis INGRID detector, a set of modules with sufficient
target mass and transverse extent to continuously monitor the interaction rate,
beam direction, and profile, and the off-axis ND280 detector,
a sophisticated set of sub-detectors that measure neutrino interaction 
products in detail.

This section describes the INGRID and ND280 detectors and the methods
used to select high purity samples of neutrino interactions.
The observed neutrino interaction rates and distributions are compared
to the predictions using the beamline and interaction models, with
nominal values for the systematic parameters.
Section~\ref{sec:BANFF} describes how ND280 data are used to improve the
systematic parameter estimates and compares the adjusted
model predictions with the ND280 measurements.

\subsection{\label{sec:INGRID} INGRID}
\subsubsection{INGRID detector}
\begin{figure}[tbp]
  \begin{center}
  \includegraphics[width=70mm]{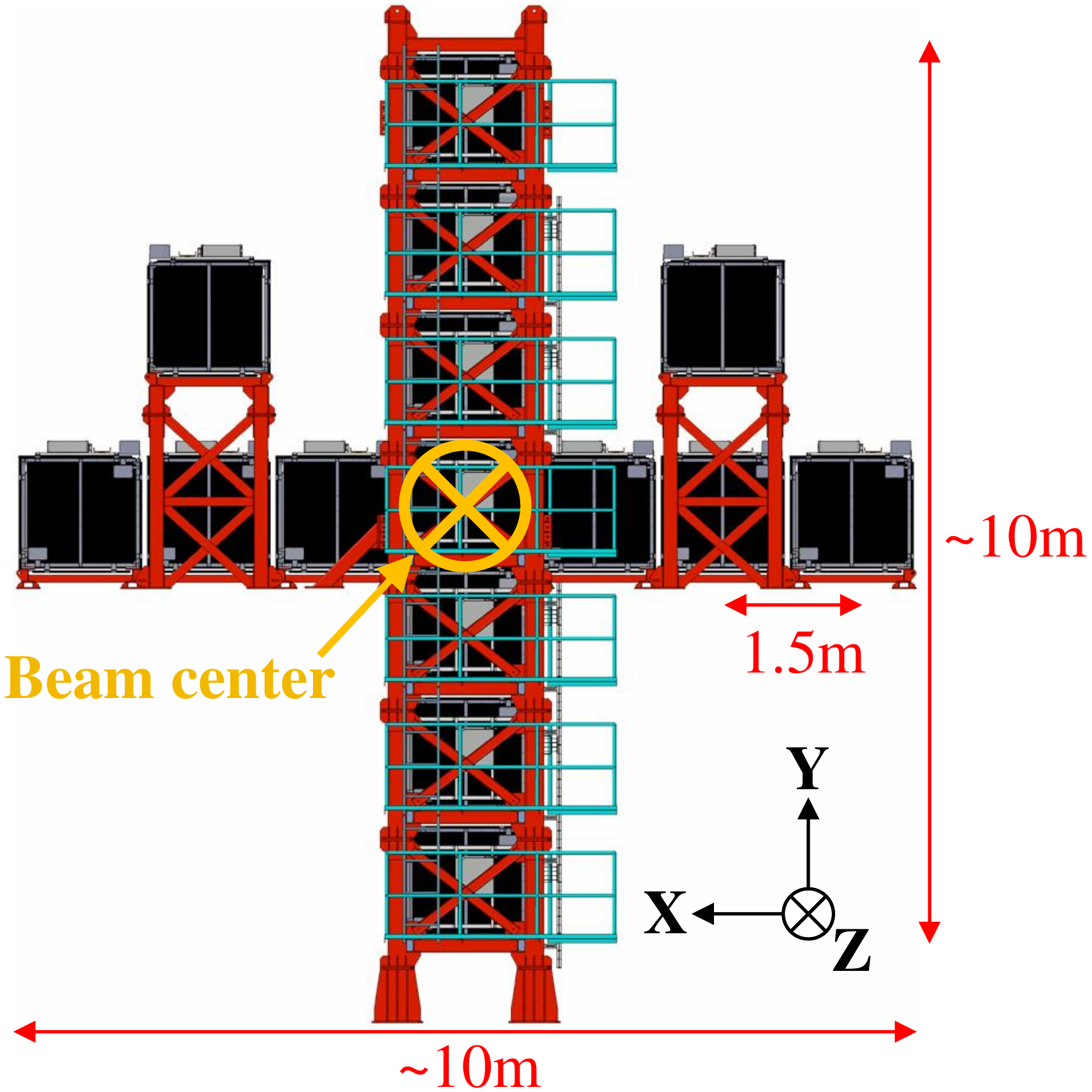}
  \caption{Overview of the INGRID viewed from beam upstream. Two separate modules are placed at off-axis positions off the main cross to monitor the asymmetry of the beam.}
  \label{ingrid_overview}
  \end{center}
\end{figure}

The main purpose of INGRID is to monitor the neutrino beam rate, profile, and center.
In order to sufficiently cover the neutrino beam profile, INGRID is designed to sample the beam in a transverse section of 
10\,m$\times$10\,m, with 14 identical modules arranged in two identical groups along the horizontal and vertical axes, as shown in Fig.~\ref{ingrid_overview}. 
Each of the modules consists of nine iron target plates and eleven tracking scintillator planes,
each made of two layers of scintillator bars (X and Y layers).
They are surrounded by veto scintillator planes to reject charged particles coming from outside of the modules.
Scintillation light from each bar is collected and transported to a photo-detector with a wavelength shifting fiber (WLS fiber) inserted in a hole through the center of the bar.
The light is read out by a Multi-Pixel Photon Counter (MPPC)~\cite{Yokoyama:2010qa} attached to one end of the WLS fiber.
A more detailed description can be found in Ref.~\cite{Abe2012}.

\subsubsection{Event selection}
Neutrino interactions within the INGRID modules are selected by first reconstructing tracks using the X and Y layers independently with an algorithm based on a cellular automaton.
Pairs of tracks in the X and Y layers with the same Z coordinates at the track ends are matched to form 3D tracks.
The upstream edges of the 3D tracks in an event are compared to form a vertex.
Events are rejected if the vertex is outside the fiducial volumes, the time is more than 100~ns from a beam pulse, or if there is a signal in the veto plane at the upstream position extrapolated from a track.

This analysis~\cite{Abe2014_ingrid_ccincl_paper}
significantly improves upon the original method established in 2010~\cite{Abe2012}.
The new track reconstruction algorithm has a higher track reconstruction efficiency and is less susceptible to MPPC dark noise.
Event pileup, defined as more than one neutrino interaction occurring in a module in the same beam pulse, occurs in as many as 1.9\% of events with interactions at the current beam intensity.
The new algorithm handles pileup events correctly as long as the vertices are distinguishable.
For the full dataset, $4.81\times10^6$ events are selected as candidate neutrino events in INGRID. The expected purity of the neutrino events in INGRID is 99.58\%.

\subsubsection{Corrections}
Corrections for individual iron target masses and the background are applied in the same way as the previous 
INGRID analysis~\cite{Abe2012}.
In addition, we apply corrections for dead channels and event pileup which can cause events to be lost.
There are 18 dead channels out of 8360 channels in the 14 standard modules and the correction factor for the dead channels is estimated from a Monte Carlo simulation.
The correction factor for the event pileup is estimated as a linear function of the beam intensity, since the event-pileup effect is proportional to the beam intensity.
The slope of the linear function is estimated from the beam data by 
combining events to simulate event pileup~\cite{Abe2014_ingrid_ccincl_paper}.
The inefficiency due to pileup is less than 1\% for all running periods.

\subsubsection{Systematic error}
Simulation and control samples are used to study potential sources of systematic error and to assign systematic uncertainties.
The sources include target mass, MPPC dark noise and efficiency, event pileup, beam-induced and cosmic background, and
those associated with the event selection criteria. 

The total systematic error for the selection efficiency, 
calculated from the quadratic sum of all the systematic errors, is 0.91\%.
It corresponds to about a quarter of the 3.73\% error from the previous analysis method~\cite{Abe2012}.
The reduction of the systematic error results from the analysis being less sensitive to MPPC dark noise and event pileup, the improved track reconstruction efficiency, and more realistic evaluations of systematic errors which had been conservatively estimated in the previous analysis.

\subsubsection{Results of the beam measurement}
\begin{figure*}[tbp]
  \begin{center}
  \includegraphics[width=160mm]{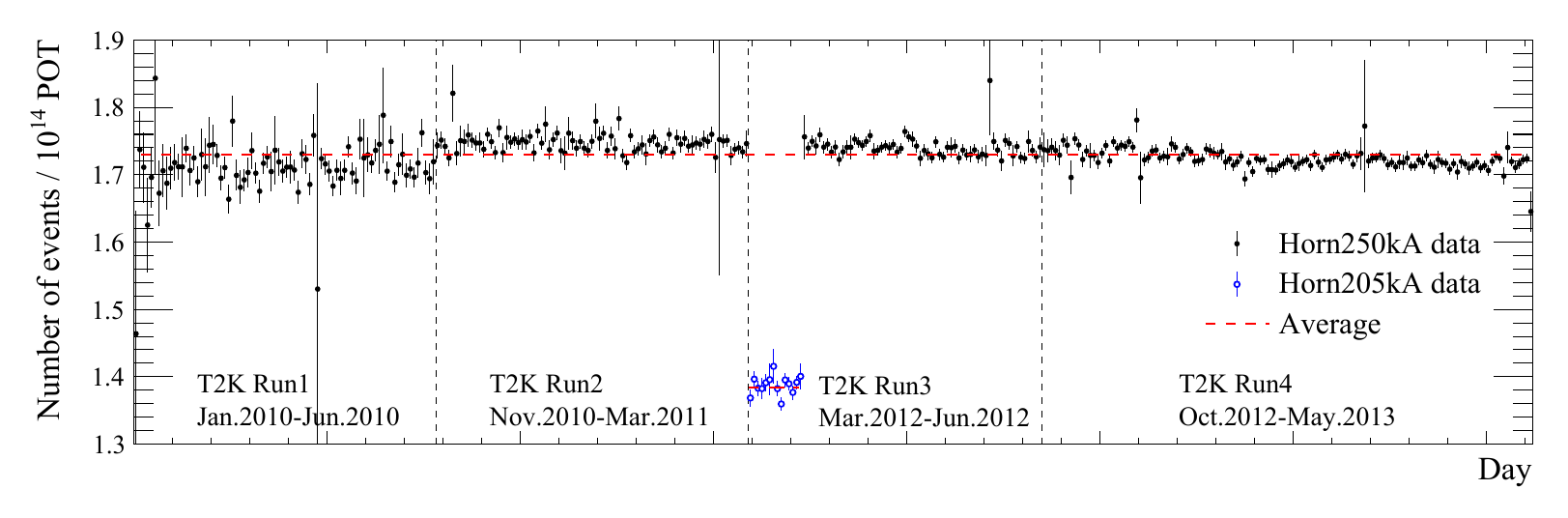}
  \caption[Daily event rate of the neutrino events normalized by protons on target.]{Daily event rate of the neutrino events normalized by protons on target. The error bars show the statistical errors. The horn current was
reduced to 205\,kA for part of Run 3.}
  \label{ingrid_evtrate}
  \end{center}
\end{figure*}

Figure~\ref{ingrid_evtrate} shows the daily rates of the neutrino events normalized by POT.
When the horn current was reduced to 205\,kA due to a power supply problem, the on-axis neutrino flux decreased because the forward focusing of the charged pions by the horns becomes weaker.
An increase by 2\% and a decrease by 1\% of event rate were observed between Run1 and Run2, and during Run4, respectively.
However, for all run periods with the horns operated at 250\,kA, the neutrino event rate is found to be stable within 2\% and the RMS/mean of the event rate is 0.7\%.

A Monte Carlo (MC) simulation that implements the beamline and neutrino interaction models described earlier, along with the INGRID detector simulation, is used to predict the neutrino event rate with the horns operating at 250\,kA and 205\,kA. 
The ratios of observed to predicted event rates, using the nominal values for the beamline and neutrino interaction systematic parameters, are:
\begin{eqnarray}
\frac{N^{\mathrm{data}}_{\mathrm{250kA}}}{N^{\mathrm{MC}}_{\mathrm{250kA}}}&=&1.014\pm 0.001(\mathrm{stat})\pm 0.009(\mathrm{det\ syst}),\\
\frac{N^{\mathrm{data}}_{\mathrm{205kA}}}{N^{\mathrm{MC}}_{\mathrm{205kA}}}&=&1.026\pm 0.002(\mathrm{stat})\pm 0.009(\mathrm{det\ syst}),
\end{eqnarray}
The uncertainties from the neutrino flux prediction and the neutrino interaction model are not included in the systematic errors.

\begin{figure*}[tbp]
  \begin{center}
  \includegraphics[width=160mm]{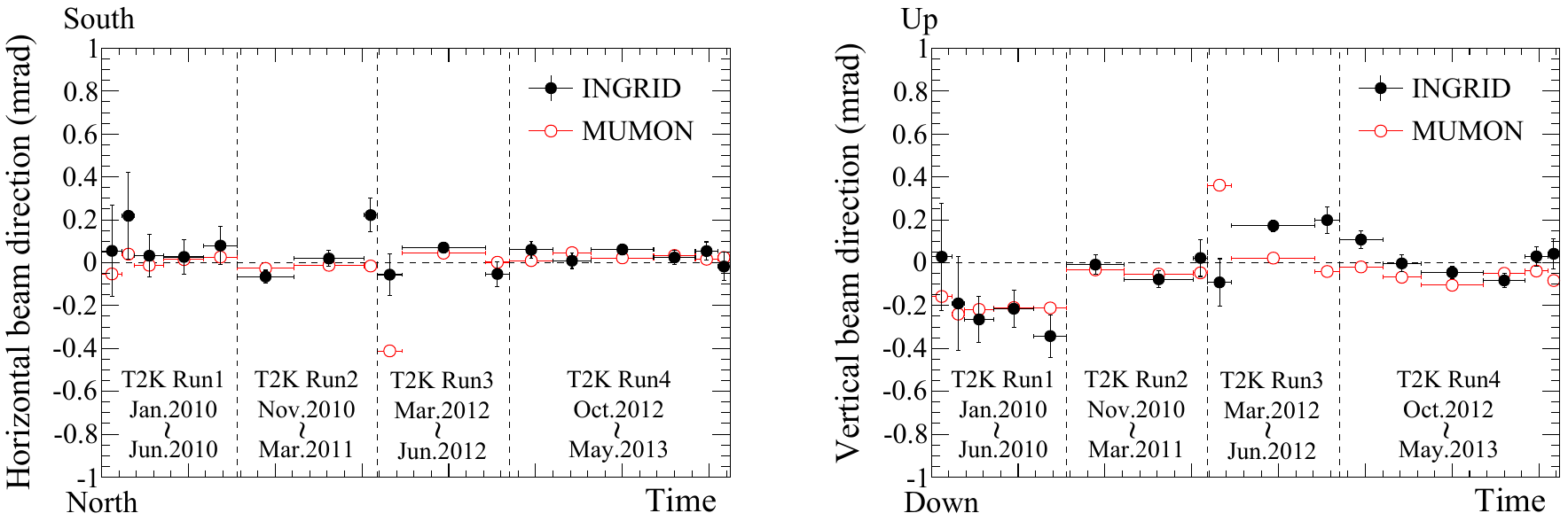}
  \caption[History of neutrino beam directions compared with the muon beam directions.]{History of neutrino beam directions for horizontal (left) and vertical (right) directions as measured by INGRID and by the muon monitor (MUMON). The zero points of the vertical axes correspond to the nominal directions. The error bars show the statistical errors.}
  \label{beam_center}
  \end{center}
\end{figure*}

\begin{figure}[tbp]
  \begin{center}
  \includegraphics[width=75mm]{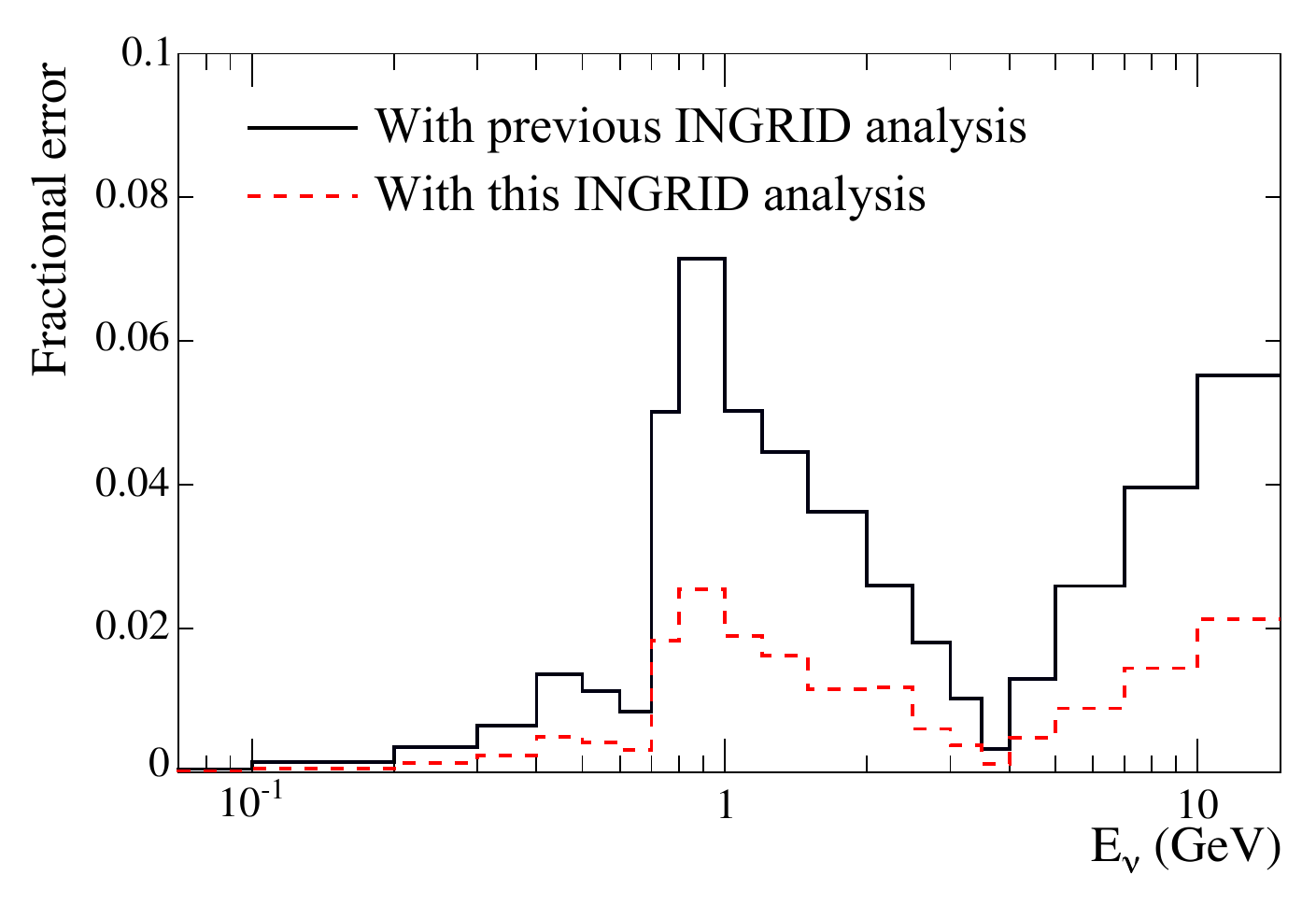}
  \caption{Fractional uncertainties of the $\nu_\mu$ flux at SK due to the beam direction uncertainty evaluated from the previous and this INGRID beam analyses. These evaluations do not include constraints from ND280.}
  \label{flux_err_oa}
  \end{center}
\end{figure}

The profiles of the neutrino beam in the horizontal and vertical directions are measured using the number of neutrino events in the seven horizontal and seven vertical modules, respectively.
The observed horizontal and vertical profiles are fitted with separate Gaussian functions and
the profile center is defined as the fitted peak positions.
Finally, the neutrino beam direction is reconstructed as the direction from the proton beam target position to the measured profile center at INGRID using the result of accurate surveys of the proton beam target and the INGRID detectors.
Figure~\ref{beam_center} shows the history of the horizontal and vertical neutrino beam directions relative to the nominal directions as measured by INGRID and by the muon monitor.
The measured neutrino beam directions are stable well within the physics requirement of 1 mrad.
A 1 mrad change in angle changes the intensity and peak energy of an unoscillated neutrino beam at SK by 3\% and 13~MeV, respectively.
Because a misalignment in the proton beamline was adjusted in November 2010, the subsequent beam centers in the vertical direction are slightly shifted toward the center.
A conservative estimate of the 
systematic error of the profile center is calculated by assuming that the detector systematic uncertainties for the neutrino event rate are not correlated between different INGRID modules. 
The average horizontal and vertical beam directions are measured as
\begin{eqnarray}
\bar{\theta}_X^{\mathrm{beam}}&=&0.030\pm 0.011(\mathrm{stat})\pm 0.095(\mathrm{det\ syst})\ \mathrm{mrad},\\
\bar{\theta}_Y^{\mathrm{beam}}&=&0.011\pm 0.012(\mathrm{stat})\pm 0.105(\mathrm{det\ syst})\ \mathrm{mrad},
\end{eqnarray}
respectively.
The neutrino flux uncertainty arising from possible incorrect modeling of the beam direction is evaluated from this result.
This uncertainty, when evaluated without ND280 data, is significantly reduced compared to the previous analysis,
as shown in Fig.~\ref{flux_err_oa}.

\begin{figure*}[tbp]
  \begin{center}
  \includegraphics[width=160mm]{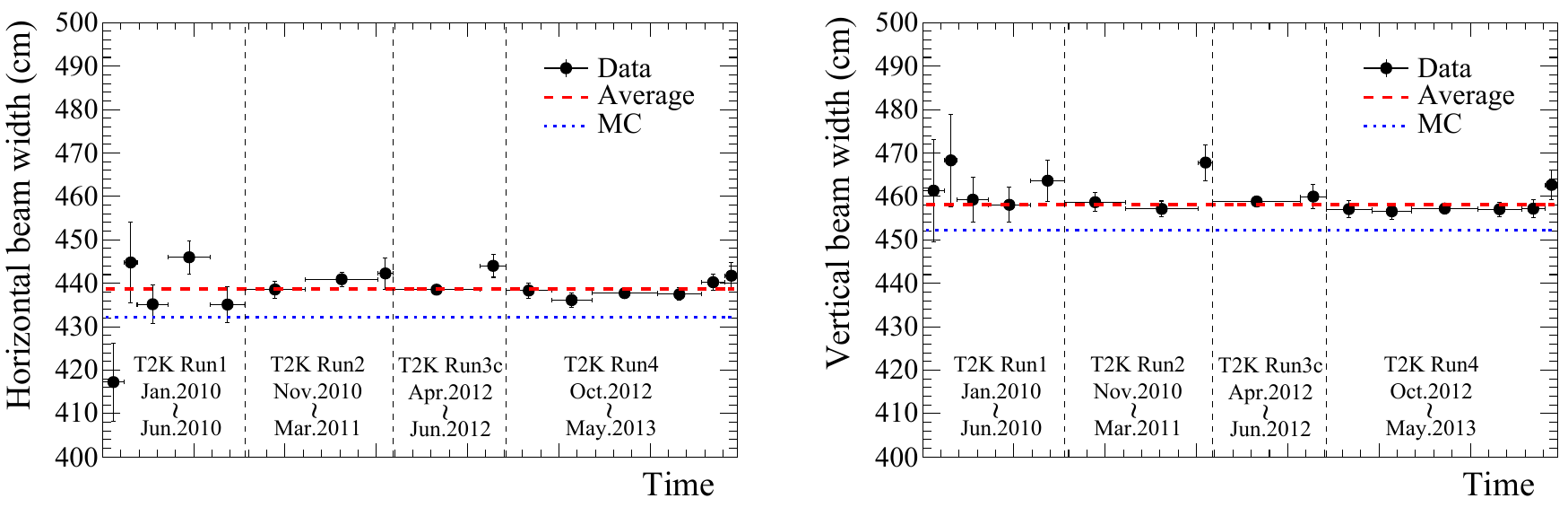}
  \caption[History of neutrino beam width.]{History of neutrino beam width for horizontal (left) and vertical (right) directions for the horn 250\,kA operation. The error bars show the statistical errors.}
  \label{beam_width}
  \end{center}
\end{figure*}

The horizontal and vertical beam width measurements are given by
the standard deviations of the Gaussians fit to the observed profiles.
Figure~\ref{beam_width} shows the history of the horizontal and vertical beam widths with the horns operating at 250\,kA which are
found to be stable within the statistical errors.
The ratios of observed to predicted widths, using nominal values for the systematic parameters, are: 
\begin{eqnarray}
\frac{W_X^{\mathrm{data}}}{W_X^{\mathrm{MC}}}&=&1.015\pm 0.001(\mathrm{stat})\pm 0.010(\mathrm{det\ syst}),\\
\frac{W_Y^{\mathrm{data}}}{W_Y^{\mathrm{MC}}}&=&1.013\pm 0.001(\mathrm{stat})\pm 0.011(\mathrm{det\ syst}),
\end{eqnarray}
for the horizontal and vertical direction, respectively.

\subsection{\label{sec:ND280} ND280}
In designing the experiment, it was recognized that
detailed measurements of neutrino interactions near the production
target and along the direction to the far detector would be necessary
to reduce uncertainty in the models of the
neutrino beam and of neutrino interactions.
To achieve this, the T2K collaboration chose to use
a combination of highly segmented scintillator targets 
and gaseous trackers in a magnetic spectrometer.
Segmented active targets allow for the neutrino interaction to be
localized and the trajectories of the charged particles to be reconstructed,
and those passing through the gaseous
trackers have their charge, momentum, and particle type measured.
The targets and gaseous trackers are surrounded by a calorimeter to detect photons and assist
in particle identification.
The refurbished UA1/NOMAD magnet was acquired and
its rectangular inner volume led to a design with
rectangular sub-detectors.
Spaces within the yoke allowed 
for the installation of outer muon detectors.

The following sections describe the ND280 detector, its simulation,
and the analyses used as input for the T2K oscillation analyses.

\subsubsection{ND280 detector}
\label{subsec:ND280_detector}
\begin{figure}[tbp]
  \begin{center}
    \includegraphics[keepaspectratio=true,width=0.48\textwidth]{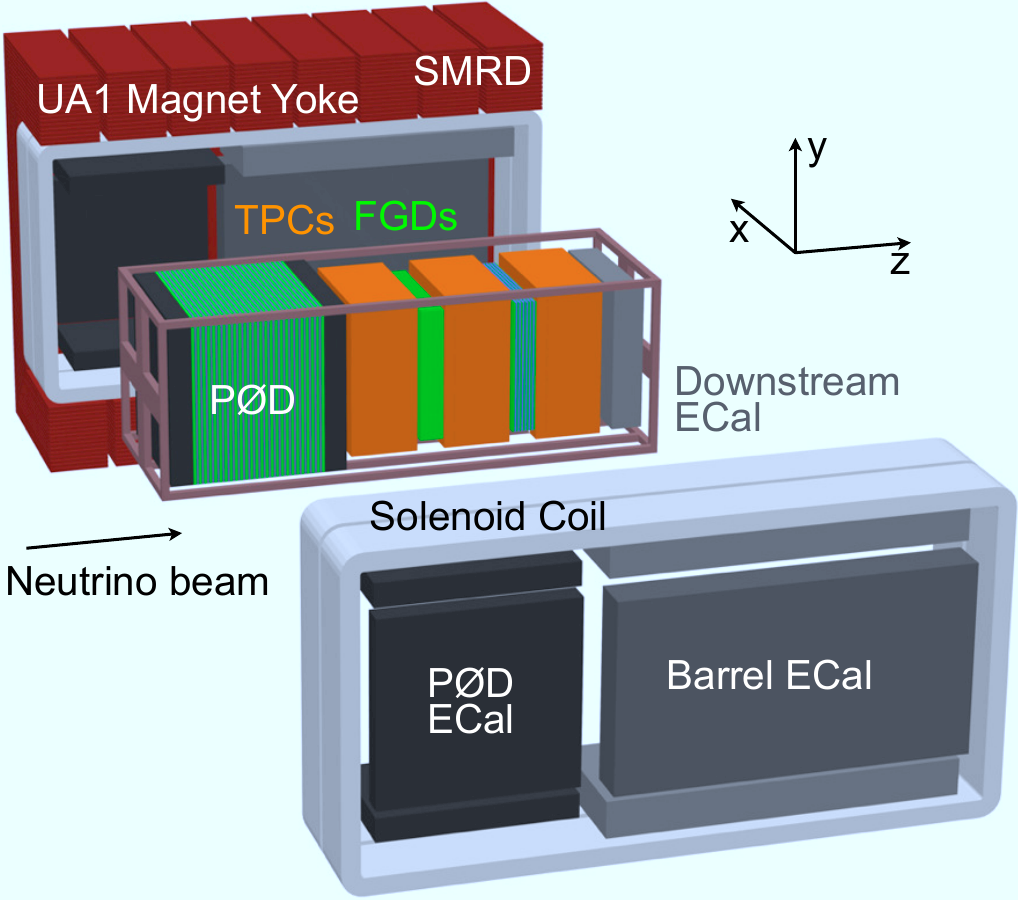}
    \caption{Sketch of the ND280 off-axis detector in an exploded view. 
             A supporting basket holds the $\pi^0$ detector (P0D) as well as the Time Projection 
             Chambers (TPCs) and Fine Grained Detectors (FGDs) that make up the 
             ND280 Tracker. Surrounding the basket is a calorimeter (ECal) and within
             the magnet yoke is the Side Muon Range Detector (SMRD).}
    \label{fig:ND280detector}
  \end{center}
\end{figure}

The ND280 detector is illustrated in
Fig.~\ref{fig:ND280detector}, 
where the coordinate convention is also indicated.
The $x$ and $z$ axes are in the horizontal plane and the $y$
axis is vertical.
The origin is at the center of the magnet
and the 0.2~T magnetic field is along the $+x$ direction. 
The $z$ axis is the direction to the far detector projected onto the horizontal plane.

The analyses presented in this paper use neutrino interactions within the ND280 tracker,
composed of two fine-grained scintillator 
bar detectors (FGDs~\cite{Amaudruz:2012pe}), 
used as the neutrino interaction target,
sandwiched between three gaseous time
projection chambers (TPCs~\cite{Abgrall:2010hi}). 

The most upstream FGD (FGD1) primarily consists of polystyrene scintillator 
bars having a square cross section, 9.6\,mm on a side,
with layers oriented alternately in the $x$ and $y$ 
directions allowing
projective tracking of charged particles.
Most of the interactions in the first FGD 
are on carbon nuclei. 
The downstream FGD (FGD2) has a similar structure but the polystyrene bars
are interleaved with water layers to allow for the measurement
of neutrino interactions on water. 
The FGDs are thin enough that most of the penetrating 
particles produced in neutrino interactions, especially muons, 
pass through to the TPCs. 
Short-ranged particles such as recoil protons can be reconstructed in the FGDs,
which have fine granularity so that individual particle tracks
can be resolved and their directions measured.

Each TPC consists of a field cage filled with Ar:CF$_4$:iC$_4$H$_{10}$ (95:3:2)
inside a box filled with CO$_2$.
The $+x$ and $-x$ walls of the field cages are each instrumented with 12 MicroMEGAS modules arranged
in two columns.
The 336\,mm $\times$ 353\,mm active area for each
MicroMEGAS is segmented into 1728 rectangular pads arranged in 48 rows and 36 
columns, providing 3D reconstruction of charged particles
that pass through the TPCs.
The curvature due to the magnetic field provides measurements of particle momenta
and charges and, when combined with ionization measurements, allows for
particle identification (PID).

The tracker is downstream of a $\pi^0$ detector (P0D~\cite{Assylbekov201248})
and all of these detectors are surrounded by electromagnetic
calorimeters (ECals~\cite{Allan:2013ofa})
and side muon range detectors (SMRDs~\cite{Aoki:2012mf}). 

Data quality is assessed weekly.
Over the entire running period, the ND280 data taking efficiency is 98.5\%.
For the analyses presented here, only data recorded with all detectors having good status
are used, giving an overall efficiency of 91.5\%.

\subsubsection{ND280 simulation}
\label{subsec:ND280_simulation}

A detailed simulation is used to interpret the data recorded by ND280.
The neutrino flux model described in Sec.~\ref{sec:beam:fluxmc} is combined with
the NEUT neutrino interaction model described in Sec.~\ref{subsec:nuintmodel} 
and a detailed material and geometrical description of the ND280 detector including the magnet,
to produce a simulated sample of neutrino interactions distributed
throughout the ND280 detector with the beam time structure.
For studies of particles originating outside of the ND280 detector, 
separate samples are produced using a description of the concrete that forms the
near detector hall and the surrounding sand.

The passage of particles through materials and the ND280 detector response 
are modeled using the GEANT4 toolkit~\cite{Agostinelli2003250}.
To simulate the scintillator detectors, including the FGDs, we use custom 
models of the scintillator photon yield, photon propagation 
including reflections and attenuation, and electronics 
response and noise~\cite{Vacheret:2011zza}.
The gaseous TPC detector simulation includes
the gas ionization, transverse and longitudinal diffusion
of the electrons, transport of the electrons to the readout
plane through the magnetic and electric field, gas amplification, and
a parametrization of the electronics response. 

Imperfections in the detector response simulation can cause the model to match the detector
performance poorly, potentially generating a systematic bias in parameter estimates.
After describing the methods to select neutrino interactions in the following
section, we quantify
the systematic uncertainty due to such effects with data/simulation comparisons in
Sec.~\ref{subsec:ND280_systematics}.

\subsubsection{ND280 $\nu_\mu$ Tracker analysis}
\label{subsec:ND280_numu}
We select an inclusive
sample of $\nu_\mu$ CC interactions in the ND280 detector in order to constrain parameters in our flux and 
cross section model. 
Our earlier oscillation analyses divided the inclusive sample into two: CCQE-like and the remainder.
New to this analysis is the division of the inclusive sample into three sub-samples, defined by 
the number of final state pions: zero (CC0$\pi$-like), one positive pion (CC$1\pi^{+}$-like),
and any other combination of number and charge (CCOther-like). 
This division has enhanced ability to constrain the CCQE and resonant 
single pion cross section parameters, which, in turn, decreases the uncertainty they contribute to
the oscillation analyses. 

The CC-inclusive selection uses the highest momentum negatively charged particle in
an event as the $\mu^-$ candidate and it is
required to start inside the 
FGD1 fiducial volume (FV) and enter the middle TPC (TPC2).
The FV begins 58~mm inward from the boundaries of the FGD1 active 
volume in $x$ and $y$ and 21~mm inward from the upstream boundary of the FGD1 active volume in $z$,
thereby excluding the first two upstream layers.
The TPC requirement has the consequence of producing a sample with predominantly forward-going $\mu^{-}$.
Additional requirements are included to reduce background in which the start of the $\mu^-$ candidate is incorrectly
assigned inside the FGD1 FV, due to a failure to correctly reconstruct a particle passing through the FGD1 (through-going veto).
The $\mu^-$ candidate is required to be consistent with a muon (muon PID requirement)  
based on a truncated mean of measurements of energy loss in the TPC gas~\cite{Abgrall:2010hi}. 
A similar PID has been developed for the FGD, which is not used for the muon selection, but is used in secondary particle identification~\cite{Amaudruz:2012pe}. 

Events passing this selection comprise the CC-inclusive sample which is then divided into three exclusive sub-samples on the basis of secondary tracks from the event vertex.
The names for these samples have the ``-like'' suffix to distinguish them 
from the corresponding topologies that are based on truth information. 
Those events with no additional TPC tracks consistent with being a pion or electron and with no additional FGD tracks consistent with being a pion, nor any time-delayed signal in the FGD which is consistent with a Michel electron, comprise the CC0$\pi$-like sample.
Those events with one positive pion candidate in a TPC and no additional negative pions, electrons or positrons comprise the CC1$\pi^+$-like sample.
The CCOther-like sample contains all other CC-inclusive events not in the CC0$\pi$-like or CC1$\pi^+$-like samples.

In the simulation we find that the CC-inclusive sample is composed of 90.7$\%$ true $\nu_\mu$ CC interactions within the FGD fiducial volume, and 89.8$\%$ of the muon candidates are muons (the rest are mainly 
mis-identified negative pions).
Table~\ref{tab:numberEvents_byCut} shows the number of events after each cut for data and simulation scaled to data POT, with systematic parameters set to their nominal values.

\begin{table*}[tbp]
\begin{center} 
\caption{Number of events at each cut step, for data and for
simulation (scaled to data POT) for the CC-inclusive sample.}
\begin{tabular}{l  c  c}
\hline\hline
Requirement \ \ \                         & Data & \ \ Simulation \ \  \\ 
\hline
$\mu^-$ candidate starts within FGD1 FV and enters TPC2  \ \ & 48731 & 47752  \\ 
passes through-going veto            & 34804 & 36833 \\
passes muon PID requirement          & 25917 & 27082 \\ 
\hline\hline
\end{tabular}
\label{tab:numberEvents_byCut}
\end{center} 
\end{table*}

Table~\ref{tab:purity_CCreaction} shows that the CC0$\pi$-like sample is significantly enhanced in CCQE interactions, the CC1$\pi^{+}$-like sample in CC resonant pion interactions, and the CCOther-like sample in CC deep inelastic scattering (DIS) interactions. This division improves the constraints on several neutrino interaction model parameters. 
As shown in Tab.~\ref{tab:purity_CCtopology}, the CC1$\pi^+$ true topology is the most difficult to isolate.
Most of the contamination in the CC1$\pi^+$-like sample comes from deep inelastic scattering events for which only one pion is detected and any other hadrons have escaped or have been lost to interactions in the surrounding material.

\begin{figure*}[tbp]
 \centering
\ifx\figstyle\bw
  \includegraphics[keepaspectratio=true,width=.5\textwidth]{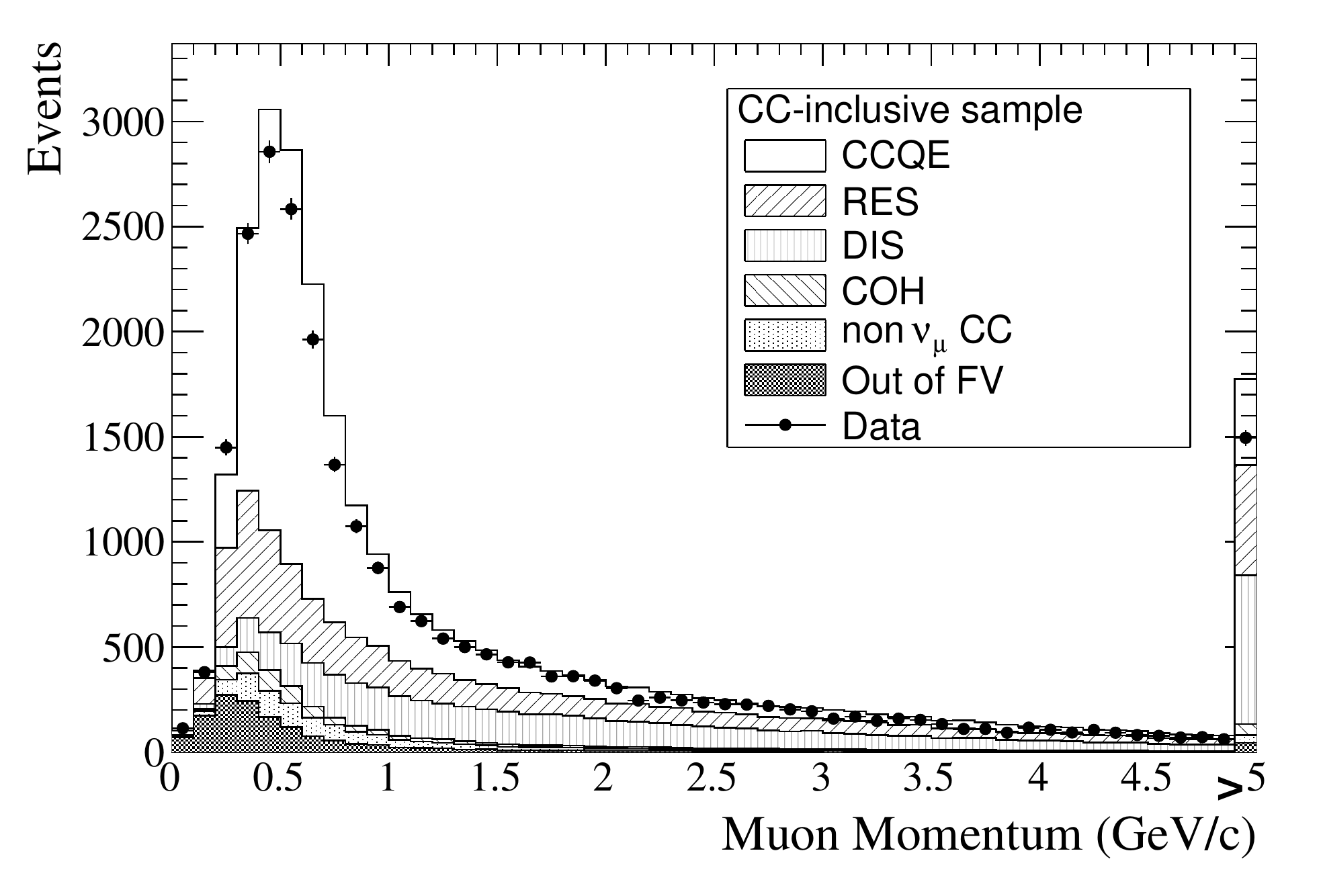} \\
  \includegraphics[keepaspectratio=true,width=.5\textwidth]{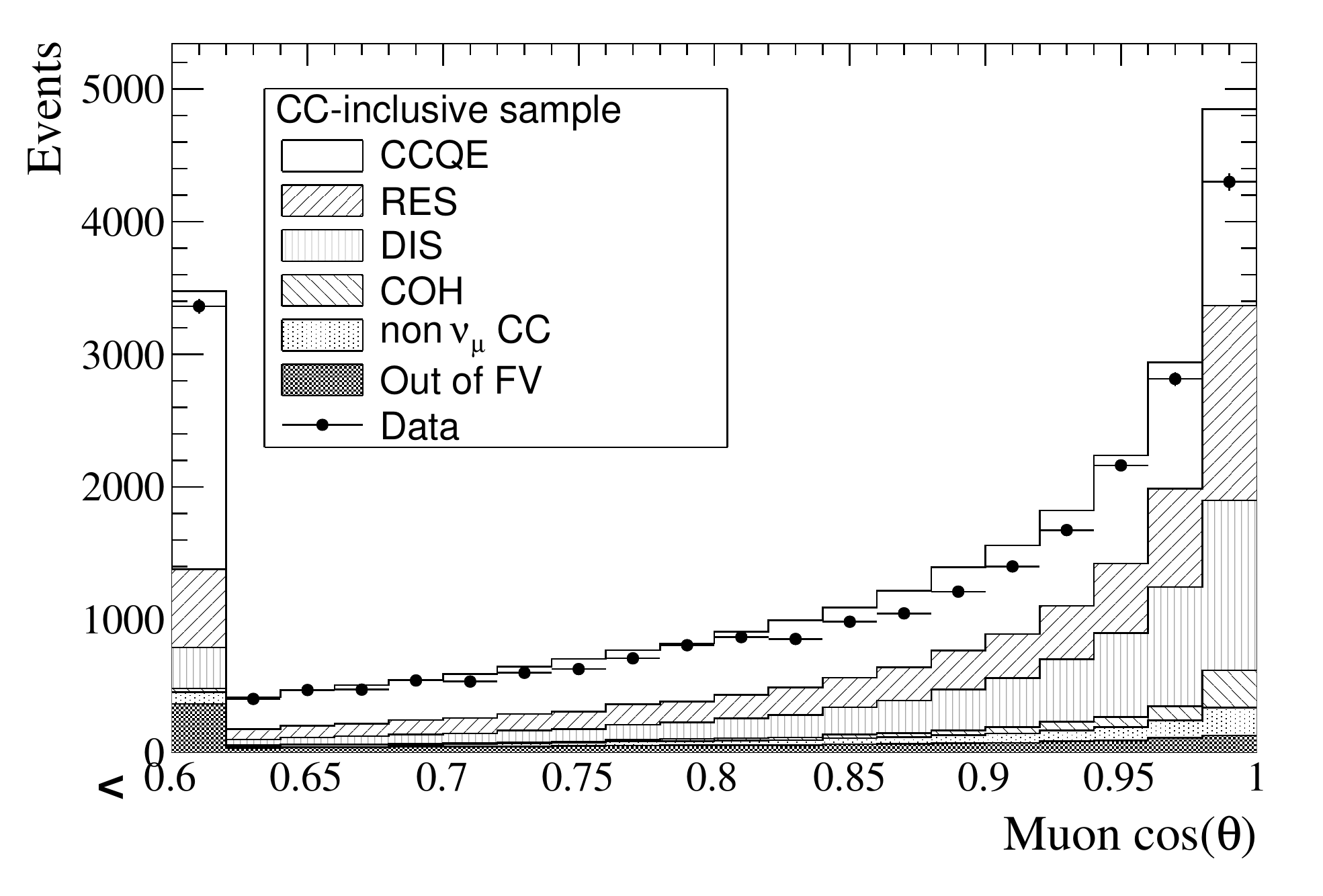}
\else
  \includegraphics[keepaspectratio=true,width=.5\textwidth]{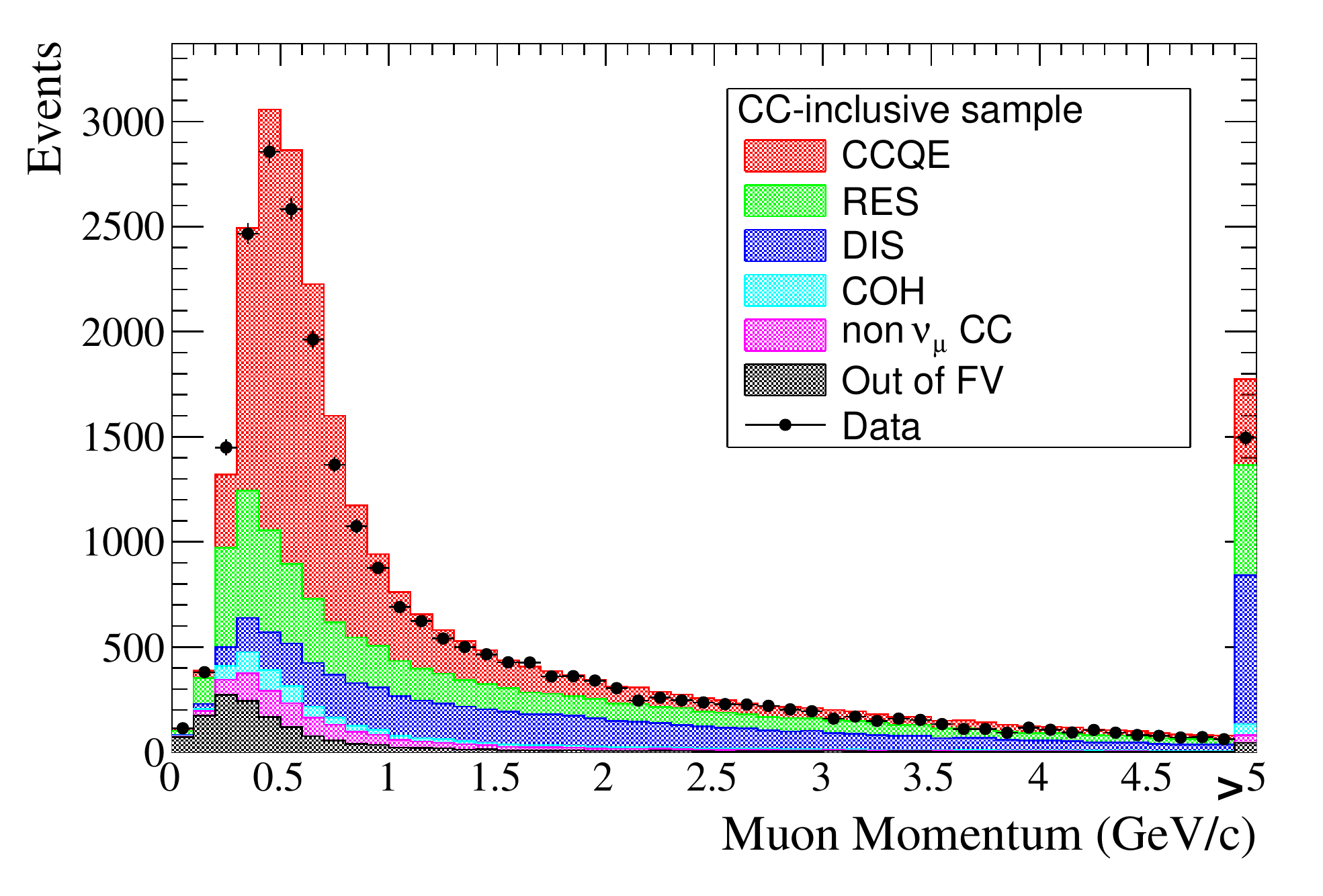} \\
  \includegraphics[keepaspectratio=true,width=.5\textwidth]{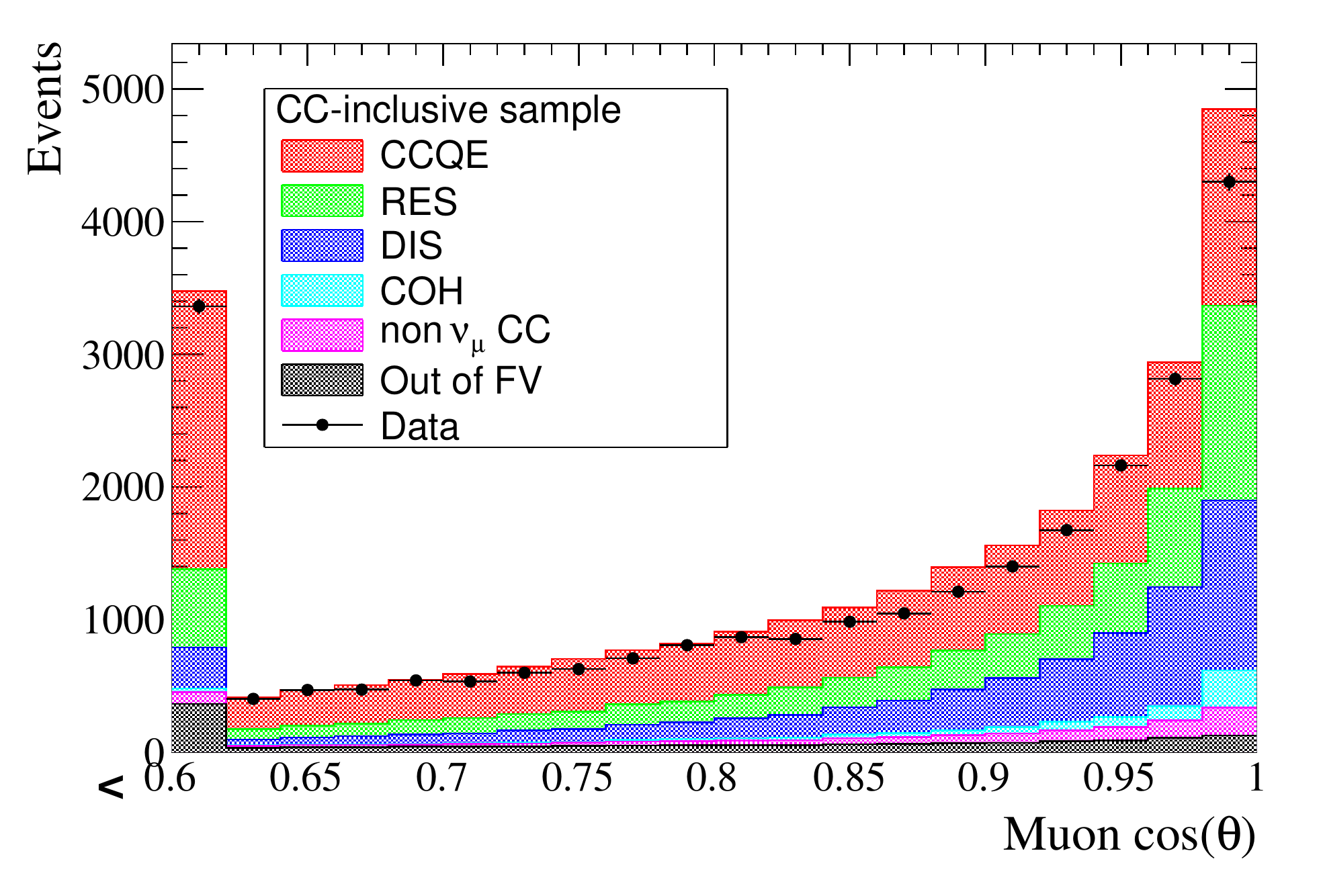}
\fi
 \caption{Muon momentum and angle distribution for the CC-inclusive sample.
These are compared to the simulation, broken down into the different reaction types shown in Tab.~\ref{tab:purity_CCreaction} and where non $\nu_\mu$ CC refers to NC, $\bar{\nu}_\mu$, and $\nu_e$ interactions.
All systematic parameters are set to their nominal values.}
 \label{fig:ND280_mu_CC}
\end{figure*}

\begin{figure*}[tbp]
 \centering
\ifx\figstyle\bw
  \includegraphics[keepaspectratio=true,width=.5\textwidth]{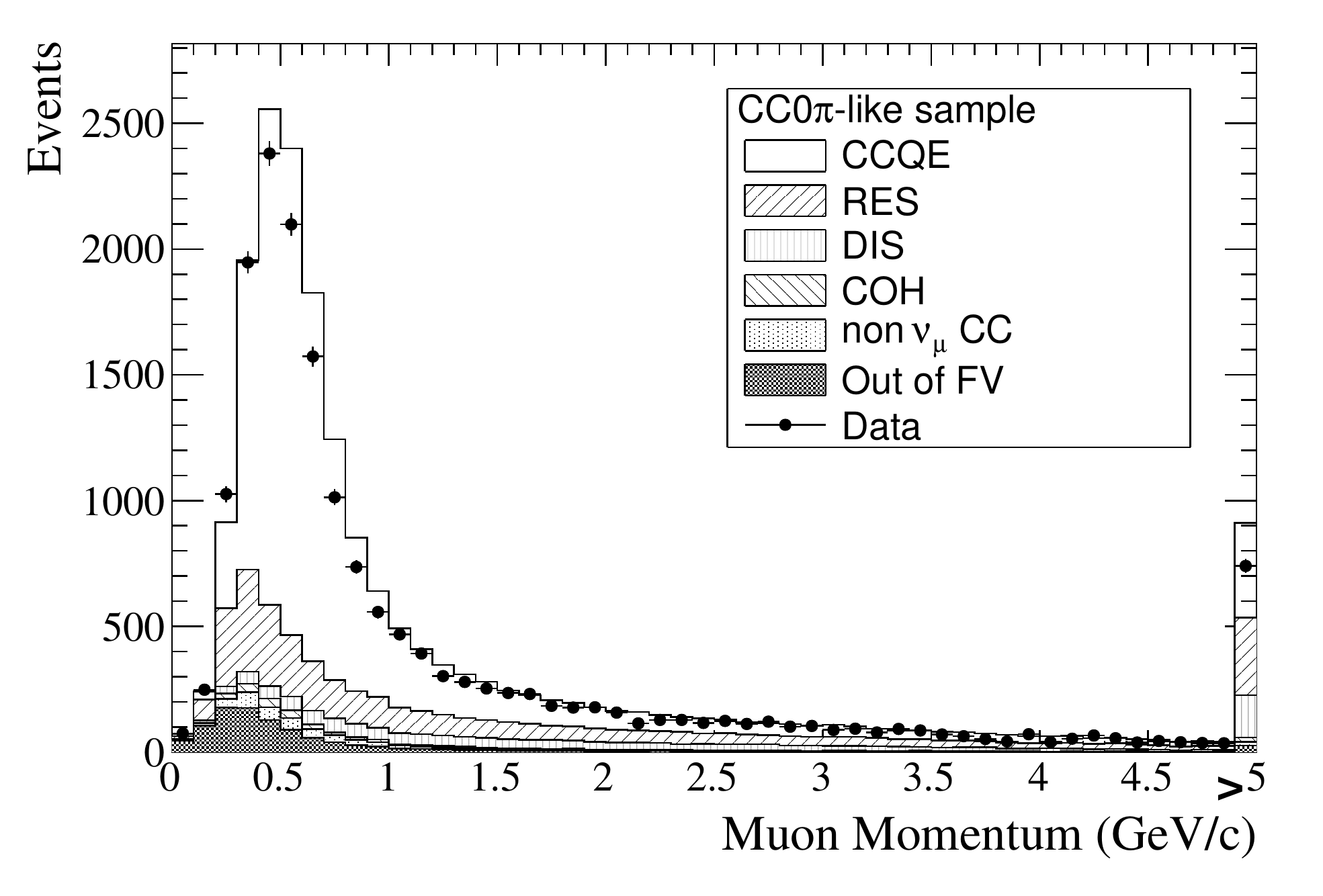} \\
  \includegraphics[keepaspectratio=true,width=.5\textwidth]{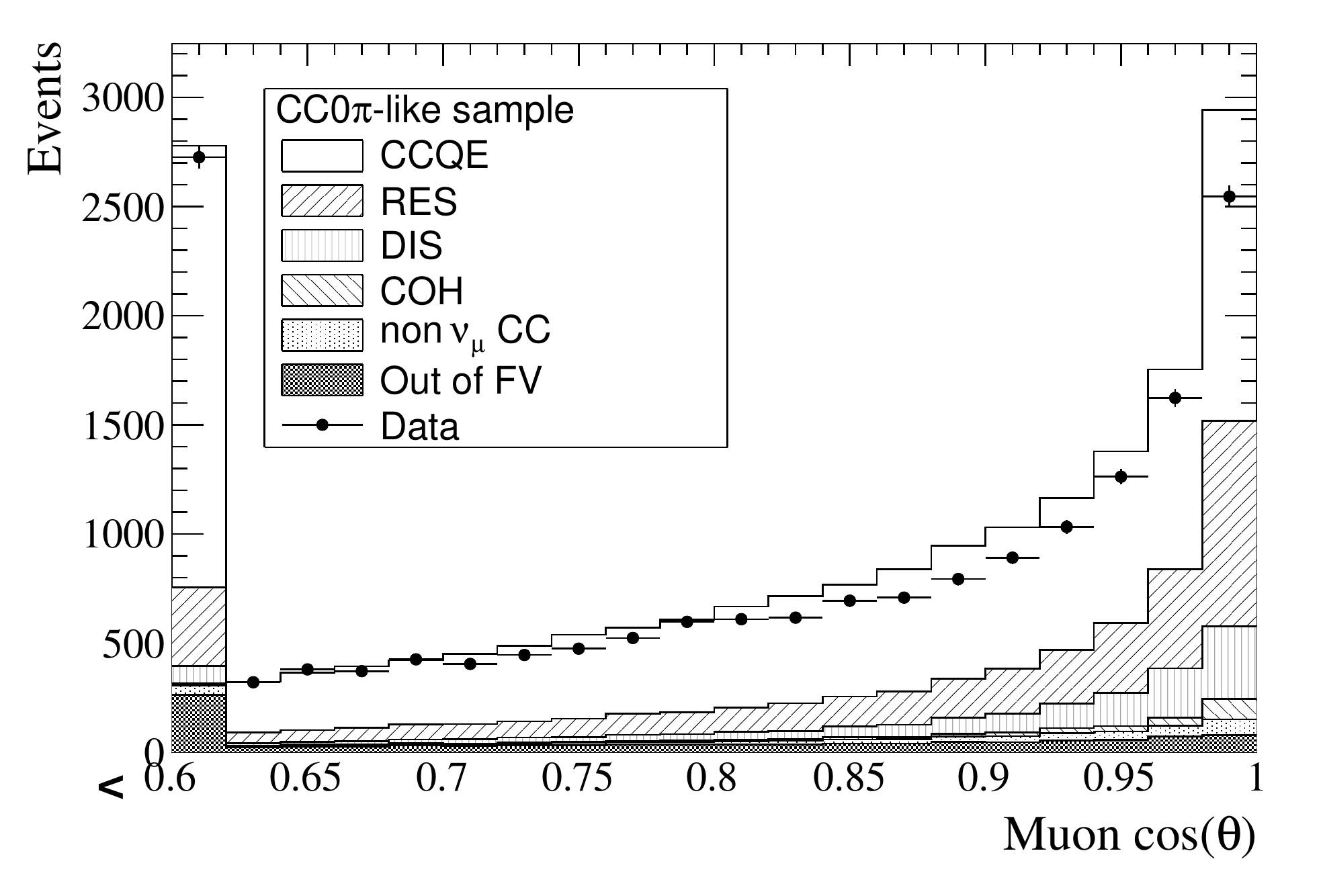} 
\else
  \includegraphics[keepaspectratio=true,width=.5\textwidth]{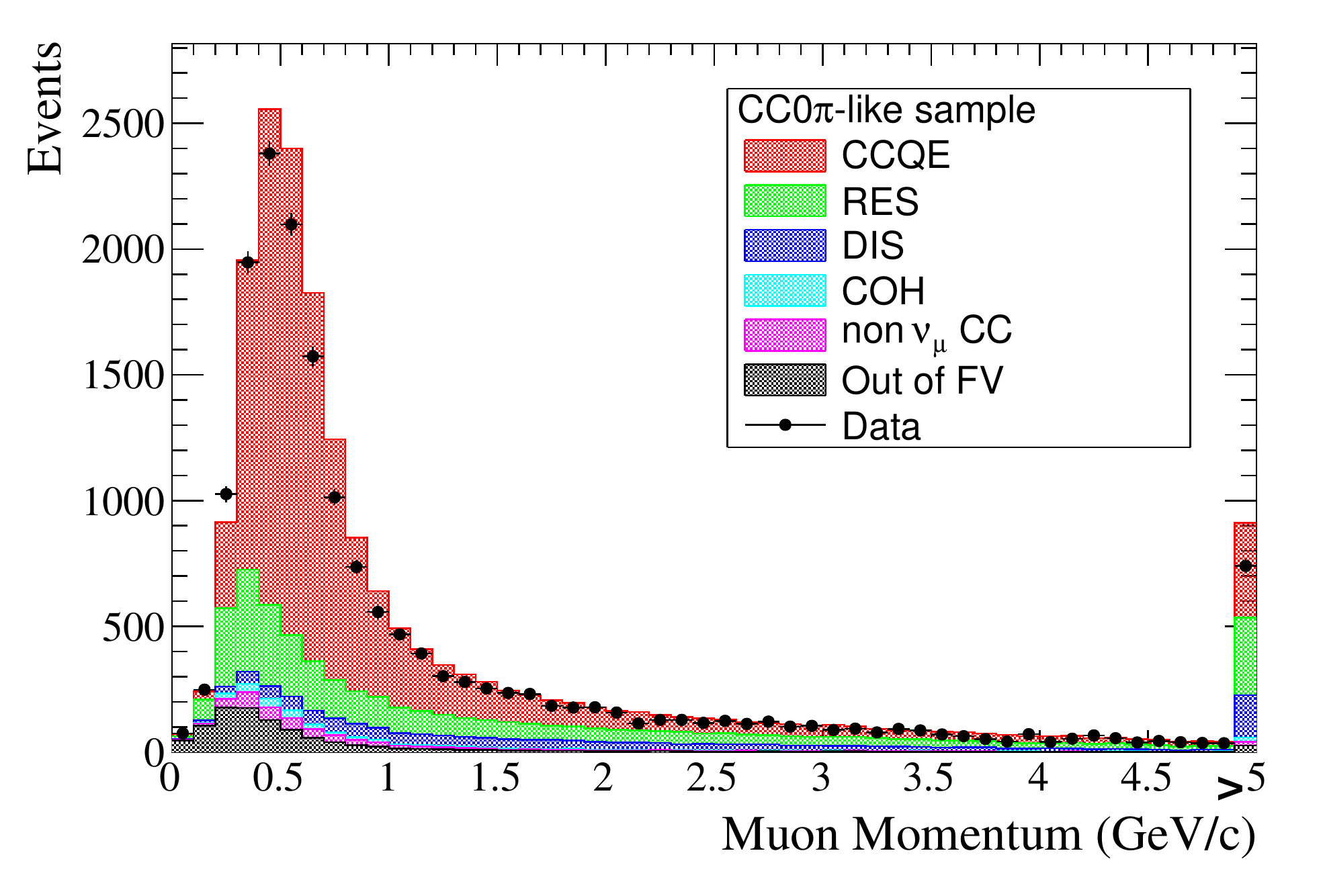} \\
  \includegraphics[keepaspectratio=true,width=.5\textwidth]{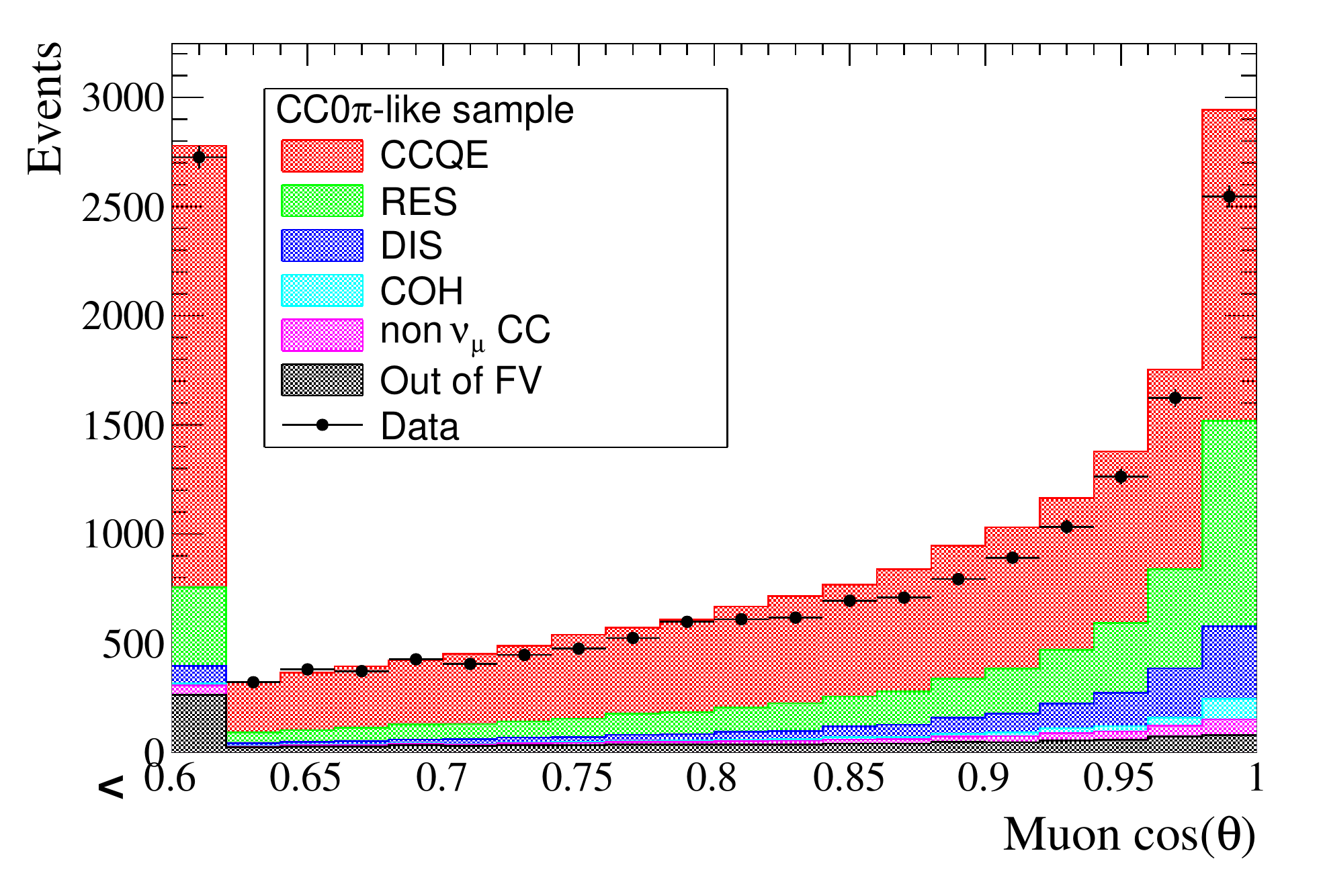} 
\fi
 \caption{Muon momentum and angle distribution for the CC0$\pi$-like sample.
These are compared to the simulation, broken down into the different reaction types, with
all systematic parameters set to their nominal values.}
 \label{fig:ND280_mu_CC0pi}
\end{figure*}

\begin{figure*}[tbp]
 \centering
\ifx\figstyle\bw
  \includegraphics[keepaspectratio=true,width=.5\textwidth]{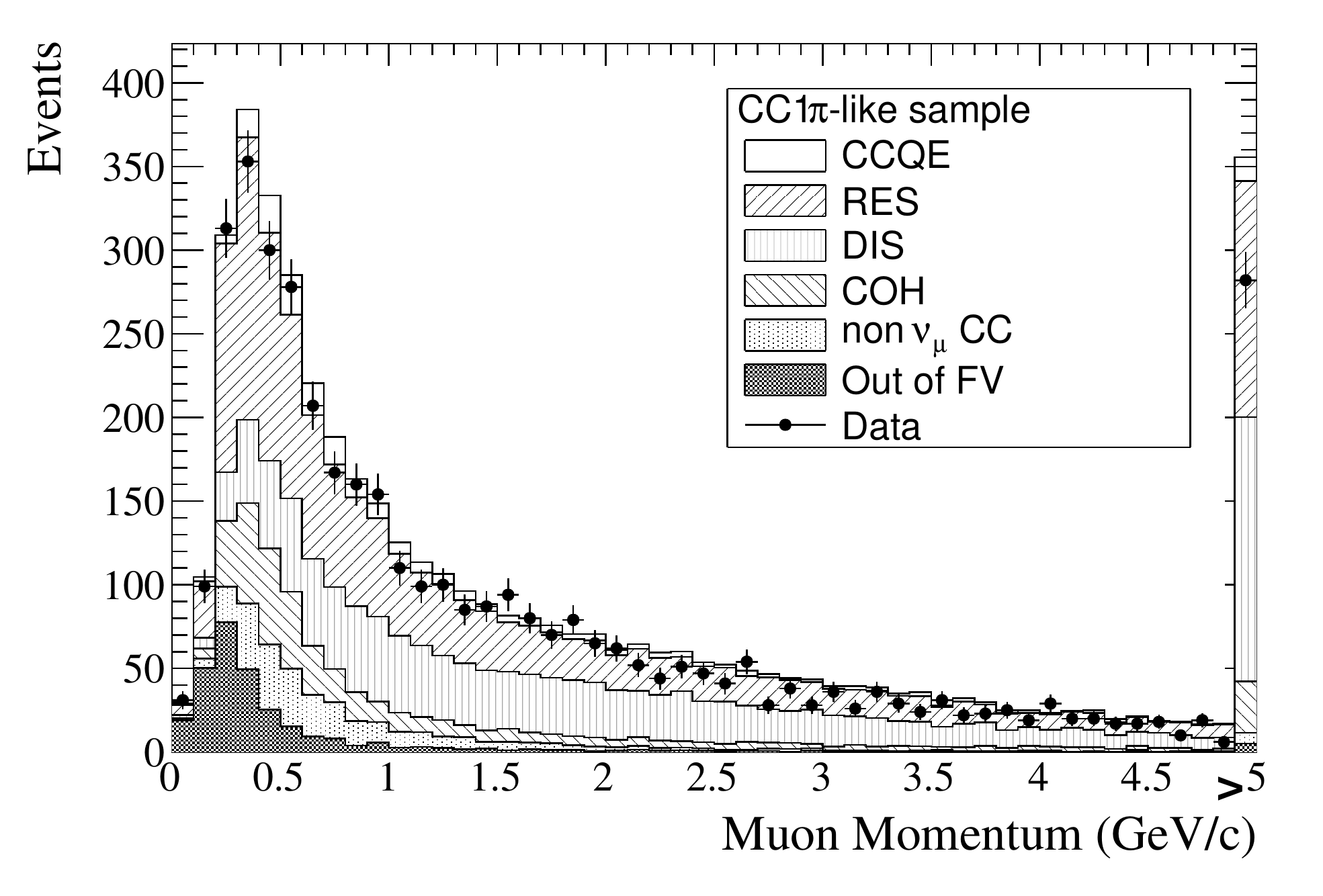} \\
  \includegraphics[keepaspectratio=true,width=.5\textwidth]{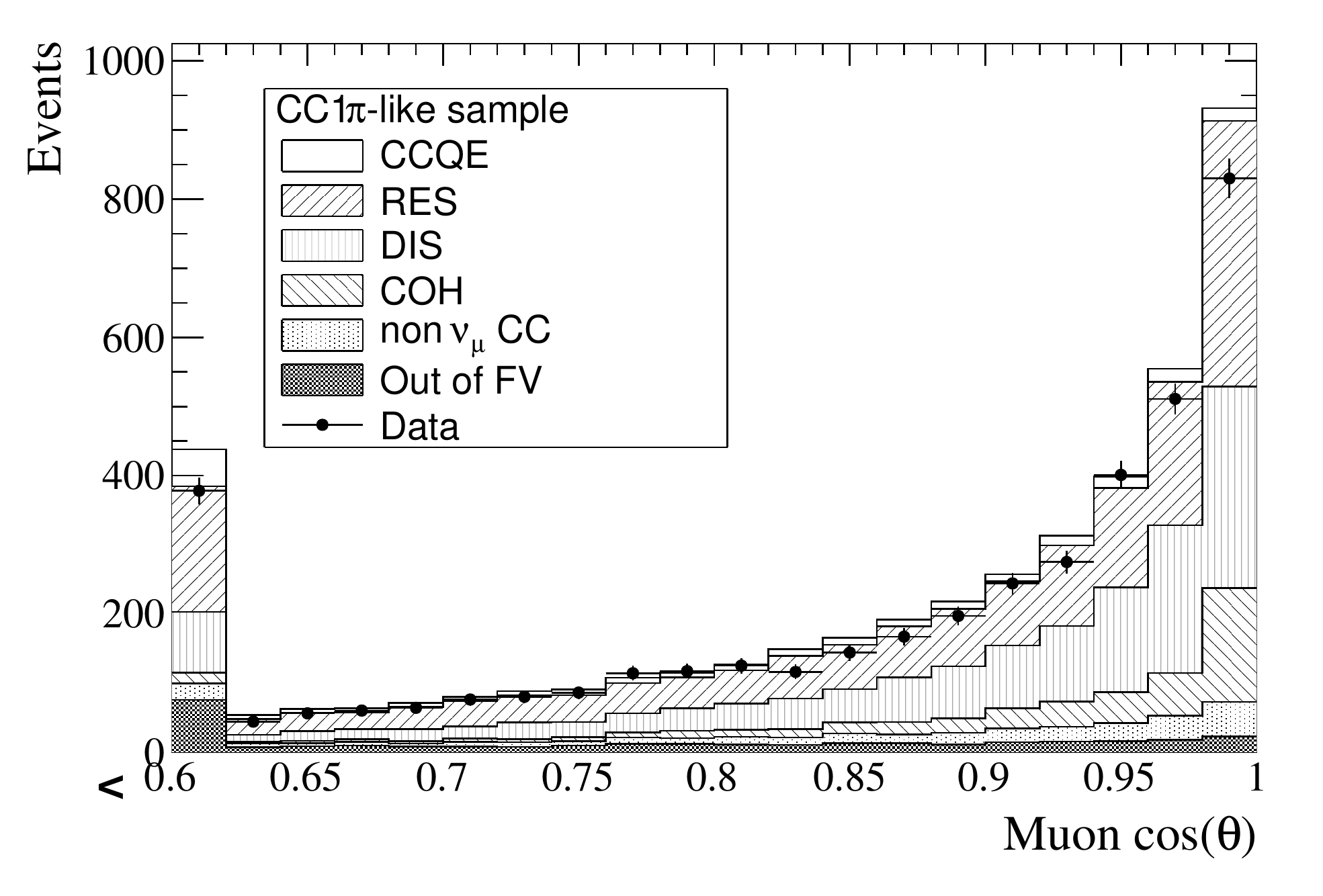}
\else
  \includegraphics[keepaspectratio=true,width=.5\textwidth]{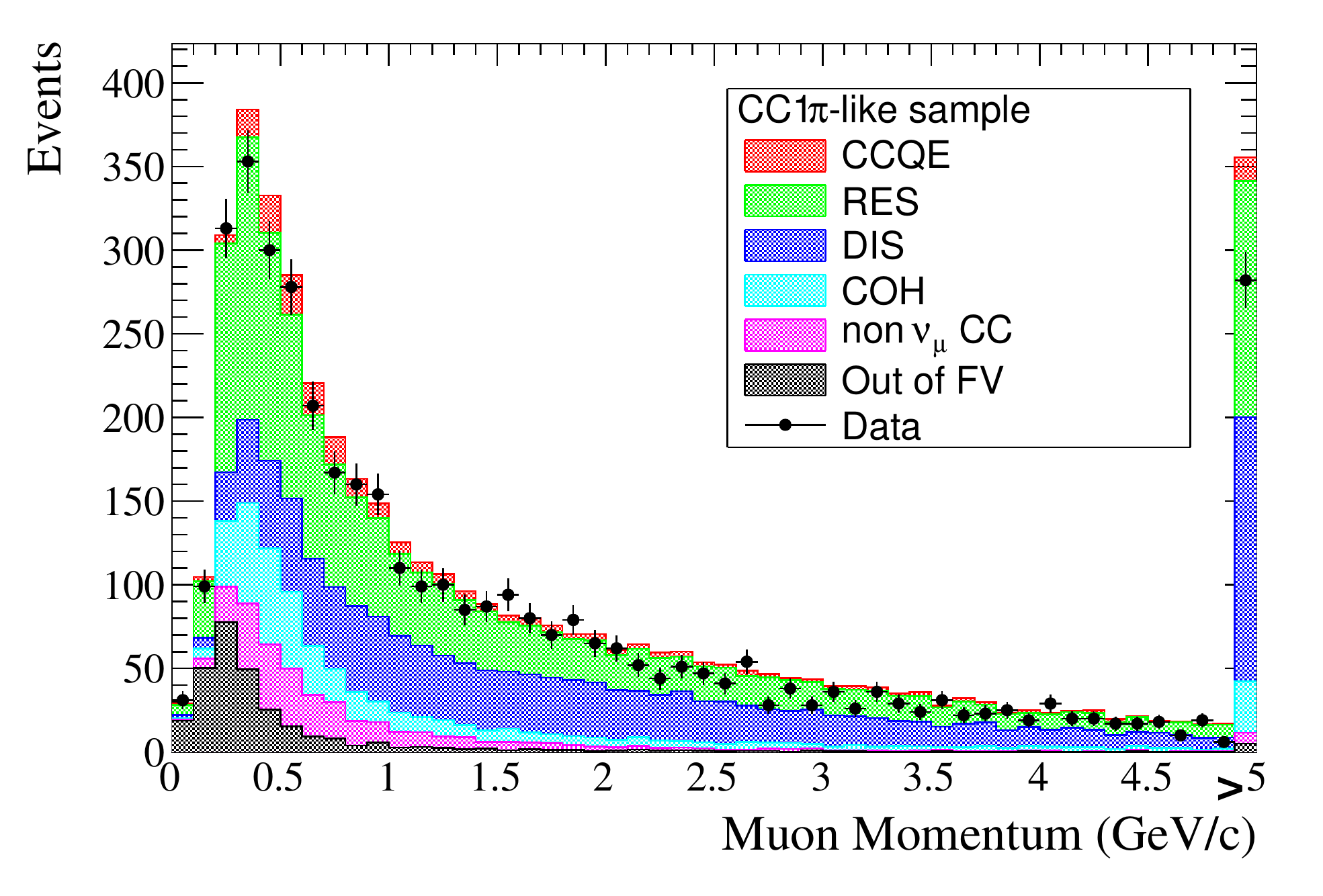} \\
  \includegraphics[keepaspectratio=true,width=.5\textwidth]{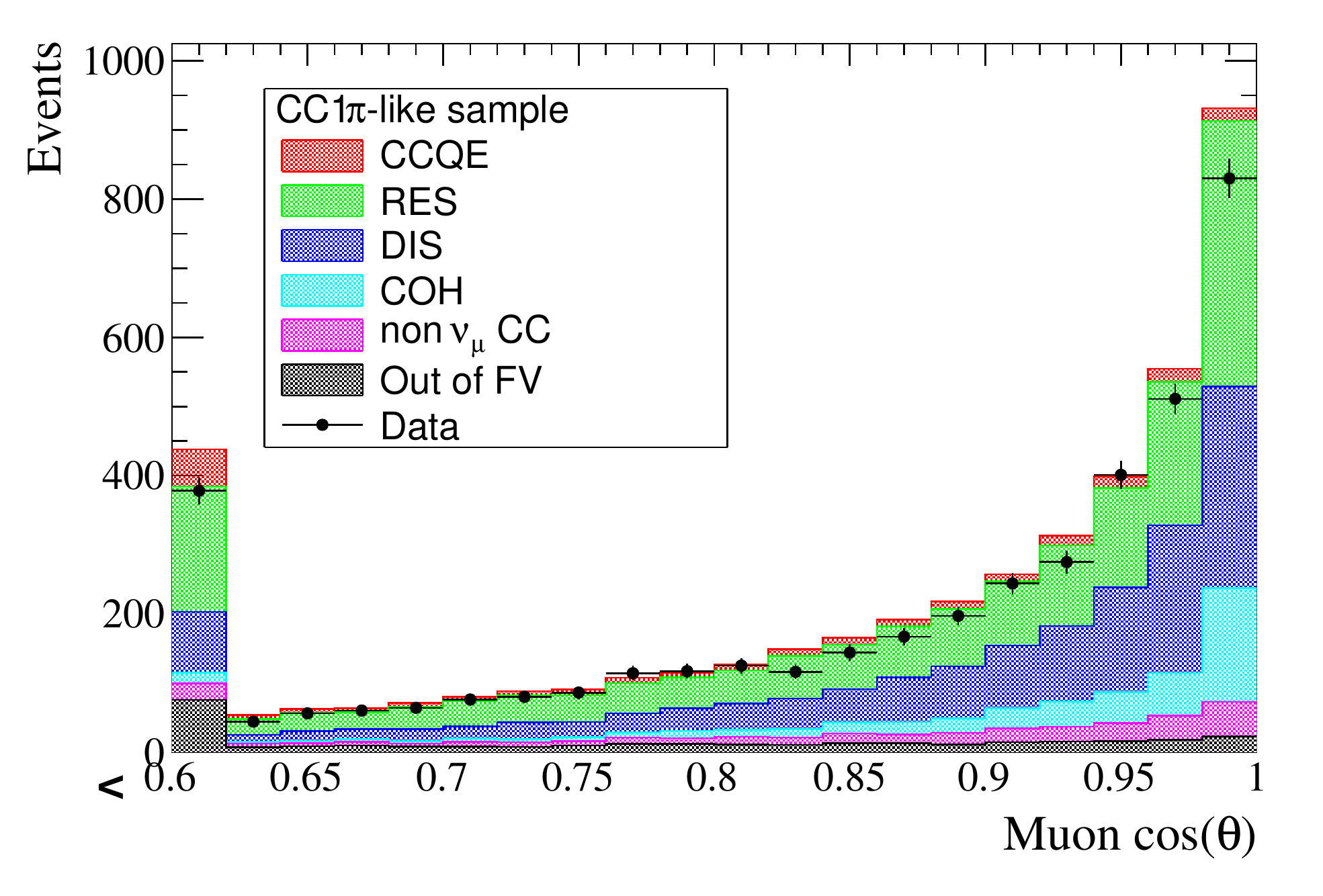}
\fi
 \caption{Muon momentum and angle distribution for the CC1$\pi^+$-like sample.
These are compared to the simulation, broken down into the different reaction types, with
all systematic parameters set to their nominal values.}
 \label{fig:ND280_mu_CC1pi}
\end{figure*}

\begin{figure*}[tbp]
 \centering
\ifx\figstyle\bw
  \includegraphics[keepaspectratio=true,width=.5\textwidth]{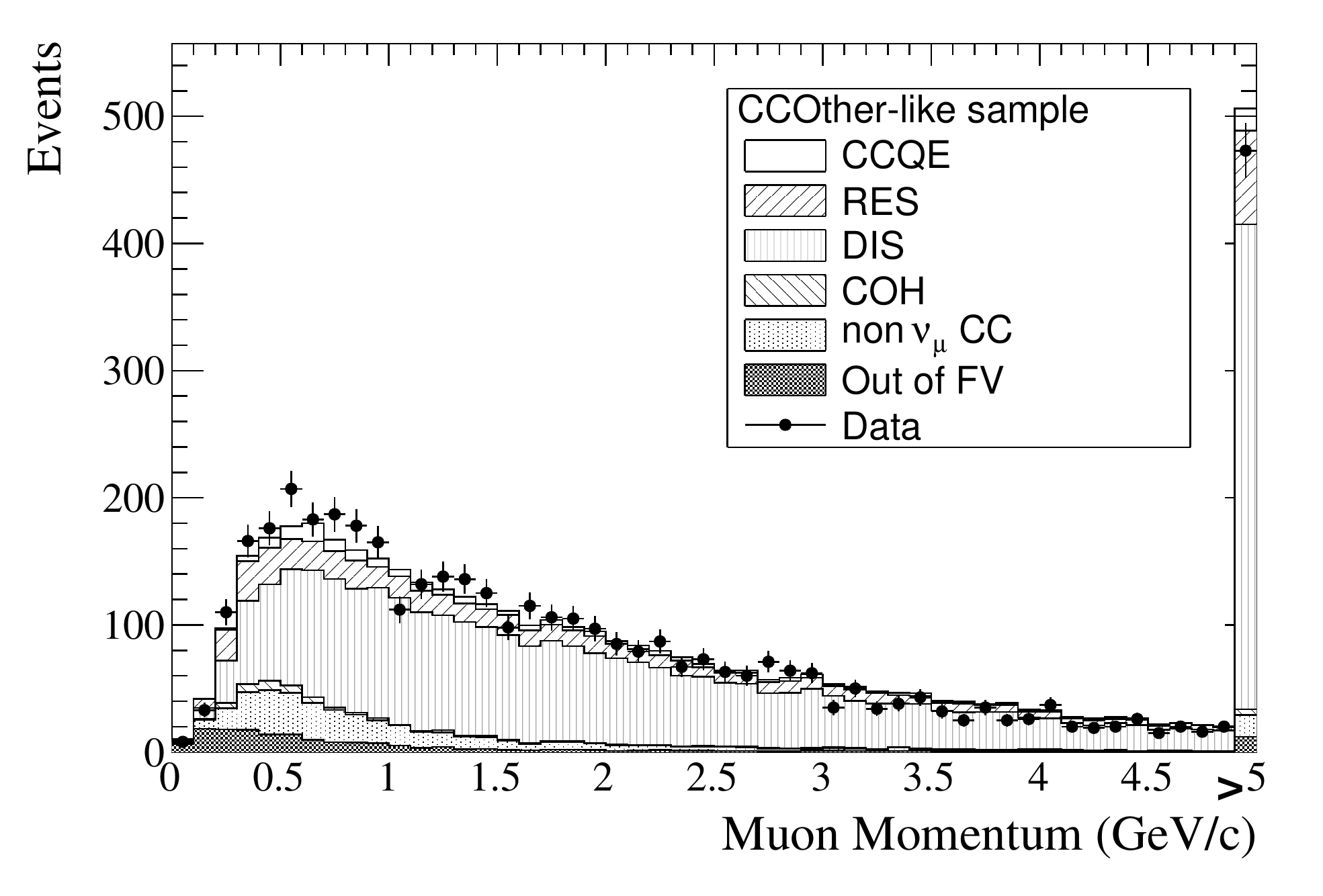} \\
  \includegraphics[keepaspectratio=true,width=.5\textwidth]{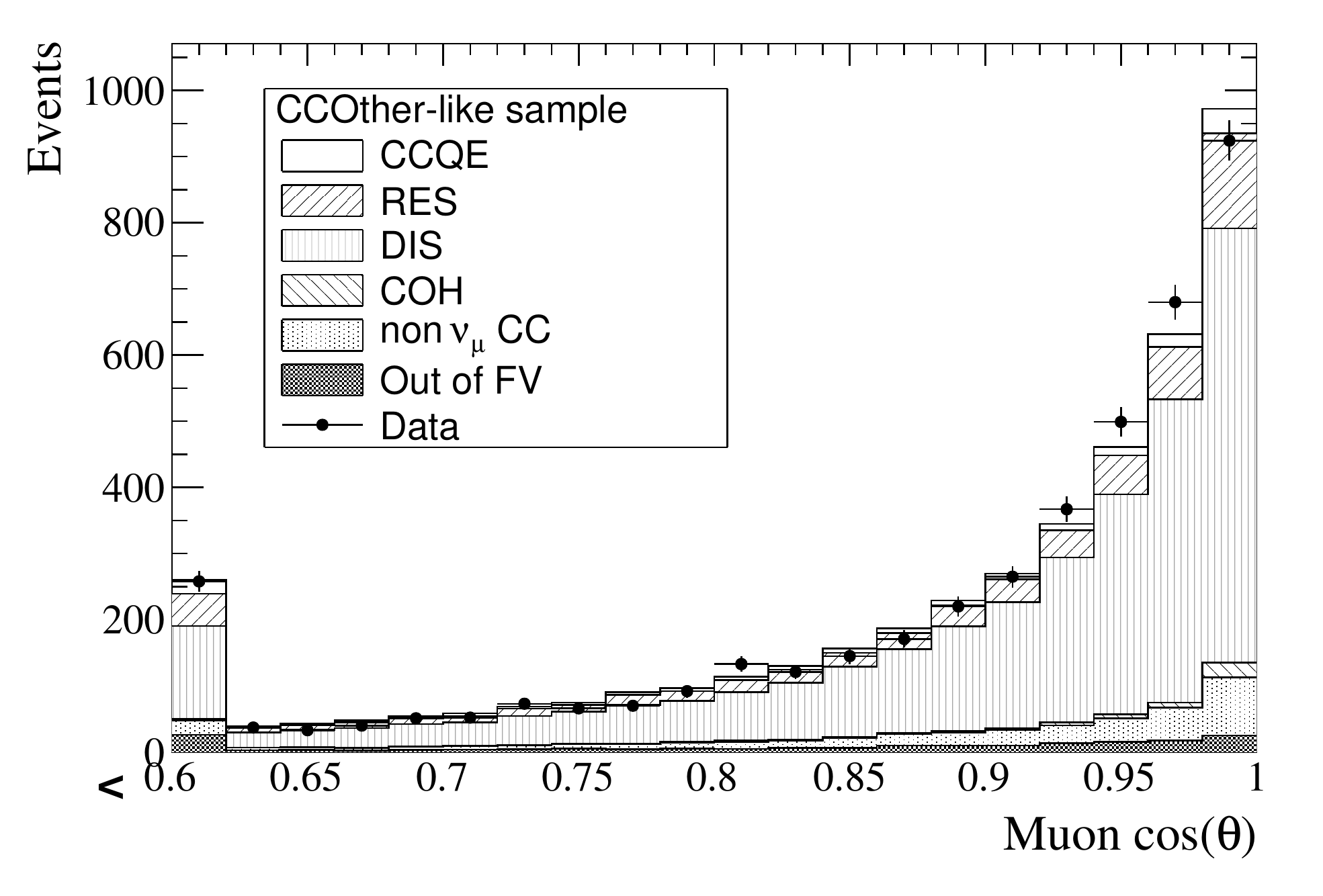}
\else
  \includegraphics[keepaspectratio=true,width=.5\textwidth]{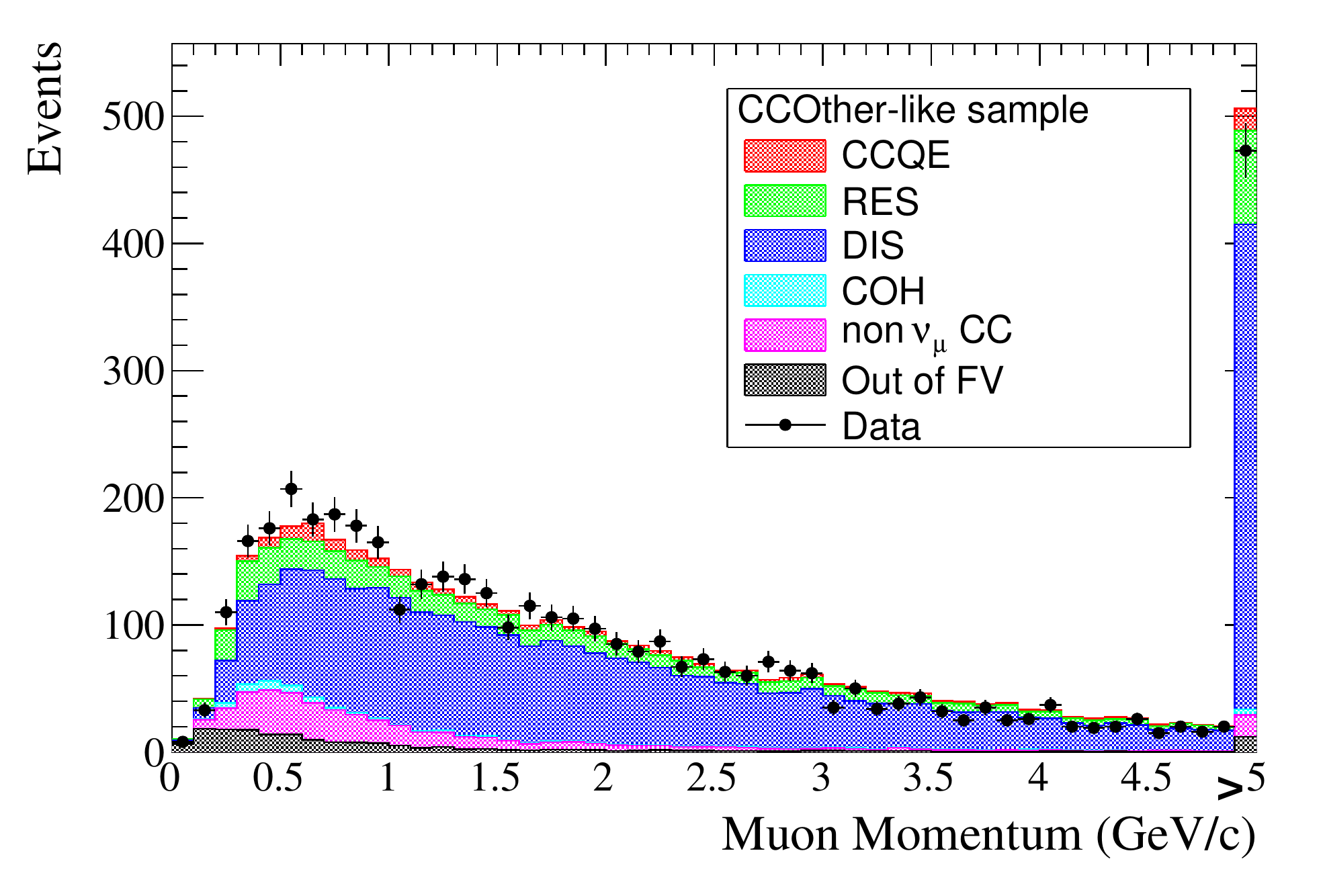} \\
  \includegraphics[keepaspectratio=true,width=.5\textwidth]{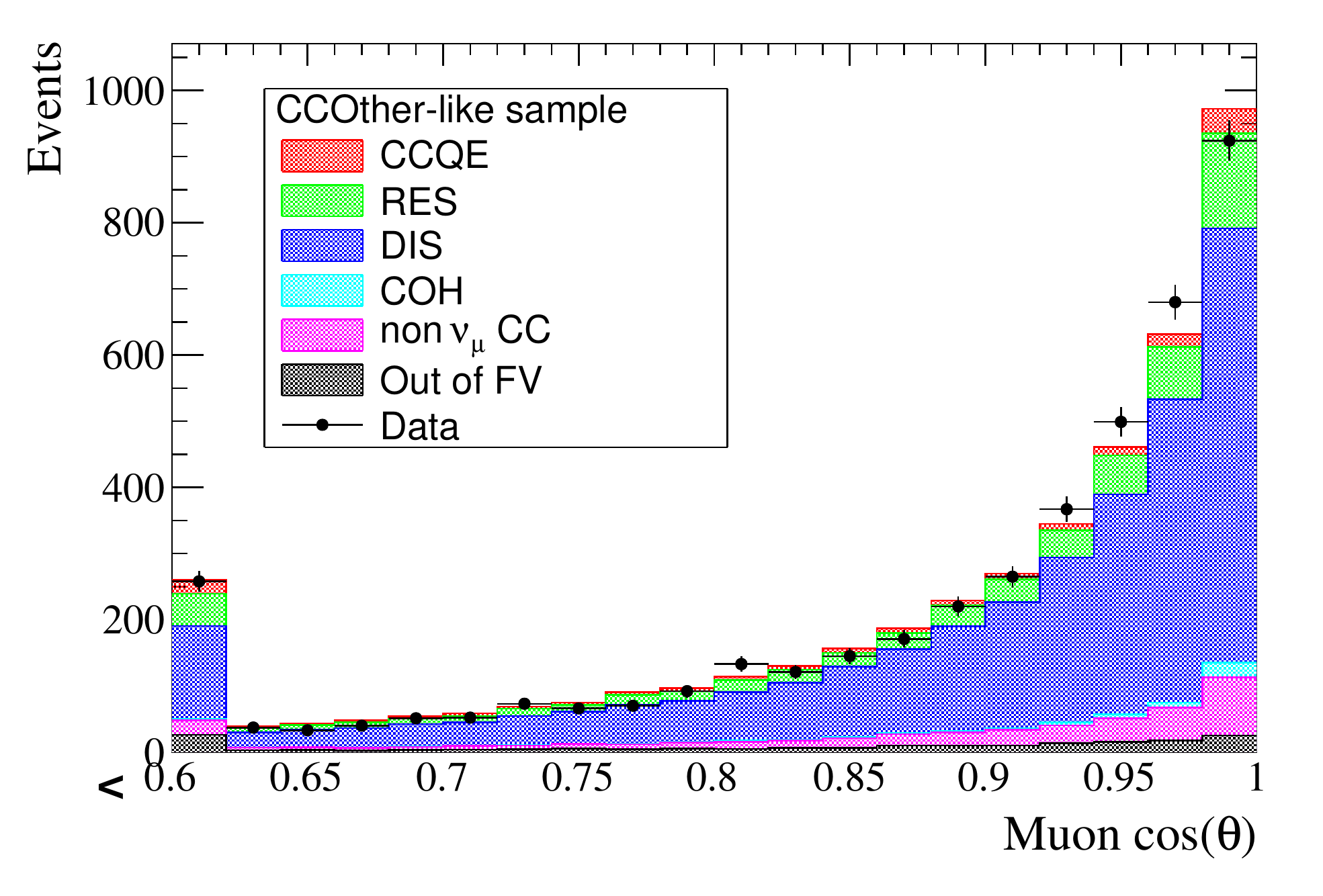}
\fi
 \caption{Muon momentum and angle distribution for the CCOther-like sample.
These are compared to the simulation, broken down into the different reaction types, with
all systematic parameters set to their nominal values.}
 \label{fig:ND280_mu_CCNpi}
\end{figure*}

Figures~\ref{fig:ND280_mu_CC}, \ref{fig:ND280_mu_CC0pi}, \ref{fig:ND280_mu_CC1pi}, and \ref{fig:ND280_mu_CCNpi} show the distributions of the muon momentum $p_\mu$ and angle $\theta_\mu$ (with respect to the $z$-axis) for the CC-inclusive sample and each sub-sample.
These are compared to the nominal simulation, broken down by true reaction type.

\begin{table*}[tbp]
\begin{center}
\caption{Composition for the selected samples (CC-inclusive, CC0$\pi$-like, CC1$\pi^+$-like, CCOther-like) according to the reaction types.} 
\begin{tabular}{l c c c c}
\hline\hline
 True Reaction         & \ \ CC-inclusive \ \ & \ \ CC0$\pi$-like \ \ & \ \ CC1$\pi^+$-like \ \ & \ \ CCOther-like \ \ \\
\hline
CCQE                   & 44.6\% & 63.3\% &  5.3\%  &  3.9\% \\
Resonant pion production \ \  & 22.4\% & 20.3\% & 39.4\%  & 14.2\% \\
Deep inelastic scattering  & 20.6\% & 7.5\%  & 31.3\%  & 67.7\% \\
Coherent pion production \ \  & 2.9\% & 1.4\%  & 10.6\%  &  1.4\% \\
NC                     & 3.1\% & 1.9\%  &  4.7\%  &  6.8\% \\
$\overline{\nu}_{\mu}$  & 0.5\% & 0.2\% &  1.7\%  &  0.9\% \\
$\nue$                 & 0.3\% & 0.2\% &  0.4\%  &  0.9\% \\
Out of FGD1 FV         & 5.4\% & 5.2\%  &  6.6\%  &  4.1\% \\
Other                  & 0.05\% & 0.03\% & 0.04\%  &  0.2\% \\
\hline\hline
\end{tabular}
\label{tab:purity_CCreaction}
\end{center} 
\end{table*}

\begin{table*}[tbp]
\begin{center} 
\caption{Composition of the selected samples (CC-inclusive, CC0$\pi$-like, CC1$\pi^+$-like, CCOther-like) divided into the true topology types. The non-$\nu_\mu$ CC topology includes $\nu_e$, $\bar{\nu}_\mu$ and NC interactions.}
\begin{tabular}{l c c c c}
\hline\hline
 True Topology  & \ \ CC-inclusive \ \ & \ \ CC0$\pi$-like \ \ & \ \ CC1$\pi^+$-like \ \ & \ \ CCOther-like \ \ \\
\hline
CC0$\pi$            & 51.5\%  & 72.4\% &  6.4\% &  5.8\% \\
CC1$\pi^+$          & 15.0\%  &  8.6\% & 49.2\% &  7.8\% \\
CCOther             & 24.2\%  & 11.5\% & 31.0\% & 73.6\% \\
non-$\nu_\mu$ CC    & 4.1\%   &  2.3\% &  6.8\% &  8.7\% \\
Out of FGD1 FV \ \    & 5.2\%   &  5.2\% &  6.6\% &  4.1\% \\
\hline\hline
\end{tabular}
\label{tab:purity_CCtopology}
\end{center} 
\end{table*}

\subsubsection{ND280 detector systematics}
\label{subsec:ND280_systematics}
In this section we explain how we use control samples to
assess uncertainty in the modeling of FGD and TPC response and of neutrino interactions outside of the fiducial volume of the FGD.

TPC systematic uncertainties are divided into three classes: selection efficiency, momentum resolution and PID.
The efficiency systematic uncertainty arises in the modeling of the ionization, cluster finding (where a cluster is defined as a set of contiguous pads in a row or column with charge above threshold), track finding, and charge assignment. This is assessed by looking for missed track components in control samples with particles that pass through all three TPCs.
The single track-finding efficiency is determined to be (99.8$^{+0.2}_{-0.4}\%$) for data and simulation for all angles, momenta and track lengths, and shows no dependence on the number of clusters for tracks with 16 clusters or more. 
The inefficiency due to the overlap from a second nearly collinear track is found to be negligible for both data and simulation, so this systematic uncertainty can be ignored.
The same control samples are used to evaluate the charge mis-identification systematic uncertainty.
This systematic uncertainty is evaluated by comparing data and simulation of the charge mis-identification probability as a function of momentum. 
This is found to be less than 1$\%$ for momenta less than 5\,\gevc. 

The momentum resolution is studied using particles crossing at least one FGD and two TPCs by evaluating the effect on the reconstructed momenta when the information from one of the TPCs is removed from the analysis.
The inverse momentum resolution is found to be better in simulations than in data, typically by 30\%,
and this difference is not fully understood. 
A scaling of the difference between true and reconstructed inverse momentum 
is applied to the simulated data to account for this.
Uncertainty in the overall magnetic field strength leads to an uncertainty on the momentum scale of 0.6$\%$, which is confirmed using the range of cosmic ray particles that stop in the FGD.

The TPC measurement of energy loss for PID is evaluated by studying high-purity control samples of electrons, muons and protons.
The muon control sample has the highest statistics and is composed of particles from neutrino interactions outside the ND280 detector that pass through the entire tracker. 
For muons with momenta below 1\,\gevc, the agreement between data and simulation is good, 
while above 1\,\gevc the resolution is better in simulation than in data. 
Correction factors are applied to the simulation to take into account this effect.

The performance for track finding in the FGD is studied separately for tracks which are connected to TPC tracks and tracks which are isolated in the FGD.
The TPC-FGD matching efficiency is estimated from the fraction of through-going muons, in which the presence of a track in the TPC upstream and downstream of the FGD implies that a track should be seen there. The efficiency is found to be 99.9\% for momentum above 200\,\mevc for both simulation and data.

The FGD-only track efficiency is computed as a function of the direction of the track using a sample of stopping protons going from TPC1 to FGD1. 
This efficiency is found to be slightly better for data than simulation when $\cos \theta_\mu< 0.9$.
A correction is applied to the simulation to account for this and the correction uncertainty is included in the overall detector uncertainty.

The FGD PID performance is evaluated by comparing the energy deposited along the track with the expected energy deposit for a given particle type and reconstructed range in the FGD. We use control samples of muons and protons tagged by TPC1 and stopping in FGD1. The pull distributions (residual divided by standard error) for specific particle hypotheses (proton, muon or pion) for data and simulation are fitted with Gaussian distributions. 
To account for the differences in the means and widths of the distributions between data and simulation, corrections are applied to simulation and the correction uncertainty is included in the overall detector uncertainty.

The Michel electron tagging efficiency is studied using a sample of cosmic rays that stop in FGD1 for which the delayed electron is detected. 
The Michel electron tagging efficiency is found to be $(61.1 \pm 1.9)\%$ for simulation
and $(58.6 \pm 0.4)\%$ for data.
A correction is applied to simulation and the correction uncertainty is included in the overall detector uncertainty.

The uncertainty on the mass of the FGD, computed using the uncertainties in the size and density of the individual components, is 0.67\%~\cite{Amaudruz:2012pe}.

There is systematic uncertainty in the modeling of pion interactions traveling through the FGD. This is evaluated from differences between external pion interaction data~\cite{ashery:piscat,levenson:piscat,ingram:piscat,jones:piscat,giannelli:piscat,ransome:piscat,miller:piscat,nakai:piscat,navon:piscat,ashery:pioncx,rowntree:piscat,fujii:piscat}
and the underlying GEANT4 simulation. The external data do not cover the whole momentum range of T2K, so some extrapolation is necessary. 
Incorrect modeling can migrate events between the three sub-samples and for some ranges of momentum this produces the largest detector systematic uncertainty.

An out-of-fiducial volume (OOFV) systematic is calculated by studying nine different categories 
of events that contribute to this background.
Examples of these categories are: a high energy neutron that creates a $\pi^-$ inside the FGD that is mis-identified as a muon,
 a backwards-going $\pi^+$ from the barrel-ECal that is mis-reconstructed as a forward-going muon, and
 a through-going muon passing completely through the FGD and the TPC-FGD matching failed in such a way that mimics a FV event.
Each of these categories is assigned a rate uncertainty (of 0 or 20\%) and a reconstruction-related uncertainty. 
The reconstruction-related uncertainty is below 40\% for all categories but one: we assign a reconstruction-related uncertainty of 150\% to the high-angle tracks category, in which matching sometimes fails to include some hits that are
outside the FGD FV.

An analysis of the events originating from neutrino interactions outside the ND280 detector (pit walls and surrounding sand) is performed using a dedicated simulation (sand muon simulation). The data/simulation discrepancy is about 10\% and is included as a systematic uncertainty on the predicted number of sand muon events in the CC-inclusive sample.

Pileup corrections are applied to account for the inefficiency due to sand muons crossing the tracker volume in coincidence with a FV event.
The correction is evaluated for each dataset separately and is always below 1.3\%; the systematic uncertainty arising from this correction is always below 0.16\%.

Table~\ref{tab:syst_model} shows the full list of base detector systematic effects 
considered and the way each one is treated within the simulated samples to propagate the uncertainty.
Normalization systematics are treated by a single weight applied to all events.
Efficiency systematics are treated by applying a weight that depends on one or more observables.
Finally, several systematics are treated by adjusting
the observables and re-applying the selection.

\begin{table*}[tbp]
\begin{center} 
\caption{List of base detector systematic effects and the way
each one is treated within the simulated samples to propagate the uncertainty.
Normalization systematics are treated with a signgle weight applied to all
events. Efficiency systematics are treated by applying a weight that depends 
on one or more observables. Observable variation systematics are treated by
adjusting the observables and re-applying the selection.
} 
\begin{tabular}{l    l}
\hline\hline
 Systematic effect           & treatment                  \\
\hline                                                               
TPC tracking efficiency      & efficiency           \\
TPC charge misassignment     & efficiency           \\
TPC momentum resolution      & observable variation    \\
TPC momentum scale           & observable variation    \\
B Field distortion           & observable variation    \\
TPC PID                      & observable variation    \\
TPC-FGD matching efficiency\ \  & efficiency           \\
FGD tracking efficiency      & efficiency           \\ 
FGD PID                      & observable variation    \\ 
Michel electron efficiency   & efficiency           \\
FGD mass                     & normalization             \\ 
Pion secondary int.          & efficiency           \\ 
Out of Fiducial Volume       & efficiency           \\ 
Sand muon                    & efficiency           \\ 
Pileup                       & normalization             \\
TPC track quality requirements \ \ & efficiency           \\
\hline\hline
\end{tabular}
\label{tab:syst_model}
\end{center} 
\end{table*}

\begin{figure}[tbp]
  \begin{center}
    \includegraphics[keepaspectratio=true,width=0.45\textwidth]{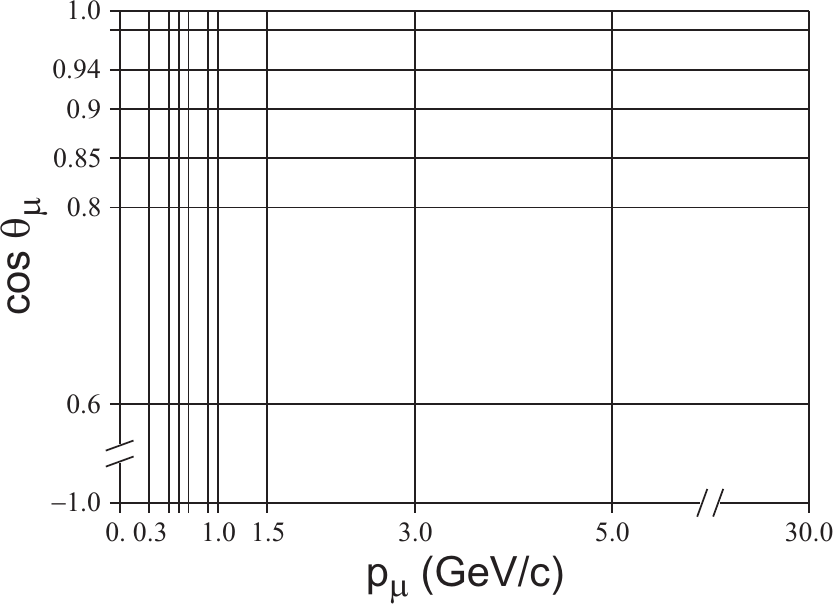}
    \includegraphics[keepaspectratio=true,width=0.45\textwidth]{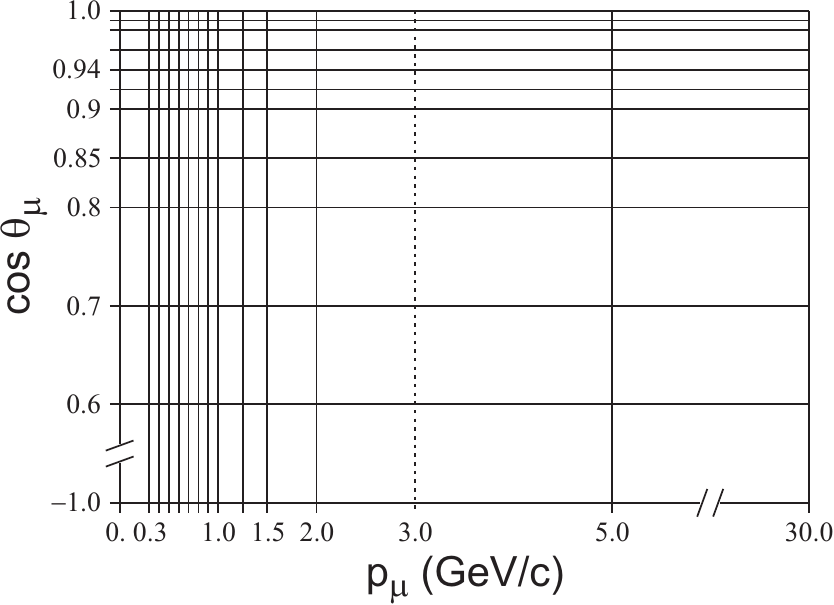}
    \caption{The $p_\mu$-$\cos\theta_\mu$ binning for the systematic parameters $\vec{d}$ that propagate the base detector systematic effects are shown in the left figure for the three event selections. The binning for the observed number of events is shown in the right figure.
For the CC1$\pi^+$-like sample, the bin division at $p_\mu=3.0$\,\gevc is not used.}
    \label{fig:ptbins}
  \end{center}
\end{figure}

The base detector systematic effects are propagated using a vector of systematic parameters $\vec{d}$
that scale the nominal expected numbers of events in bins of 
$p_\mu$-$\cos\theta_\mu$ for the three selections, with the binning illustrated in Fig.~\ref{fig:ptbins}.
When a base systematic parameter is adjusted, $d_i$ is the ratio of the modified to nominal
expected number of events in bin $i$.
The covariance of $\vec{d}$ due to the variation of each base systematic parameters is evaluated
and the full covariance of $\vec{d}$, $V_d$, is found by adding the individual covariances together.
This covariance, and the observed number of events in the three samples in bins of $p_\mu$-$\cos\theta_\mu$, shown in Fig.~\ref{fig:ptbins}, are used by the
subsequent analyses in order to constrain neutrino flux and interaction systematic parameters.

\clearpage
\section{\label{sec:BANFF} Near Detector  Analysis}
In this section we explain how we use the large and detailed samples from
ND280 in conjunction with models for the beam, neutrino interactions, and
the ND280 detector to improve our predictions of the flux at SK and some
cross section parameters.
The systematic parameters for the beam model ($\vec{b}$), binned in energy as shown in Fig.~\ref{fig:flux_at_sk}, the cross section model ($\vec{x}$), listed in Tab.~\ref{tbl:xsecpar}, and detector model ($\vec{d}$), illustrated in Fig.~\ref{fig:ptbins}, are used to describe the systematic uncertainties in the analysis.
We use the three \num CC samples described in Sec.~\ref{sec:ND280} and external data discussed in Sec.~\ref{subsec:externalconstraints}
and summarize our knowledge of the neutrino cross section parameters and unoscillated neutrino flux parameters with a covariance matrix, assuming that a multivariate Gaussian is an appropriate description.

\subsection{ND280 Likelihood}
The three \num CC samples are binned in the kinematic variables $p_{\mu}$ and $\cos\theta_{\mu}$, 
as shown in Fig.~\ref{fig:ptbins}, and the observed and predicted number of events in the bins are used to
define the likelihood,
\begin{equation}
\begin{split}
\mathcal{L}(\vec{b},\vec{x},\vec{d}) =& \prod_i^{N_{bins}} p\left(N^d_{i}|N^{p}_{i}(\vec{b},\vec{x},\vec{d})\right)\\
=& \  c \prod_i^{N_{bins}} \left(N^{p}_{i}(\vec{b},\vec{x},\vec{d})\right)^{N^d_i} e^{-N^{p}_{i}(\vec{b},\vec{x},\vec{d})}\\
\label{eq:Lratio}
\end{split}
\end{equation}
where $N^p_{i}$ is the number of unoscillated MC predicted events and $N^d_i$ is the number of data events in the $i$th bin of the CC samples, the second line assumes the Poisson distribution, and $c$ is a constant. The number of MC predicted events, $N^p_{i}(\vec{b},\vec{x},\vec{d})$, is a function of the underlying beam flux $\vec{b}$, cross section $\vec{x}$, and detector $\vec{d}$ parameters, and these parameters are constrained by external data as described in the previous sections. We model these constraints as multivariate Gaussian likelihood functions and use the product of the above defined likelihood and the constraining likelihood functions as the total likelihood for the near detector analysis. This total likelihood is maximized to estimate the systematic parameters and evaluate their covariance. In practice, the quantity $-2\ln\mathcal{L}_{total}$ is minimized. Explicitly, this quantity is:
\begin{equation}
\begin{split}
& -2\ln\mathcal{L}_{total} = {\mathrm{constant\ }} +\\
&  2 \sum_{i=1}^{N_{bins}}N^{p}_{i}(\vec{b},\vec{x},\vec{d})-N^{d}_{i}\ln[N^{p}_{i}(\vec{b},\vec{x},\vec{d})] \\
& +\sum_{i=1}^{N_{b}}\sum_{j=1}^{N_{b}}(b^0_{i}-b_{i})(V_{b}^{-1})_{i,j}(b^0_{j}-b_{j})  \\
&  +\sum_{i=1}^{N_{x}}\sum_{j=1}^{N_{x}}(x^{0}_{i}-x_{i})(V^{-1}_{x})_{i,j}(x^{0}_j-x_{j}) \\
& +\sum_{i=1}^{N_{d}}\sum_{j=1}^{N_{d}}(d^0_{i}-d_{i})(V^{-1}_{d})_{i,j}(d^0_{j}-d_{j}) \\
\end{split}
\label{eq:Ltotal}
\end{equation}
where $\vec{b}^0$, $\vec{x}^0$, and $\vec{d}^0$ are the nominal values (best estimates prior to the ND280 analysis) and $V_{b}$, $V_{x}$, and $V_{d}$ are the covariance matrices of the beam, cross section, and detector systematic parameters. 

\subsection{Fitting methods}
A reference Monte Carlo sample of ND280 events is generated using the models described in the previous sections and the nominal values for the systematic parameters. Predicted distributions for adjusted values of the systematic parameters are calculated by weighting each event of the Monte Carlo sample individually. For the flux parameters, the true energy and flavor of each MC event determine the normalization weight appropriate for that event. For the detector parameters, the reconstructed momentum and angle of the muon candidate are used. For cross section scaling parameters (e.g., $x^{QE}_1$), weights are applied according to the true interaction mode and true energy. For other cross section parameters (e.g., $M_{A}^{QE}$), including the FSI parameters,
the ratio of the adjusted cross section to the nominal cross section 
(calculated as a function of the true energy, interaction type, and lepton kinematics)
is used to weight the event. The FSI parameters are constrained by a covariance matrix constructed by using representative points on the 1-$\sigma$ surface for the parameters in Table~\ref{tab:fsi_parsets}. 

The fit is performed by minimizing $-2\ln\mathcal{L}_{total}$ using MINUIT~\cite{James:1975dr}. Parameters not of interest to the oscillation analyses (e.g. ND280 detector systematic uncertainties) are treated as nuisance parameters. 

\subsection{Results}
\begin{figure}[tbp]
   \centering
   \includegraphics[width=0.5\textwidth]{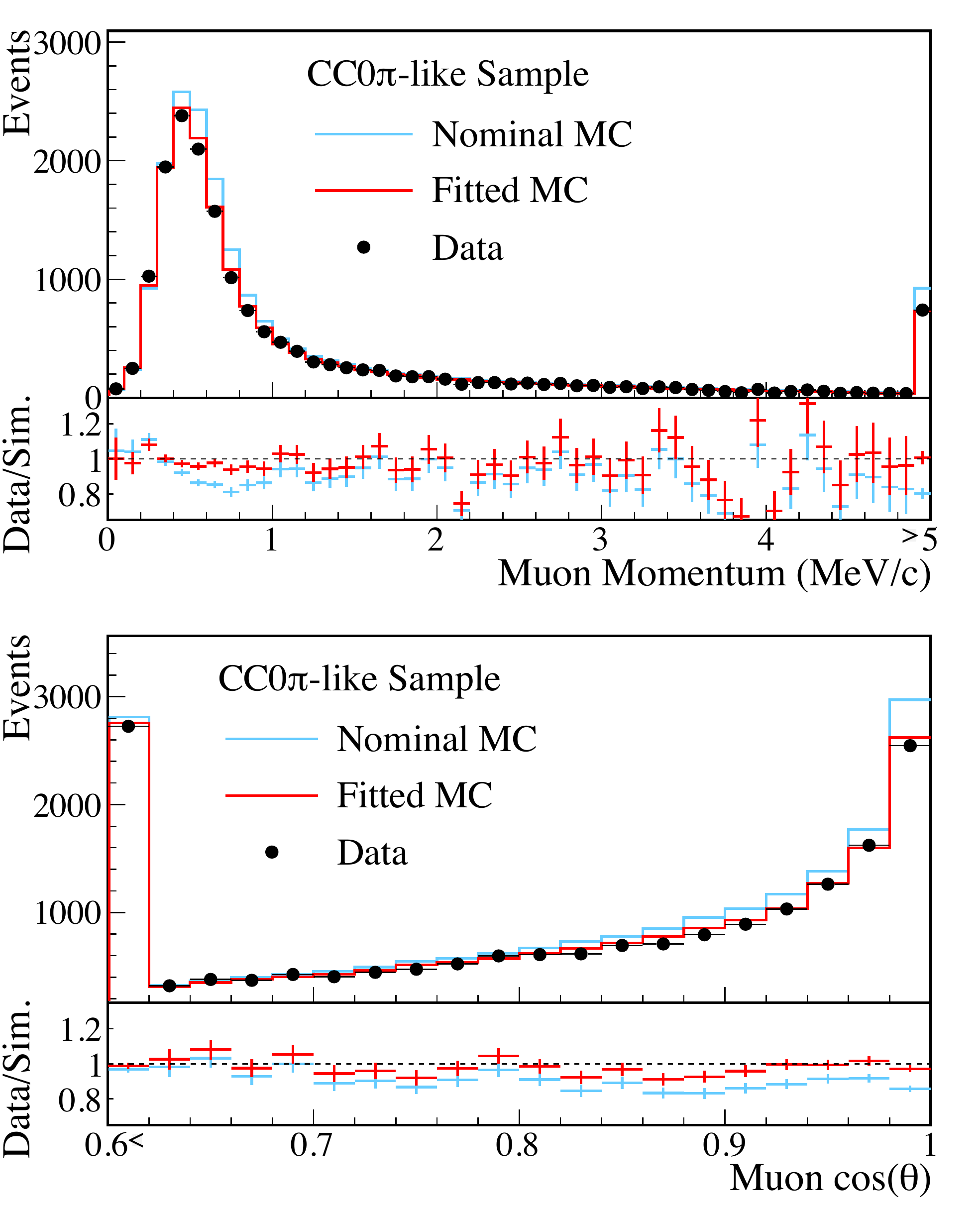} 
   \caption{Comparison of the data and Monte Carlo distributions for muon momentum (top) and angle (bottom) in the CC0$\pi$-like sample, 
using the nominal and fitted values for the systematic parameters.}
   \label{fig:BANFFCC0pi}
\end{figure}

\begin{figure}[tbp]
   \centering
   \includegraphics[width=0.5\textwidth]{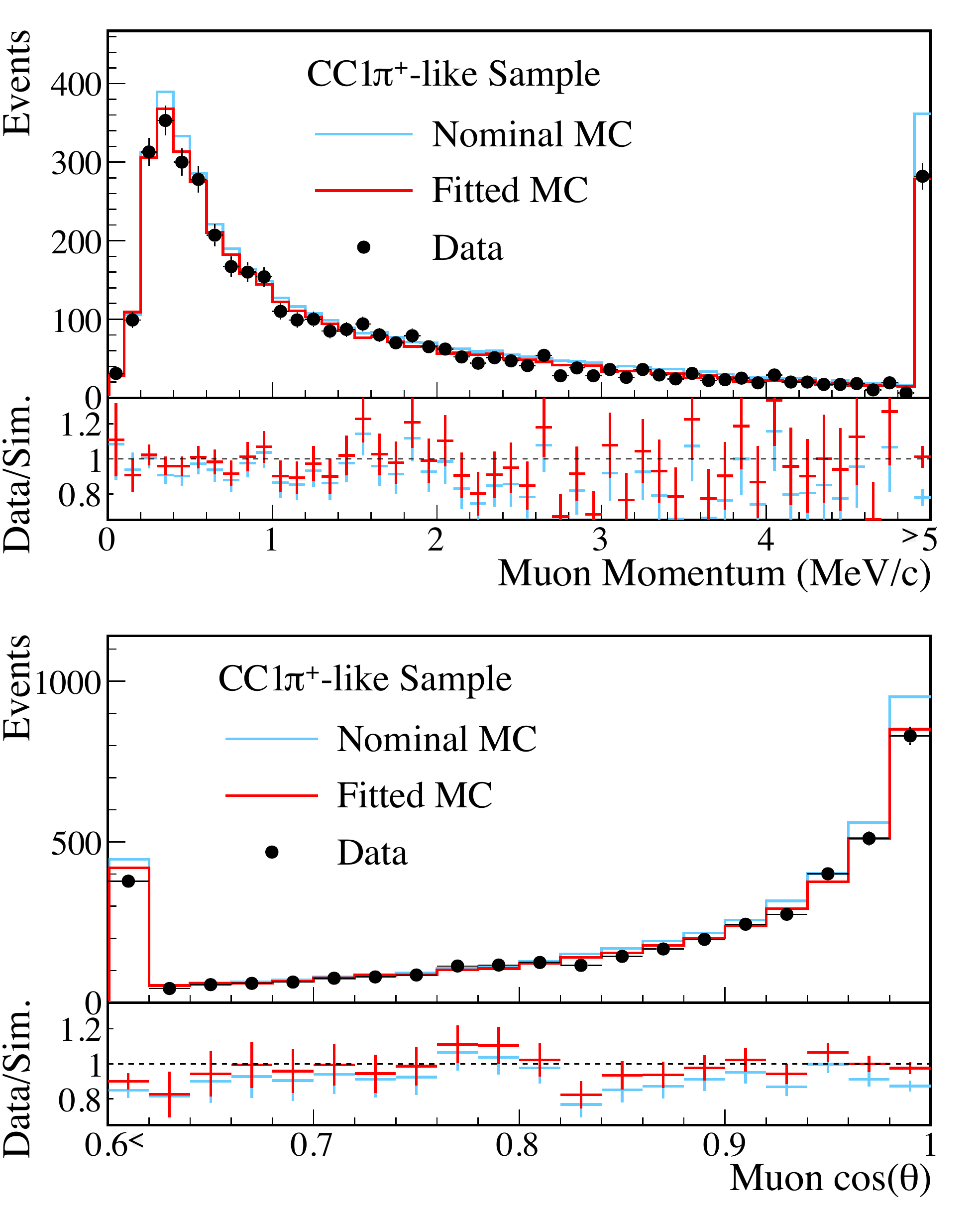} 
   \caption{Comparison of the data and Monte Carlo distributions for muon momentum (top) and angle (bottom) in the CC1$\pi^+$-like sample, 
using the nominal and fitted values for the systematic parameters.}
   \label{fig:BANFFCC1pi}
\end{figure}

\begin{figure}[tbp]
   \centering
   \includegraphics[width=0.5\textwidth]{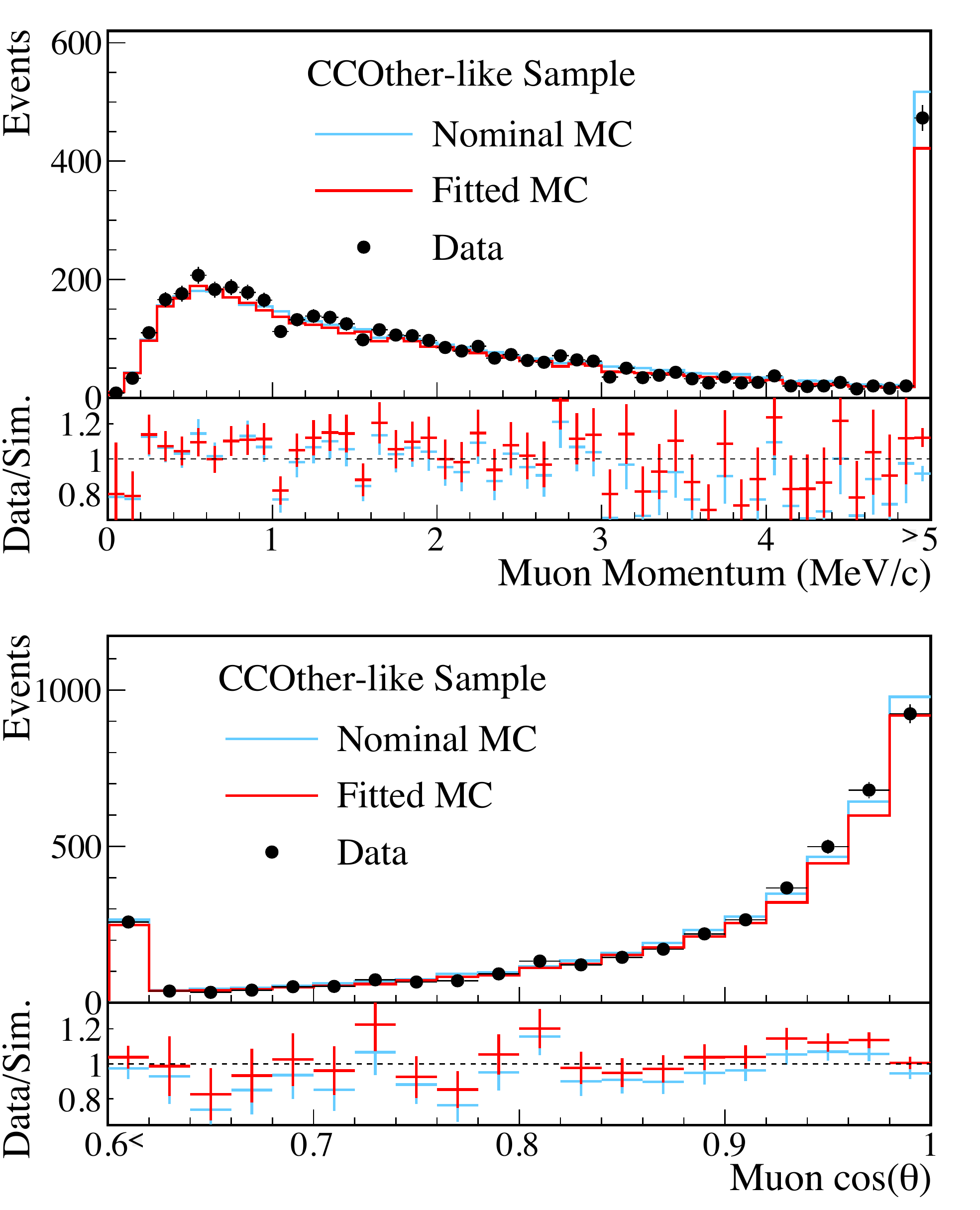} 
   \caption{Comparison of the data and Monte Carlo distributions for muon momentum (top) and angle (bottom) in the CCOther-like sample, 
using the nominal and fitted values for the systematic parameters.}
   \label{fig:BANFFCCOth}
\end{figure}

The result of this analysis is a set of point estimates ($\vec{g}$) and covariance ($V_g$) for the 
systematic scaling factors for the
unoscillated neutrino flux at SK in bins of energy and flavor ($\vec{b}_{s}$) and the cross section parameters which are constrained by ND280 data ($\vec{x}_{n}$). 
Figures~\ref{fig:BANFFCC0pi},~\ref{fig:BANFFCC1pi}, and~\ref{fig:BANFFCCOth} show the projected kinematic variable distributions of the three ND280 samples used in this analysis, comparing the data to the MC prediction for the two cases of using nominal values of the systematic parameters and using the best-fit values of the parameters. The MC distributions show better agreement with the data when using the best-fit values for the parameters, especially decreasing the prediction near the momentum peak and in the forward direction ($\cos\theta_{\mu}$ close to 1). 

\begin{figure*}[tbp]
  \begin{center}
  \includegraphics[width=0.6\textwidth]{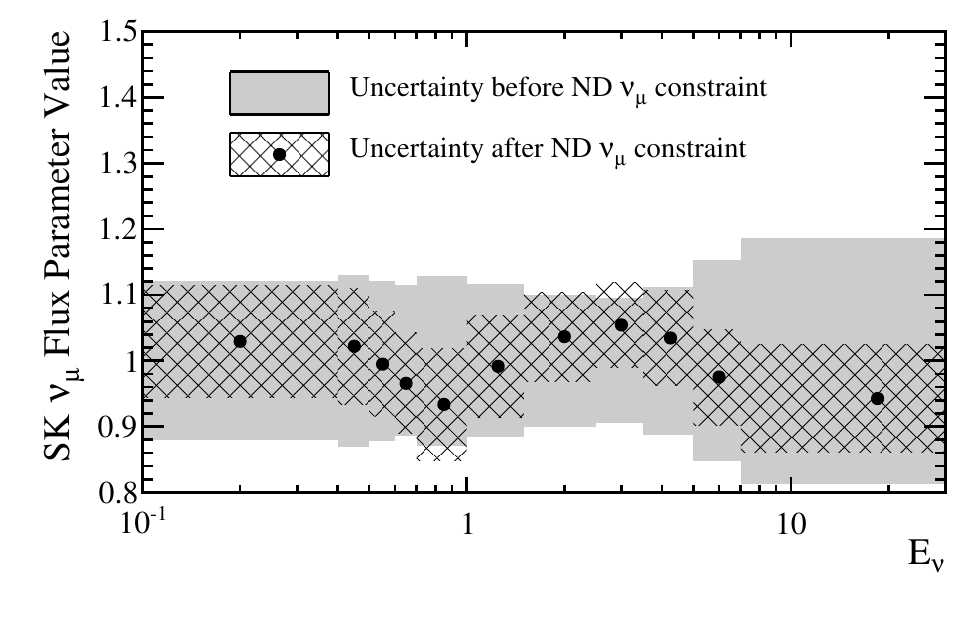}
  \includegraphics[width=0.6\textwidth]{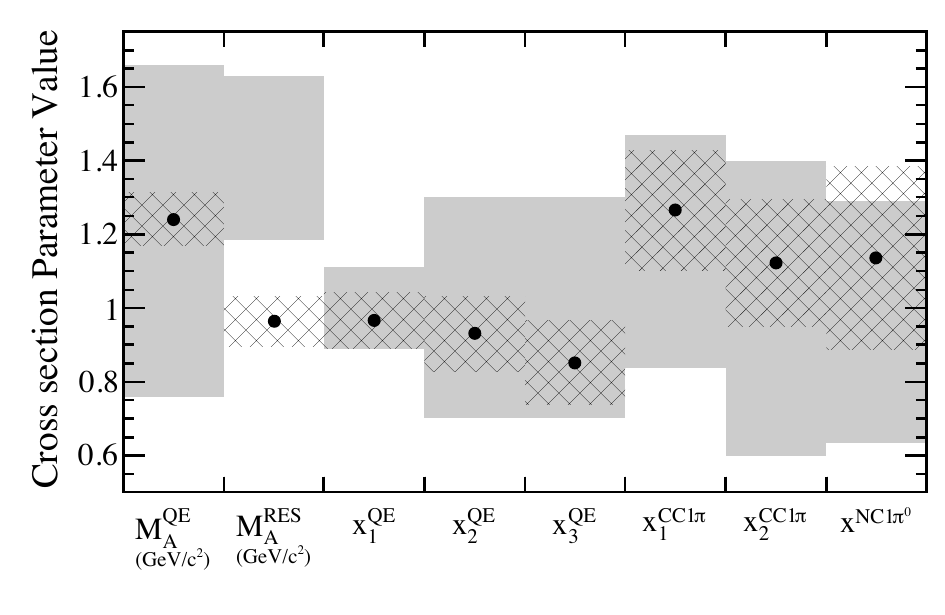}
  \caption{Prior and fitted values and uncertainties for the SK $\nu_{\mu}$ flux parameters (upper figure) and cross section parameters (lower figure) constrained by the near detector analysis for the oscillation analyses. Uncertainties are calculated as the square root of the diagonal of the relevant covariance matrix. The value of $M_{A}^{QE}$ and $M_{A}^{RES}$ are given in units of \gevcc, and all other parameters are multiplicative corrections. }
  \label{fig:BANFFuncertainties}
  \end{center}
\end{figure*}

Figure~\ref{fig:BANFFuncertainties} shows the values of the \num flux and cross section parameters that are constrained by the near detector analysis for the oscillation analyses; Table~\ref{tab:propagatedparameters} lists the flux parameters and Table~\ref{tab:propagatedparametersxsec} lists the values of the cross section parameters. These tables contain all of the point estimates in $\vec{g}$ as well as the errors calculated as the square root of the diagonal of the covariance $V_g$. One of the interesting features of the best-fit parameters is the dip in the flux parameters just below 1~GeV, which is near the peak of the T2K beam flux. This is particularly important, as this is the region of interest for oscillation analyses, and an incorrect prediction of the flux in this region can bias estimates of oscillation parameters. Another interesting point is the value of $M_{A}^{RES}$, which is pulled to a much lower value than the external data constraint used in the fit. This highlights both the power of the ND280 data, and the importance of the CC1$\pi^{+}$-like sample, which is dominant in determining this parameter. This selection is new to the ND280 analysis for the set of oscillation analyses reported in this paper, and provides an improved ability to use T2K data to constrain resonant interaction parameters.

 The predicted event rate at SK is given by the product of the flux, cross section, and detector efficiency, and the typical uncertainties of the flux and cross section parameters constrained by ND280 are 7-10\%. The estimators of these flux and cross section parameters have a strong negative correlation, however, because they use the rate measurements in the near detector. As a result, their contribution to the SK event rate uncertainty is less than 3\%, significantly smaller than the individual flux and cross section parameter uncertainties.

A cross-check to this analysis is performed by studying a selection of electron neutrino interactions in ND280~\cite{intrinsicnumeasurement2014}, and finds that the relative rate of selected 
electron neutrino events to that predicted by MC using the best-fit parameter values from this analysis
is $R(\nue) = 1.01\pm0.10$. 

\begin{table}[tbp]
   \centering
      \caption{Prior and fitted values and uncertainties for the near-detector-constrained SK flux parameters. All parameters are multiplicative corrections, and the uncertainties are calculated as the square root of the diagonal of the covariance matrix.}
   \begin{tabular}{l c c}
   \hline\hline
   Parameter & \ \ Prior Value \ \ & \ \ Fitted Value\ \ \\
   \hline
   \num 0.0--0.4 GeV\ \ & 1.00$\pm$0.12&1.03$\pm$0.09\\
   \num 0.4--0.5 GeV&1.00$\pm$0.13&1.02$\pm$0.09\\
   \num 0.5--0.6 GeV&1.00$\pm$0.12&0.99$\pm$0.08\\
   \num 0.6--0.7 GeV&1.00$\pm$0.11&0.97$\pm$0.08\\
   \num 0.7--1.0 GeV&1.00$\pm$0.13&0.93$\pm$0.08\\
   \num 1.0--1.5 GeV&1.00$\pm$0.12&0.99$\pm$0.08\\
   \num 1.5--2.5 GeV&1.00$\pm$0.10&1.04$\pm$0.07\\
   \num 2.5--3.5 GeV&1.00$\pm$0.09&1.05$\pm$0.06\\
   \num 3.5--5.0 GeV&1.00$\pm$0.11&1.03$\pm$0.07\\
   \num 5.0--7.0 GeV&1.00$\pm$0.15&0.98$\pm$0.07\\
   \num $>$7.0 GeV&1.00$\pm$0.19&0.94$\pm$0.08\\
\hline
   \numb 0.0--0.7 GeV&1.00$\pm$0.13&1.03$\pm$0.10\\
   \numb 0.7--1.0 GeV&1.00$\pm$0.12&1.01$\pm$0.09\\
   \numb 1.0--1.5 GeV&1.00$\pm$0.12&1.01$\pm$0.09\\
   \numb 1.5--2.5 GeV&1.00$\pm$0.12&1.03$\pm$0.10\\
   \numb $>$2.5 GeV&1.00$\pm$0.12&1.01$\pm$0.11\\
\hline
   \nue 0.0--0.5 GeV&1.00$\pm$0.13&1.03$\pm$0.10\\
   \nue 0.5--0.7 GeV&1.00$\pm$0.13&1.01$\pm$0.09\\
   \nue 0.7--0.8 GeV&1.00$\pm$0.14&0.98$\pm$0.11\\
   \nue 0.8--1.5 GeV&1.00$\pm$0.11&1.00$\pm$0.07\\
   \nue 1.5--2.5 GeV&1.00$\pm$0.10&1.02$\pm$0.07\\
   \nue 2.5--4.0 GeV&1.00$\pm$0.12&1.00$\pm$0.07\\
   \nue $>$4.0 GeV&1.00$\pm$0.17&0.95$\pm$0.08\\
\hline
   \nueb 0.0--2.5 GeV&1.00$\pm$0.19&1.01$\pm$0.18\\
   \nueb $>$2.5 GeV&1.00$\pm$0.14&0.96$\pm$0.08\\
   \hline\hline
   \end{tabular}

   \label{tab:propagatedparameters}
\end{table}

\begin{table}[tbp]
   \centering
      \caption{Prior and fitted values and uncertainties for the near-detector-constrained cross section model parameters. The value of $M_{A}^{QE}$ and $M_{A}^{RES}$ are given in units of \gevcc and all other parameters are multiplicative corrections. The uncertainties are calculated as the square root of the diagonal of the covariance 
matrix.}
   \begin{tabular}{l c c c}
   \hline\hline
   Parameter & \ \ units \ \ & \ \ Prior Value \ \ & \ \ Fitted Value \ \ \\
   \hline
   $M_{A}^{QE}$ & \gevcc &1.21$\pm$0.45&1.24$\pm$0.07\\
   $M_{A}^{RES}$ & \gevcc & 1.41$\pm$0.22&0.96$\pm$0.07\\
   $x^{QE}_1$ & & 1.00$\pm$0.11&0.97$\pm$0.08\\
  $ x^{QE}_2$ & &1.00$\pm$0.30&0.93$\pm$0.10\\
   $x^{QE}_3$ & &1.00$\pm$0.30&0.85$\pm$0.11\\
   $x^{CC1\pi}_1$ & &1.15$\pm$0.32&1.26$\pm$0.16\\
   $x^{CC1\pi}_2$ & &1.00$\pm$0.40&1.12$\pm$0.17\\
   $x^{NC\pi^0}$ & &0.96$\pm$0.33&1.14$\pm$0.25\\
   \hline\hline
   \end{tabular}
   \label{tab:propagatedparametersxsec}
\end{table}

\clearpage
\section{\label{sec:SK}Far Detector}
Precision measurements of neutrino oscillation by T2K rely on the capabilities of the far detector,
most notably, its large target volume and acceptance and efficient discrimination between the primary leptons produced in $\nu_{\mu}$ and  $\nu_{e}$ CC interactions.  
Additionally, since CCQE scattering interactions are expected to dominate at the energies below 1 GeV, 
accurate reconstruction of the parent neutrino energy is reliant upon accurate estimation of the lepton kinematics.
Finally, the suppression of backgrounds, particularly those from NC and single-pion production processes, is needed. 
Here we discuss the performance of SK in this context, focusing on the event selections 
and the estimation of systematic uncertainties in the modeling of SK. 

Super-Kamiokande is a 50 kton water Cherenkov detector located in the Kamioka Observatory, Gifu, Japan. 
It is divided into two concentric cylinders, an inner detector (ID) with 11,129 inward-facing 
20-inch photomultiplier tubes (PMTs) and an outer detector (OD), used primarily as a 
veto, which has 1885 outward-facing eight-inch PMTs.
The ID PMTs view a 32 kton target volume and the OD collects light within a 2-m wide cylindrical shell surrounding the ID. 
The photocathode coverage of the ID is 40\% and the space between PMTs 
is covered with a black plastic sheet to reduce reflection.  
To overcome its reduced photocathode coverage, reflective Tyvek$^\circledR$ lines the inner and outer surfaces of the OD and each PMT is coupled to a $60 \times 60 \mbox{\ cm}^{2}$ wavelength-shifting plate to improve light collection.  

Cherenkov radiation from charged particles traversing the detector produces ring patterns recorded by the ID PMTs and is the primary 
tool for particle identification (PID). 
Due to their relatively large mass, muons passing through the detector are often unscattered and thereby produce clear ring patterns. 
Electrons, in contrast, scatter and produce electromagnetic showers, resulting in a diffuse ring edge. 
These differences in conjunction with estimation of the Cherenkov opening angle enable efficient 
discrimination between leptons. 
The probabilities to misidentify a single electron as a muon or a single muon as an electron are 0.7\% and 0.8\%, respectively,
for typical lepton energies in T2K events.
Since the recoil proton from CC interactions at T2K is usually below Cherenkov threshold, a single lepton 
is the dominant topology for beam-induced events at SK. 
For such isolated electrons (muons) the momentum and angular resolutions are estimated to be $0.6\% + 2.6\%/\sqrt{P[{\mathrm{GeV/c}}]}$ $(1.7\% + 0.7\%/\sqrt{P[{\mathrm{GeV/c}}]})$
and $3.0^\circ$ $(1.8^\circ)$, respectively.
Since the start of T2K, SK has operated with upgraded electronics which provide lossless acquisition of all PMT hits above threshold.
As a result the efficiency for tagging electrons from muon decays within the ID is $89.1\%$, an essential element of removing backgrounds containing 
sub-threshold muons or charged pions. 
Further details of the detector and its calibration may be found in~\cite{Fukuda:2002uc,Abe:2011ks,Abe:2013gga}.

Due to its large size, SK observes roughly ten atmospheric neutrino interactions per day within its fiducial volume.
These neutrinos serve as control samples for the estimation of systematic errors. 
Similarly, although the detector is located at a depth of 2700 meters water equivalent, cosmic ray muons traverse the detector at approximately $3\mbox{\ Hz}$ and 
together with their decay electrons provide an additional sample for systematic error evaluation. 
Details of these and other control samples are presented in the following subsections. 

\subsection{Event Selection and Data Quality}
\label{sec:SK_event_selection}
We define a sample of fully contained (FC) events whose Cherenkov light is deposited 
exclusively in the ID.  PMTs in the OD that register light above threshold are referred to as
``hit PMTs'' and are grouped with 
neighboring hit PMTs to form clusters.  If the largest such cluster contains more 
than 15 PMTs the event is rejected from the FC sample
and included in the OD sample.  Low energy (LE) events are 
removed by requiring that the total charge from the ID PMT hits in a 300\ns window 
be greater than 200 photoelectrons (p.e.), which corresponds to the charge observed 
from a 20\mev electromagnetic shower.  Events are also removed if a single ID PMT hit 
constitutes more than half of the total p.e.\ observed, in order to reject events due to noise.  
The final criterion rejects events 
that occur due to light from a discharge at the dynode of a PMT, known as ``flasher'' events.  
Such events have a 
broader timing distribution than neutrino interactions and tend to form a repeated pattern 
of light.  A total of 18 events were rejected as flashers from all run periods, although from 
event timing information and visual scans we are confident that all are in fact due 
to beam neutrino interactions.  Nevertheless, these events are discarded and the resulting 
selection inefficiency is taken into account.

\begin{figure}[tbp]
\begin{center}
\includegraphics[width=0.45\textwidth]{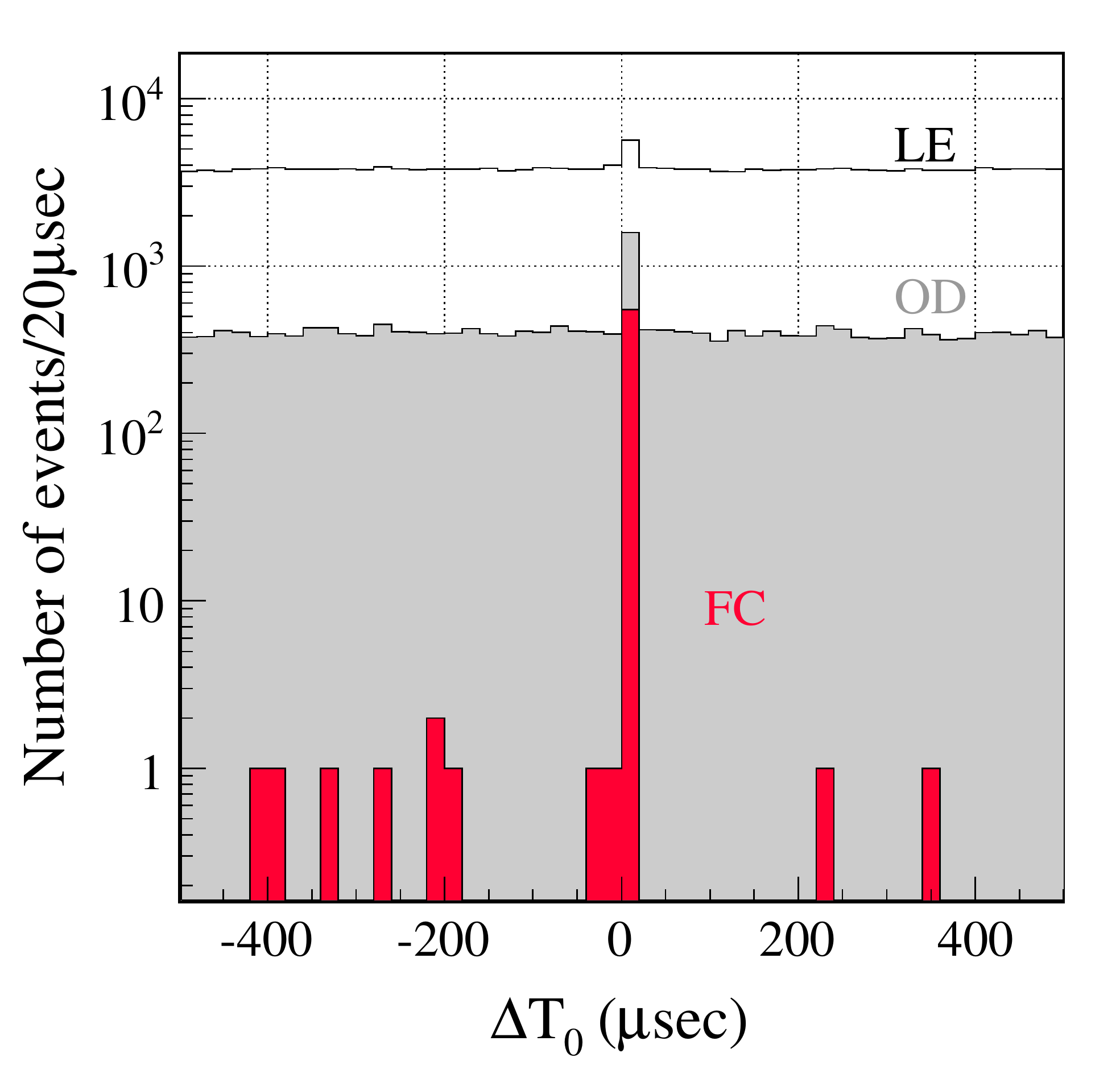}
\caption{$\Delta T_0$ distribution of all FC, OD, and LE events within $\pm$500\mus of the 
expected beam arrival time.  The histograms are stacked in 
that order.}
\label{fig:SK_dT0_offtiming}
\end{center}
\end{figure}

\begin{figure}[tbp]
\begin{center}
\includegraphics[width=0.45\textwidth]{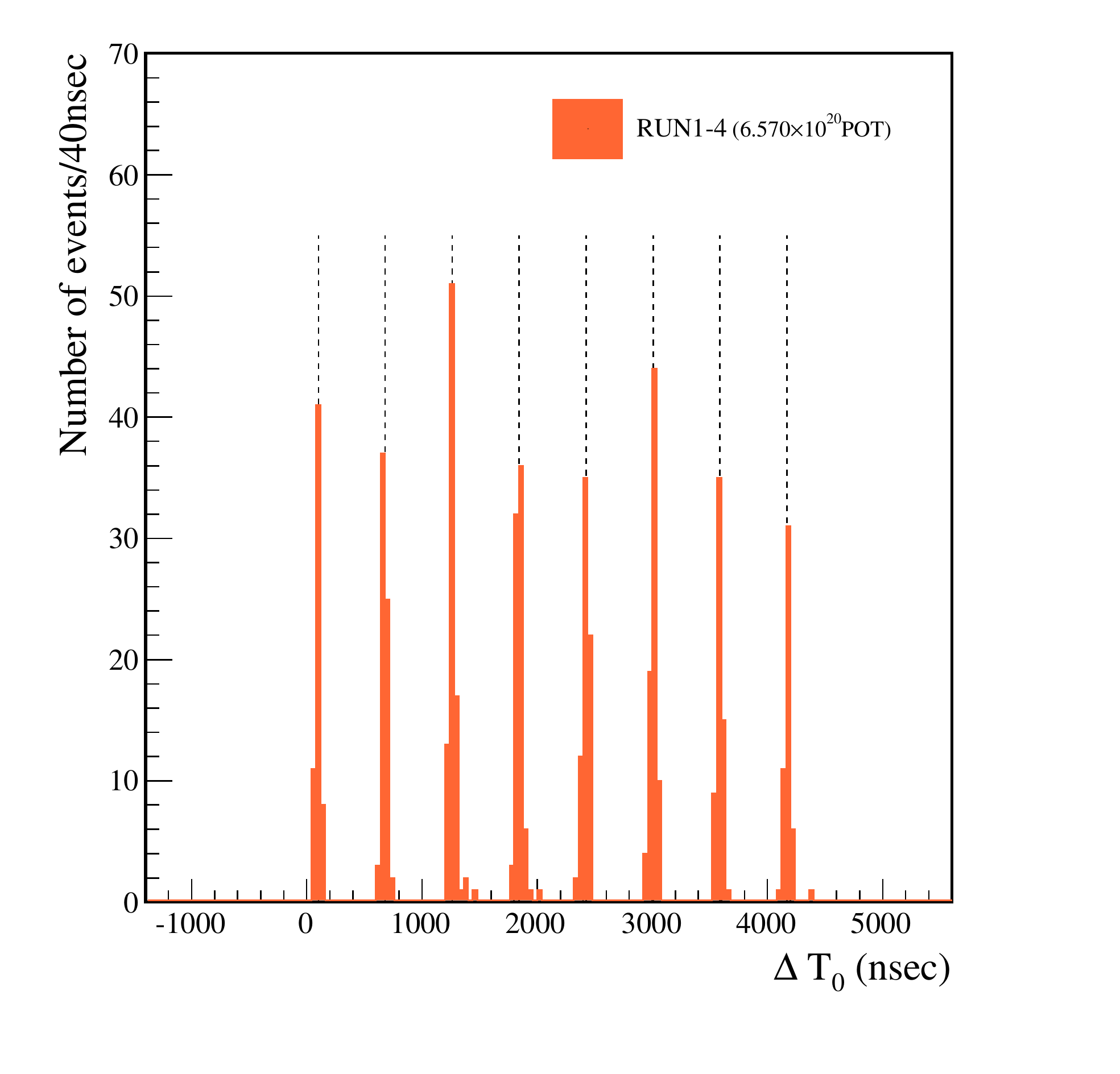}
\caption{$\Delta T_0$ distribution of all FC events within the beam spill window.}
\label{fig:SK_dT0_ontiming}
\end{center}
\end{figure}

Events are timed with respect to the leading edge of the beam spill, taking into account 
the time of flight of the neutrino and myriad other sources of delay~\cite{Abe:2013gga,Abe:2015gna}.
Figure~\ref{fig:SK_dT0_offtiming} shows the event timing ($\Delta T_0$) distribution for all 
ID, OD, and LE events within $\pm500\mus$ of the beam arrival time.  There is a 
clear peak near $\Delta T_0 = 0$ for the FC sample. Eleven FC events have been observed 
outside the spill window.  Using data collected with no beam we estimate the expected 
number of these events to be 5.85, mainly low energy events.  $\Delta T_0$ 
is corrected to take into account the neutrino interaction vertex position and the 
photon time-of-flight from the vertex to the PMTs.  FC events within the spill window 
can be seen in Fig.~\ref{fig:SK_dT0_ontiming} where the beam structure with eight bunches 
is clearly visible.  The dotted lines represent the fitted bunch center times with 
a fixed bunch interval of 581\ns.
For an event to be incorporated into the analysis, $\Delta T_0$ must lie between $-2$ to 10\mus.

A fiducial volume is defined within the ID, 2\,m away from the detector wall, with a fiducial mass of 22.5\,kton. 
Events whose vertex is reconstructed within this volume and with 
visible energy ($E_{\mathrm{vis}}$) greater than 30\mev are selected into the fully contained fiducial 
volume sample (FCFV).  Visible energy is defined as the energy of an electromagnetic shower that
produces the observed amount of Cherenkov light. We observe 377 events classified as FCFV.  
The expected number of background events from non-beam related sources in accidental coincidence is estimated to be 0.0085.

Charged current interactions $(\nu + N \rightarrow l^- + X)$ in the narrow energy range of the T2K 
beam tend to produce single ring events at SK because most of the particles produced, except for the 
primary lepton, do not escape the nucleus or are below detection threshold.  The energy 
of the incoming neutrino can be calculated assuming the kinematics of a CCQE interaction 
and neglecting Fermi motion:
\begin{equation}
\label{eq:SK_Erec}
E^{\mathrm{rec}}_{\nu} = \frac{m^2_p - (m_n - E_b)^2 - m^2_l + 2(m_n - E_b)E_l}{2(m_n - E_b - E_l + p_l\cos\theta_l)}
\end{equation}
where $E^{\mathrm{rec}}_{\nu}$ is the reconstructed neutrino energy, $m_p$ is the proton mass, 
$m_n$ the neutron mass, $m_l$ the lepton mass 
and $E_b=27$\mev is the binding energy of a nucleon inside $^{16}\mathrm{O}$ nuclei. 
$E_l$, $p_l$ and $\theta_l$ are the reconstructed lepton energy, momentum, and 
angle with respect to the beam, respectively.  
The selection criteria for both \nue CC and \num CC events
were fixed using MC studies before being applied to data.  Events are determined 
to be $e$-like or $\mu$-like based on the PID of the brightest Cherenkov ring.  The PID of each ring 
is determined by a likelihood incorporating information on the charge distribution and the opening 
angle of the Cherenkov cone.

We select \nue CC candidate events using the criteria listed in Tab.~\ref{tab:SK_nue_events}.
The $E_{\mathrm{vis}}$ requirement removes low energy NC interactions and electrons from the 
decay of unseen parents that are below Cherenkov threshold or fall outside the 
beam time window.  The $\pi^0$-like event rejection uses an independent 
reconstruction algorithm which is described in Sec.~\ref{sec:SK_fiTQun}.
We require 
$E^{\mathrm{rec}}_{\nu} < 1.25\gev$ since above this energy the intrinsic beam 
\nue background is dominant.
The numbers of events remaining after successive selection criteria for a simulation sample 
produced with a nominal set of oscillation parameter values are shown in Tab.~\ref{tab:SK_nue_events}.
After all cuts 28 events remain in the \nue CC candidate sample.  A Kolmogorov-Smirnov 
(KS) test of the accumulated events with accumulated POT 
is compatible with a constant rate with a p-value of 0.7.

\begin{table*}[t]
\begin{center}
\caption{
Event reduction for the \nue CC selection at the far detector.  The numbers of 
expected MC events divided into four categories are shown after each 
selection criterion is applied.  The MC expectation is based upon three-neutrino 
oscillations for $\sin^{2}2\theta_{23}=1.0$, $\Delta m^2_{32}=2.4\times10^{-3}\evvcccc$,
$\sin^{2}2\theta_{13}=0.1$, $\delta_{CP}=0$ and normal mass hierarchy
(parameters chosen without reference to the T2K data).
}
\begin{tabular}{ll}
(1) & There is only one reconstructed Cherenkov ring \\
(2) & The ring is $e$-like \\
(3) & The visible energy, $E_{\mathrm{vis}}$, is greater than 100 \mev \\
(4) & There is no reconstructed Michel electron \\
(5) & The reconstructed energy, $E_{\nu}^{\mathrm{rec}}$, is less than 1.25 \gev \\
(6) & The event is not consistent with a $\pi^0$ hypothesis \\
\end{tabular}
\vskip 6mm
\begin{tabular}{lccccc}
\hline
\hline
&                 & \ \ $\num+\numb$ \ \ & \ \ $\nue+\nueb$ \ \ & \ \ $\nu+\bar{\nu}$ \ \ & \ \ $\num\rightarrow\nue$ \ \ \\
& \ \ MC total\ \ & CC & CC & NC & CC \\
\hline
interactions in FV                 & 656.83 & 325.67 & 15.97 & 288.11 & 27.07 \\
FCFV              & 372.35 & 247.75 & 15.36 & 83.02  & 26.22 \\
(1) single ring                    & 198.44 & 142.44 & 9.82  & 23.46  & 22.72 \\
(2) electron-like                  & 54.17  & 5.63   & 9.74   & 16.35 & 22.45 \\
(3) $E_{\rm vis}>100{\rm \mev}$      & 49.36  & 3.66   & 9.68   & 13.99 & 22.04 \\
(4) no Michel election            & 40.03  & 0.69   & 7.87   & 11.84 & 19.63 \\
(5) $E_{\nu}^{\rm rec}<1250{\rm \mev}$& 31.76  & 0.21   & 3.73   & 8.99 & 18.82 \\
(6) not $\pi^{0}$-like             & 21.59  & 0.07   & 3.24   & 0.96  & 17.32  \\
\hline
\hline
\end{tabular}
\label{tab:SK_nue_events}
\end{center}
\end{table*}

We select \num CC candidate events using the selection criteria shown in Tab.~\ref{tab:SK_numu_events}.
The momentum cut rejects charged pions and misidentified electrons from the decay of unobserved muons 
and pions.  We require fewer than two Michel electrons to reject events with additional unseen muons or pions.
After all cuts are applied, 120 events remain in the \num CC candidate sample.

\begin{table*}[t]
\begin{center}
\caption{
Event reduction for the \num CC selection at the far detector.  The numbers of 
expected MC events divided into four categories are shown after each 
selection criterion is applied.  The MC expectation is based upon three-neutrino 
oscillations for $\sin^{2}2\theta_{23}=1.0$, $\Delta m^2_{32}=2.4\times10^{-3}\evvcccc$ 
and normal mass hierarchy (parameters chosen without reference to the T2K data).
}
\begin{tabular}{ll}
(1) & There is only one reconstructed Cherenkov ring \\
(2) & The ring is $\mu$-like \\
(3) & The reconstructed momentum, $p_{\mu}$, is greater than 200 \mevc \\
(4) & There are less than two reconstructed Michel electrons \\
\end{tabular}
\vskip 5mm
\begin{tabular}{lccccc}
\hline
\hline
 &             & \ \ $\num+\numb$ \ \ & $\num+\numb$ & \ \ $\nue+\nueb$ \ \ & \ \ $\nu+\bar{\nu}$ \ \ \\
 & MC total\ \ & CCQE  & \ \ CC nonQE\ \  & CC & NC \\
\hline
interactions in FV          & 656.83 & 111.71 & 213.96 & 43.05 & 288.11 \\
FCFV                        & 372.35 & 85.55  & 162.20 & 41.58 & 83.02 \\
(1) single ring             & 198.44 & 80.57  & 61.87  & 32.54 & 23.46 \\
(2) muon-like               & 144.28 & 79.01  & 57.80  & 0.35  & 7.11 \\
(3) $p_{\mu}>200 \mevc$      & 143.99 & 78.84  & 57.77  & 0.35  & 7.04 \\
(4) $N_{\rm Michel-e} \leq 1$  & 125.85 & 77.93  & 40.78  & 0.35  & 6.78 \\
\hline
\hline
\end{tabular}
\label{tab:SK_numu_events}
\end{center}
\end{table*}

\begin{figure}[tbp]
\begin{center}
\subfloat[]{
\includegraphics[width=0.4\textwidth]{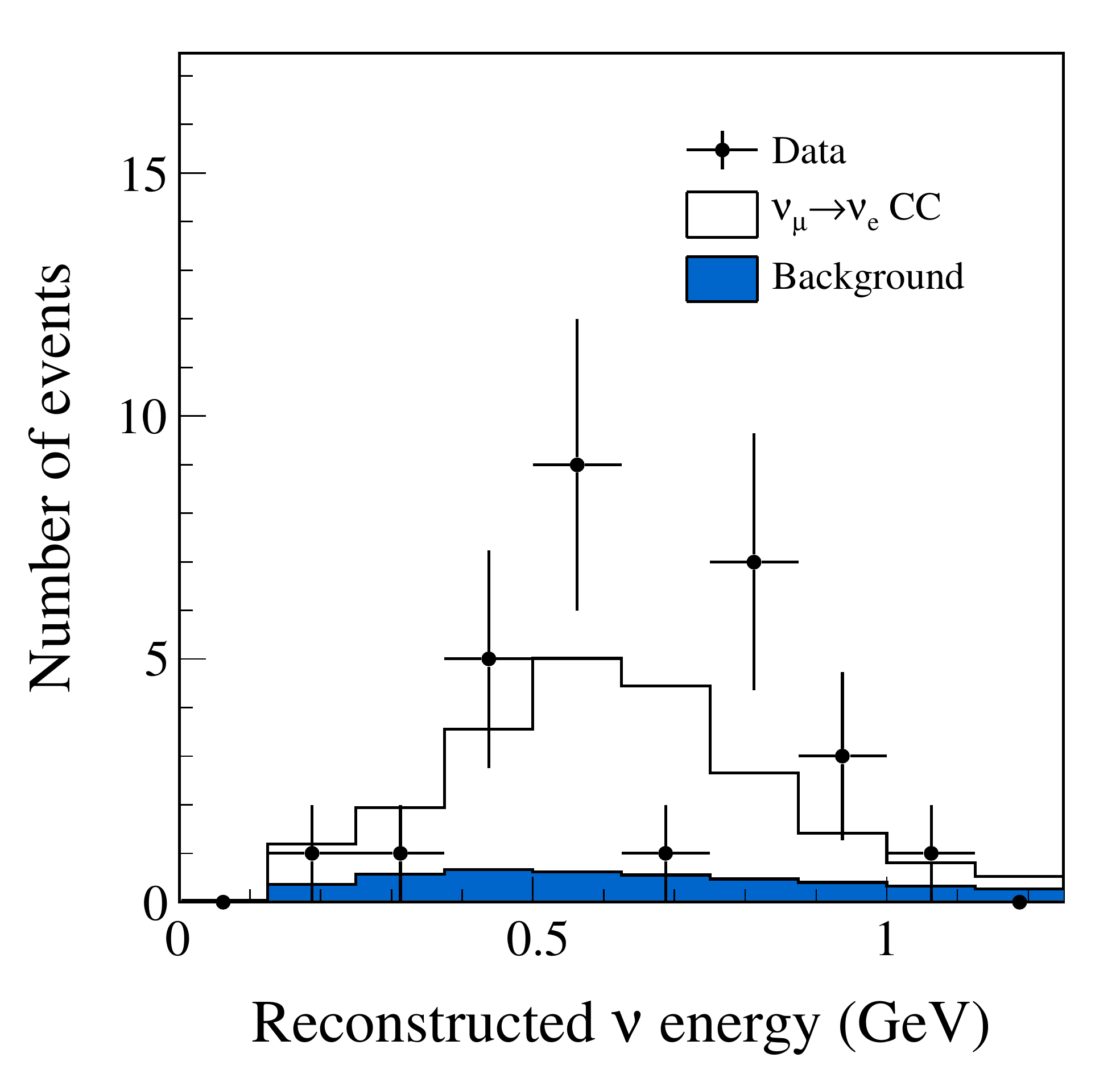}
}
\subfloat[]{
\includegraphics[width=0.4\textwidth]{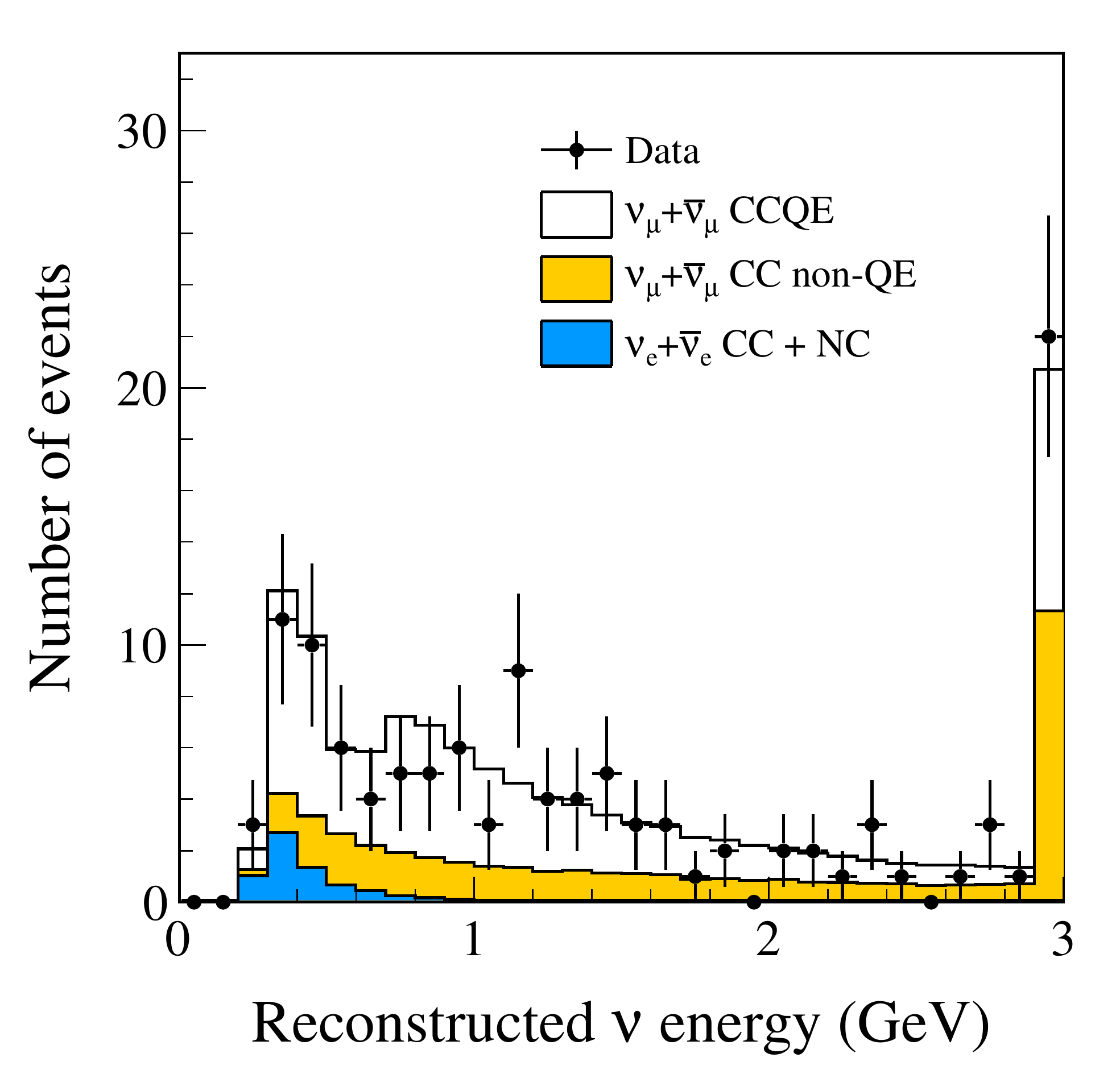}
}
\caption{The reconstructed energy spectra 
of the observed \nue (a) and \num (b) CC candidate events assuming CCQE interaction kinematics.  
The data are shown as points with statistical error bars and the shaded, stacked 
histograms are the MC predictions, and the rightmost bin includes overflow.
The expectation is based on the following oscillation parameters: 
$\sin^{2}2\theta_{13}=0.1$, $\sin^{2}2\theta_{23}=1.0$, $\delta_{CP}=0$,
$\Delta m^2_{23}=2.4\times10^{-3}\evvcccc$ and normal mass hierarchy.}
\label{fig:SK_enurec_plots}
\end{center}
\end{figure}

\begin{figure}[tbp]
\begin{center}
\subfloat[]{
\includegraphics[width=0.4\textwidth]{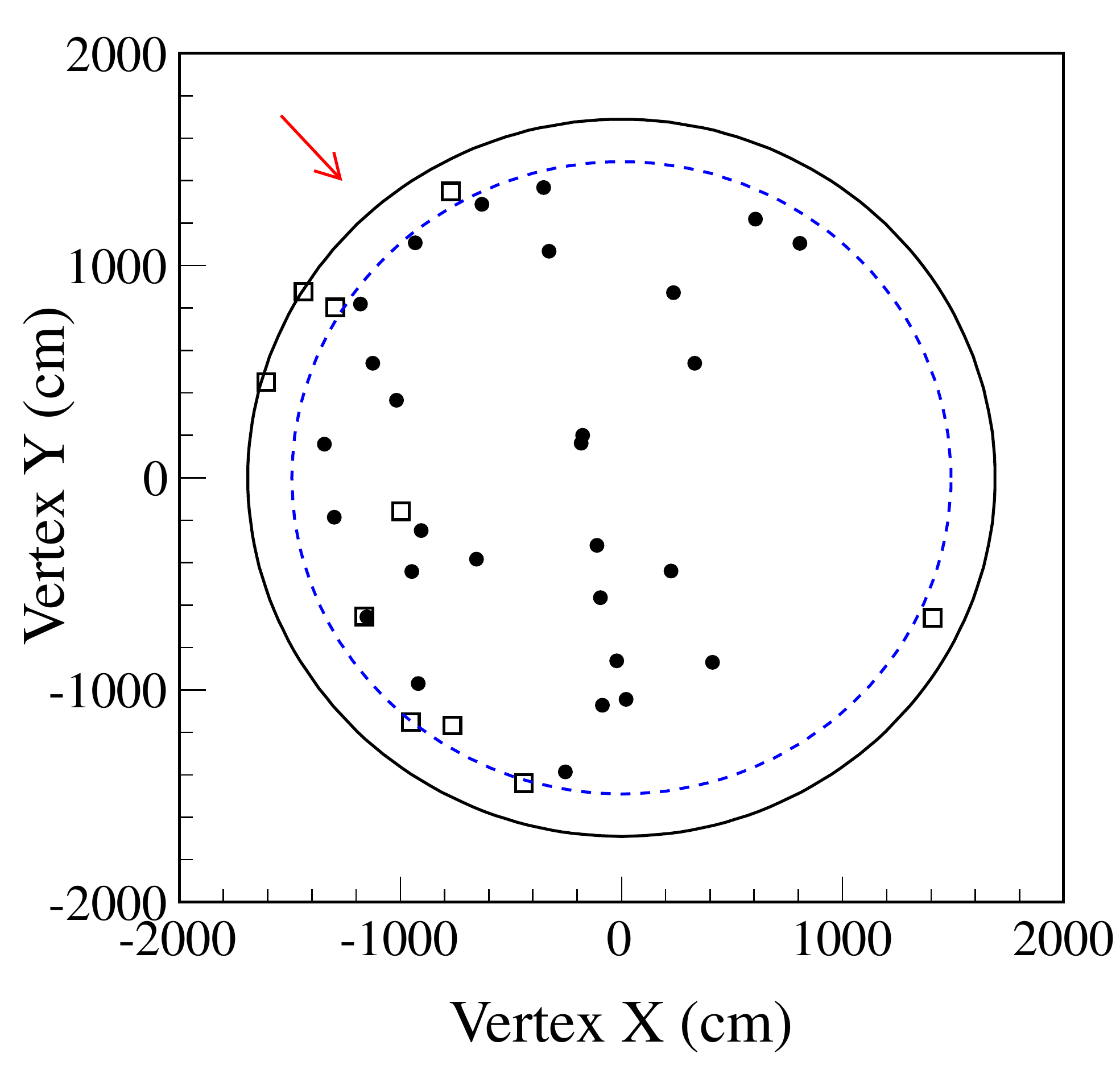}
}
\subfloat[]{
\includegraphics[width=0.4\textwidth]{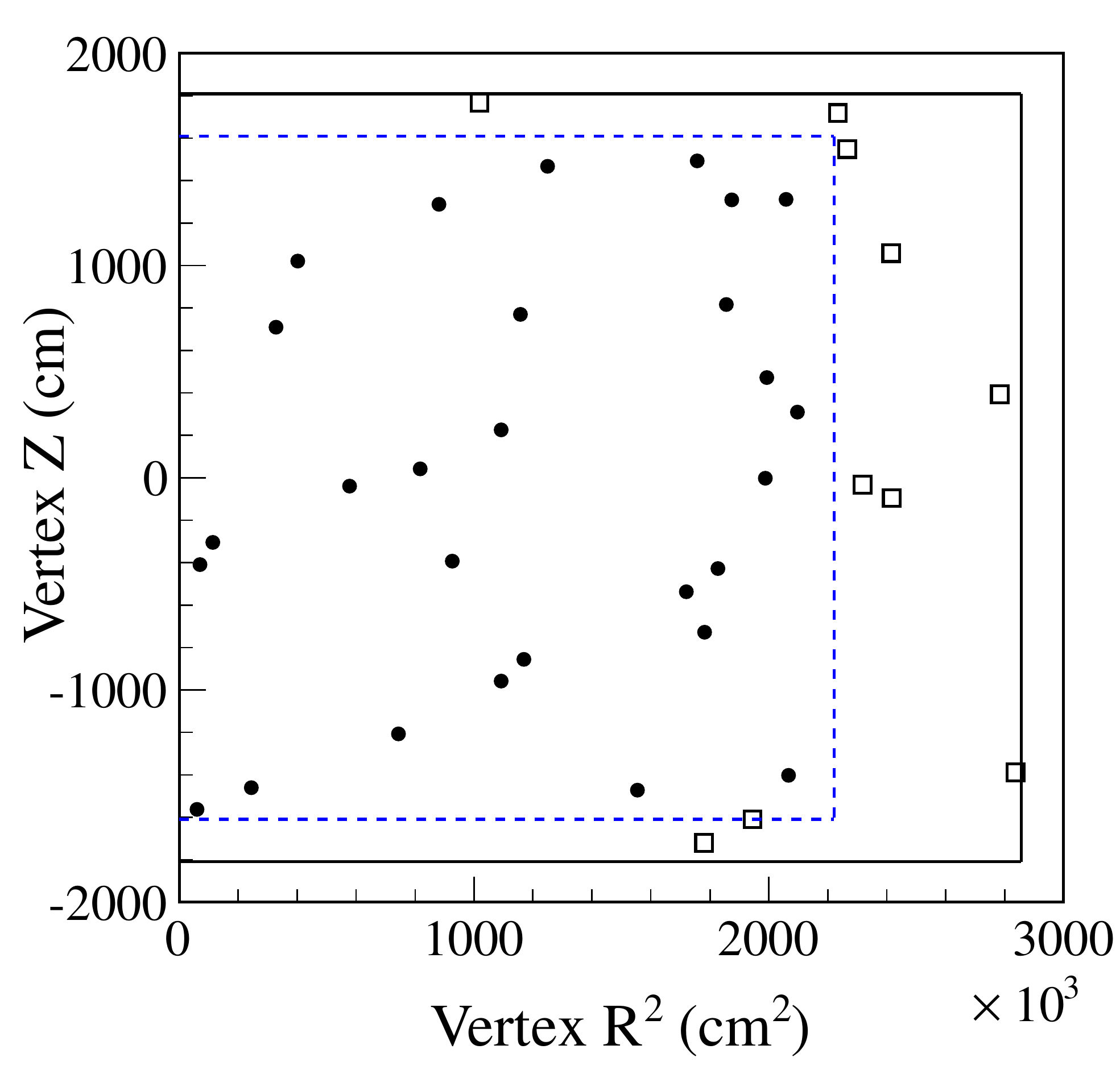}
}
\caption{Two-dimensional vertex distributions of the observed \nue CC candidate events in 
$(x,y)$ and $(r^2 = x^2 + y^2,z)$.  The arrow indicates the neutrino beam direction and 
the dashed (blue online) line indicates the fiducial volume boundary.  
Events indicated 
by open square markers passed all of the \nue selection cuts except for the fiducial 
volume cut.}
\label{fig:SK_nue_vtx}
\end{center}
\end{figure}

Figure \ref{fig:SK_enurec_plots} shows the candidate event spectra for the appearance (\nue) and 
disappearance (\num) channels.
We monitor the vertex distributions of the candidate event samples for 
signs of bias that might suggest background contamination.  Figure~\ref{fig:SK_nue_vtx} 
shows the vertex distribution of the \nue CC candidate events 
in the SK tank coordinate system.  We observe no unexpected clustering and 
combined KS tests for uniformity in $r^2$ and $z$ yields a p-value of 0.6.

\subsection{$\pi^0$ Rejection with the New Event Reconstruction Algorithm}
\label{sec:SK_fiTQun}
As mentioned in the previous section, in order to select $\nu_{e}$ CC events, we require that only one electron-like ring is reconstructed. The $\nu_{e}$ CC selection criteria 1-5 in Tab.~\ref{tab:SK_nue_events} are based on the information provided by SK event reconstruction software which has been used at SK for atmospheric neutrino and nucleon decay analyses\cite{Ashie:2005ik} and, as shown in the 
table, we reject most of the background events by these selection cuts. The $\nu_e$ appearance signal purity is 59.3\% and the selection efficiency for the signal is 71.8\%.
The remaining backgrounds are predominantly NC single $\pi^{0}$ events, as one of the two decay $\gamma$s from a $\pi^{0}$ is occasionally missed and the other $\gamma$ forms an electron-like ring.

In order to reject such $\pi^{0}$ events, we employ a new event reconstruction algorithm which is based on the methods developed by MiniBooNE~\cite{MBRecon}. The new algorithm adopts a maximum likelihood method to reconstruct particle kinematics in the SK detector. For a given event, we construct a likelihood function which uses the observed charge and time information from the PMTs:
\begin{equation}
\begin{split}
\mathcal{L}(\bm{x})\equiv&\prod_{j}^{\rm unhit}P_{j}({\rm unhit}|\bm{x})\\
&\times\prod_{i}^{\rm hit}\{1-P_{i}({\rm unhit}|\bm{x})\}f_{q}(q_{i}|\bm{x})f_{t}(t_{i}|\bm{x}).
\label{eq:likelihood_orig}
\end{split}
\end{equation}
In the equation, $\bm{x}$ represents particle track parameters such as the vertex, direction, and momentum which are to be estimated. The first index $j$ runs over the PMTs which do not register a hit, and for each of such PMTs the conditional probability $P_{j}({\rm unhit}|\bm{x})$ of not registering a hit given $\bm{x}$ is evaluated. For each PMT which does register a hit, in addition to the hit probability, we calculate the probability density $f_{q}(q_{i}|\bm{x})$ of observing charge $q_{i}$ as well as the probability density $f_{t}(t_{i}|\bm{x})$ of the hit occurring at time $t_{i}$. The estimated track parameters, $\bm{x}$, are those that maximize the likelihood function. For every event we construct and maximize the likelihood assuming several different particle hypotheses, and particle identification is done using ratios of the maximum likelihoods for the different hypotheses.

In this analysis, we use a single electron hypothesis and a $\pi^{0}$ hypothesis for $\pi^{0}$ rejection. The single electron hypothesis has seven parameters which are the initial vertex position, time, direction, and momentum. Since a $\pi^{0}$ decays into two $\gamma$s and produces two electron-like Cherenkov rings, the $\pi^{0}$ hypothesis is constructed by combining the charge and time contributions from two electron tracks which point back to a common vertex. 
In addition to the common vertex and the directions and momenta of the two $\gamma$ tracks, each track has an additional free parameter which shifts its origin along its direction in order to account for photon conversion points. The $\pi^{0}$ hypothesis therefore has twelve parameters. 

\begin{figure}[tbp]
\begin{center}
\includegraphics[width=0.49\textwidth]{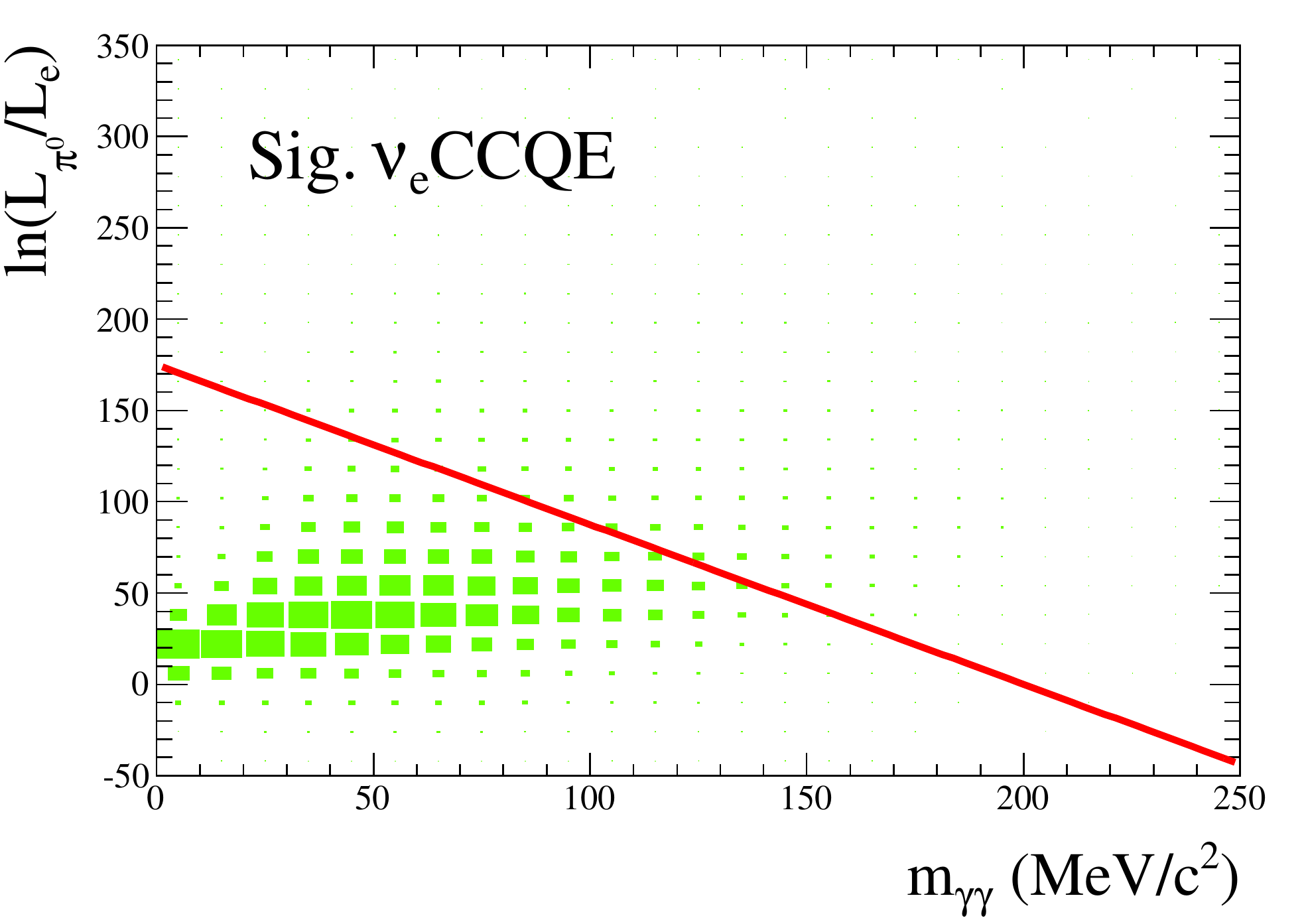}
\includegraphics[width=0.49\textwidth]{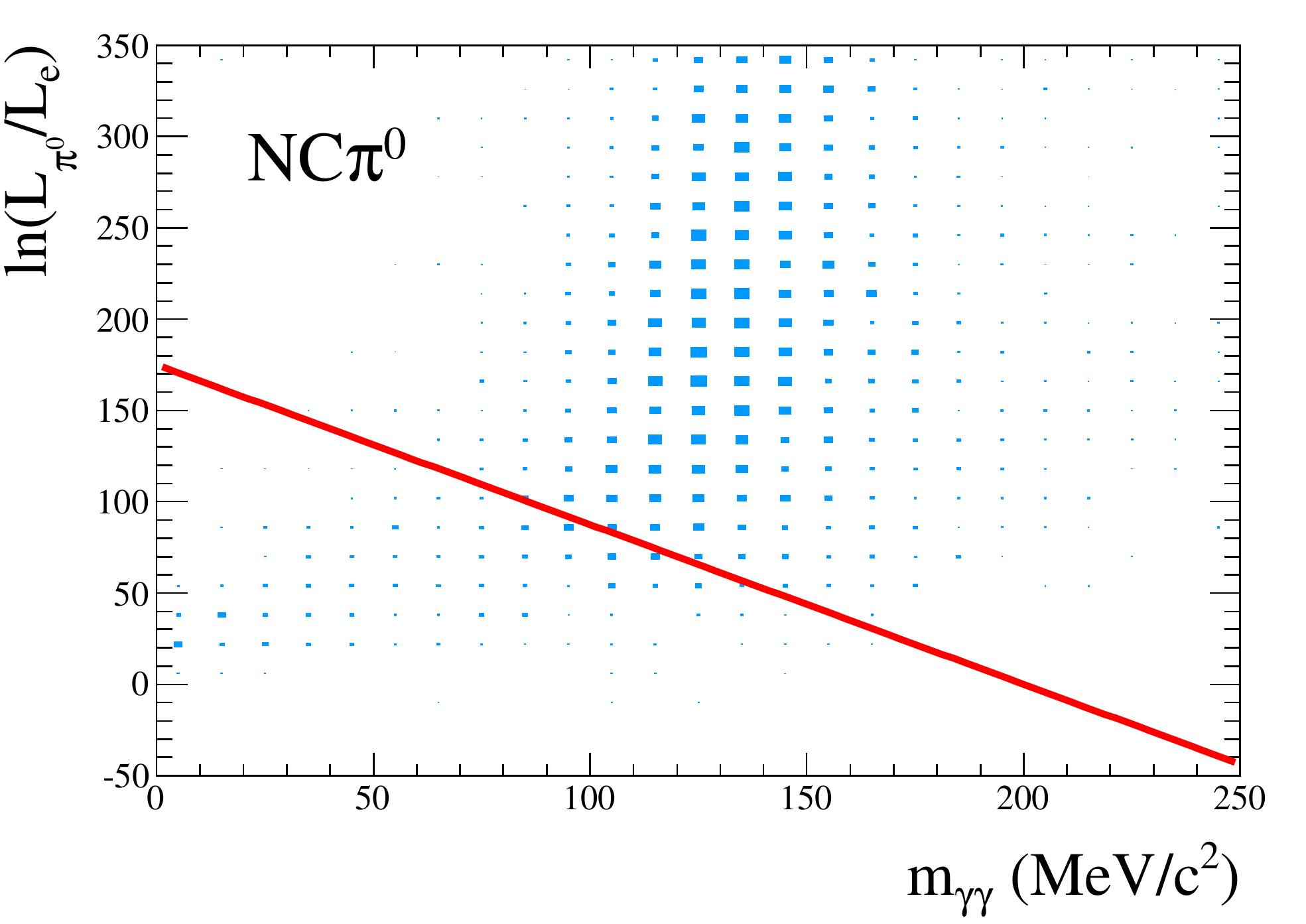}
\caption{2D distributions of the logarithm of the likelihood ratio $\ln(L_{\pi^{0}}/L_{e})$ vs. the reconstructed invariant mass $m_{\gamma\gamma}$, for signal $\nu_{e}$ CCQE(left) and background NC $\pi^{0}$(right) events. The diagonal line indicates the $\pi^{0}$ rejection criterion, and events lying above the line are rejected as $\pi^{0}$ background. The size of each box is proportional to the number of events the bin. The two figures use the same scale for representing the number of events and are normalized to the same POT.}
\label{fig:SK_pi02dcut}
\end{center}
\end{figure}

In order to distinguish signal $\nu_{e}$ CC events from $\pi^{0}$ background events, we use the maximum likelihood values of the electron hypothesis $L_{e}$ and the $\pi^{0}$ hypothesis $L_{\pi^{0}}$ as well as the reconstructed invariant mass $m_{\gamma\gamma}$ obtained from the $\pi^{0}$ hypothesis. 
Figure~\ref{fig:SK_pi02dcut} shows the 2D distributions of the logarithm of the likelihood ratio $\ln(L_{\pi^{0}}/L_{e})$ vs. $m_{\gamma\gamma}$ for signal $\nu_{e}$ CCQE and background NC $\pi^{0}$ events which satisfy the $\nu_{e}$ selection criteria 1-5, produced by MC. We see a clear separation between the two event types, and we accept an event as a $\nu_{e}$ CC candidate if it satisfies $\ln(L_{\pi^{0}}/L_{e})<175-0.875\times m_{\gamma\gamma}[\mevcc]$, which is indicated by the diagonal line in the plots. As shown in Tab.~\ref{tab:SK_nue_events}, the remaining NC background is reduced by roughly a factor of nine by introducing the $\pi^{0}$ rejection cut.
After the cut, the purity and the selection efficiency for the $\nu_e$ appearance signal are 80.2\% and 66.1\% respectively.

\begin{figure}[tbp]
\begin{center}
\includegraphics[width=0.6\textwidth]{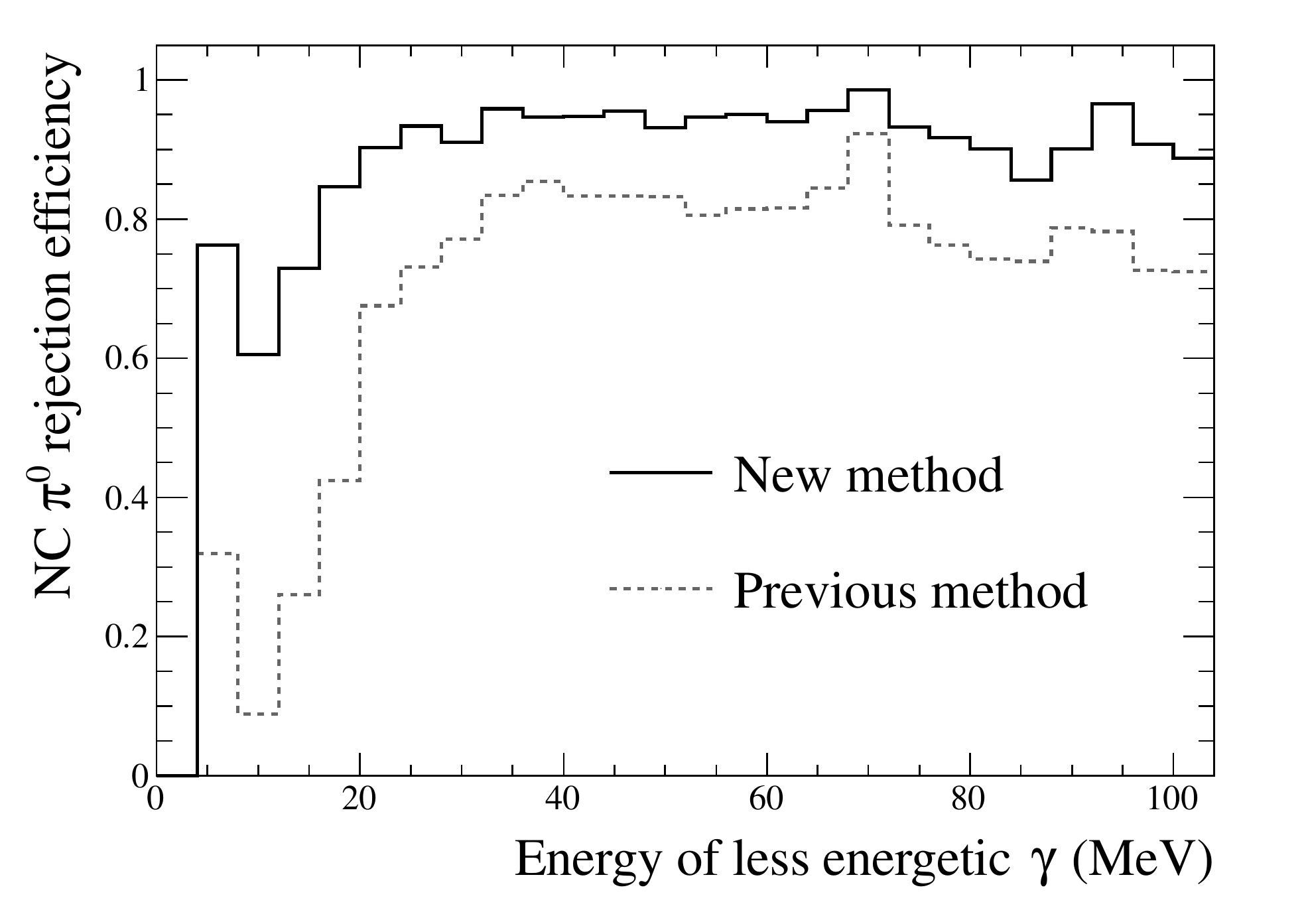}
\caption{Efficiencies for rejecting NC $\pi^{0}$ events for the previous and the new $\pi^{0}$ rejection methods, plotted in bins of the energy of the less energetic $\gamma$.}
\label{fig:SK_pi0eff_POLfQ}
\end{center}
\end{figure}

In earlier published T2K $\nu_{e}$ appearance analysis results~\cite{PhysRevLett.107.041801,Abe:2013xua}, we used a $\pi^{0}$ rejection method which is different from what is described above~\cite{Barszczak:2005sf}. To demonstrate the improvement over the previous method, Fig.~\ref{fig:SK_pi0eff_POLfQ} shows the efficiency for rejecting NC $\pi^{0}$ events for the two methods, plotted as a function of the energy of the less energetic $\gamma$. In calculating the efficiencies, only the events which satisfy the $\nu_{e}$ selection criteria 1-5 are included. As the figure indicates, the rejection efficiency by the new method remains high even in cases where the energy of one of the two $\gamma$s is low. By employing the new method, we have reduced the $\pi^{0}$ background remaining in the final $\nu_{e}$ CC candidate event sample by 69$\%$ relative to the previous method.

\subsection{Systematic uncertainty}
\label{sec:SK_detector_errors}
This section describes the studies and treatment of uncertainty in modeling
the SK detector that lead to systematic uncertainty in estimating the 
selection efficiency and background for the oscillation samples.
We use SKDETSIM~\cite{Abe:2013xua,Abe:2011ks}, 
a GEANT3-derived simulation of the SK detector, 
to model the propagation of particles produced by neutrino interactions.  The GCALOR
physics package is used to simulate hadronic interactions in water owing to its 
ability to reproduce pion interaction data around 1\gevc.  However for pions with 
momentum below 500\mevc, custom routines are employed based on the cascade model used by NEUT 
to simulate interactions of final state hadrons.  SKDETSIM incorporates
the propagation of photons in water, subject to absorption, Rayleigh scattering, 
and Mie scattering.  
The simulation of these processes is tuned using laser calibration sources in situ~\cite{Fukuda:2002uc}.

Control samples that are not related to the T2K beam spills
are used to assess systematic uncertainty, including
muons and neutrinos produced from cosmic ray interactions with the atmosphere (cosmic ray muons and atmospheric
neutrinos) and combinations of simulated and cosmic ray data (hybrid-$\pi^0$ sample).
As described below, cosmic ray muons are used to evaluate the systematic
uncertainty due to the fully-contained (FC), fiducial-volume, and decay-electron requirements.
Atmospheric neutrinos are used to assess uncertainty from the
ring counting, particle identification, and $\pi^0$ rejection.
The hybrid-$\pi^0$ sample is used to study the SK response to $\pi^0$'s.
The uncertainties due to energy scale, modeling of pion final state interactions (FSI) and
secondary interactions (SI) are evaluated separately.

Cosmic ray muon samples are used to estimate uncertainties related to the
FC, fiducial-volume and decay-electron requirements, for the selections of both 
\nue and \num CC candidates.
The error from the initial FC event selection is 1\% and is dominated by 
the event-by-event flasher rejection cut.
The uncertainty in the fiducial volume
is estimated to be 1\% using the vertex distribution of cosmic ray muons which 
have been independently determined to have stopped inside the ID.
The uncertainty due to the Michel electron tagging efficiency is estimated by comparing cosmic ray 
stopped-muon data and MC.  This uncertainty is applied based on the fraction of events with true 
Michel electrons in the T2K beam MC.  The rate of falsely identified
Michel electrons is estimated from MC and 100\% uncertainty in that rate is assumed. 
Overall, the event rate uncertainty related to the decay-electron requirements is small.
For the \nue CC candidate sample, it
is 0.2\% for $\nu_e$ CC events and 0.4\% for $\nu_\mu$ CC and NC events.  For the \num 
CC candidate sample it is 1.0\%.  

Other studies of systematic uncertainty in SK modeling divide simulated events into
categories according to their final state (FS) topologies, with the criteria shown in Tab.~\ref{tab:FSmodes}.
These topologies do not correspond exactly with true interaction modes due to
subsequent interactions 
within the nucleus or with neighboring nuclei or because one or more particles 
are produced below Cherenkov threshold.

\begin{table}[tbp]
 \centering
 \caption{Criteria for categorization of simulated events by final state topology for systematic studies.
 $N_x$ is the number of particles of type $x$ and  
 the number of charged pions ($N_{\pi^{\pm}}$) and protons ($N_{P}$)
 only includes those particles produced with momentum above Cherenkov threshold set at 156.0\,\mevc
 and 1051.0\,\mevc respectively.}
 \begin{tabular}{ll}
   \hline
   \hline
   Event type & MC truth selection criteria \\
   \hline
   CC 1$e$  & $\nu_{e}$ CC and $N_{\pi^{0}}=0$ and $N_{\pi^{\pm}}=0$ and $N_{P}=0$ \\
   CC $e$ other            & $\nu_{e}$ CC and not $\nu_{e}$ CC1$e$ \\
   CC 1$\mu$    & $\nu_{\mu}$ CC and $N_{\pi^{0}}=0$  and $N_{\pi^{\pm}}=0$ and $N_{P}=0$\\
   CC $\mu$ other                & $\nu_{\mu}$ CC and $N_{\pi^{0}}=0$ \\
   CC $\mu$ $\pi^{0}$ other\ \ \ \ & $\nu_{\mu}$ CC and $N_{\pi^{0}}>0$ \\
   NC 1$\pi^{0}$                 & NC and not NC 1$\gamma$ and $N_{\pi^{0}}=1$ and $N_{\pi^{\pm}}=0$ and $N_{P}=0$\\
   NC $\pi^{0}$ other            & NC and not NC 1$\gamma$ and $N_{\pi^{0}}\geq1$ and not NC 1$\pi^{0}$\\
   NC 1$\gamma$                 & NEUT truth\\
   NC 1$\pi^{\pm}$               & NC and not NC 1$\gamma$ and $N_{\pi^{0}}=0$ and $N_{\pi^{\pm}}=1$ and $N_{P}=0$\\
   NC other                     & NC and not NC 1$\gamma$ and not NC 1$\pi^{0}$ and not NC 1$\pi^{\pm}$ 
                                and not NC $\pi^{0}$ other \\
   \hline
   \hline
 \end{tabular}
 \label{tab:FSmodes}
\end{table}

Atmospheric neutrino data are used to assess
possible mismodeling of the ring counting (RC), particle identification, and $\pi^0$ rejection for the
first four FS topologies shown in Tab.~\ref{tab:FSmodes}.
Atmospheric neutrino samples fully 
contained within the fiducial volume and with $E_{\mathrm{vis}}>30$ \mev are divided into 
CCQE and CC non-quasi-elastic (CCnQE) enriched samples using
the number of Michel electrons and the visible energy. These samples 
are further split into ``core'' samples of events which pass all of the 
requirements and tail samples of events which fail only one requirement.  An 
additional background sample is included, enhanced in NC $\pi^{0}$.  These samples, 
13 in total, are summarized in Tab.~\ref{tab:ControlSample}
and are binned in $E_{\mathrm{vis}}$, for $E_{\mathrm{vis}}<30$\gev.  

\begin{table}[htbp]
 \caption{
    SK atmospheric neutrino control samples. The parent sample is
    defined to be fully contained and in the fiducial volume.
    This parent sample is divided into four sets of core and tail
    samples and one background (BG) control sample. The
    main difference between CCQE and CCnQE is the
    number of decay-$e$ $N_{dcy-e}$ cut, which is based on the hit time
    distribution. The distance from the
    expected muon stopping point to the nearest decay-$e$,
    $D_{dcy-e}$, is used to select high
    purity \num CCQE and CCnQE samples.
    The BG sample is enriched in NC $\pi^{0}$ to constrain the NC
    normalization.
}
\label{tab:ControlSample}
 \begin{center}
  \begin{tabular}{ccllp{1.5cm}p{2cm}l}
    \hline
    \hline
       \multicolumn{2}{c}{\multirow{2}{4.4cm}{Type of control sample}} &\multicolumn{5}{c}{Branch of control sample} \\ \cline{3-7}
\multicolumn{2}{c}{} & \multicolumn{2}{c}{ sample     } & RC cut & PID cut & $\pi^{0}$ / $D_{dcy-e}$ cut \\ 
    \hline
       \multirow{4}{2.2cm}{\nue CCQE enriched} & \multirow{4}{2.2cm}{$N_{dcy-e} = 0$ \& $E_{vis} > 100$} \ \  & \multicolumn{2}{l}{Core} & 1R &\& $e$-like & \& not $\pi^{0}$-like \\ \cline{3-7}
        & & & RC   tail \ \ & $>1$R  & \& $e$-like   &\& not $\pi^{0}$-like  \\ 
        & & & PID  tail & 1R & \& $\mu$-like &\& not $\pi^{0}$-like  \\ 
        & & & $\pi^{0}$ tail & 1R & \& $e$-like   &\& $\pi^{0}$-like  \\
    \hline
        \multirow{4}{2.2cm}{\nue CCnQE enriched} & \multirow{4}{2.2cm}{$N_{dcy-e} \ge 1$ \& $E_{vis} > 100$}   & \multicolumn{2}{l}{Core} & 1R &\& $e$-like & \& not $\pi^{0}$-like  \\ \cline{3-7}
        & & & RC   tail & $>1$R  & \& $e$-like   & \& not $\pi^{0}$-like  \\ 
        & & & PID  tail & 1R & \& $\mu$-like & \& not $\pi^{0}$-like  \\ 
        & & & $\pi^{0}$ tail & 1R & \& $e$-like   & \& $\pi^{0}$-like  \\ 
    \hline
      \multirow{2}{2.2cm}{\num CCQE enriched} & \multirow{2}{2.2cm}{$N_{dcy-e} = 1$} & \multicolumn{2}{l}{Core} & 1R & \& $\mu$-like & \& $D_{dcy-e} < 80$~cm \\ \cline{3-7}
      & & & RC   tail & $>1$R  & \& $\mu$-like   &\& $D_{dcy-e} < 80$~cm  \\ \cline{1-7}

       \multirow{2}{2.2cm}{\num CCnQE enriched} & \multirow{2}{2.2cm}{$N_{dcy-e}\ge 2$} & \multicolumn{2}{l}{Core} & 1R & \& $\mu$-like & \& $D_{dcy-e} < 160$~cm  \\ \cline{3-7}
       &  & & RC   tail & $>1$R  & \& $e$-like   &\& $D_{dcy-e} < 160$~cm  \\ \cline{1-7}
 \hline
          BG enriched & $N_{dcy-e} = 0$  & \multicolumn{2}{c}{ NC $\pi^{0}$ } & $>1$R & \& $e$-like & \& $\pi^{0}$-like \\ 
     \hline
     \hline
  \end{tabular}
 \end{center}
\end{table}

\begin{figure}[tbp]
\begin{center}
\includegraphics[width=0.6\textwidth]{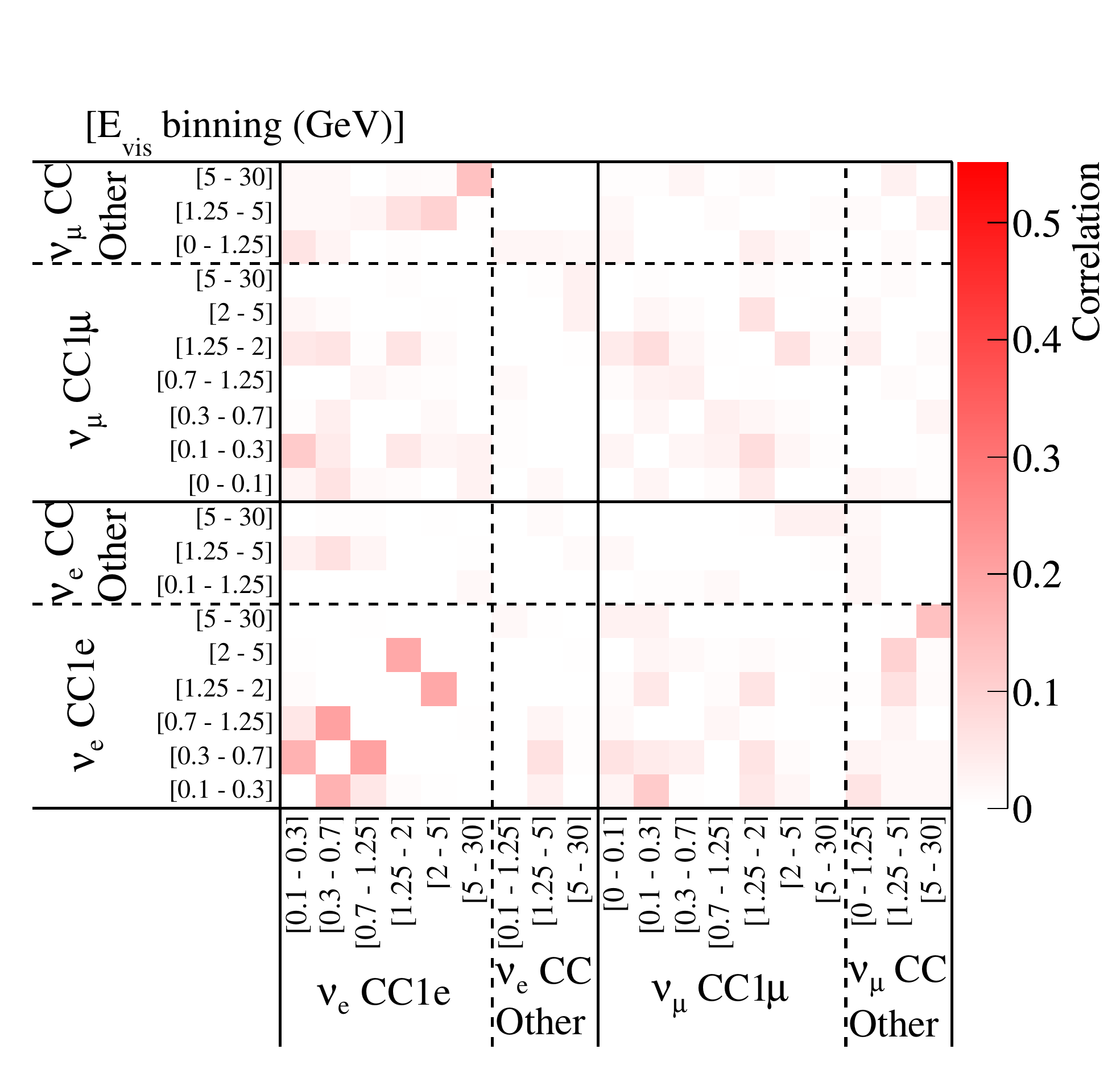} 
\includegraphics[width=0.6\textwidth]{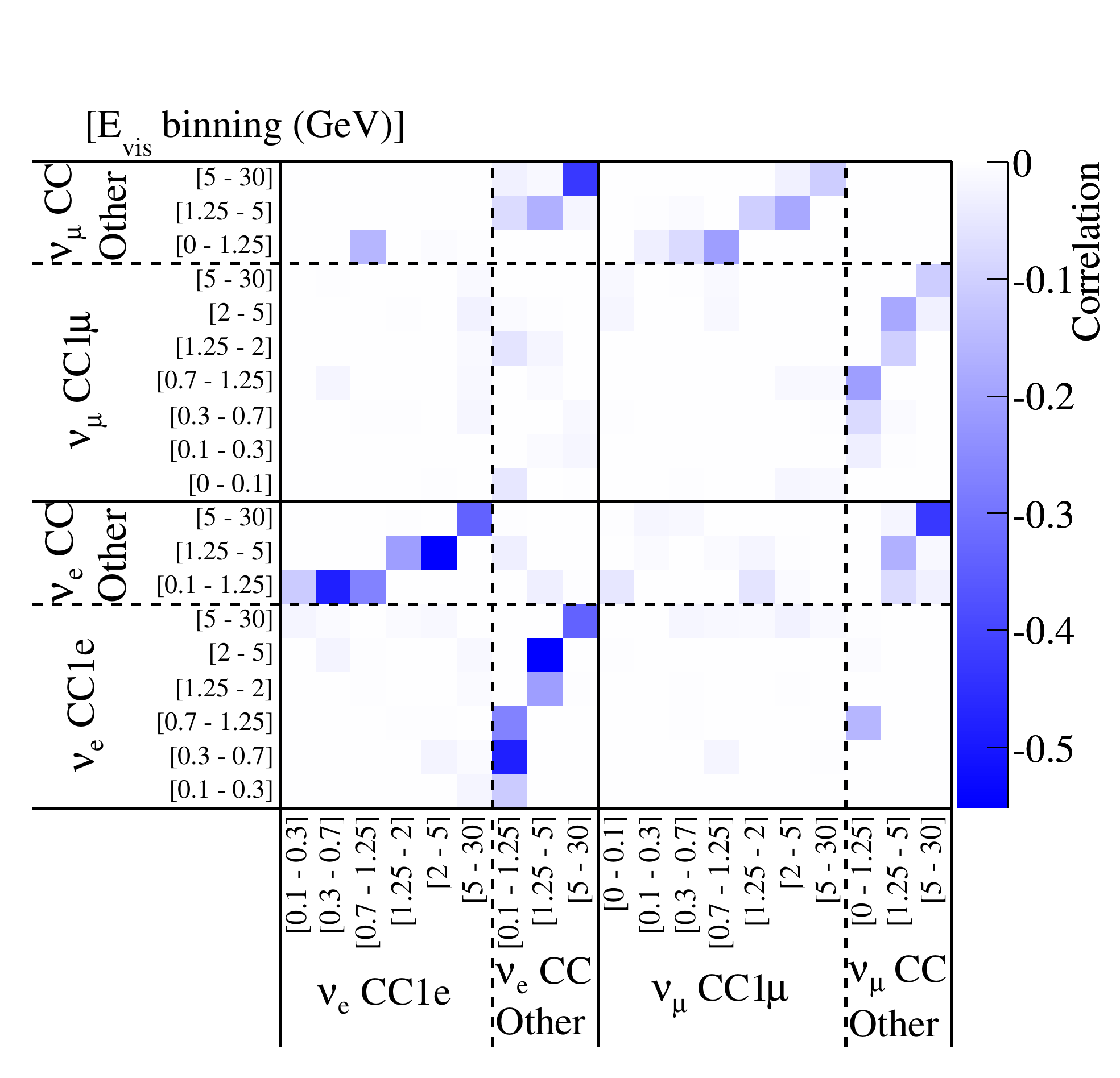}
\caption{The correlations between the estimated efficiencies for the final state topologies
CC 1$e$ and CC $e$ other for the \nue CC event selection 
and CC 1$\mu$ and CC $\mu$ other for the \num CC event selection.
The upper figure shows the combinations with positive correlation, and the lower with negative correlation. 
The diagonal correlations (correlation $= 1$) are not shown. 
}
\label{fig:SK_err_cov}
\end{center}
\end{figure}

In order to adjust the modeling of ring counting, particle identification, and $\pi^0$ rejection,
a set of parameters is defined to alter the cut values for these three classifiers.
Separate parameters are used for the first four FS topologies in Tab.~\ref{tab:FSmodes} and
for each visible energy bin within those topologies.
By adjusting these parameters, simulated events, generated according to models of the atmospheric neutrino flux,
migrate between the branches in Tab.~\ref{tab:ControlSample} thus 
changing the efficiency for true CC 1$e$ and CC $e$ other 
(CC 1$\mu$ and CC $\mu$ other) events in the \nue (\num) core samples.
Using the observed numbers of core and tail data events in each visible energy bin,
a likelihood function is defined and marginalized over the
neutrino flux, neutrino interaction systematic parameters and cut adjustment parameters, 
using a Markov Chain Monte Carlo.
The marginalized likelihood is used to estimate corrected efficiencies 
for the four FS topologies in bins of $E_{\mathrm{vis}}$ and their covariance.
The observed differences between the nominal and corrected efficiencies may indicate mismodeling of the
detector response, so additional covariance is included, with the
diagonal elements being the square of these differences and the off-diagonal terms calculated
by assuming full correlation.
The correlations between the estimated efficiencies are shown in Fig.~\ref{fig:SK_err_cov}.

To evaluate the systematic uncertainty in modeling $\pi^0$'s in SK, 
we construct a set of ``hybrid-$\pi^0$'' control samples. These events are constructed
by overlaying one electron-like ring from the SK atmospheric neutrino or cosmic ray muon 
samples with one simulated photon ring. The simulated photon ring kinematics are chosen 
such that the momenta and opening angle between the two rings follow the decay kinematics 
of NC $\pi^0$ events from the T2K MC. Hybrid-$\pi^0$ MC samples with both 
rings from the SK MC are produced to compare with the
hybrid-$\pi^0$ data samples and the difference in the fractions that pass the \nue\ selection criteria
is used to assign the systematic error.
The difference could be due to incorrect modeling of
scattered or reflected light 
from the higher energy ring which obscures the lower energy ring.
In order to investigate this, we compare
hybrid-$\pi^0$ samples in which the electron constitutes the higher energy 
ring from the $\pi^0$ decay with 
hybrid-$\pi^0$ samples in which it constitutes the lower energy ring. 
For events with additional particles in the final state,
we add a MC ring to the existing hybrid-$\pi^0$ samples and
assume the dominant source of error comes from the detection of the lower energy photon 
from the $\pi^0$ decay. Relative uncertainties on the efficiency, calculated 
for 17 bins in reconstructed electron momentum and angle, are in the range 2-60\%. 
Relative statistical errors, in the range 15-50\%, are applied assuming no correlation between bins.

Neutral-current interactions can produce a final state containing just a 
single photon via radiative decays of $\Delta$ resonances (NC 1\g).  This is a 
background in the \nue CC candidate sample because photons and electrons 
produce very similar charge patterns in the SK detector.  The uncertainty in 
the efficiency of selecting NC 1\g events is determined by comparing the efficiency of 
a single photon MC sample with that of a single electron MC.  The difference is 
no more than 1\%.  This error is added in quadrature to the uncertainty for the CC 1$e$ 
FS topology described above to give the total uncertainty on the NC 1\g 
background in the \nue CC candidate sample.

Muon decay-in-flight events make up a small background in the \nue CC candidate sample.
Such events can be 
misidentified because the decay electron is boosted in the direction of the 
parent muon and thus their Cherenkov rings can overlap.  MC studies indicate that 
such events make up 19\% of the background from \num interactions and 
its rate is assigned a 16\% selection uncertainty.  
The remaining background from \num interactions are assigned a conservative 150\% error.
A conservative 100\% uncorrelated error is assigned to the NC 1$\pi^\pm$ 
and the NC other.

For the \num CC candidate sample, the dominant NC backgrounds are NC 1$\pi^\pm$ and 
events with just a single proton (NC other).
The relative uncertainty in this background, due to systematic uncertainties in
ring counting and particle identification, is found to be 59\%.
The background from \nue interactions in the \num CC candidate sample is assigned a 
conservative 100\% error.

All aspects of SK detector simulation that can affect the modeling of the SK candidate event selection 
described above are propagated using a vector of systematic parameters, $\vec{s}$,
which scale the nominal expected number of events in bins of
the observable kinematic variables $E_{\mathrm{rec}}$ or $p-\theta$ for 
the true $\nu$ interaction mode categories.  
The binning is shown in Tab.~\ref{tab:SK_sbin}.
The covariance of these parameters, $V_s$, is used 
to propagate the uncertainties in the detector simulation
to the oscillation analyses.

The energy scale uncertainty is estimated by comparing data with simulated
samples spanning the momentum range 30\mevc to 6\gevc.  
Starting at the lowest energy, we use the reconstructed momentum spectrum 
of electrons produced by the decay of cosmic ray muons, the reconstructed 
mass of neutral pions from atmospheric neutrino interactions and cosmic ray 
muons that stop within the SK tank. The final uncertainty is 2.4\%, independent of $E_\nu$.

\begin{table}[tbp]
   \centering
   \caption{Binning for the vector of SK detector systematic parameters $\vec{s}$. 
   Two schemes are defined since analyses use either 
   $E_{rec}$ or $p-\theta$ binning for the $\nu_e$ appearance channel.} 
   \begin{tabular}{ll}
     \hline
     \hline
     \multicolumn{2}{l}{$\nu_e$ appearance: \ \ $E_{rec}$ (\gev)}  \\
     \hline
     Osc.$\nu_e$CC & 0--0.35--0.8--1.25 (3 bins) \\
     $\nu_{\mu}$CC & 0--0.35--0.8--1.25 (3 bins) \\
     $\nu_e$CC & 0--0.35--0.8--1.25 (3 bins) \\
     NC & 0--0.35--0.8--1.25 (3 bins) \\
     \hline
     \multicolumn{2}{l}{$\nu_{\mu}$ disappearance: \ \ $E_{rec}$ (\gev)} \\   
     \hline
     $\nu_{\mu}$CCQE & 0--0.4--1.1--30.0 (3 bins) \\
     $\nu_{\mu}$CCOther & 0--30.0 (1 bin) \\
     $\nu_e$ & 0--30.0 (1 bin) \\
     NC & 0--30.0 (1 bin) \\
     \hline
     \hline
   \end{tabular}
\vskip 8mm
 \begin{tabular}{cll}
   \hline
   \hline
    {$\nu_e$ appearance} & $p$ (\gevc) & $\theta$ (degree) \\
   \hline
    \multirow{3}{*}{Osc.$\nu_e$CC:$\nu_{\mu}$CC:$\nu_e$CC:NC 
   $\left\{ \begin{array}{l} \\ \\ \end{array} \right.$ } & 0--0.3   & 0--40--60--80--100--120--140--180 (7 bins)  \\
                              & 0.3--0.7   & 0--40--60--80--180 (4 bins) \\
                              & 0.7--       & 0--40--180 (2 bins)\\
   \hline
   \hline
   $\nu_{\mu}$ disappearance & \multicolumn{2}{l}{$E_{rec}$ (\gev)} \\
     \hline
   $\nu_{\mu}$CCQE & \multicolumn{2}{l}{0--0.4--1.1--30.0 (3 bins)} \\
   $\nu_{\mu}$CCOther & \multicolumn{2}{l}{0--30.0 (1 bin)} \\
   $\nu_e$ & \multicolumn{2}{l}{0--30.0 (1 bin)} \\
   NC & \multicolumn{2}{l}{0--30.0 (1 bin)} \\
   \hline
   \hline
 \end{tabular}
 \label{tab:SK_sbin}
\end{table}

Systematic uncertainties in pion interactions in the target nucleus 
(FSI uncertainties) and SK detector 
(SI uncertainties) are evaluated by varying pion 
interaction probabilities in the NEUT cascade model.
In the NEUT sample we store the information necessary to recompute 
the pion cascade using modified interaction probabilities 
to weight each event.  
Altered CC sample distributions are produced using 16 representative points 
$\vec{x}^{FSI}_k$
on the 1-$\sigma$ 
surface for the parameters.
 The covariance 
matrix $V$, which describes the variations in the number of 
events in the binned observables ($N_i$) due to the variation in 
$\vec{x}^{FSI}$, is given by
\begin{equation}
V_{ij} = \frac{1}{16}\sum_{k=1}^{16}(N_i(\vec{x}^{FSI}_k) - N_i)(N_j(\vec{x}^{FSI}_k) - N_j)
\end{equation}

The binning of this matrix is chosen to match that of the detector error 
covariance matrix shown in Table \ref{tab:SK_sbin}.

A simulation of photo-nuclear (PN) interactions is incorporated into 
the SK MC. The model allows for the absorption 
of photons based on the measured cross section and assumes that there 
is no subsequent emission above Cherenkov threshold.
A systematic uncertainty of 100\% is assumed for the normalization of the PN cross section.

\clearpage
\section{\label{sec:OA} Oscillation model and parameter estimation}
The previous sections have described the T2K experiment and the way we model all elements of the experiment and
neutrino interactions which are necessary to interpret our data, and how we use internal and external data to
improve our models.
In this section, we turn our attention to general aspects of estimating 
neutrino oscillation parameters from our data.
The oscillation model is given along with the predictions 
for the probability for muon neutrino disappearance and electron neutrino appearance, the key
observables for our experiment.
We explain how we use external data for some of the oscillation parameters
and the general approaches we use to estimate the remaining parameters.
Finally, we characterize the importance of the different sources of systematic uncertainty.
Sections~\ref{sec:numu}--\ref{sec:jointbayes} describe the individual analyses and their results in detail.

\subsection{\label{sec:OA:oscmodel}Oscillation model}
The Pontecorvo-Maki-Nakagawa-Sakata (PMNS) matrix, $U$, defines the mixture of the
mass eigenstates ($\nu_1$, $\nu_2$, and $\nu_3$) that make up each flavor state:
\begin{equation}
\left(
\begin{array}{c}
\nu_{e} \\
\nu_{\mu} \\
\nu_{\tau} 
\end{array} \right) = 
{U}
\left(
\begin{array}{c}
\nu_{1} \\
\nu_{2} \\
\nu_{3} 
\end{array} \right)
\end{equation}
and it has become standard to parametrize this matrix, ignoring the Majorana phases, as:
\begin{equation}
\begin {split}
&U = \left(
\begin{array}{ccc}
1 & 0 & 0 \\
0 & c_{23} & s_{23} \\
0 & -s_{23} & c_{23} 
\end{array} \right)
\left(
\begin{array}{ccc}
c_{13} & 0 & s_{13}e^{-i\delta} \\
0 & 1 & 0 \\
-s_{13}e^{i\delta} & 0 & c_{13}
\end{array} \right)
\left(
\begin{array}{ccc}
c_{12} & s_{12} & 0 \\
-s_{12} & c_{12} & 0 \\
0 & 0 & 1
\end{array} \right)
\label{eq:PMNSmatrix}
\end{split}
\end{equation}
where $s_{ij} = \sin\theta_{ij}$, $c_{ij} = \cos\theta_{ij}$, and $\delta = \dcp$ is the CP-violating phase. 

The \num-survival probability for a neutrino with energy $E$ traveling a distance $L$ is:
\begin{equation}
\begin{array}{l}
P(\nu_{\mu}\rightarrow\nu_{\mu}) = \\
1 -4\left( s_{12}^{2} c_{23}^{2} +  s_{13}^{2}  s_{23}^{2}  c_{12}^{2} + 2  s_{12}s_{13}s_{23} c_{12}c_{23}\cos\delta\right)s_{23}^{2}c_{13}^{2}\sin^{2}\phi_{31}\\
- 4\left( c_{12}^{2} c_{23}^{2} +  s_{13}^{2}  s_{23}^{2}  s_{12}^{2} - 2  s_{12}s_{13}s_{23} c_{12}c_{23}\cos\delta\right)s_{23}^{2}c_{13}^{2}\sin^{2}\phi_{32}\\
-4\left( s_{12}^{2} c_{23}^{2} +  s_{13}^{2}  s_{23}^{2}  c_{12}^{2} + 2  s_{12}s_{13}s_{23} c_{12}c_{23}\cos\delta\right)\left( c_{12}^{2} c_{23}^{2} +  s_{13}^{2}  s_{23}^{2}  s_{12}^{2} - 2  s_{12}s_{13}s_{23} c_{12}c_{23}\cos\delta\right) \sin^{2}\phi_{21}
\end{array}
\label{eq:numusurv3f}
\end{equation}
where
\begin{equation}
\phi_{ij} = \frac{\Delta m_{ij}^{2} L}{4E}
\label{eq:phiDef}
\end{equation}
in natural units and $\Delta m_{ij}^{2} = m_i^2 - m_j^2$ is the difference in the squares of masses of eigenstates.

The $\nue$-appearance probability, to first order approximation in matter effects, can be written as:
\begin{equation}
\begin{array}{l}
P(\nu_{\mu}\rightarrow\nu_{e}) = \\
4c^{2}_{13}s^{2}_{13}s^{2}_{23}\sin^{2}\phi_{31}\left(1+\frac{2a}{\Delta m^{2}_{31}}(1-2s^2_{13})\right)\\
+8c^{2}_{13}s_{12}s_{13}s_{23}\left(c_{12}c_{23}\cos\delta-s_{12}s_{13}s_{23}\right)\cos\phi_{23}\sin\phi_{31}\sin\phi_{21}\\
-8c^{2}_{13}c_{12}c_{23}s_{12}s_{13}s_{23}\sin\delta\sin\phi_{32}\sin\phi_{31}\sin\phi_{21}\\
+4s^{2}_{12}c^{2}_{13}\left(c^{2}_{12}c^{2}_{23}+s^{2}_{12}s^{2}_{23}s^{2}_{13}-2c_{12}c_{23}s_{12}s_{23}s_{13}\cos\delta\right)\sin^{2}\phi_{21}\\
-8c^{2}_{13}s^{2}_{13}s^{2}_{23}\left(1-2s^{2}_{13}\right)\frac{aL}{4E_{\nu}}\cos\phi_{32}\sin\phi_{31}
\end{array}
\end{equation}
The effect on oscillation due to the density, $\rho$, of matter through which the neutrinos travel
is included with the terms, $a[\evvcccc]=7.56\times10^{-5} \rho[$g/cm$^3]$E$_\nu[$GeV$]$.
The corresponding $\nueb$-appearance probability is calculated by changing the sign of $a$ and $\dcp$.
Our analyses use
the complete formulas, without approximating matter effects, 
to compute the oscillation probabilities.

Since the neutrino mass hierarchy (MH) is not yet known, we parametrize the large mass splitting by $\Dmsq=\dmsq$ for normal hierarchy (NH, where $m_3$ is
the largest mass)
and $\Dmsq=\dmsqo$ for inverted hierarchy (IH, where $m_3$ is the smallest mass).

\begin{figure}[tbp]
\begin{center}
\includegraphics[width=0.75\textwidth]{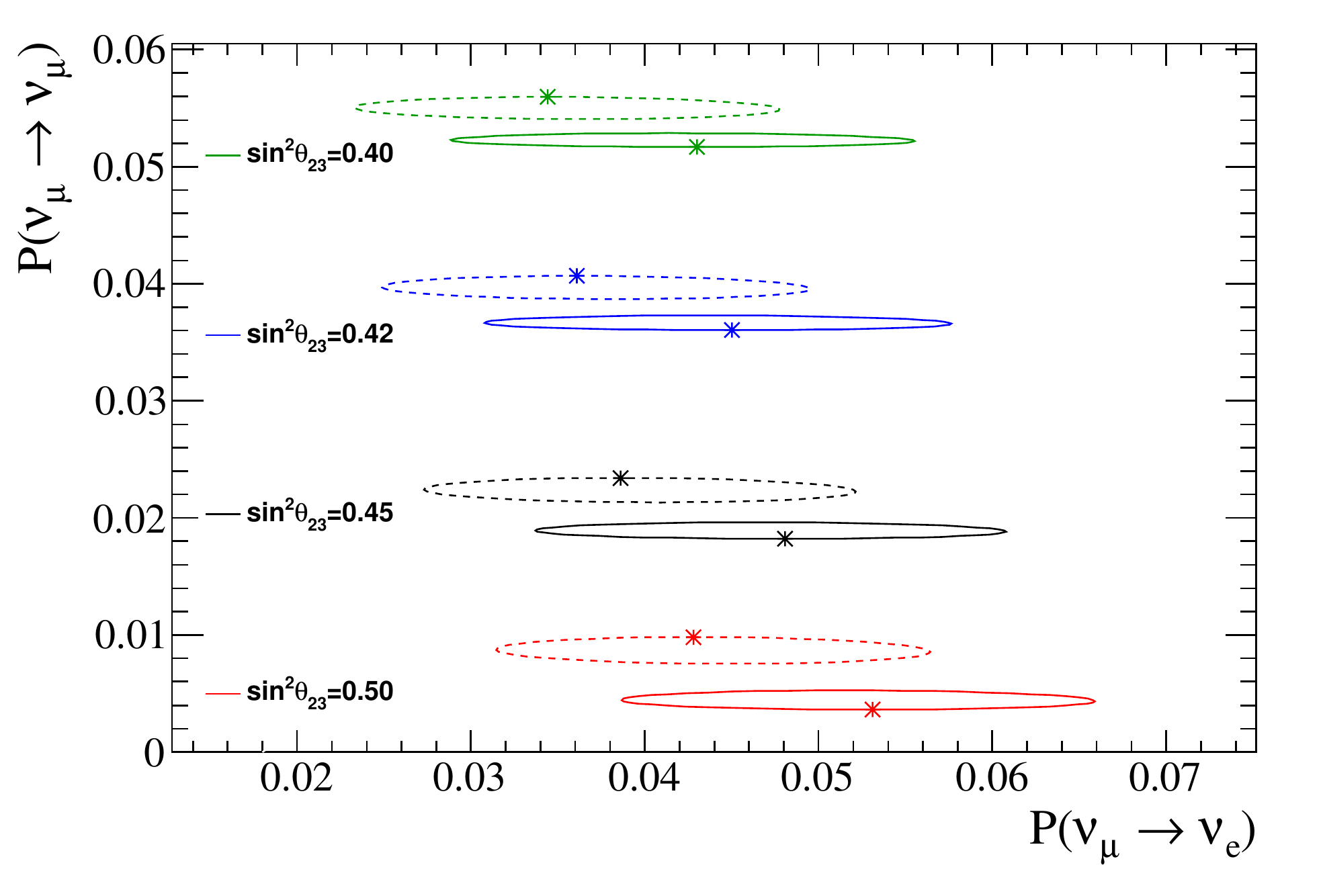}
\caption{The $P(\nu_{\mu}\rightarrow\nu_{\mu})$ survival probability and $P(\nu_{\mu}\rightarrow\nu_{e})$ appearance 
probability for different values of $\stt$ and for $\dcp$ in the interval $[-\pi,\pi]$ for normal (solid) and inverted (dashed) mass hierarchy.
The highlighted dot on each ellipse is the point for $\dcp=0$ and $\dcp$ increases clockwise (anti-clockwise) for normal (inverted) mass hierarchy.
The other oscillation parameter values are fixed (solar parameters as established in Sec.~\ref{sec:OA:osc} and $\sot = 0.0243$) and the neutrino energy is fixed to 0.6~GeV.
}
\label{fig:oscprob}
\end{center}
\end{figure}

It is not possible to estimate all of the oscillation parameters using only our measurements of $\num$-disappearance and $\nue$-appearance.
Instead, we estimate the four oscillation parameters, \Dmsq, \stt, \sot, \dcp, and the mass hierarchy, and use external measurements
for the solar oscillation parameters, \stso and \dmsqso, as we have negligible sensitivity to those.
Figure~\ref{fig:oscprob} illustrates how our key observables depend on the two parameters, \stt\ and \dcp, for the two mass hierarchies.
In this figure the neutrino energy is at the oscillation maximum (0.6~GeV), and the other oscillation parameters are fixed 
(solar parameters as established in Sec.~\ref{sec:OA:osc} and $\sot = 0.0243$).
To a good approximation, with our current dataset, $\num$-disappearance can be treated on its own to estimate $\theta_{23}$.
The oscillation parameter dependence on $\nue$-appearance cannot be factorized, however. 
In order to estimate the full set of oscillation parameters and properly account for all uncertainties, it is necessary 
to do a joint analysis of $\num$-disappearance and $\nue$-appearance.

\subsection{\label{sec:OA:osc} External input for oscillation parameters}
Since our experiment is insensitive to the solar oscillation parameters, we
fix them to the values
$\stso = 0.306$ and $\dmsqso = 7.5\times 10^{-5}$\evvcccc from~\cite{PDG2012}.
As a check, the Bayesian analysis presented in Sec.~\ref{sec:jointbayes} applies Gaussian priors 
with standard deviations (0.017 and $0.2\times10^{-5}$\evvcccc) and finds
that the uncertainties in these parameters do not affect the intervals of the other oscillation parameters.

When combining the results for the T2K joint oscillation analyses in Secs.~\ref{sec:jointfreq} and~\ref{sec:jointbayes}
with the results from the reactor experiments, we use the 
weighted average of the results from the three reactor experiments Daya Bay, RENO, and Double Chooz
which is: $(\stot)_{reactor} = 0.095\pm0.01$~\cite{PDG2013}.
In terms of the parametrization that we use in this paper, $(\sot)_{reactor} = 0.0243\pm0.0026$.

\subsection{\label{sec:OA:fit} Oscillation parameter estimation}
Sections~\ref{sec:numu}--\ref{sec:jointbayes} describe analyses 
which use T2K and external data
to estimate oscillation parameters and provide frequentist confidence intervals or Bayesian credible intervals.
Using the disappearance channel alone, the atmospheric oscillation parameters are studied using frequentist approaches.
The disappearance and appearance channels are used in combination to study a larger set of oscillation parameters,
using frequentist and Bayesian approaches.
This section describes general methods that are applied in these analyses.

The oscillation analyses compare the event rate and distribution of the reconstructed neutrino energies for the observed $\nu_\mu$ CC and $\nu_e$ CC candidate events
recorded by the far detector, 
selected as described in Sec.~\ref{sec:SK_event_selection},
with model predictions.
The overall number of predicted events for typical oscillation parameter values 
and without oscillations are shown in Tab.~\ref{tab:jointfreq:prediction:all_extra_templates}.

\begin{table}[tbp]
\center
\caption{Predicted number of $\nu_\mu$ CC candidates and $\nu_e$ CC candidates for an exposure of 6.57 $\times 10^{20}$ POT
with and without oscillations and with oscillations using the typical parameter values: 
$\sin^{2}\theta_{12} = 0.306$, $\Delta m_{21}^{2} = 7.5 \times 10^{-5}$\evvcccc,
$\sin^{2}\theta_{23} =  0.5$, $\Delta m_{32}^{2}= 2.4 \times 10^{-3}$\evvcccc, $\sin^{2}\theta_{13} =  0.0243$, $\delta_{CP} = 0$ and normal mass hierarchy.
The total numbers are broken down into the intrinsic beam components (those without an arrow) and oscillated components.
}
\begin{tabular}  {l c c c c c}
\toprule
  & \multicolumn{2}{c}{$\nu_\mu$ CC} & \ \ \ & \multicolumn{2}{c}{$\nu_{e}$ CC} \\
  & Osc. & No osc. & & Osc. & No osc. \\
\hline
$\nu_{\mu}$  & 116.46 & 431.77 & & 0.94 & 1.38\\
$\nu_{e} \rightarrow \nu_{\mu}$  & 0.16  & 0 & & 0.00 & 0\\
$\bar{\nu}_{\mu}$  & 7.81 & 13.92 & & 0.05 & 0.06\\
$\nu_{e}$  & 0.26 & 0.27 & & 3.13 & 3.38\\
 $ \nu_{\mu} \rightarrow \nu_{e}$ & 0.26 & 0  & & 16.55  & 0\\
$\bar{\nu}_{e}$  & 0.02 & 0.02 & & 0.15 & 0.16\\
$\bar{\nu}_{\mu} \rightarrow \bar{\nu_{e}}$  & 0.00 & 0 & & 0.22 & 0\\
\hline
Total  & 124.98 & 445.98 & & 21.06 & 4.97 \\
\botrule
\end{tabular}
\label{tab:jointfreq:prediction:all_extra_templates}
\end{table}

Point estimates for the oscillation parameters are those that maximize a likelihood function
(or the posterior probability density for Bayesian analyses)
that accounts for T2K-SK data, as well as internal control samples and external data.
The observed numbers of events in SK are treated as outcomes of Poisson distributions.
Systematic uncertainties are encapsulated by the systematic parameters and their covariance matrices, defined in Sec.s~\ref{sec:beam}-\ref{sec:SK}.
These provide a convenient mechanism to connect the separate analyses of the neutrino beamline, neutrino interactions, near detectors, and far detector 
to the full oscillation analyses.
The analyses use different approaches to deal with the large number of oscillation and nuisance parameters, and
report intervals based on either frequentist or Bayesian methods.

With the large number of oscillation and nuisance parameters involved, it is not possible to calculate confidence intervals
for a subset of the parameters with a method that guarantees frequentist coverage\footnote{coverage demands 
that in an ensemble of repeated experiments, $\alpha$\% of the $\alpha$\% confidence intervals contain the true parameter(s).
Coverage in presence of systematic uncertainty is difficult to define, in part due to the definition of an appropriate ensemble.}
for any possible values of the remaining parameters.
Instead, a pragmatic approach is followed by reducing the high dimensionality of the likelihood functions through either profiling or marginalization.
The profile likelihood, a function of only the subset of parameters of interest, is the likelihood maximized over the remaining parameters.
The marginal likelihood is found by integrating the product of the likelihood function and priors over all parameters, except those of interest.
In the case of linear parameter dependence and where the nuisance parameters appear in a Gaussian form,
the profile and marginal likelihood functions will be identical and can be used to produce intervals with correct frequentist coverage.
For the neutrino oscillation analysis, the parameter dependence is non-linear and as a result the profile and marginal likelihoods
differ, and frequentist coverage is not guaranteed.

When practical, we use the Neyman approach of constructing $\alpha$\% confidence intervals
whereby, for any value of the parameter(s) of interest,
$\alpha$\% of possible data outcomes are accepted on the basis of a statistic. 
In our analyses, they are accepted if the likelihood ratio is larger than a critical value. 
The confidence interval is the set of all values for the parameter(s) for which the data are accepted.
When physical boundaries or non-linearities appear in the parametrization, as in the case for the oscillation parameters,
they can cause confidence intervals to be empty or misleadingly small.
In order to reduce the chance of producing such confidence intervals, 
we use the likelihood ratio recommended by Feldman and Cousins~\cite{PhysRevD.57.3873} to form the interval.
When producing joint intervals for two oscillation parameters, this approach is not always computationally practical, and instead
approximate intervals are shown using contours of the likelihood ratio, sometimes referred to as the constant $\Delta\chi^2$ method.

To construct Bayesian credible intervals, the posterior probability density function of the oscillation and nuisance parameters
is calculated as the product of the likelihood function for the SK data with prior probability functions for the parameters.
The Markov-Chain Monte Carlo (MCMC) method using the Metropolis-Hastings algorithm~\cite{hastings1970monte}
is used to efficiently produce a set of points that populate the full parameter space proportional
to the posterior probability density function.
The chain is the set of accepted stepping points in a random walk through parameter space, 
in which a proposed step from point $A$ to a point $B$ with lower density
is accepted with a probability equal to the ratio of the densities $f(B)/f(A)$ and is always accepted when the density increases. 
When a step is not accepted, the last point in the chain is repeated, and another random step from that point is proposed.
With the chain, consisting typically of millions of points,
$\alpha$\% highest-posterior-density (HPD) credible intervals~\cite{chen1999monte}
are constructed by selecting the region of highest
density that contain $\alpha$\% of all the points.
HPD intervals are constructed such that no point in parameter space outside the interval has a higher probability 
density than any point inside the interval.
This is done for one or two parameters of interest, and the values of the remaining parameters are ignored in the process,
equivalent to producing a set of points distributed according to the marginalized posterior probability density function.
Unlike the frequentist approaches used, for which coverage is approximate, 
there are no approximations necessary to produce the credible intervals.

The prior probability densities are, by default, uniform for the oscillation parameters over a large bounded region
in the standard oscillation parametrization (\Dmsq, \stt, \sot, \dcp), 
multidimensional Gaussians for the nuisance parameters, and the prior probabilities 
for the two mass hierarchies are set to 0.5.
As a result, the posterior probability density is proportional to the likelihood functions used for the frequentist analyses.
Checks are made for alternative priors which are uniform in the oscillation angles, and
the resulting interval boundaries are not strongly affected.

\subsection{\label{sec:OA:syst} Characterizing systematic uncertainty}
The systematic parameters considered for the oscillation analyses can be grouped into three different categories: i) SK flux parameters and cross section parameters in common with ND280, ii) independent cross section parameters and iii) SK efficiencies, final state and secondary interactions (FSI+SI) and photo-nuclear (PN) parameters.
The first category includes the systematic uncertainties related to the neutrino flux at SK and some cross sections, which are constrained by the near detector data as explained in Sec.~\ref{sec:BANFF}. 
The values and uncertainties of these parameters used in the oscillation analyses are summarized in Tabs.~\ref{tab:propagatedparameters} and~\ref{tab:propagatedparametersxsec}.
The independent cross section parameters, described in Sec.~\ref{sec:nuint}, are related to the nuclear model, therefore independent between the near and far detector as they contain different nuclei, or those which are common between the near and far detector but for which the near detector is insensitive. 
Table~\ref{tbl:xsecpar} in Sec.~\ref{sec:nuint} summarizes the values and uncertainties of the independent cross section parameters used for the SK oscillation analyses. 
Finally, the far detector efficiencies and uncertainties on final state, secondary and photo-nuclear interactions are described in Sec.~\ref{sec:SK}. 
A covariance matrix is computed for the uncertainties in this group; however, the uncertainty on the SK reconstructed energy scale, estimated to be 2.4\%, is not included in the calculation of the covariance matrix, but considered as an independent systematic parameter.

The effects of the systematic uncertainties on the predicted event rate are summarized in 
Tab.~\ref{tab:systematics:nsk_table_summary}
for the typical values of the oscillation parameters. In this table, the effects are presented as percentage uncertainties
computed by throwing $10^{6}$ toy experiments, varying only the systematics in the selected category
(fixing the rest to their nominal values) and finding the RMS/mean of the distribution of number
of events.

\begin{table}[tbp]
\caption{
  Relative uncertainty (1$\sigma$) on
  the predicted rate of $\nu_\mu$ CC and $\nu_e$ CC candidate events.}
    \begin{tabular}{lcc} \toprule
  { Source of uncertainty }  & \ \ $\nu_\mu$ CC \ \ &  \ \ $\nu_e$ CC \ \ \\
\hline
Flux and common cross sections & & \\
(w/o ND280 constraint)       &    21.7\%   &  26.0\% \\
(w ND280 constraint)      &  2.7\%   &  3.2\% \\
\hline
Independent cross sections    &   5.0\%   &  4.7\% \\
\hline
SK    &    4.0\%   &  2.7\% \\
FSI+SI(+PN) & 3.0\% & 2.5\% \\
\hline
{{Total}} & & 	\\ 
{{(w/o ND280 constraint) }}  & {{23.5\%}}   &  {{26.8\%}}  \\
{{(w ND280 constraint) }}     &   {{7.7\%}}   &  {{6.8\%}} \\
\botrule
\end{tabular}
\label{tab:systematics:nsk_table_summary}
\end{table}

\begin{figure}[tbp]
\begin{center}
\includegraphics[width=0.47\textwidth]{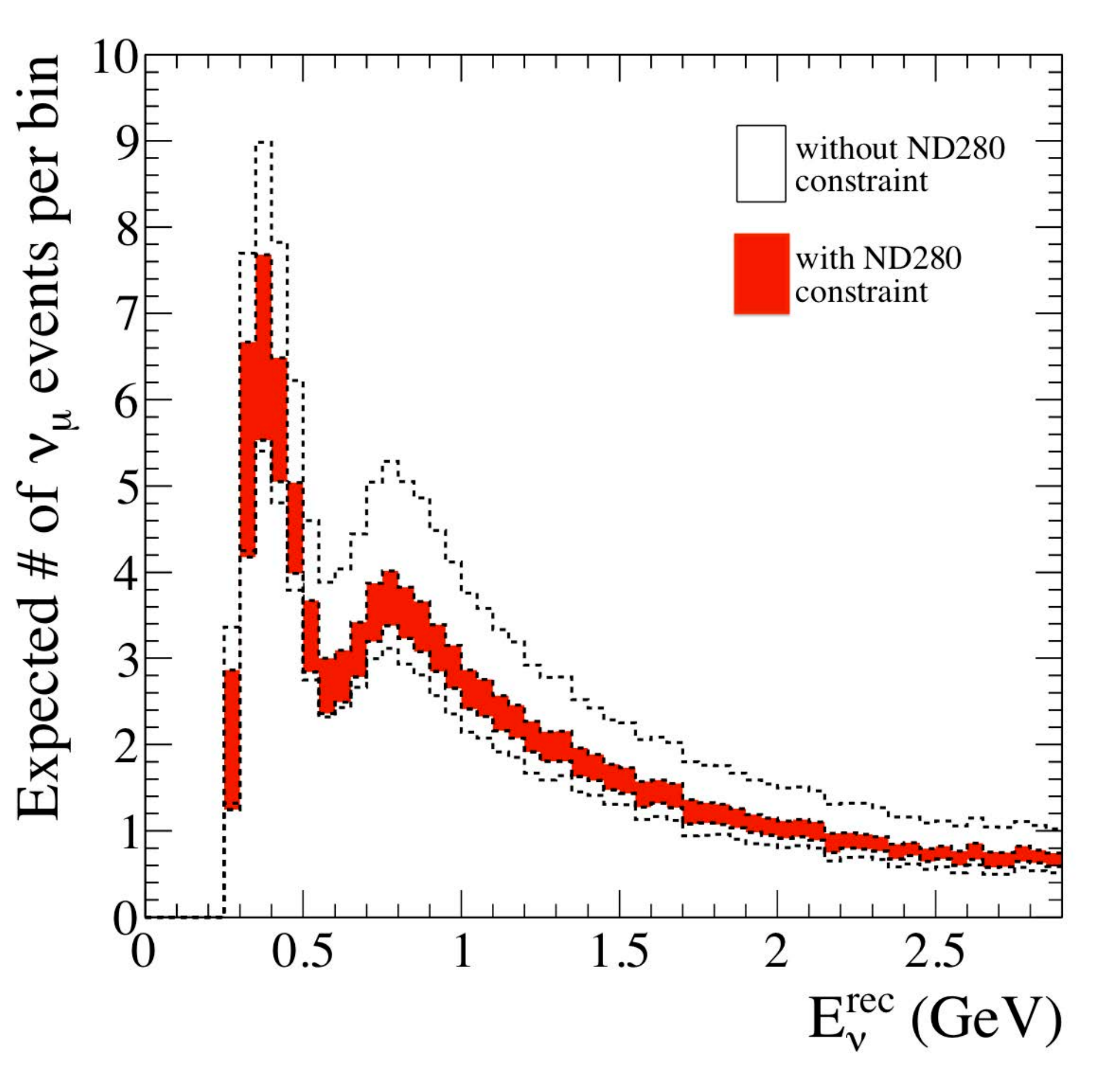}
\includegraphics[width=0.47\textwidth]{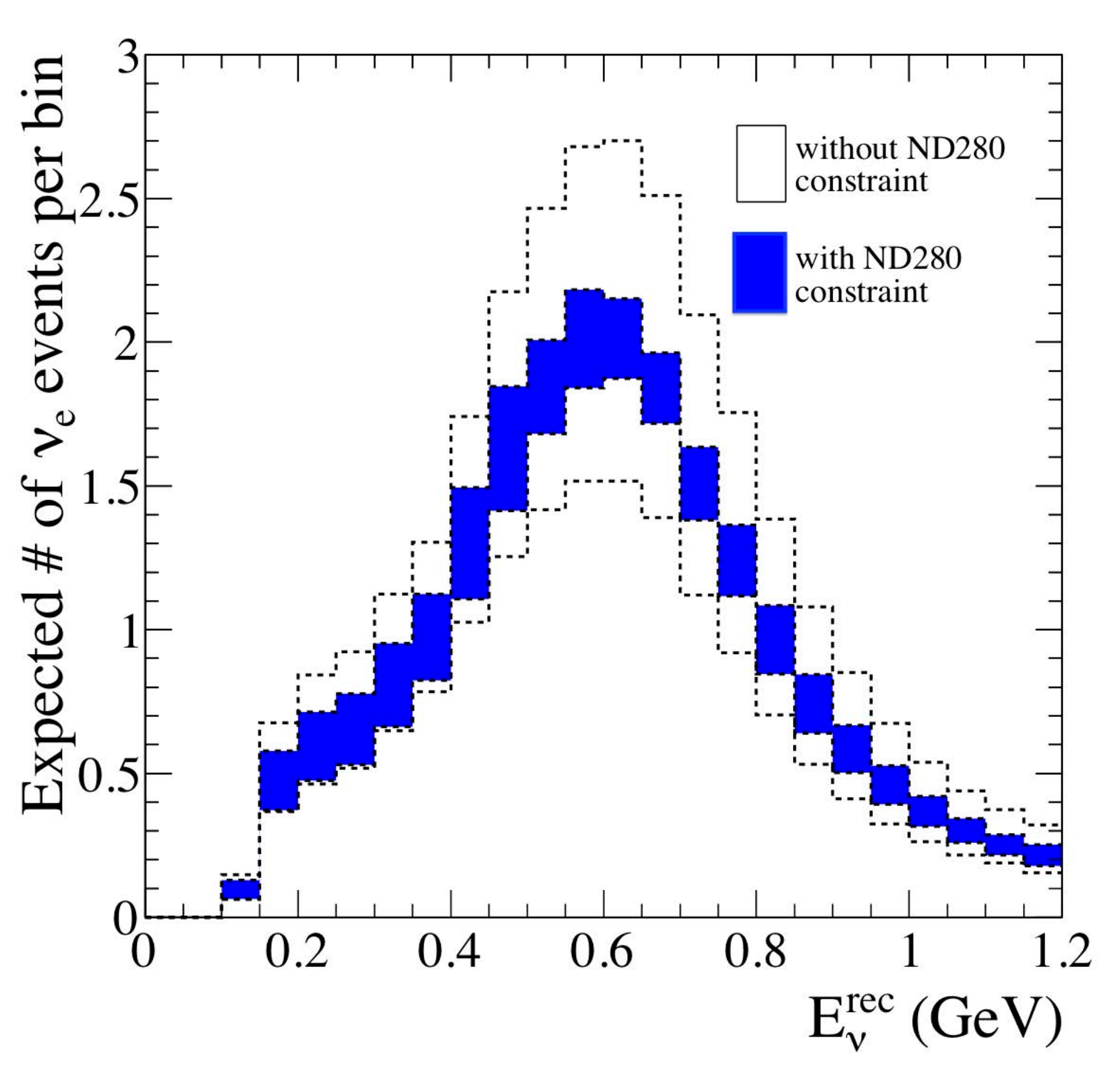}
\caption {
Total error envelopes for the reconstructed energy distributions of
$\nu_\mu$ CC (left) and $\nu_e$ CC (right) candidate events, 
using typical oscillation parameter values,
with and without the ND280 constraint applied.}
\label{fig:systematics:error_envelope}
\end{center}
\end{figure}

Figure~\ref{fig:systematics:error_envelope} shows the total error envelope combining all systematic uncertainties, calculated as the RMS from $10^6$ toy MC experiments generated with randomized systematic parameters, taking into account all correlations between them, with and without the constraint from the ND280 data, showing a clear reduction of the error envelope when the constraint is applied. 

\clearpage
\section{\label{sec:numu} $\num \rightarrow \num$ Analysis}
T2K has published several measurements of muon
neutrino disappearance~\cite{PhysRevD.85.031103,abe2013measurement,PhysRevLett.112.181801}.
These measurements were performed
within the framework of the PMNS 
oscillation model described in Section~\ref{sec:OA:oscmodel} and provided
best-fit estimates and frequentist confidence intervals for the values
of the mixing parameter \stt and the mass-squared splitting
 \dmsq\ (\dmsqo) in the case of the normal (inverted) hierarchy.
Each successive measurement analyzed a larger dataset, and the most recent measurement
provides the world's strongest constraint on \stt~\cite{Abe:2014ugx}.
This section gives a more detailed description of that analysis
and the study of multi-nucleon effects.
Reducing the
uncertainty on the values of these two parameters is important for
measuring CP violation in neutrino oscillations by T2K
and other current and future experiments. Furthermore, precise
measurements of \stt\ could constrain models of neutrino mass
generation~\cite{King:2013iu,Albright:2010kl,Altarelli:2010jl,Ishimori:2010uu,Albright:2006hr,Mohapatra:940213}.

\subsection{Method}
\label{sec:numu:method}

The \num-disappearance analysis is performed by comparing the
rate and spectrum of reconstructed neutrino energies, Eq.~(\ref{eq:SK_Erec}), 
in the $\nu_\mu$ CC candidate event sample with
predictions calculated from Monte Carlo simulation.
The predicted spectrum is calculated by applying
the survival probability in Eq.~(\ref{eq:numusurv3f}) to a prediction for the unoscillated
rate and spectrum. These predictions are derived from our models of the total expected neutrino flux
at the detector (explained in Sec.~\ref{sec:beam}) and the cross section predictions
for neutrino-nucleus interactions on water (described in
Sec.~\ref{sec:nuint}), 
which are constrained by near detector data (described in Sec.~\ref{sec:BANFF}),
and a GEANT3 model of particle
interactions and transport in the \sk detector.  The models of the flux,
interaction physics, and detector include systematic parameters, whose
uncertainties are accounted for in the analysis by using their corresponding
covariance matrices. 

The oscillation parameters 
are estimated using two independent maximum likelihood fits to the reconstructed energy
spectrum.  The fits use different likelihoods and software
in order to serve as cross-checks to each other.  One analysis uses an extended
unbinned likelihood (M1), while 
the other uses a binned likelihood (M2). The
log-likelihood definitions, ignoring constant terms, are:
\begin{itemize}
\item {\it M1 likelihood},
\begin{equation}
\begin{split}
-2\ln\mathcal{L}(\vec{\theta},\vec{g},\vec{x}_s,\vec{s}) = 
& -2\sum^{N^d}_{i=1}\ln f(E_{\nu,i}^{rec}|\vec{\theta},\vec{g},\vec{x}_s,\vec{s}) \\
& + 2\left(N^{p}(\vec{\theta},\vec{g},\vec{x}_s,\vec{s})
-N^d \ln N^{p}(\vec{\theta},\vec{g},\vec{x}_s,\vec{s})\right) \\
& + \Delta\vec{g}^T V_g^{-1} \Delta\vec{g}
+ \Delta\vec{x}_s^T V_{xs}^{-1} \Delta\vec{x}_s
+ \Delta\vec{s}^T V_s^{-1} \Delta\vec{s} \ ,
\end{split}
\label{eq:numu_M1_likelihood}
\end{equation}
and
\item {\it M2 likelihood},
\begin{equation}
\begin{split}
-2\ln\mathcal{L}(\vec{\theta},\vec{g},\vec{x}_s,\vec{s}) =
& - 2\sum_{j=1}^{N_{bins}} N^d_j \ln N^p_j(\vec{\theta},\vec{g},\vec{x}_s,\vec{s}) \\
& + 2N^p(\vec{\theta},\vec{g},\vec{x}_s,\vec{s}) \\
& + \Delta\vec{g}^T V_g^{-1} \Delta\vec{g}
+ \Delta\vec{x}_s^T V_{xs}^{-1} \Delta\vec{x}_s
+ \Delta\vec{s}^T V_s^{-1} \Delta\vec{s} \ .
\end{split}
\label{eq:numu_M2_likelihood}
\end{equation}
\end{itemize}
In both definitions, $N^d$ and $N^p$ are the total number of data and predicted
events respectively; $\vec{\theta}$ represents a vector of the
PMNS oscillation parameters (Sec.~\ref{sec:OA:oscmodel}); $\vec{g}$ is a vector containing the
values of the systematic parameters
constrained by the near detector 
(Tabs.~\ref{tab:propagatedparameters},\ref{tab:propagatedparametersxsec}), 
$\vec{x}_s$ are the cross section parameters not constrained
by the near detector (Tab.~\ref{tbl:xsecpar}), 
and $\vec{s}$ are the \sk\ detector systematic parameters (Sec.~\ref{sec:SK_detector_errors}).
$\Delta$ designates the difference between the systematic parameters and their nominal values, and
$V$ designates the covariance for the systematic parameters.

For the M1 likelihood, $f(E_{\nu,i}^{rec}|\vec{\theta},\vec{g},\vec{x}_s,\vec{s})$ is the
probability density of observing an event with reconstructed energy,
$E_{\nu,i}^{rec}$, given values for the oscillation and systematic
parameters. The value of $f(E_{\nu,i}^{rec}|\vec{\theta},\vec{g},\vec{x}_s,\vec{s})$ is
calculated with a linear interpolation between the bins of a histogram
of the normalized energy spectrum. For the M2 likelihood, 
the number of data and
predicted events in the $j^{th}$ reconstructed energy bin, 
$N^d_j$ and $N^p_j$ respectively, are used instead.

Both \num-disappearance fits consider a total of 48
parameters: 6 oscillation parameters, 16 flux parameters, 20 neutrino interaction
parameters and 6 parameters related to the response of \sk.
In order to find the
best-fit values and confidence intervals for \stt\ and \Dmsq, the
profiled likelihood is maximized.  Separate fits are performed for the different
neutrino mass hierarchy assumptions. 

\subsection{Determining Confidence Intervals}
As explained in Sec.~\ref{sec:OA:fit}, the Neyman method with the 
approach
recommended by Feldman and Cousins (FC) was used to calculate confidence
intervals for the two oscillation parameters, \stt and \dmsq (\dmsqo), for the
normal (inverted) hierarchy. The constant-$\Delta\chi^{2}$ method does not
provide correct coverage due to the physical boundary near \sttt$=1$ and
because of the non-linear parametrization. Critical values of the FC
statistic were determined on a fine grid of the two oscillation
parameters of interest using 10,000 toy datasets at each point.
Each toy dataset had a set of values of the systematic parameters sampled
from a multi-dimensional Gaussian having means at the nominal values, and covariances $V$.
Each oscillation parameter, \stso, \dmsqso, and \sot, is sampled from a
Gaussian with mean and sigma values listed in Sec.~\ref{sec:OA:osc}.
The values of \dcp\ are sampled uniformly between $-\pi$ and
$+\pi$. The systematic parameters and these additional oscillation 
parameters are removed from the likelihood function by profiling. In
order to calculate an interval of just one oscillation parameter (\stt\ or $\Delta m^2$), 
we determine the critical values by
marginalizing over the second oscillation parameter. The
marginalization assumes that the probability is proportional to the
likelihood using T2K data.

\subsection{Results}
\begin{figure}[tbp]
  \centering
  \includegraphics[width=0.7\textwidth]{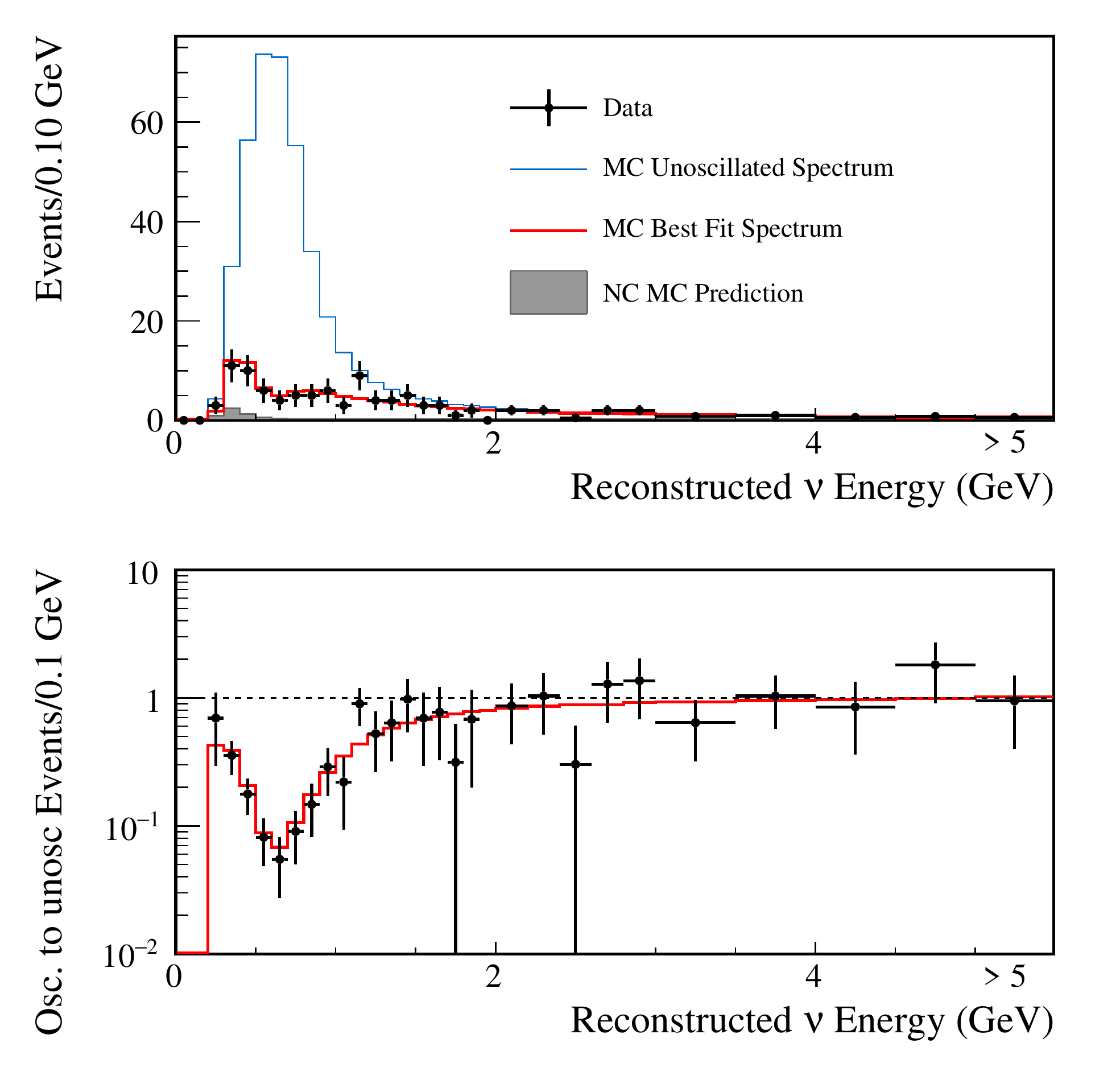}
  \caption{Top: Reconstructed neutrino energy spectrum for data, 
    best-fit prediction, and unoscillated prediction.
    Bottom: Ratio of oscillated to unoscillated events as a function of neutrino energy
    for the data and the best-fit spectrum.}
  \label{fig:numu_spectrum_result}
\end{figure}

\begin{figure}
  \centering
  \includegraphics[width=0.48\textwidth]{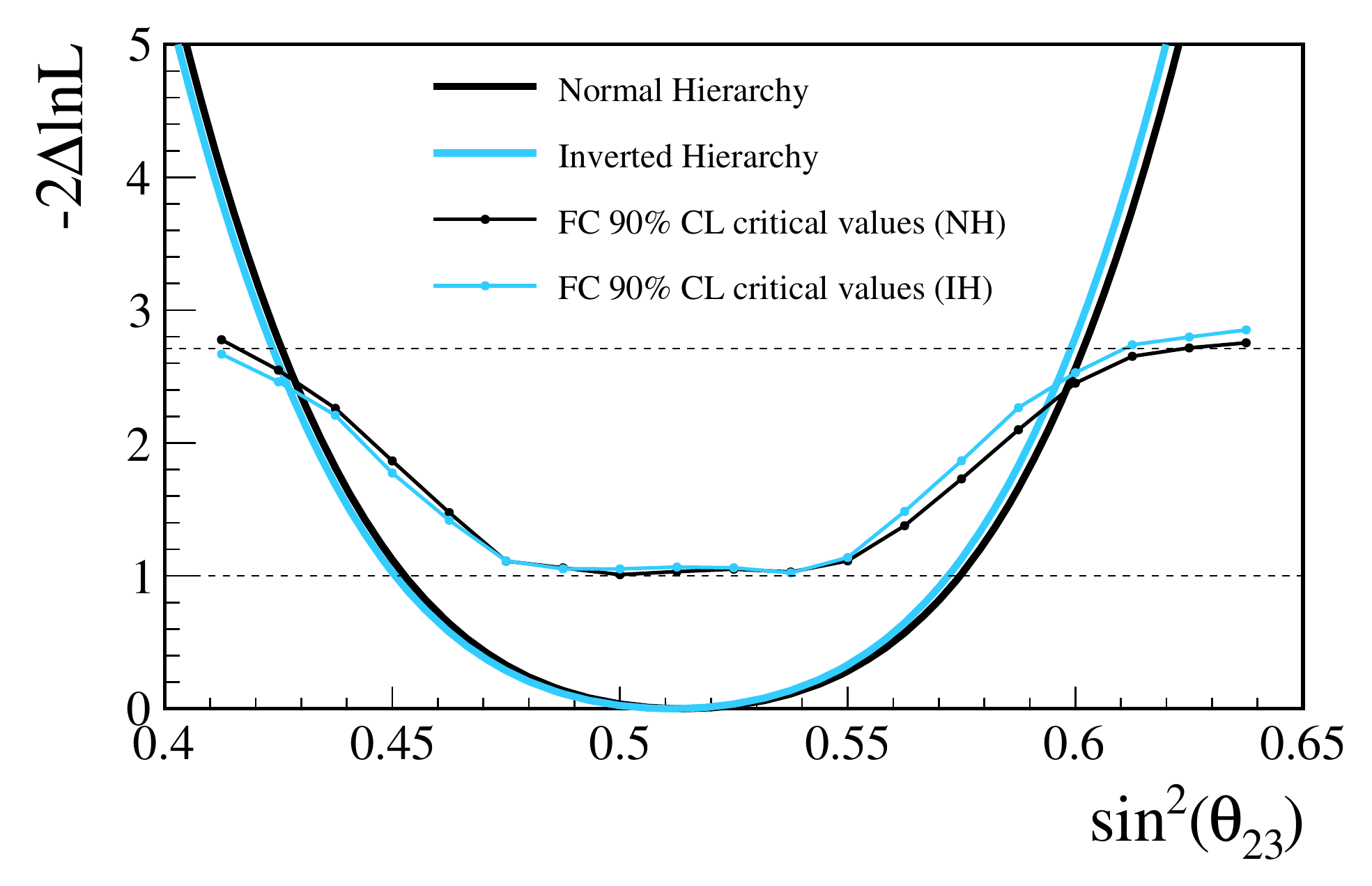}
  \includegraphics[width=0.48\textwidth]{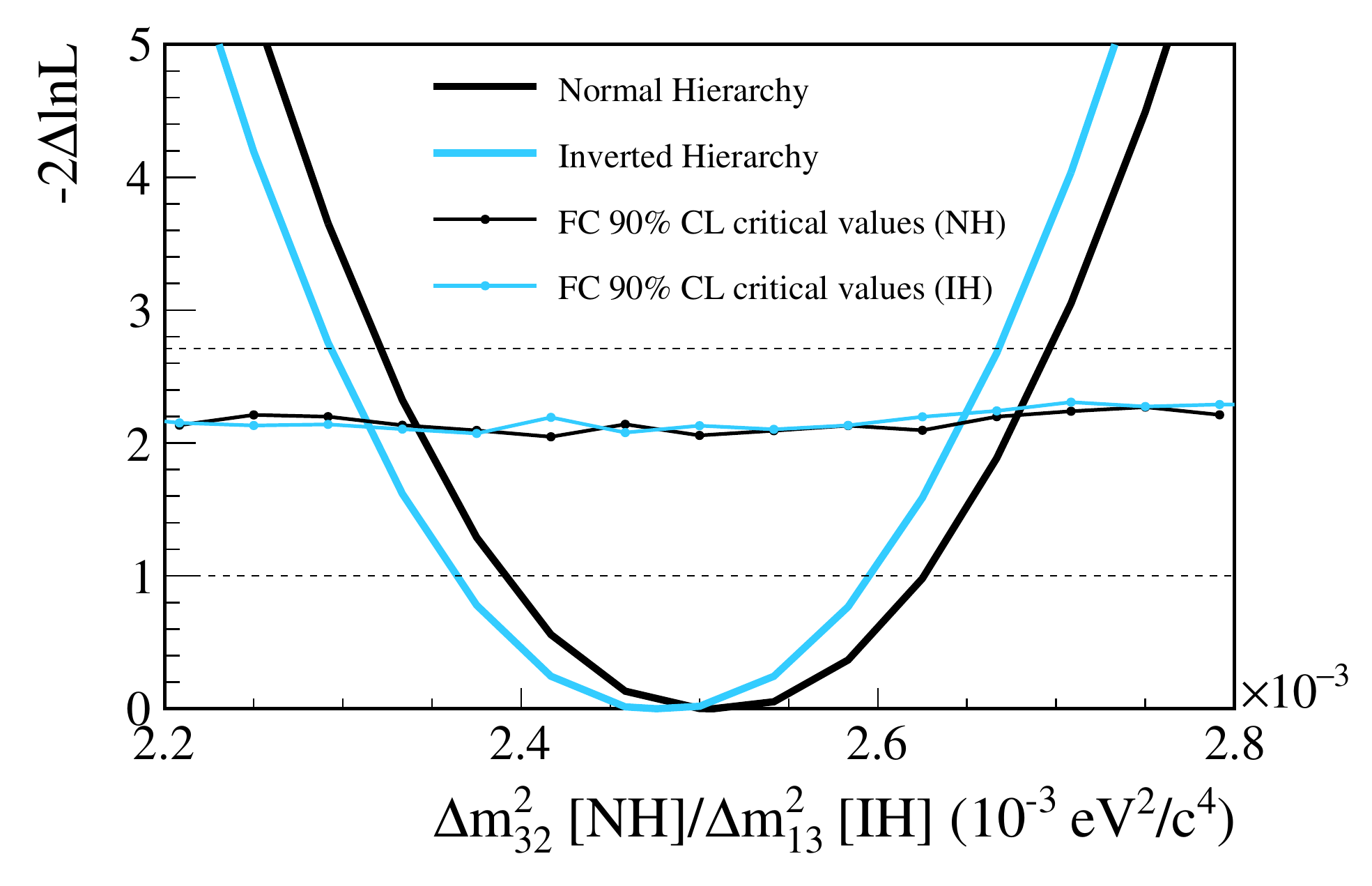}
  \caption{Profiled -2$\Delta\ln$L as a function of
    \stt\ (left) and \Dmsq\ (right), for the normal and inverted mass hierarchy assumptions.
    	The 90\% CL critical values are indicated by the lines with points.  }
  \label{fig:fc_1d_s23_normal}
\end{figure}

Both the M1 and M2 analyses find the point estimates
$\stt=0.514$ and $\dmsq=2.51\times10^{-3}$\,\evvcccc when assuming the normal mass
hierarchy and $\stt=0.511$ and $\dmsqo=2.48\times10^{-3}$\,\evvcccc when
assuming the inverted mass hierarchy. Table~\ref{tab:numu:summary}
summarizes these results from the M1 and M2 analyses. 
Likewise, the confidence intervals produced by M1 and M2 are similar. 
Since the M1 and M2
analyses are consistent with each other, only results from M1 are given below.
Figure~\ref{fig:numu_contour_nh_wsens} shows the best-fit values of the oscillation parameters, the
2D confidence intervals calculated using the Feldman and Cousins method, assuming normal and inverted hierarchy, and the
sensitivity at the current exposure.  The size of the confidence interval found by the fit to
the data is smaller than the sensitivity.  
This arises because the best-fit point is at the physical boundary corresponding to maximum disappearance probability.
The amount by which the region is smaller is not unusual in an ensemble of toy MC experiments produced under
the assumption of maximal disappearance.
The best-fit spectrum from the normal hierarchy fit compared to the observed spectrum is shown in
Figure~\ref{fig:numu_spectrum_result}, showing as well the
ratio of the number of observed events to the predicted number of
events with \stt$=0$. 
The observed oscillation dip is significant and well fit by simulation.
The calculated 1D Feldman and Cousins confidence intervals are given in Table~\ref{tab:numu:confidence_intervals}.
Figure~\ref{fig:fc_1d_s23_normal} shows the -2$\Delta\ln\mathcal{L}$
distributions for \stt\ and \Dmsq\ from the data, along with the
90\% CL critical values.

\begin{figure}[tbp]
  \includegraphics[width=0.65\textwidth]{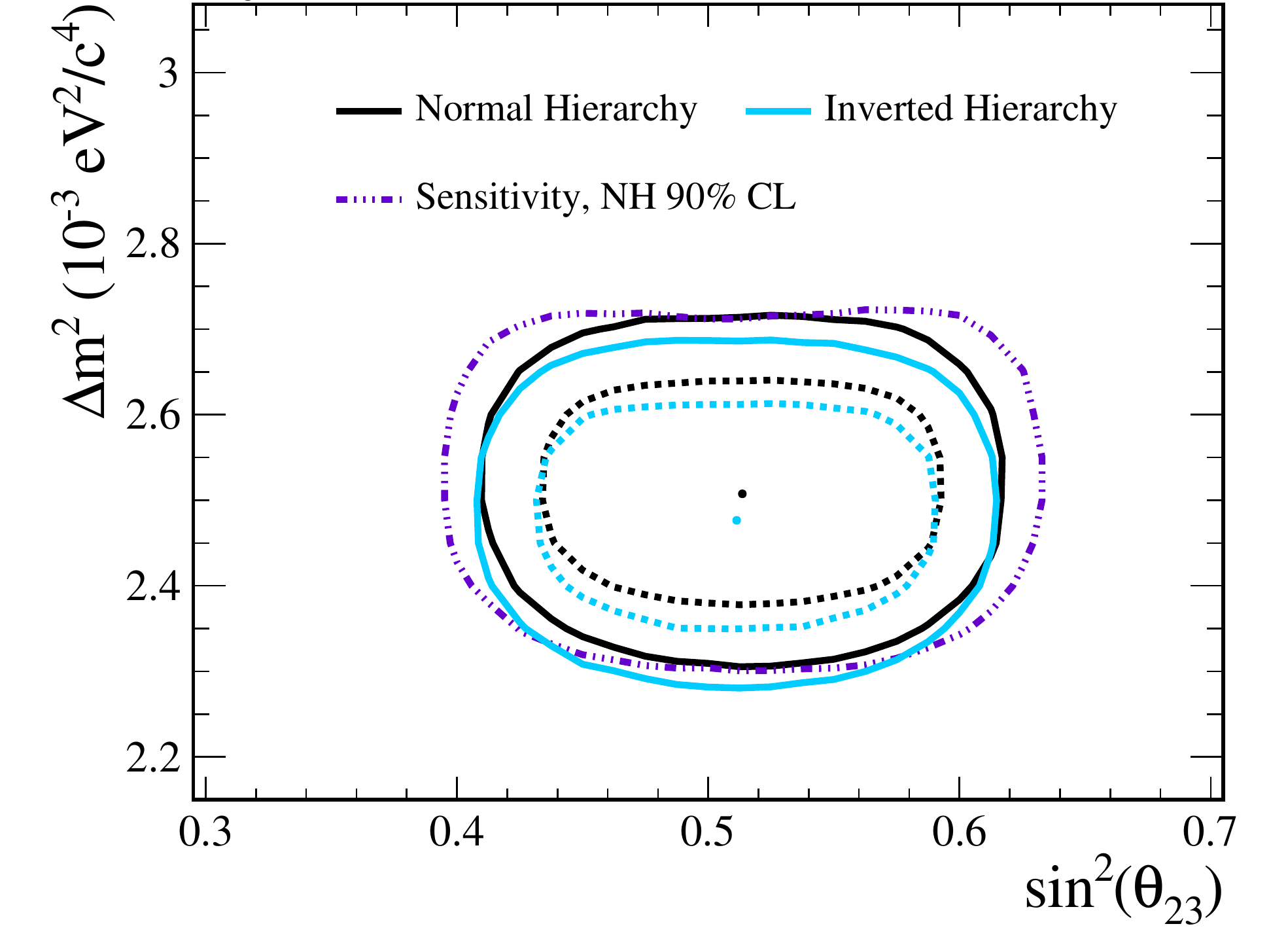}
  \caption{The 68\% (dashed) and 90\% (solid) CL intervals
    for the M1 \num-disappearance analysis assuming normal 
    and inverted  mass hierarchies. 
    The 90\% CL sensitivity contour for the normal
    hierarchy is overlaid for comparison. }
  \label{fig:numu_contour_nh_wsens}
\end{figure}

\begin{table}[tbp]
\centering
\caption{
Summary of the point estimates from the two independent 3-flavor muon neutrino disappearance oscillation frequentist analyses.
 }
\begin{tabular}{ c  c  c  c  c  }
\hline\hline
\ \ Analysis \ \ & \ \ MH \ \ & \ \ \dmsq or \dmsqo \ \ & \ \ \stt \ \  & \ \ N$_{exp}^{1R\mu}$ \ \ \\
              &        & $(10^{-3} \evvcccc)$   &         &      \\
\hline
M1 &  NH  & $2.51$ & $0.514$ & 121.4  \\
M1 &   IH  & $2.48$ & $0.511$  & 121.4  \\
\hline 
M2 &  NH  & $2.51$ & $0.514$ & 121.5  \\
M2 &   IH  & $2.48$ &  $0.511$ & 121.4  \\
\hline\hline
\end{tabular}
\label{tab:numu:summary}
\end{table}

\begin{table}[tbp]
\centering
\caption{68\% and 90\% confidence level intervals for the \num-disappearance analysis.}
\begin{tabular}{ c  c  c c }
\hline\hline
         & MH  &    68\% CL  & 90\% CL \\
\hline
\stt &  NH  & [0.458, 0.568] & [0.428, 0.598] \\
\stt &  IH  & [0.456, 0.566] & [0.427, 0.596]   \\
\hline 
\ \ \dmsq ($10^{-3}$\evvcccc) \ \ &  \ \ NH \ \ & \ \ [2.41, 2.61] \ \ &  \ \ [2.34, 2.68] \ \ \\
 \dmsqo ($10^{-3}$\evvcccc) &  IH  & [2.38, 2.58] & [2.31, 2.64]  \\
\hline\hline
\end{tabular}
\label{tab:numu:confidence_intervals}
\end{table}

\subsection{Multi-Nucleon Effects Study}

Recently, experimental~\cite{mb-ccqe, AguilarArevalo:2013hm, 
Fields:2013zhk, Fiorentini:2013ezn} and theoretical~\cite{Marteau:1999kt, Martini:2009, Carlson:2001mp, Shen:2012xz,Bodek:2011, Martini:2010, Martini:2013sha, Nieves:2005rq, Benhar:1994hw,Gran:2013kda, Nieves:2012yz, Lalakulich3, Martini2, Martini3, Meloni, Nieves:2012}
results have suggested that the
charged-current neutrino-nucleus scattering cross section at T2K
energies could contain a significant multi-nucleon component. Such
processes are known to be important in describing electron-nucleus
scattering (for a review, see~\cite{RevModPhys.80.189}), but have not yet been included in the model of
neutrino-nucleus interactions in our muon neutrino disappearance
analyses. If such multi-nucleon effects are important, their omission
could introduce a bias in the oscillation analyses. Since low energy
nucleons are not detected in \sk, such events can be selected in the QE
sample and assigned incorrect neutrino energies. 

\begin{figure}
  \includegraphics[width=0.45\textwidth]{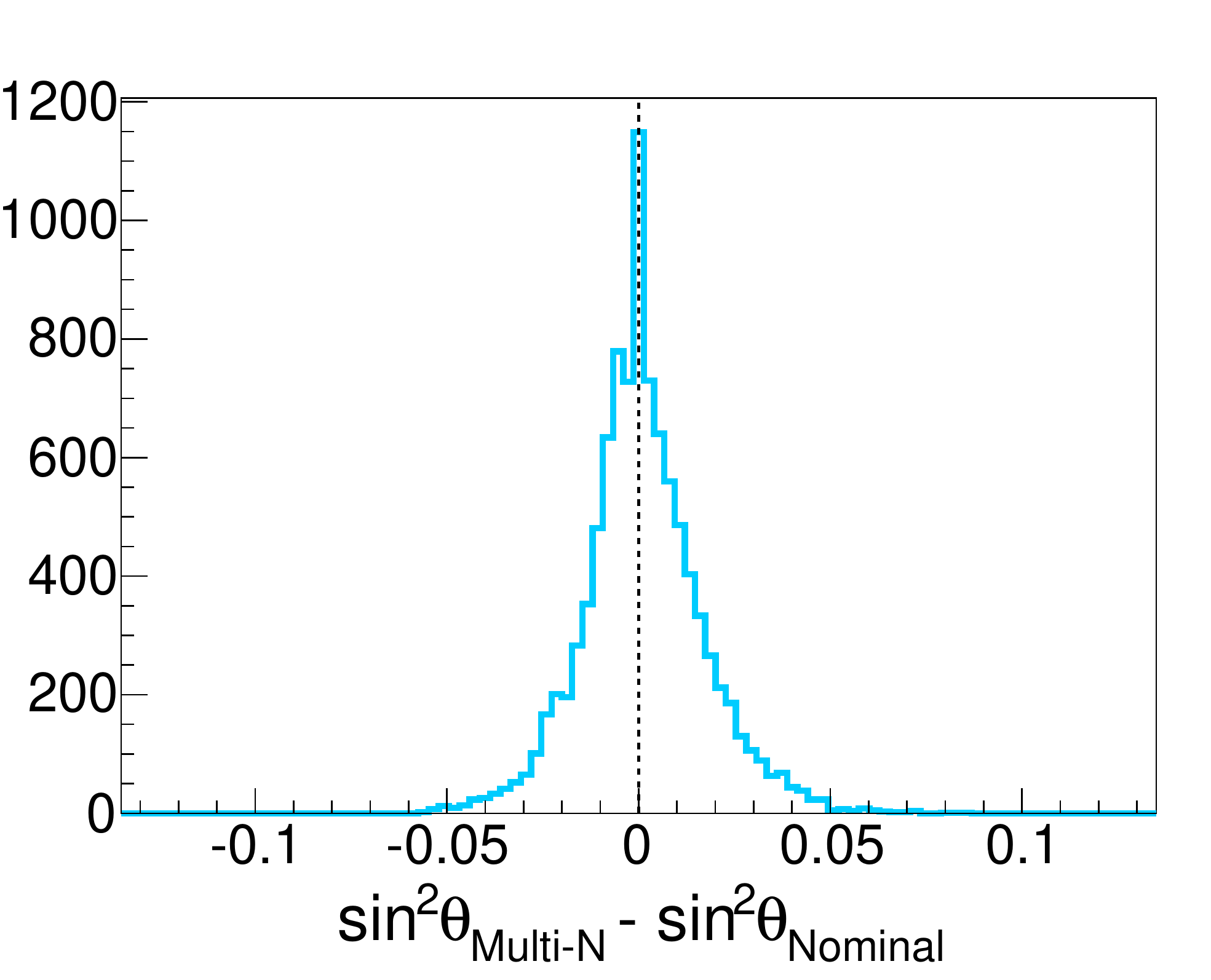}
  \includegraphics[width=0.45\textwidth]{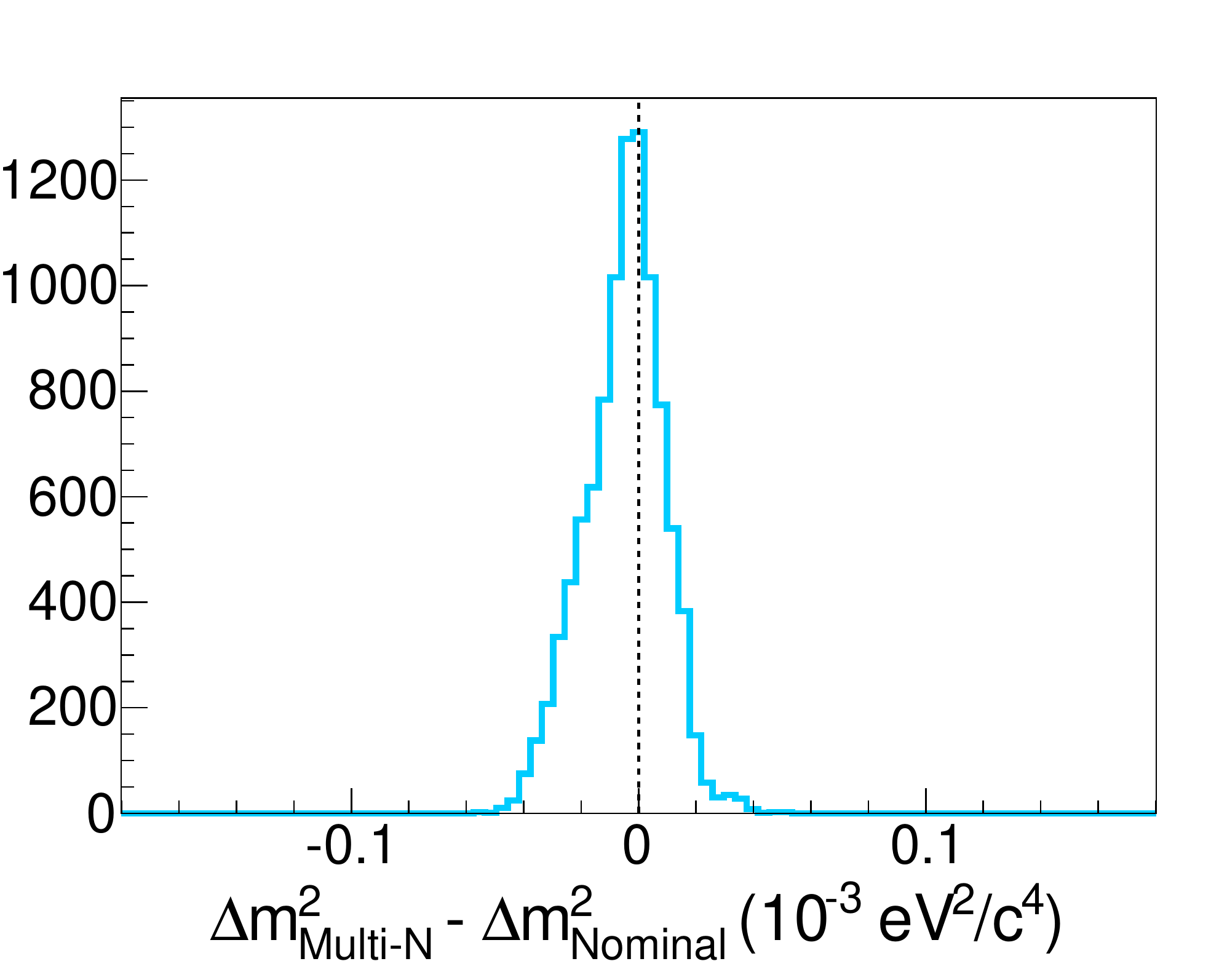}
  \caption{Difference in the point estimates of \stt (left) and \Dmsq (right)
    between pairs of toy MC datasets with and without including 
    multi-nucleon effects.}
  \label{fig:nieves_nom}
\end{figure}

A Monte Carlo study was performed in order to explore the sensitivity
of the analysis to multi-nucleon effects. The nominal interaction
model includes pion-less delta decay (PDD), which can be considered
to be a multi-nucleon effect. As an alternative, we turn off PDD and
use a model by Nieves~\cite{Nieves:2012} 
 to simulate multi-nucleon interactions for neutrino energies below 1.5~GeV.
Pairs of toy Monte Carlo experiments including both near and far
detector data were generated, one
with the nominal and one with the alternative model. Each dataset in
a pair 
was produced by using the same distribution of interacting neutrinos,
in order to reduce
statistical fluctuations in the comparison. 
Each pair of experiments used a different distribution of interacting neutrinos and a different set of systematic parameters sampled from
multivariate Gaussian distributions.
The complete analysis with near and far detector data is performed, 
assuming the nominal model in all cases.
In so doing, the study
properly accounts for the reduction in sensitivity to mis-modeling neutrino 
interactions when using near detector data to constrain flux and cross section parameters.
The differences in
the point estimates for the oscillation parameters for the two samples
in each pair are shown in Figure~\ref{fig:nieves_nom}. The overall bias for both
is negligible, compared to the precision obtained for the
parameters. However, the additional variation in \stt\ is about
3\%, comparable to the size of other systematic uncertainties. 
The bias was evaluated at $\stt=0.45$ to avoid the
the physical boundary at maximal disappearance which could reduce the size of the apparent bias.
For the
present exposure, the effect can be ignored, but future analyses will need
to incorporate multi-nucleon effects in their model of
neutrino-nucleus interactions.

\clearpage
\section{\label{sec:jointfreq} Joint \num disappearance and \nue appearance analysis using a frequentist approach.}
This section describes the joint 3-flavor oscillation analysis performed 
by combining the \num disappearance and \nue appearance channels
using a frequentist approach. 
The oscillation parameters,
$\vec{\theta}=\Dmsq$, \stt, \sot, and \dcp, described in Sec.~\ref{sec:OA:oscmodel},  are simultaneously determined. 
This is done by comparing the reconstructed energy spectra of the $\nu_\mu$ CC and $\nu_e$ CC candidate
events observed at SK, selected as described in Sec.~\ref{sec:SK}, with the predicted 
reconstructed energy spectra.
Point estimates 
of the oscillation parameters are found by minimizing the negative log-likelihood
\begin{equation}
\begin {split}
\displaystyle    
\chi^2 =  -2 \, \ln\mathcal{L} (\vec{\theta},\vec{g},\vec{x}_s,\vec{s}) = \ 
& 2N^p_\mu(\vec{\theta},\vec{g},\vec{x}_s,\vec{s})
- 2\sum_{i=1}^{N_{\mu\ \mathrm{bins}}} N^d_{\mu,i} \ln N^p_{\mu,i}(\vec{\theta},\vec{g},\vec{x}_s,\vec{s}) \\
+ \ & 2N^p_e(\vec{\theta},\vec{g},\vec{x}_s,\vec{s})
- 2\sum_{i=1}^{N_{e\ \mathrm{bins}}} N^d_{e,i} \ln N^p_{e,i}(\vec{\theta},\vec{g},\vec{x}_s,\vec{s}) \\
+ \ & \Delta\vec{g}^T V_g^{-1} \Delta\vec{g}
+ \Delta\vec{x}_s^T V_{xs}^{-1} \Delta\vec{x}_s
+ \Delta\vec{s}^T V_s^{-1} \Delta\vec{s} \ .
\label{eq:jointfreq:chisquare}
\end{split}
\end{equation} 
where $N^d_{\mu,i}$ ($N^d_{e,i}$) is the observed number of $\nu_\mu$ CC ($\nu_e$ CC) candidate events in the $i^{th}$ reconstructed energy bin, and $N^p_{\mu,i}$ ($N^p_{e,i}$)
is the corresponding predicted number of events, calculated as a function of the oscillation parameters $\vec{\theta}$ and 
the vectors of systematic parameters, $\vec{g},\vec{x}_s,\vec{s}$, as described for
Eq.~(\ref{eq:numu_M2_likelihood}).

The negative log-likelihood function is minimized using MINUIT. 
As explained in Sec.~\ref{sec:OA}, the solar oscillation parameters are kept fixed for this analysis.
To combine our measurement with the reactor measurements, we add the term, 
\begin{equation}
\chi^2_{reactor} = \Bigg( \frac{\sot - (\sot)_{reactor} }{\sigma_{reactor}} \Bigg)^2  \ ,
\label{eq:jointfreq:reactor}
\end{equation}
where  (\sot)$_{reactor}$ and $\sigma_{reactor}$ are given in Sec.~\ref{sec:OA:osc}.

When maximizing the likelihood, the systematic parameters are allowed to vary in a wide range
[-5$\sigma$, +5$\sigma$] (where $\sigma$ is the square root of the corresponding diagonal element in the covariance matrix), with the exception of the spectral function parameter which is constrained to lie between 0 (RFG) and 1 (SF).
A total of 64 systematic parameters, representing uncertainties in the far detector efficiencies,
the reconstructed neutrino energy scale, final state and secondary interactions, the flux prediction,
and the relevant neutrino interaction models, are considered. As with the disappearance analyses, the fit to the ND280 near detector data described in Sec.~\ref{sec:BANFF}
is applied as a multivariate Gaussian penalty term to constrain the flux uncertainties and cross sections common to the near and far detectors. 

The 1-dimensional limits and 2-dimensional confidence regions reported in this analysis are constructed
using the constant \Dchisq method~\cite{PDG2012} with respect to a 4-dimensional best-fit point obtained by
minimizing Eq.~(\ref{eq:jointfreq:chisquare}).  
An exception is the (\sot, \dcp) space without the reactor measurement,
as that analysis has little power to constrain \dcp.
For that case, a best-fit value of \sot\ is found for fixed values of \dcp\ in the interval [-$\pi$, $\pi$] (divided into 51 bins), resulting in 1-dimensional confidence
regions for different values of \dcp\ with respect to a line of best-fit points.
For the T2K data fit combined with the reactor constraint, described in Sec.~\ref{sec:jointfreq:results_reactor},
the Feldman and Cousins method~\cite{PhysRevD.57.3873}
is used to produce confidence intervals, by finding critical values of \Dchisq\ as a function of \dcp\ and we report excluded regions for \dcp.

\subsection{\label{sec:jointfreq:results} Results}
Point estimates for the oscillation parameters
and the expected number of events are summarized in Tab.~\ref{tab:jointfreq:bestfit}.  Notably, the value obtained for \sot\ by T2K is larger
than the value found by the reactor experiments, the best-fit value of \stt\ is consistent with maximal disappearance, and the difference in \Dchisq\ between the solutions 
for each mass hierarchy is negligible. 

\begin{table}[tbp]
\centering
\caption{
  Point estimates of the oscillation parameters for the joint 3-flavor oscillation frequentist analysis.
 }
\begin{tabular}{ c c  c  c  c c c c  }
\toprule
\ \ MH \ \ &
\ \ \dmsq\ or \dmsqo \ \ & \ \ \stt \ \ & \ \ \sot \ \ & \ \ \dcp \ \ &  \ \ N$_{exp}^{1R\mu}$ \ \ & \ \ N$_{exp}^{1Re}$ \ \ & \ \ \Dchisq \ \ \\
& $(10^{-3} \evvcccc)$ & & & & & & \\
\hline
  NH  & 2.51 & 0.524 & 0.0422 & 1.91  & 119.9  & 28.00 &  0.01\\
  IH  & 2.49 & 0.523 & 0.0491 & 1.01  & 119.9  & 28.00 &  0.00\\
\botrule
\end{tabular}
\label{tab:jointfreq:bestfit}
\end{table}

\begin{figure}[tbp]
\begin{center}
\includegraphics[width=0.9\textwidth]{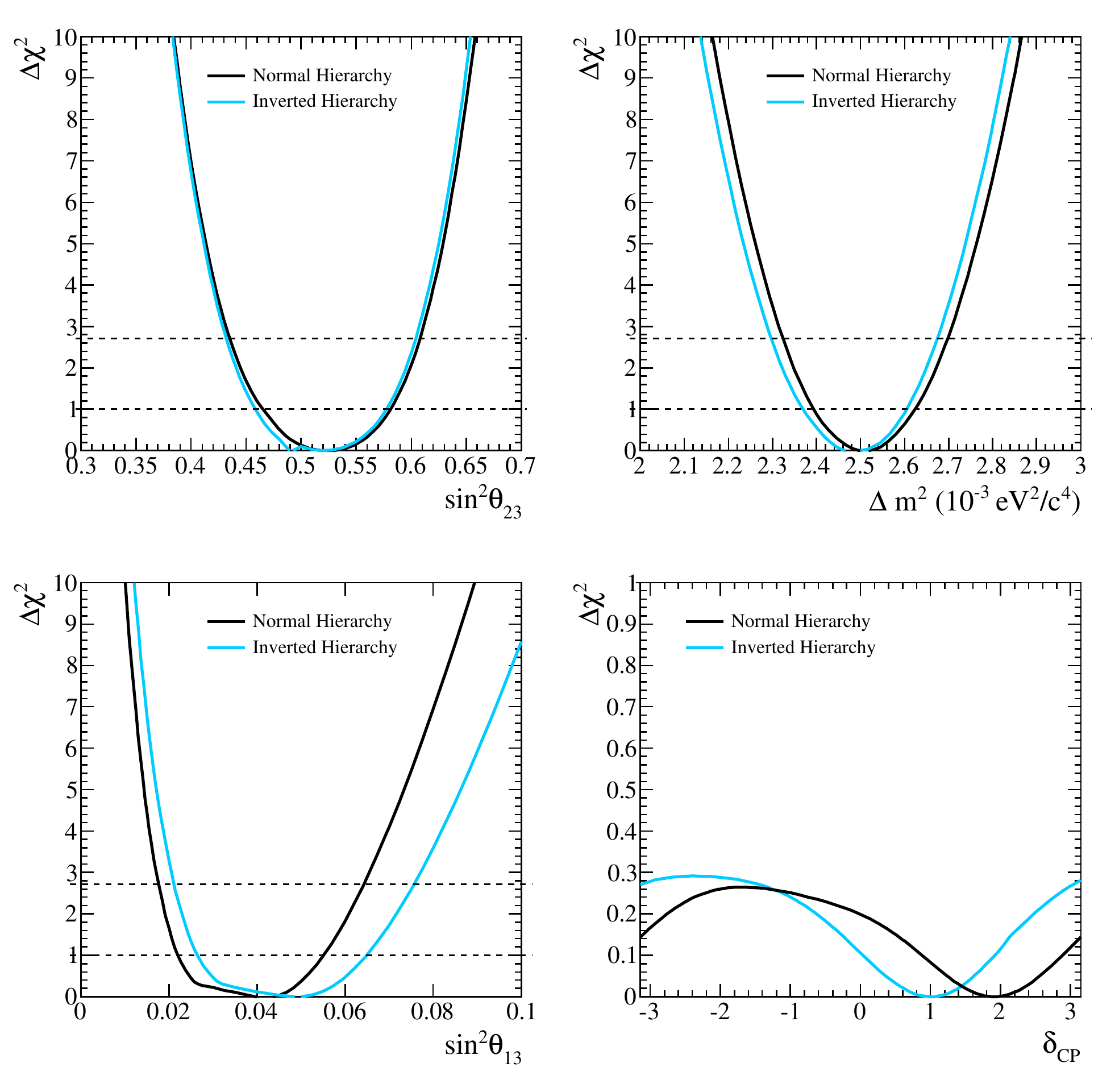}
\caption {
  Profiled \Dchisq  
 for the joint 3-flavor oscillation analysis without using reactor data. 
The parameter $\Dmsq$ represents \dmsq\ or \dmsqo\ for normal and inverted mass hierarchy assumptions respectively.
The horizontal lines show the critical \Dchisq values for one dimensional fits at the 68 \% and 90 \% CL (\Dchisq = 1.00 and 2.71 respectively).
}
\label{fig:jointfreq:1D}
\end{center}
\end{figure}

The profiled \Dchisq of each oscillation parameter was obtained by
minimizing the negative log-likelihood with respect to the systematic parameters and other three oscillation parameters using MINUIT.
Figure \ref{fig:jointfreq:1D} presents the profiled \Dchisq of each oscillation parameter, 
comparing the results for the normal and the inverted mass hierarchy.
From these figures, the 1$\sigma$ intervals estimated using the $\Dchisq=1$ criterion are:
\begin{center}
$\stt = 0.524^{+0.057}_{-0.059}$ (NH) \ \ 
$\stt = 0.523^{+0.055}_{-0.065}$ (IH)

$\sot = 0.042^{+0.013}_{-0.021}$ (NH) \ \ 
$\sot = 0.049^{+0.015}_{-0.021}$ (IH)

$\dmsq = 2.51^{+0.11}_{-0.12}$ ($10^{-3}\evvcccc$, NH) \ \ 
$\dmsqo = 2.49^{+0.12}_{-0.12}$ ($10^{-3}\evvcccc$, IH).
\end{center}

\begin{figure*}
\centering
\includegraphics[width=0.5\textwidth]{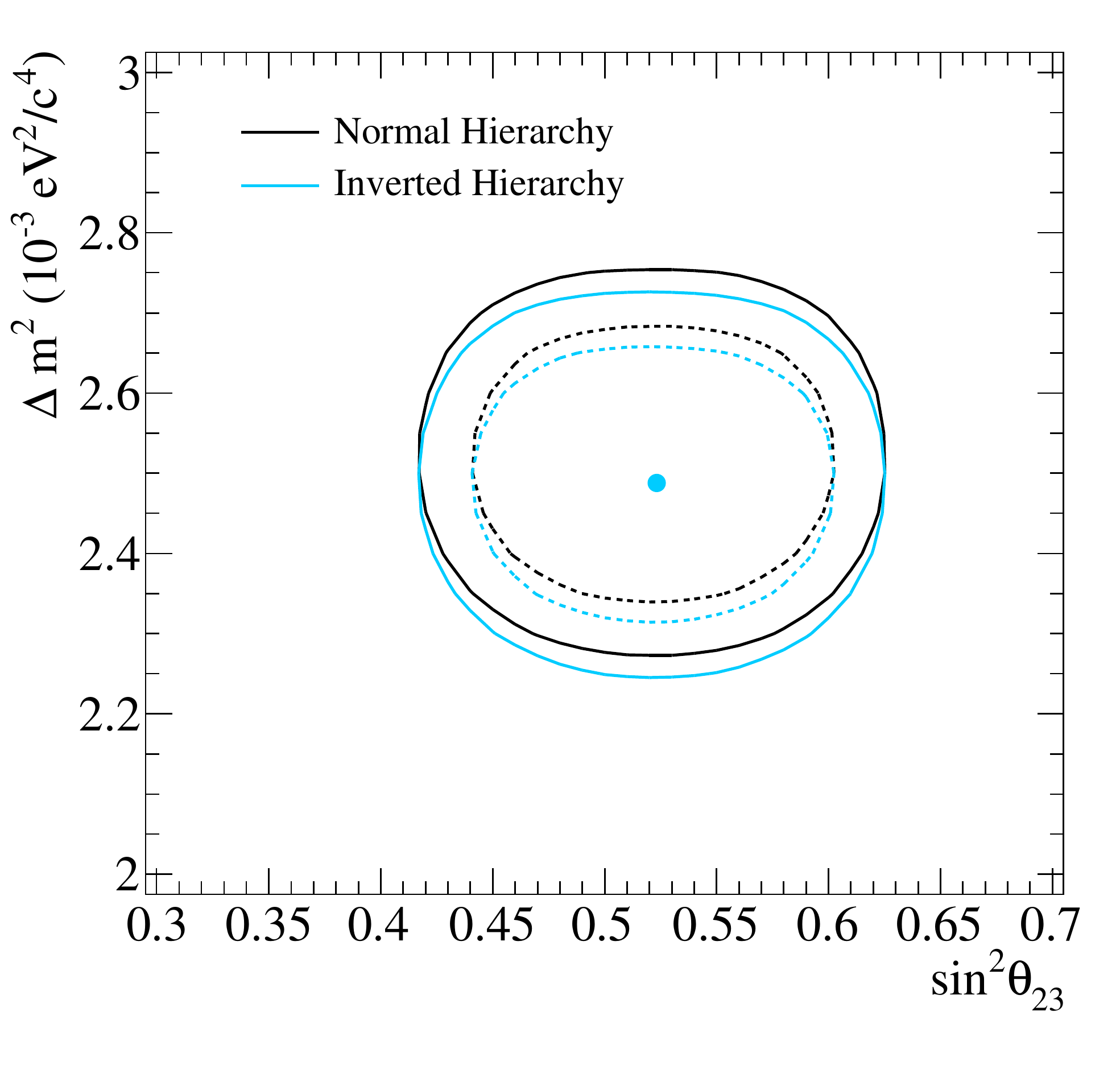}\includegraphics[width=0.5\textwidth]{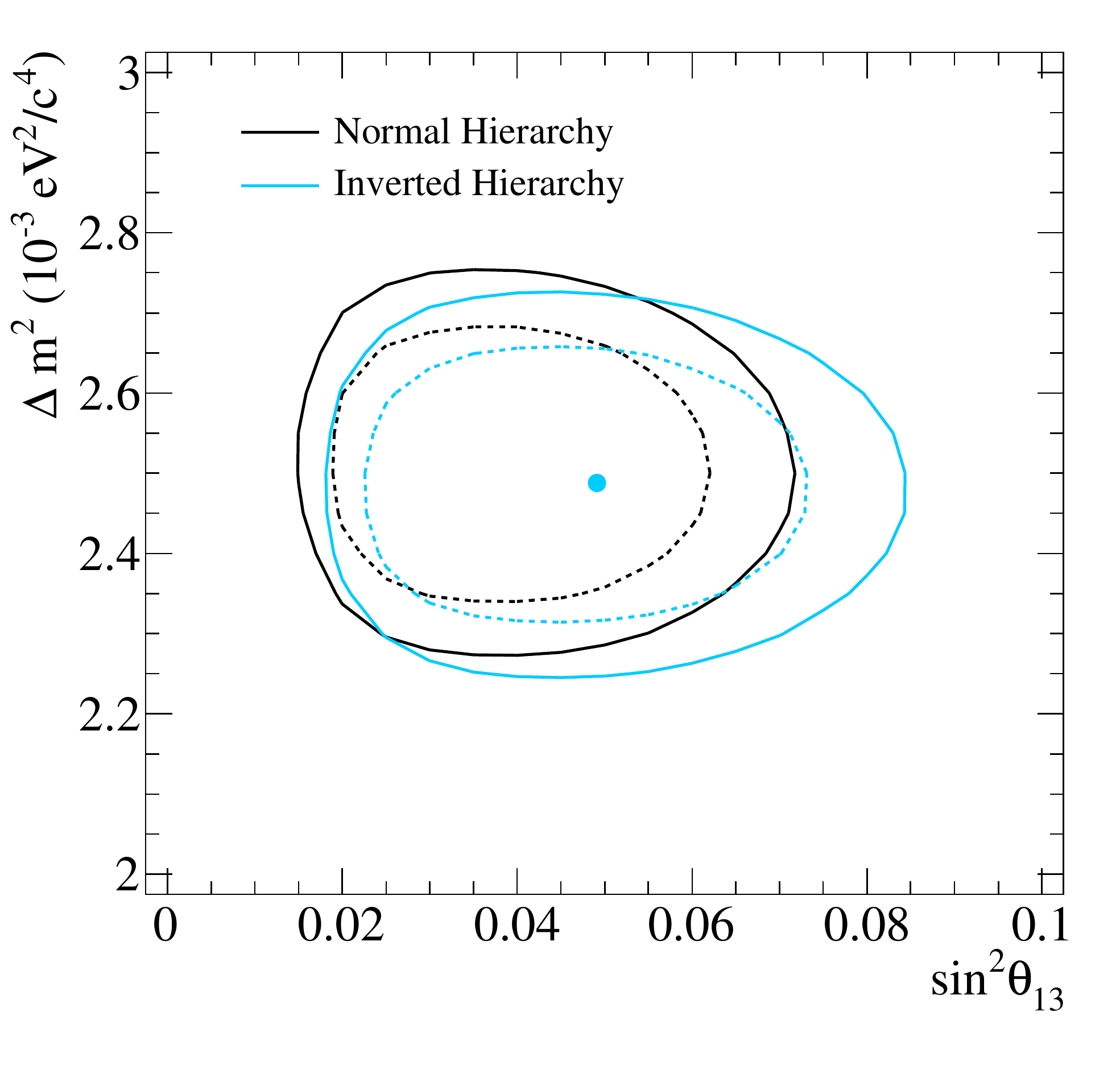}
\includegraphics[width=0.5\textwidth]{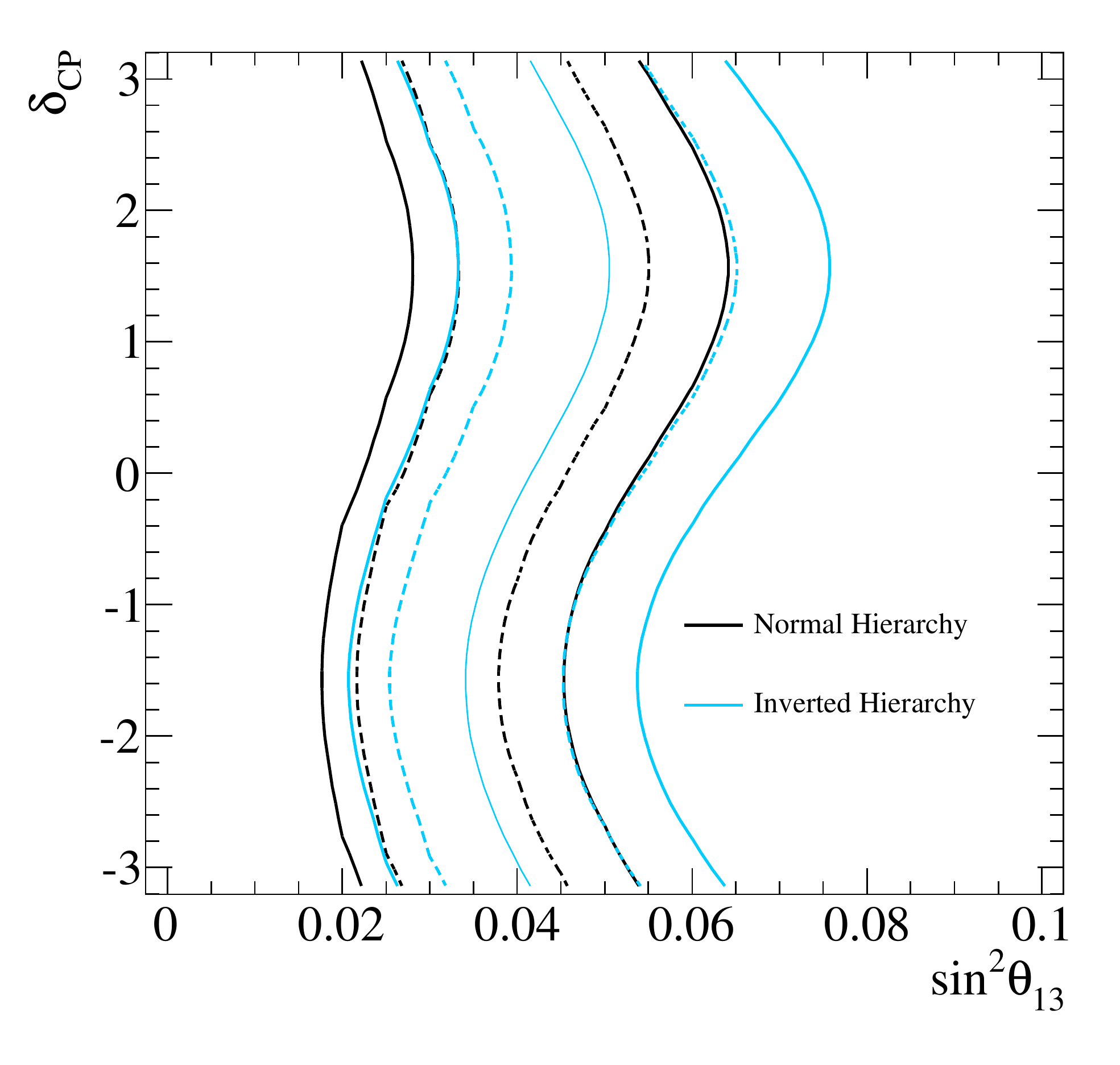}\includegraphics[width=0.5\textwidth]{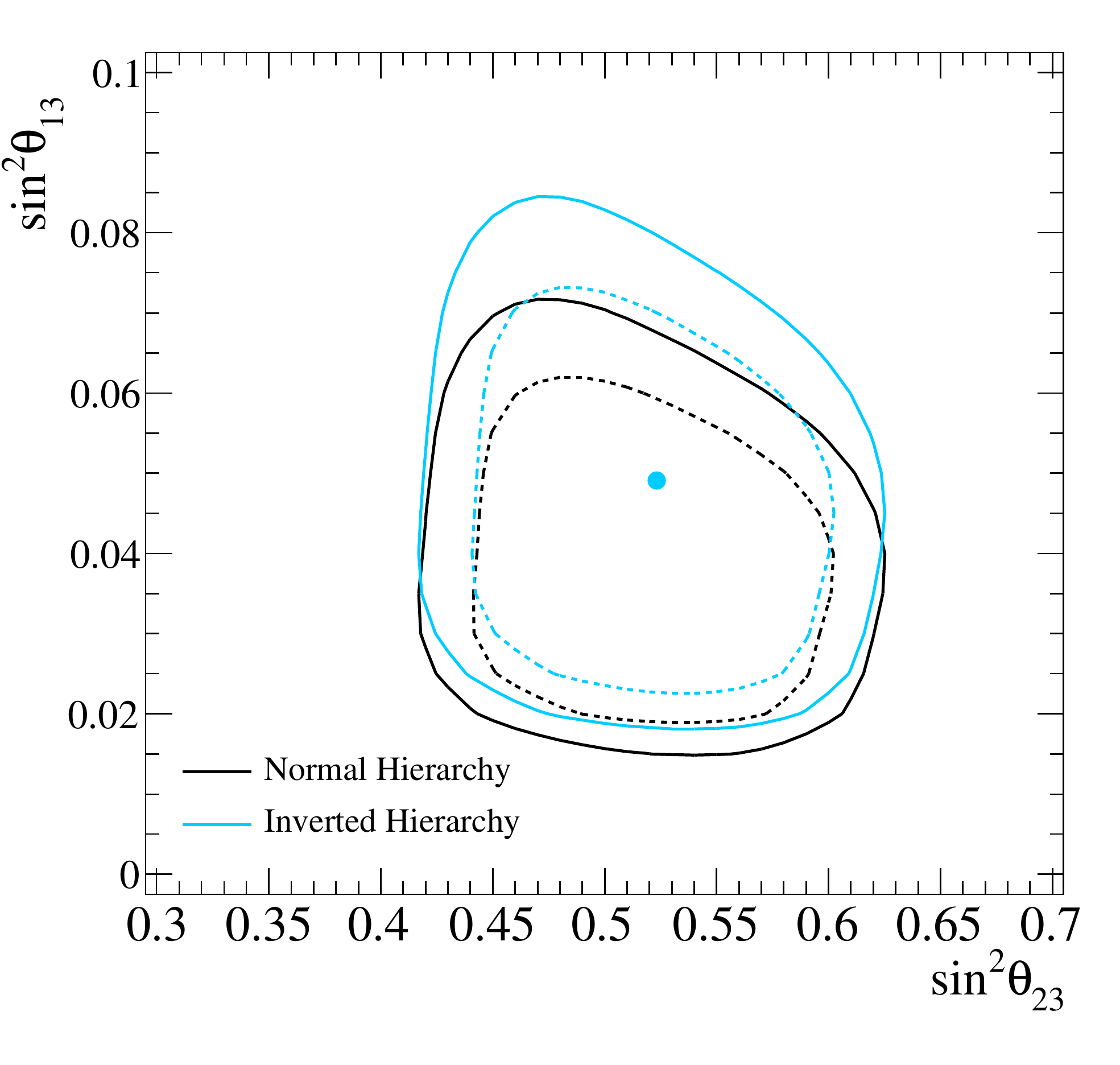}
\caption {
68\% (dashed) and 90\% (solid) CL regions, from the analysis without using reactor data,
with different mass hierarchy assumptions using \Dchisq\ with respect to the
best-fit point -- that from the inverted hierarchy.
The parameter $\Dmsq$ represents \dmsq\ or \dmsqo\ for normal and inverted mass hierarchy assumptions respectively.
The lower left plot shows 1D confidence intervals in \sot\ for different values of \dcp.
}
\label{fig:jointfreq:2D}
\end{figure*}

Figure~\ref{fig:jointfreq:2D} presents
the 68\% and 90\% CL regions 
for the two mass hierarchy assumptions 
in the four 2-dimensional oscillation 
parameter spaces
(\stt, \dmsq),
(\sot, \dmsqo),
(\sot, \dcp), and
(\stt, \sot),
constructed using constant \Dchisq\ with respect to the inverted hierarchy best-fit point.

\subsection{\label{sec:jointfreq:results_reactor} Results for T2K combined with the reactor experiment result}
The point estimates for the oscillation parameters and the predicted number of events, when the reactor measurements are included in the likelihood function, are given in Tab.~\ref{tab:jointfreq:bestfit_reactor}. 
The estimate for \sot\ is smaller than the result obtained with T2K data only, shown in Tab.~\ref{tab:jointfreq:bestfit}. 
The likelihood is maximum for normal mass hierarchy and for $\dcp=-\pi/2$, where the appearance probability is largest, as shown in Fig.~\ref{fig:oscprob}.

\begin{table}[tbp]
\centering
\caption{
  Point estimates of the oscillation parameters for the joint 3-flavor oscillation frequentist analysis
  combined with the results from reactor experiments.
 }
\begin{tabular}{c c  c  c  c c c c  }
\toprule
\ \ MH \ \ &
\ \ \dmsq\ or \dmsqo \ \ & \ \ \stt \ \ & \ \ \sot \ \ & \ \ \dcp \ \ &  \ \ N$_{exp}^{1R\mu}$ \ \ & \ \ N$_{exp}^{1Re}$ \ \ & \ \ \Dchisq \ \ \\
& $(10^{-3} \evvcccc)$ & & & & & & \\
\hline
  NH  & 2.51 & 0.527 & 0.0248 & -1.55  & 120.4  & 25.87 &  0.00\\
  IH  & 2.48 & 0.533 & 0.0252 & -1.56  & 121.2  & 23.57 &  0.86 \\
\botrule
\end{tabular}
\label{tab:jointfreq:bestfit_reactor}
\end{table}

\begin{figure}[tbp]
\begin{center}
\includegraphics[width=0.9\textwidth]{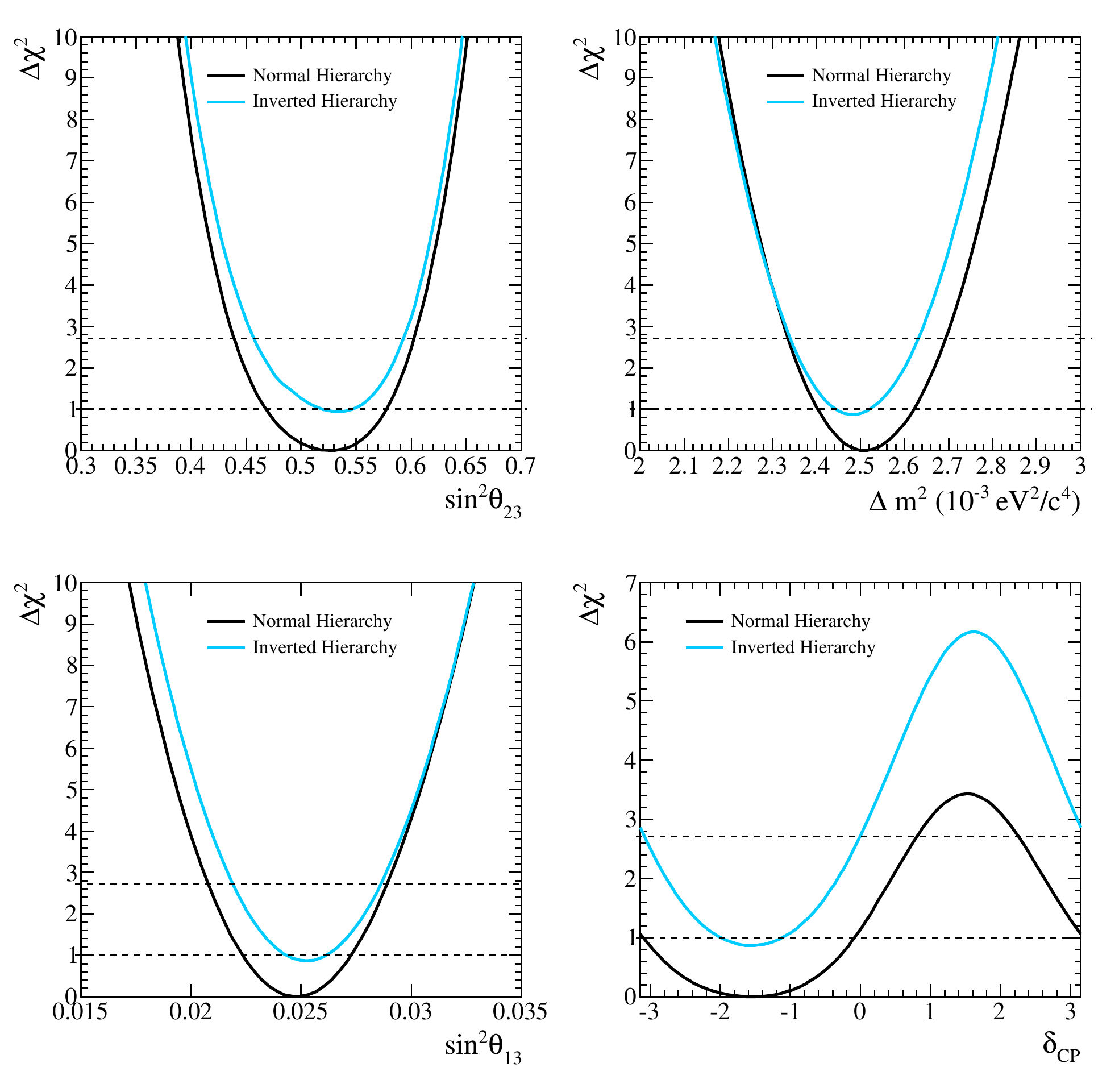}
\caption {
  Profiled \Dchisq  
 for the joint 3-flavor oscillation analysis 
 combined with the results from reactor experiments.
The parameter $\Dmsq$ represents \dmsq\ or \dmsqo\ for normal and inverted mass hierarchy assumptions respectively.
The horizontal lines show the critical \Dchisq\ values for one dimensional fits at the 68 \% and 90 \% CL (\Dchisq = 1.00 and 2.71 respectively).
}
\label{fig:jointfreq:1D_reactor}
\end{center}
\end{figure}

\begin{figure*}[tbp]    
\centering
\includegraphics[width=0.5\textwidth]{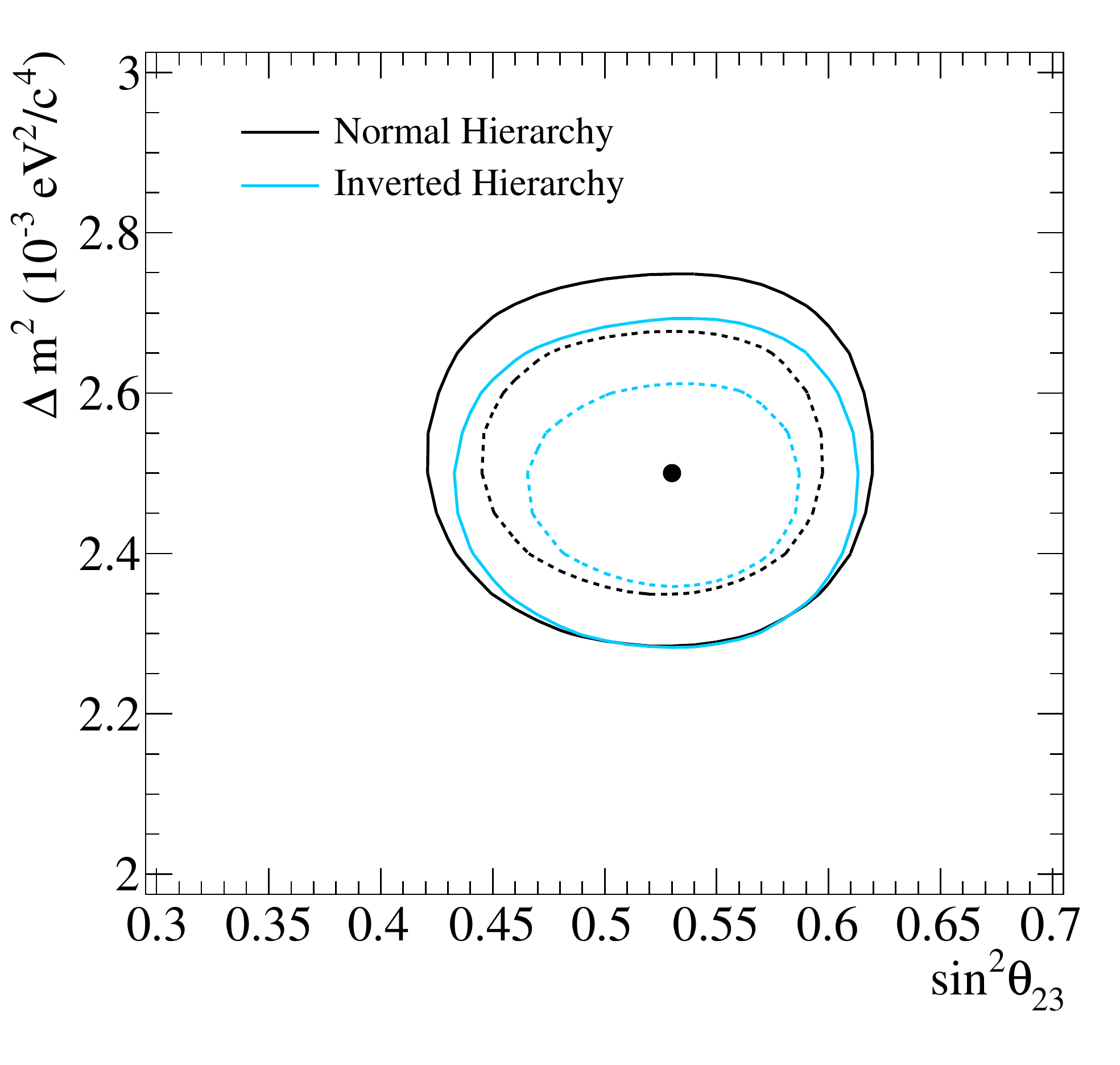}\includegraphics[width=0.5\textwidth]{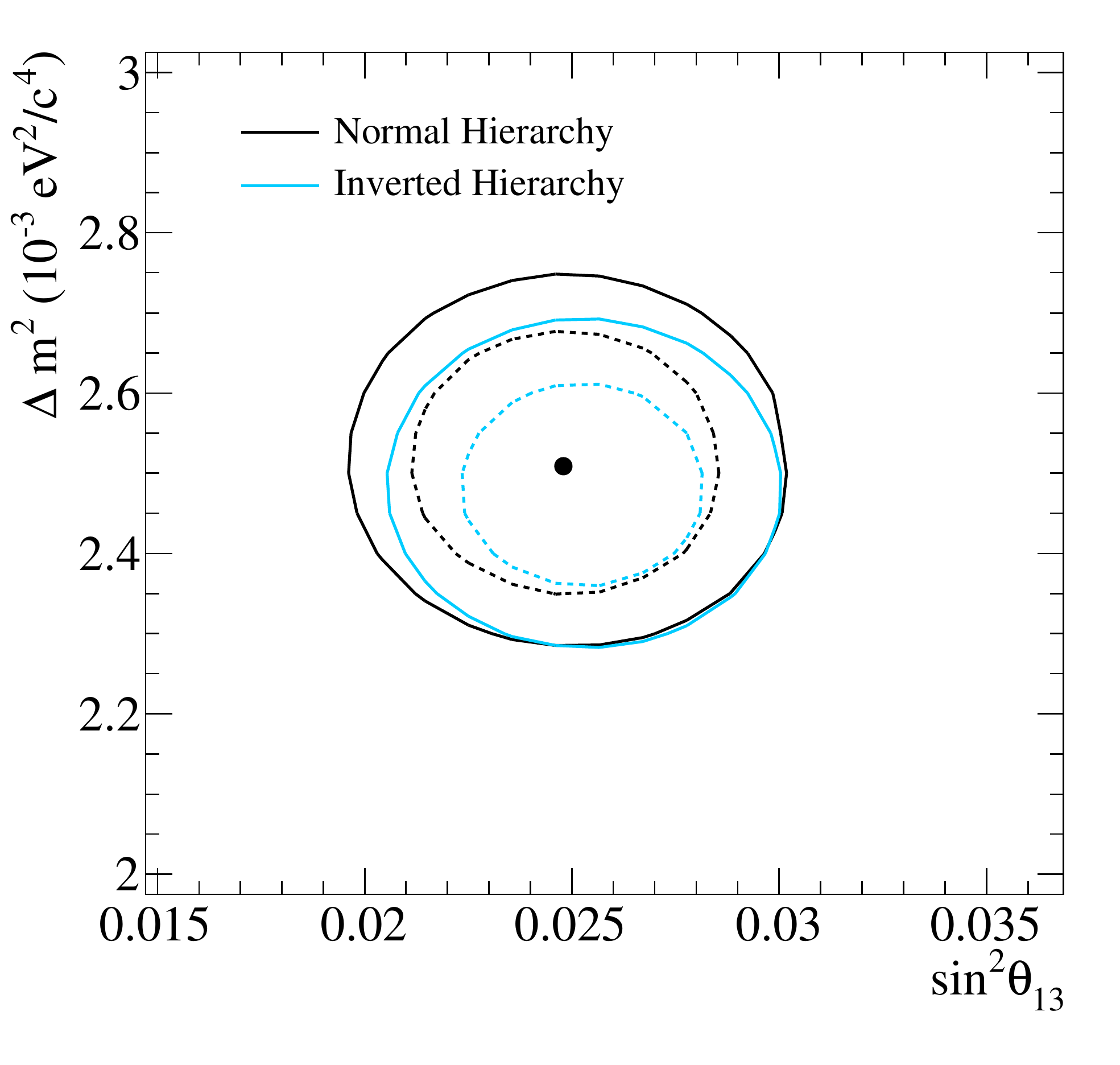}
\includegraphics[width=0.5\textwidth]{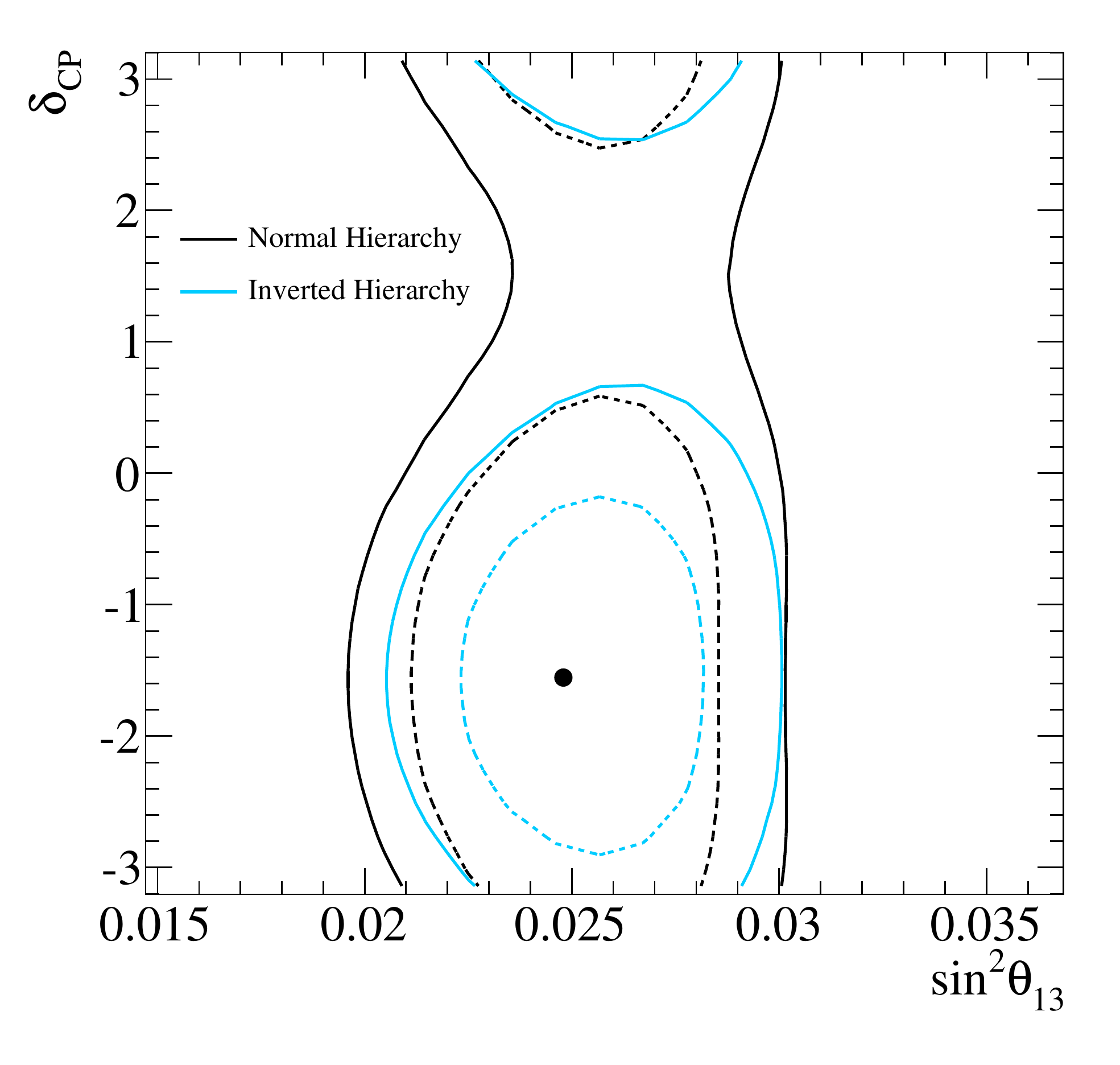}\includegraphics[width=0.5\textwidth]{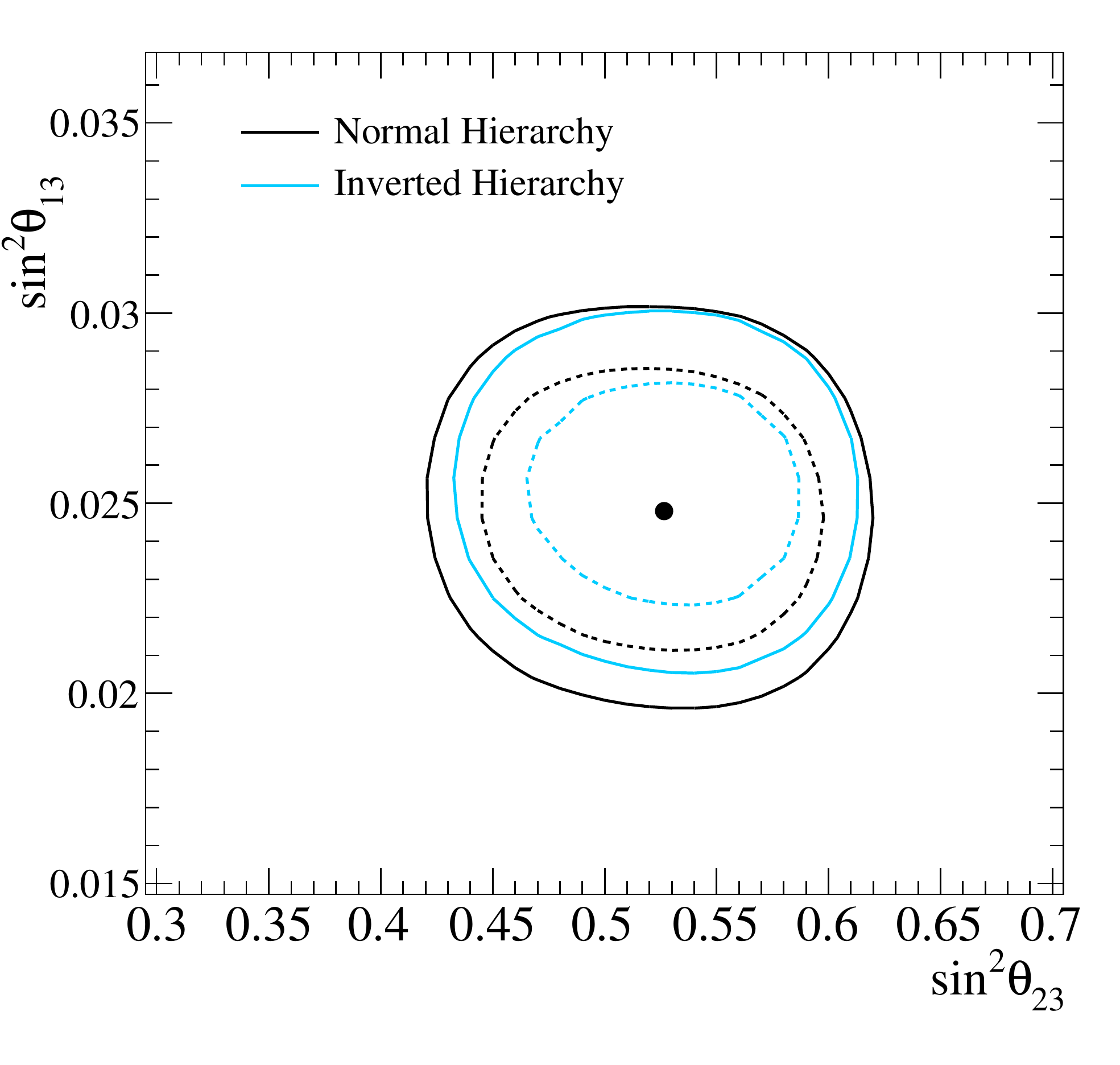}
\caption {
68\% (dashed) and 90\% (solid) CL regions from the analysis that includes
results from reactor experiments
with different mass hierarchy assumptions using \Dchisq\ with respect to the best-fit point, the one from the fit with normal hierarchy.
The parameter $\Dmsq$ represents \dmsq\ or \dmsqo\ for normal and inverted mass hierarchy assumptions respectively.
}
\label{fig:jointfreq:2D_reactor}
\end{figure*}

\begin{figure}[tbp]
\begin{center}
\includegraphics[width=0.5\textwidth]{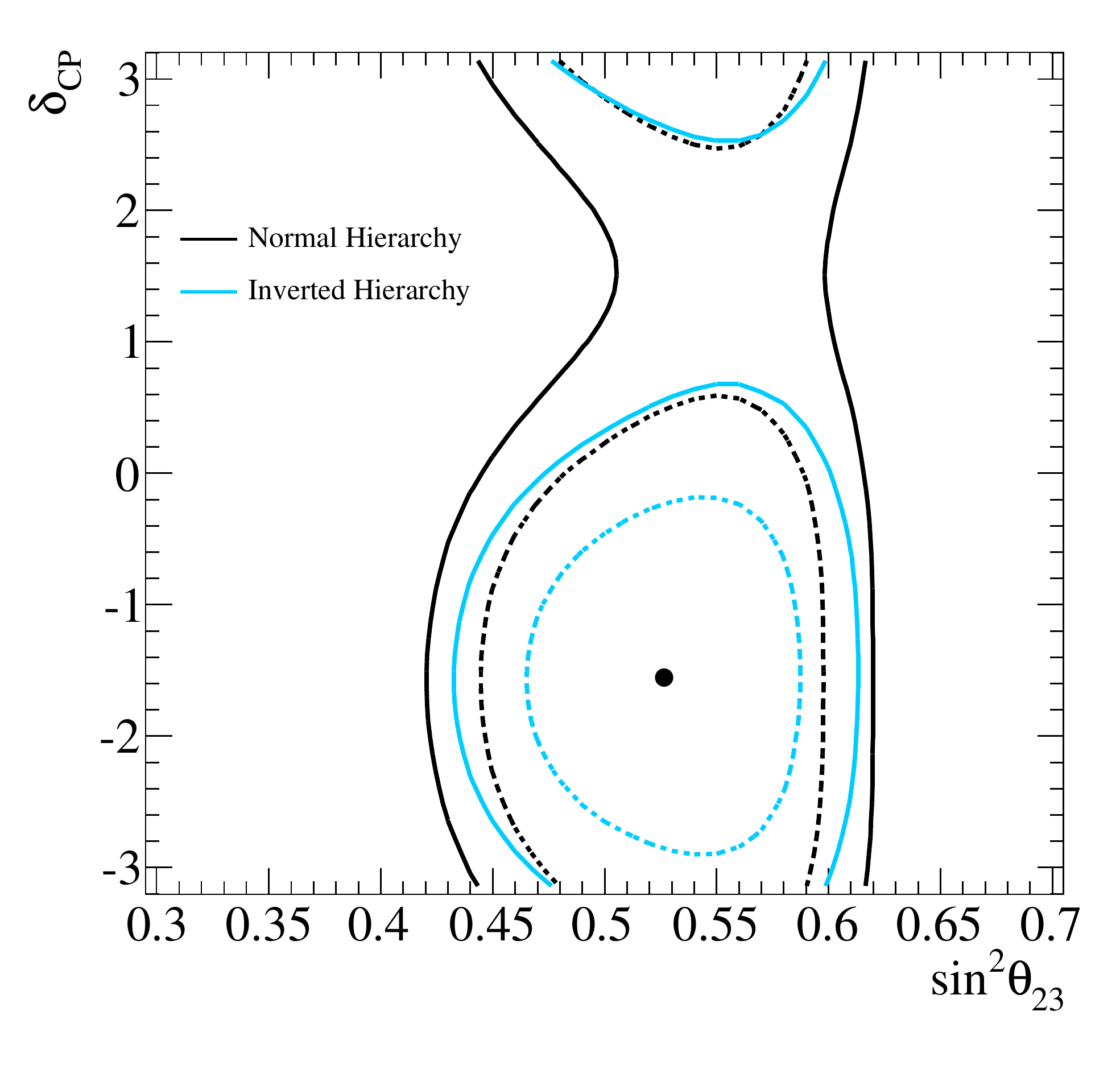}\includegraphics[width=0.5\textwidth]{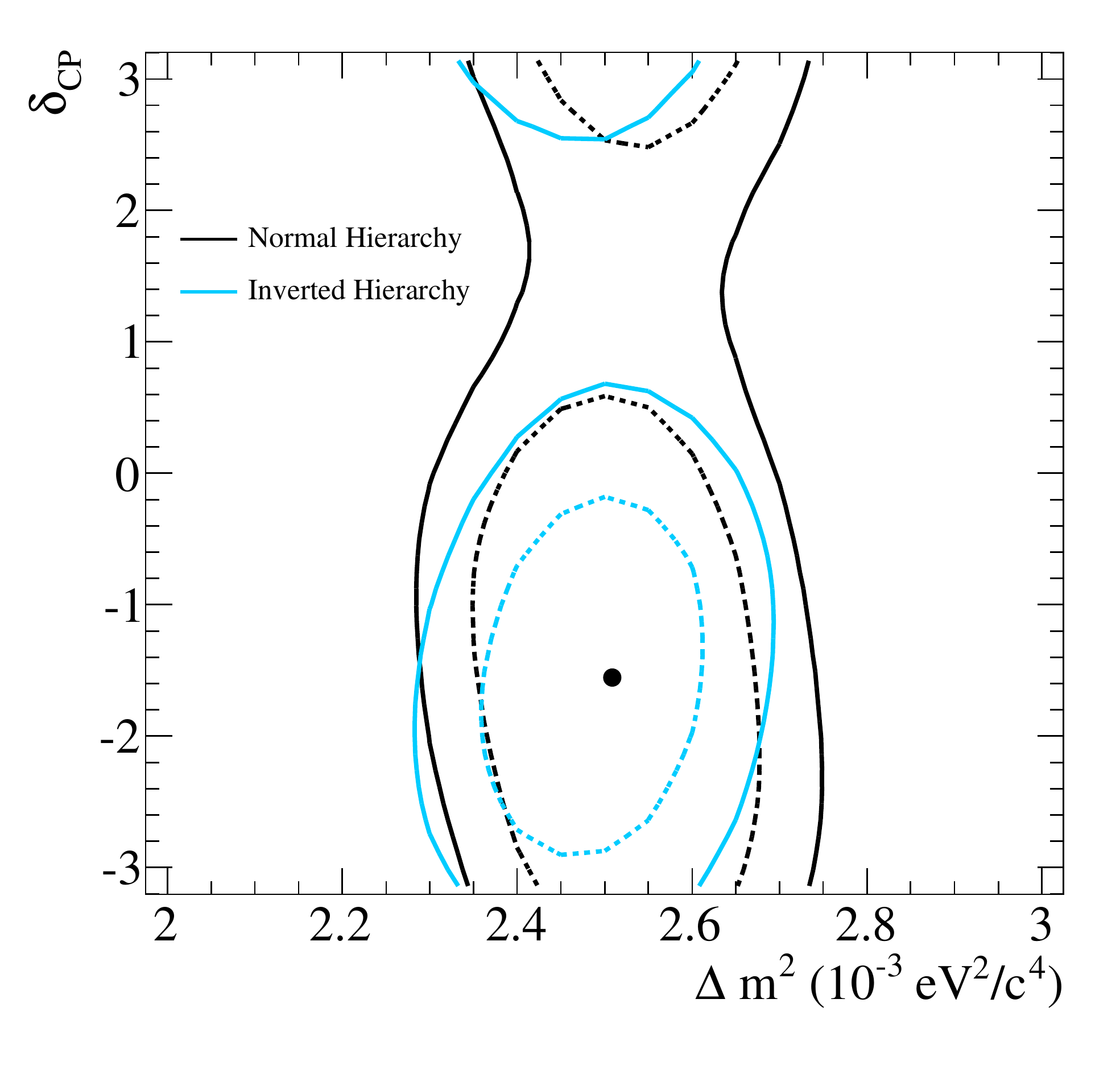}
\caption {
Comparison of 68\% (dashed) and 90\% (solid) CL regions
combined with the results from reactor experiments
with different mass hierarchy assumptions using \Dchisq\ with respect to the best-fit point, the one from the fit with normal hierarchy.
The parameter $\Dmsq$ represents \dmsq\ or \dmsqo\ for normal and inverted mass hierarchy assumptions respectively.
}
\label{fig:jointfreq:2D_reactor2}
\end{center}
\end{figure}

The profiled \Dchisq as a function of each oscillation parameter are presented in Fig.~\ref{fig:jointfreq:1D_reactor},
and the 68\% and 90\% CL regions for the two mass hierarchies constructed using \Dchisq\ with respect to the best-fit point, the one for the normal hierarchy,
are presented in Figs.~\ref{fig:jointfreq:2D_reactor} and \ref{fig:jointfreq:2D_reactor2}.

\begin{figure}[tbp]
\begin{center}
\includegraphics[width=0.6\textwidth]{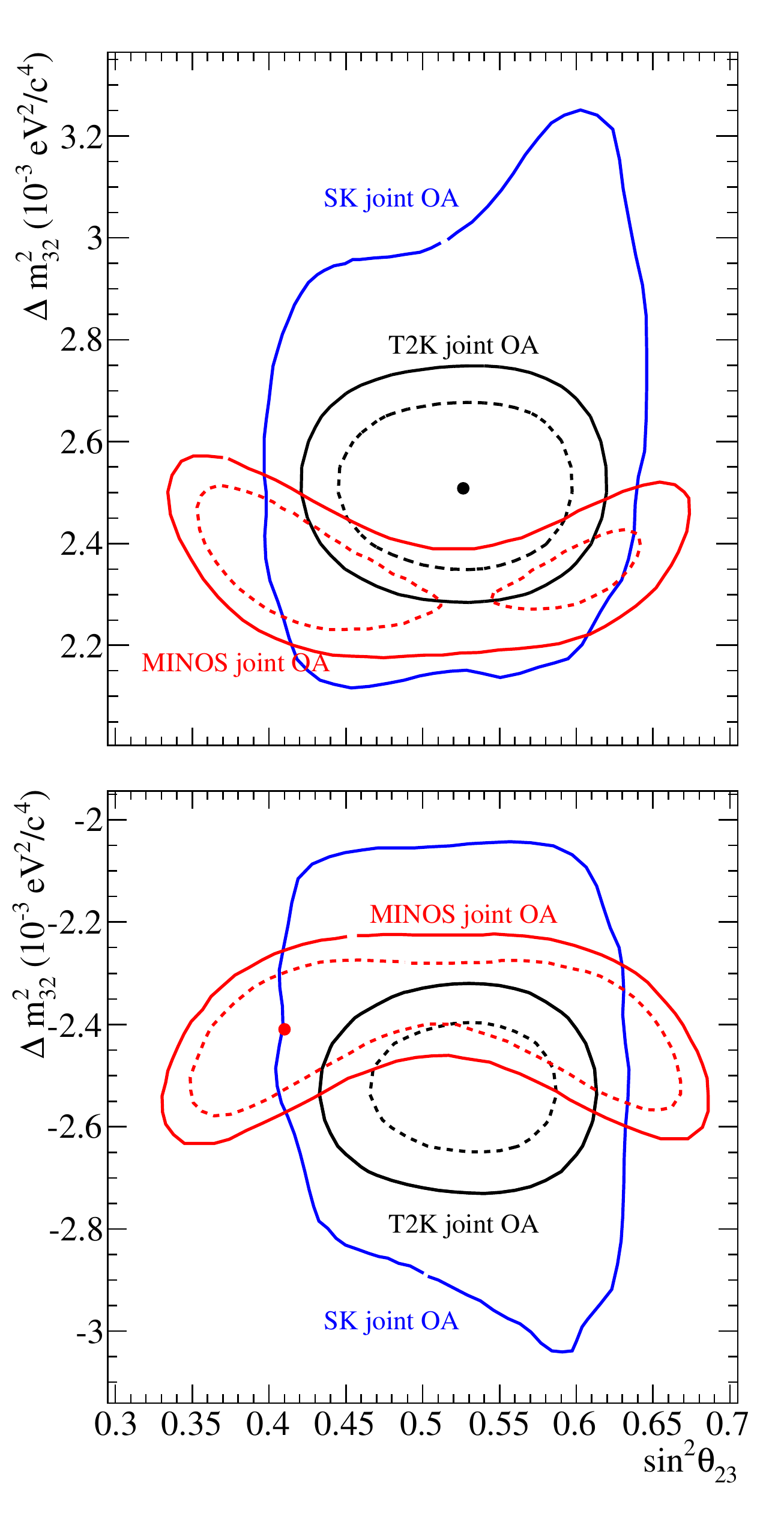}
\caption {
68\% (dashed) and 90\% (solid) CL regions for normal (top) and inverted (bottom) mass hierarchy
combined with the results from reactor experiments in the (\stt, \dmsq) space compared
to the results from the Super-Kamiokande~\cite{Himmel:2013jva} and MINOS~\cite{Adamson:2014vgd} experiments.
}
\label{fig:jointfreq:SKMINOS}
\end{center}
\end{figure}

The confidence regions obtained in the (\stt, \Dmsq) space are compared with the results
from Super-Kamiokande~\cite{Himmel:2013jva} and the MINOS~\cite{Adamson:2014vgd} experiments in Fig.~\ref{fig:jointfreq:SKMINOS}.
The results from T2K and MINOS used the latest value of \sot\ from~\cite{PDG2013}
to fit this parameter whereas the result from SK has \sot\ fixed to the previous reactor value in~\cite{PDG2012}. In the three
analyses \dcp\ was removed by profiling. 

\begin{figure}[tbp]
\begin{center}
\includegraphics[width=0.8\textwidth]{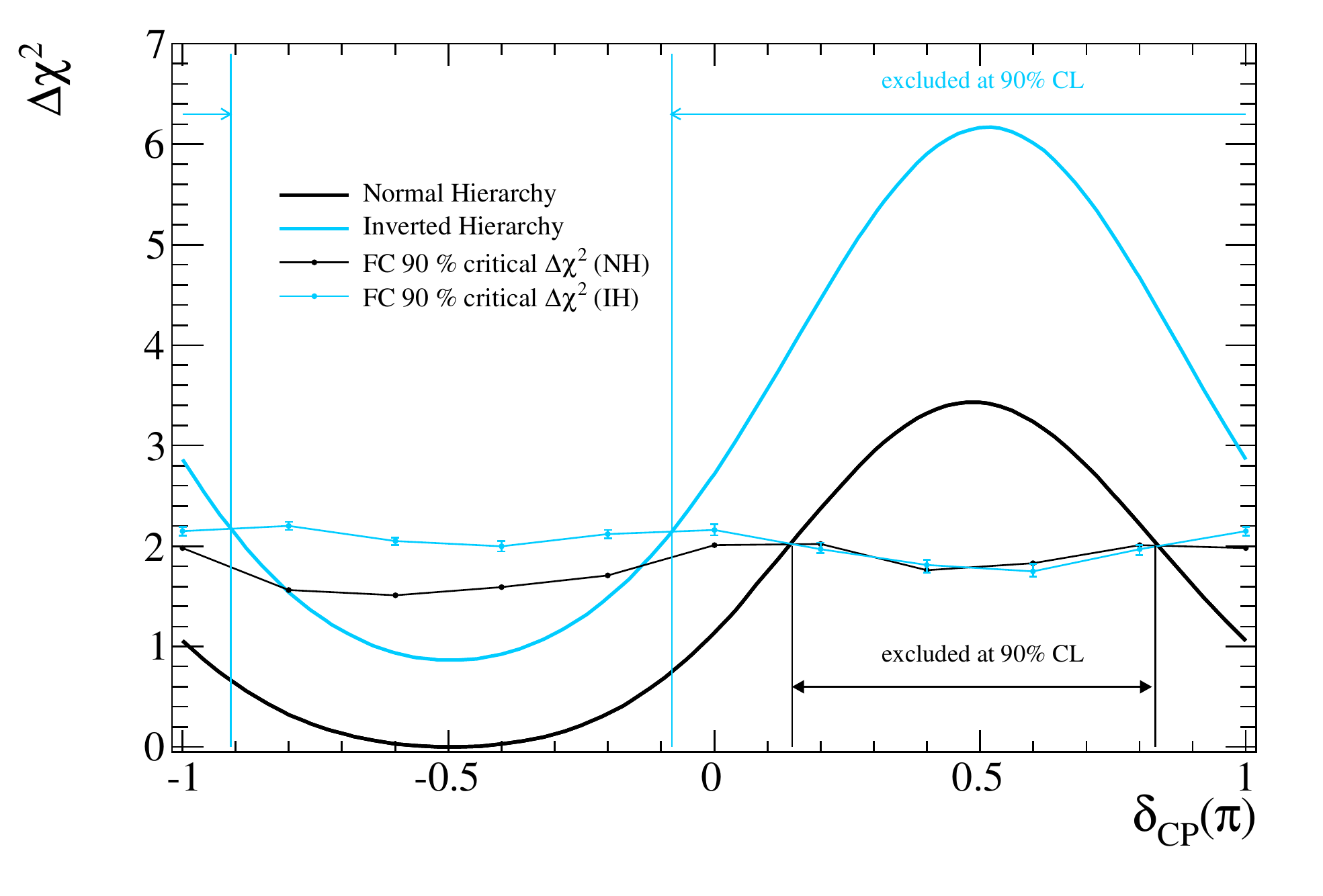}
\caption {
Profiled \Dchisq as a function of \dcp\ with the results of the critical \Dchisq
values for the normal and inverted hierarchies for the 
joint fit with reactor constraint,
with the excluded regions found overlaid.}
\label{fig:jointfreq:FC}
\end{center}
\end{figure}

An analysis using the Feldman and Cousins method was performed for the 
measurement of \dcp\ including a reactor constraint by
creating 4000 toy MC experiments at fixed values of \dcp\ in the interval [-$\pi$, $\pi$] (divided into 51 bins),
taking into account statistical fluctuations and systematic variations.
The other three oscillation parameters are removed by profiling following
the 3-dimensional \Dchisq surface obtained as a result of the joint 
fit with the reactor constraint. The values of the critical \Dchisq calculated using these toy 
experiments are overlaid with the curve of \Dchisq as a function of \dcp\ in Fig.~\ref{fig:jointfreq:FC}, and give the following 
excluded regions for \dcp\ at the 90\% C.L: \dcp = [0.15,0.83]$\pi$ for normal hierarchy and \dcp = [$-$0.08,1.09]$\pi$ for inverted hierarchy.

In order to thoroughly cross-check the analysis described above, 
an alternate frequentist joint fit analysis was performed which differs in the treatment of the systematic errors.
This originated as part of an effort to simplify and reduce the computing power needed for the analysis and to perform a study of the future sensitivity of the experiment~\cite{Abe:2014tzr}.
A new set of systematic parameters is used; they multiply the nominal expected number of \num\ or \nue\ events, with one parameter for each reconstructed energy bin. 
Results from the alternate analysis agree with the results presented in Secs.~\ref{sec:jointfreq:results} and~\ref{sec:jointfreq:results_reactor}.

\clearpage
\section{\label{sec:jointbayes} Joint $\num \rightarrow \num$ and $\num \rightarrow \nue$ Bayesian Analysis}
This section describes a complementary approach to the analysis detailed in Sec.~\ref{sec:jointfreq}, which uses Bayesian techniques to extract most probable values of oscillation parameters and their uncertainties. Bayesian inference analysis methods construct posterior probabilities of a hypothesis given the data observed by combining prior information with the likelihood function. This technique allows one to naturally include prior information about systematic parameters and external experimental data in the interpretation of the results of the experiment. Another distinguishing feature for this analysis is the fact that full marginalization of systematic parameters is achieved intrinsically, without the assumption that the observables are linear functions of the systematic parameters, taking into account the actual dependencies on the nuisance parameters.

The posterior distribution, produced using Bayes' theorem, is too difficult to compute analytically. We use two numerical methods to perform the high-dimensional integral necessary when computing the posterior distribution: a Markov Chain Monte Carlo (MCMC) in Sec.~\ref{sec:mcmc_analysis} and a sampling method in Sec.~\ref{sec:crosscheck_analysis} which is used as a cross-check.

\subsection{Joint Near-Far Markov Chain Monte Carlo Analysis}
\label{sec:mcmc_analysis}
\subsubsection{Point estimates}
To extract information about the point estimate of oscillation parameters from the posterior distribution generated by the MCMC, the density of points in 4-dimensional space was estimated using a kernel density estimator (KDE)~\cite{mills2011efficient,cranmer2001kernel}. A KDE estimates a PDF by smearing the discrete points of a MCMC in the 4 dimensions of interest. The Gaussian width of the smearing was set to be variable, and inversely proportional to the local density of MCMC points; this technique counters potential under-smoothing in low density regions and potential over-smoothing in high density regions. The maximum of the PDF produced by the KDE was then maximized using MINUIT to find the most probable value. In the case of using only T2K data, there is little sensitivity to the $\delta_{CP}$ parameter, and so a line of most probable values was created by finding the 3-dimensional density of the MCMC at a series of values of $\delta_{CP}$.

\subsubsection{Samples}

Unlike the frequentist analyses described above, the joint near-far analysis does not use the covariance matrix produced by the ND280 analysis described in Sec.~\ref{sec:BANFF}. Instead, this analysis is performed simultaneously with the three ND280 \num CC samples, and the SK $\nu_{\mu}$ CC, and SK $\nu_e$ CC samples. By fitting all samples simultaneously, this analysis avoids any error coming from neglecting non-linear dependencies of the systematic parameters constrained by ND280 analysis on the oscillation parameters. 

The systematic uncertainties used for the ND280 samples are nearly identical to those in Sec.~\ref{sec:BANFF} with the following exceptions: the uncertainties on the cross section ratios $\sigma_{\nue}/\sigma_{\num}$ and $\sigma_{\bar{\nu}}/\sigma_{\nu}$ are applied and the NC normalization uncertainties are divided into NC1$\pi^0$, NC$1\pi^{\pm}$, NC coherent, and NCOther for all samples. Additionally, the number of bins in the ND280 detector systematic covariance matrix is reduced to 105, in order to reduce the total number of parameters. There are no differences in the systematic uncertainties for the SK samples. 
Ignoring constant terms, the negative log of the posterior probability is 
given by,

\begin{equation}
\begin{split}
-\ln(P) = & \sum_{i}^{ND280bins}N^{p}_{i}(\vec{b},\vec{x},\vec{d})
-N^{d}_{i} \ln N^{p}_{i}(\vec{b},\vec{x},\vec{d}) \\
& +  \sum_{i}^{N_{\mu\ \mathrm{bins}}}N^{p}_{\mu,i}(\vec{\theta}, \vec{b},\vec{x},\vec{s})
-N^{d}_{\mu,i} \ln N^{p}_{\mu,i}(\vec{\theta}, \vec{b},\vec{x},\vec{s}) \\
& +  \sum_{i}^{N_{e\ \mathrm{bins}}}N^{p}_{e,i}(\vec{\theta},\vec{b},\vec{x},\vec{s})
-N^{d}_{e,i} \ln N^{p}_{e,i}(\vec{\theta}, \vec{b},\vec{x},\vec{s}) \\
& + {\textstyle \frac{1}{2}}\, \Delta\vec{b}^T V_b^{-1} \Delta\vec{b}
+ {\textstyle \frac{1}{2}}\, \Delta\vec{x}^T V_x^{-1} \Delta\vec{x}
+ {\textstyle \frac{1}{2}}\, \Delta\vec{d}^T V_d^{-1} \Delta\vec{d} \\
& + {\textstyle \frac{1}{2}}\, \Delta\vec{s}^T V_s^{-1} \Delta\vec{s} 
+ {\textstyle \frac{1}{2}}\, \Delta\vec{\theta}_{sr}^T V_{\theta sr}^{-1} \Delta\vec{\theta}_{sr} \ \ .\\
\end{split}
\label{eq:likelihood}
\end{equation}
The vector $\vec{\theta}_{sr}$ contains the solar oscillation parameters and for combined fits with reactor data $\sin^2 2\theta_{13}$, with priors described in Sec.~\ref{sec:OA:osc}.
The priors on the other oscillation parameters of interest are uniform in $\sin^2 \theta_{13}$ between 0 and 1,  $\sin^2 \theta_{23}$ between 0 and 1, $|\Delta m^2_{32}|$ between 0.001 and 0.005 \evvcccc, and $\delta_{CP}$ between $-\pi$ and $\pi$. Additionally, the prior probability of the normal hierarchy and inverted hierarchy are each 0.5. Priors for the systematic parameters are the multivariate Gaussian terms shown, with the exception of the cross section spectral function parameters which are given a uniform prior between 0 and 1. 

In this analysis, both ND280 and SK MC sample events are weighted individually for all parameters in the analysis. This means that each PDF is rebuilt from the MC at every iteration of the MCMC. This has the advantage of retaining shape information within each bin of the PDF, especially desirable for the oscillation parameters, and also allows a more natural treatment of certain parameters such as the SK energy scale uncertainty which may cause events to migrate between bins. The increase in computational load was offset by performing certain calculations on GPUs, including the event-by-event calculation of oscillation probability~\cite{calland2014accelerated}.

\subsubsection{Results}
The MCMC was run with $5.6\times10^7$ steps using only T2K data, and for $1.4\times10^8$ steps for T2K data combined with reactor experiment results. The most probable values for the oscillation parameters for both analyses are shown in Table~\ref{tab:data_results}. For the T2K-only analysis, the values are shown for \dcp=0, as the analysis has little sensitivity to the value of \dcp. The 68\% 1D credible intervals, marginalized over all other parameters, including mass hierarchy, for each of the parameters except $\dcp$ are shown in Table~\ref{tab:data_CI}. 

\begin{table}[tbp]
\caption{Most probable values for oscillation parameters from Bayesian analysis.}
    \begin{tabular}{llcccc} \toprule
           & Hierarchy \ \ & $|\Delta m^{2}_{32} |$ & \ \ $\sin^{2}\theta_{23}$ \ \ & \ \ $\sin^{2}\theta_{13}$ \ \ & \ \ $\dcp$ \ \ \\ 
  Analysis &           &   \ \ $10^{-3}$ \evvcccc \ \     &                       &                       &   \\ \hline
    T2K-only &Inverted  & 2.571 & 0.520 & 0.0454 & \ \ 0 (fixed) \ \ \\ 
        T2K+reactor \ \ & Normal  & 2.509 & 0.528 & 0.0250 & -1.601 \\ \botrule
    \end{tabular}
\label{tab:data_results}
\end{table}

\begin{table}[tbp]
\caption{68\% Bayesian credible intervals for oscillation parameters.} 
    \begin{tabular}{lcccc} \toprule
              & $|\Delta m^{2}_{32} |$ & $\sin^{2}\theta_{23}$ & $\sin^{2}\theta_{13}$ \\ 
     Analysis \ \ & \ \  $10^{-3}$ \evvcccc \ \    &                       &         \\ \hline
    T2K-only  &[2.46, 2.68] & \ \ [0.470, 0.565] \ \ & \ \ [0.0314 ,0.0664] \ \  \\ 
    T2K+reactor &[2.40, 2.62] & [0.490, 0.583] & [0.0224, 0.0276]   \\ 
    \botrule
    \end{tabular}
\label{tab:data_CI}
\end{table}

\begin{figure}
\includegraphics[width=0.7\textwidth]{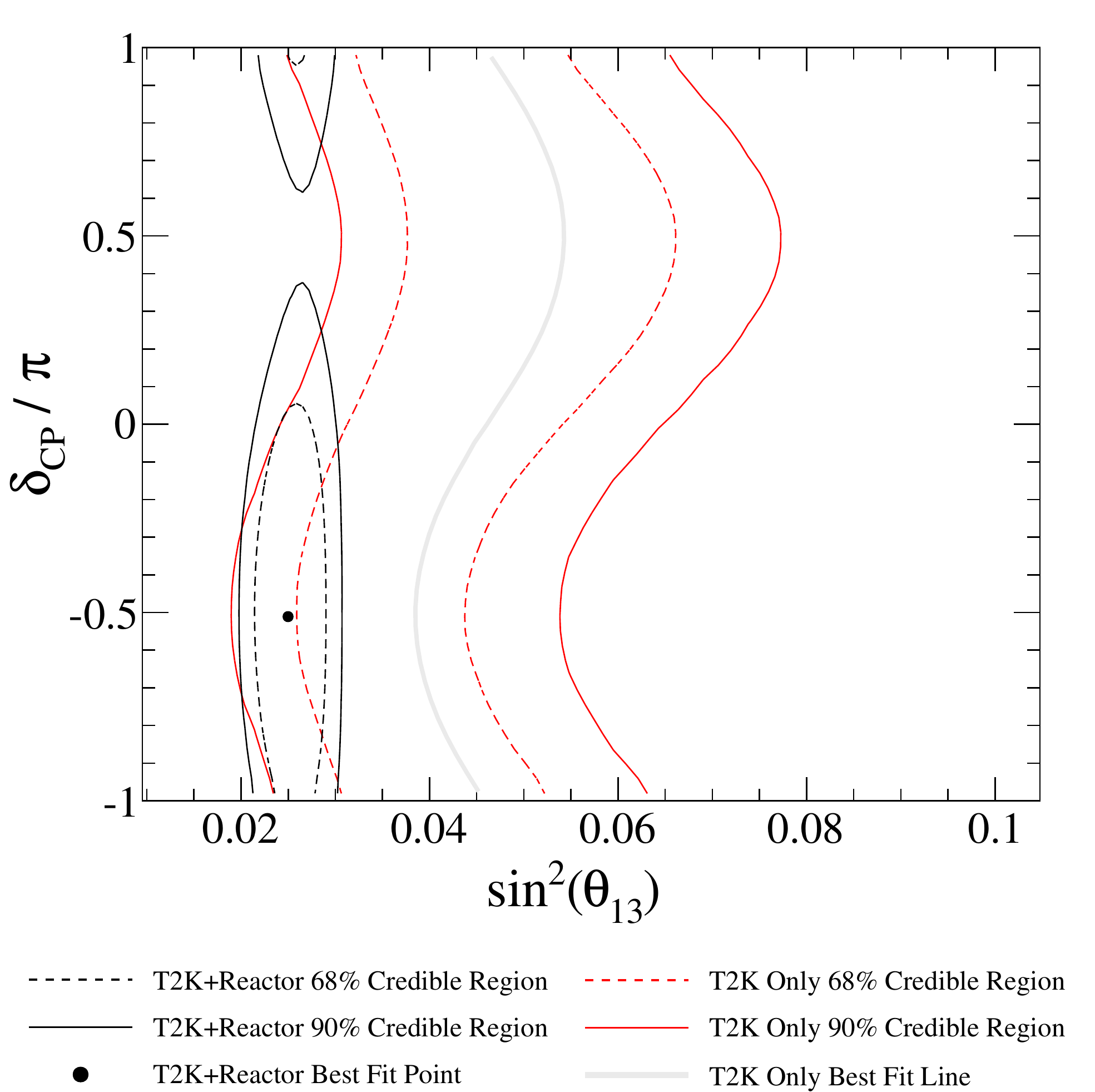}
\caption{Credible regions for $\sin^2\theta_{13}$ and $\delta_{CP}$ for T2K-only and T2K+reactor combined analyses. These are constructed by marginalizing over both mass hierarchies.
For the T2K-only analysis, the best fit line is shown instead of the best fit point because the analysis has little sensitivity to \dcp.}
\label{fig:th13_dcp_reactor}
\end{figure}

\begin{figure}
\includegraphics[width=0.7\textwidth]{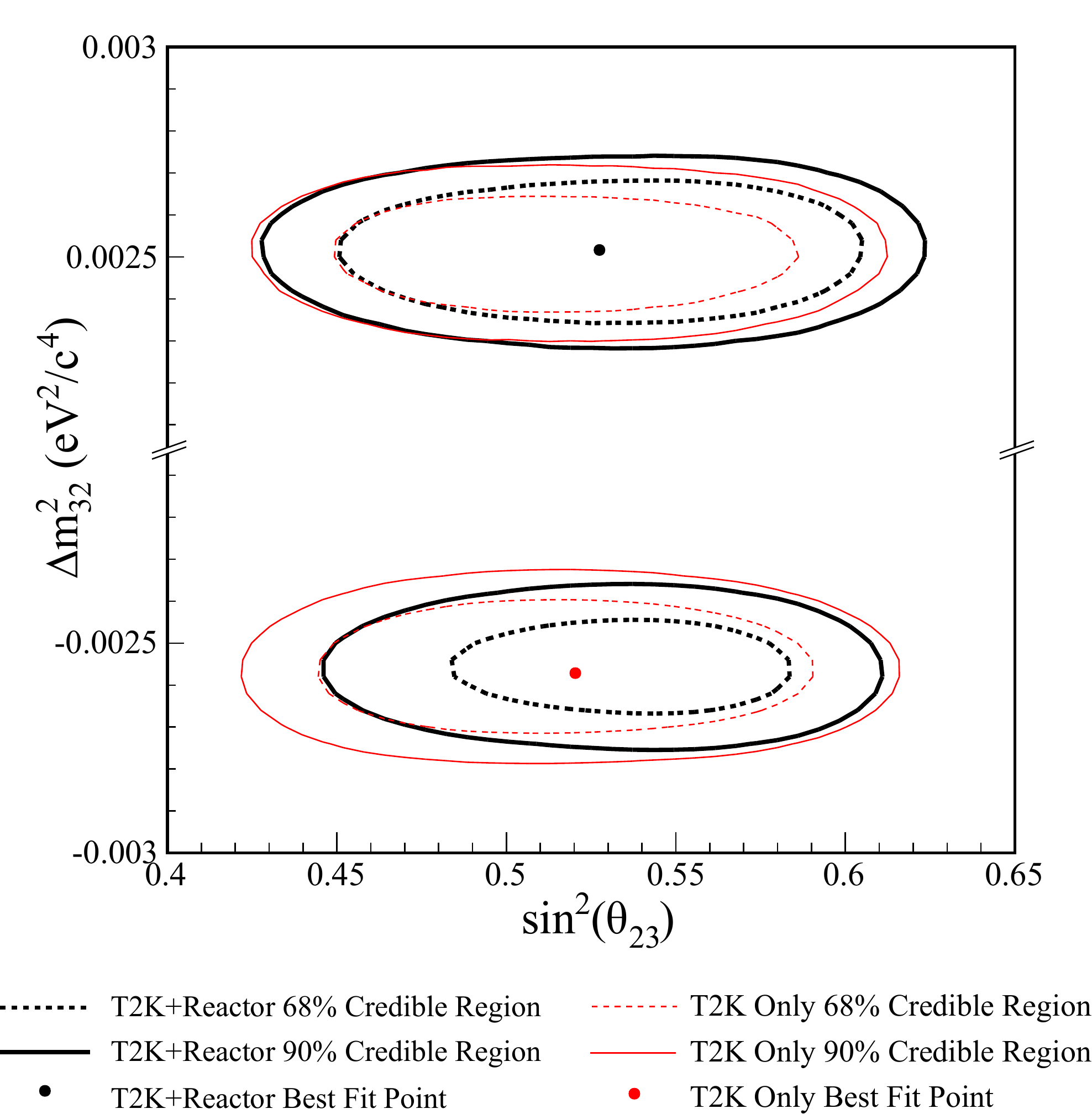}
\caption{Credible regions for $\sin^2\theta_{23}$ and $\Delta m^2_{32}$ for T2K-only and T2K+reactor combined analyses. The normal hierarchy corresponds to positive values of $\Delta m^2_{32}$ and the inverted hierarchy to negative values.  }
\label{fig:th23_dm32_both}
\end{figure}

\begin{figure}
\includegraphics[width=0.5\textwidth]{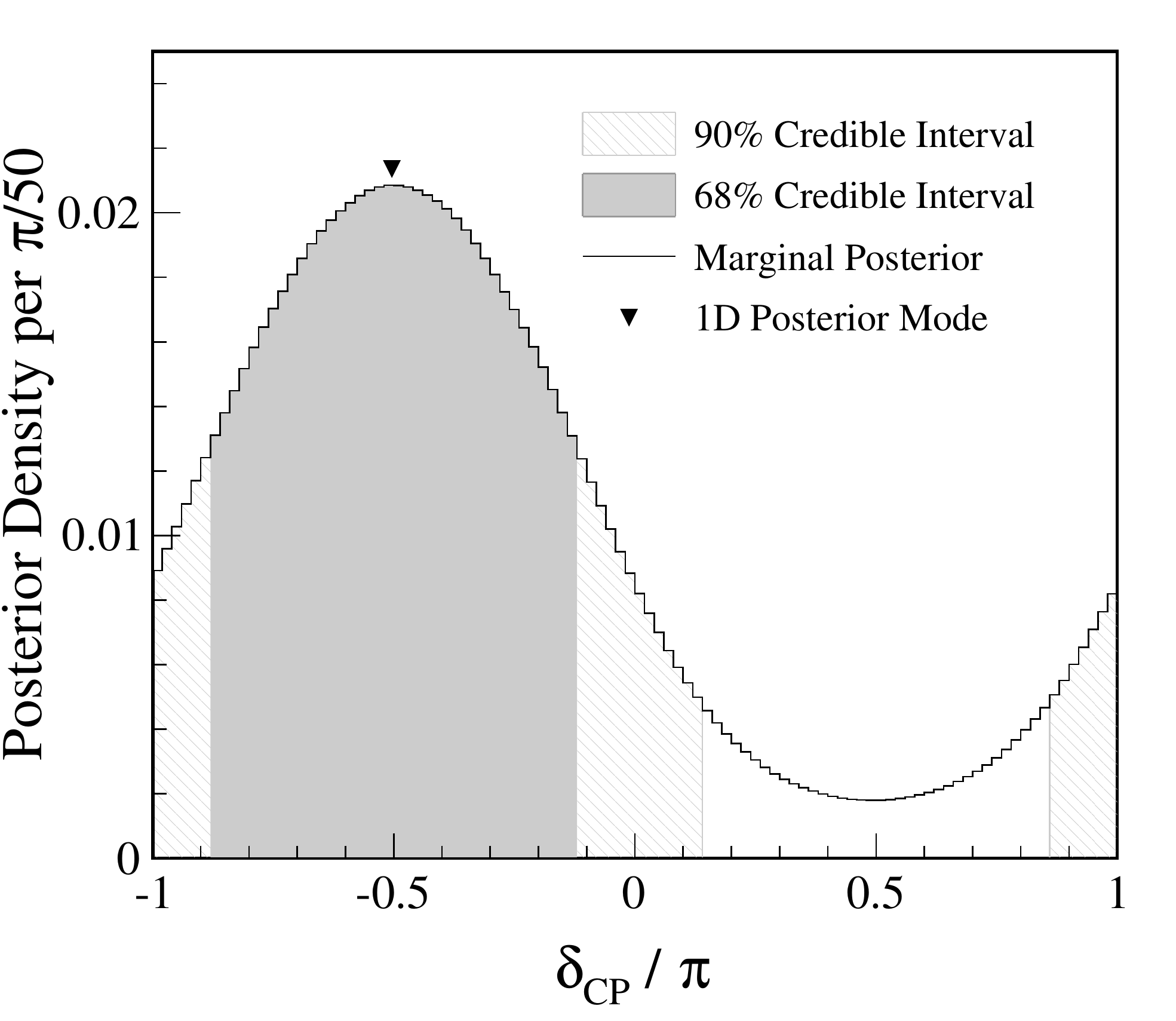}
\caption{The posterior probability for \dcp, marginalized over all other parameters, including mass hierarchy, for the T2K+reactor combined analysis. }
\label{fig:dcp_marg_MCMC}
\end{figure}

\begin{figure*}
\centering
\includegraphics[width=1.0\textwidth]{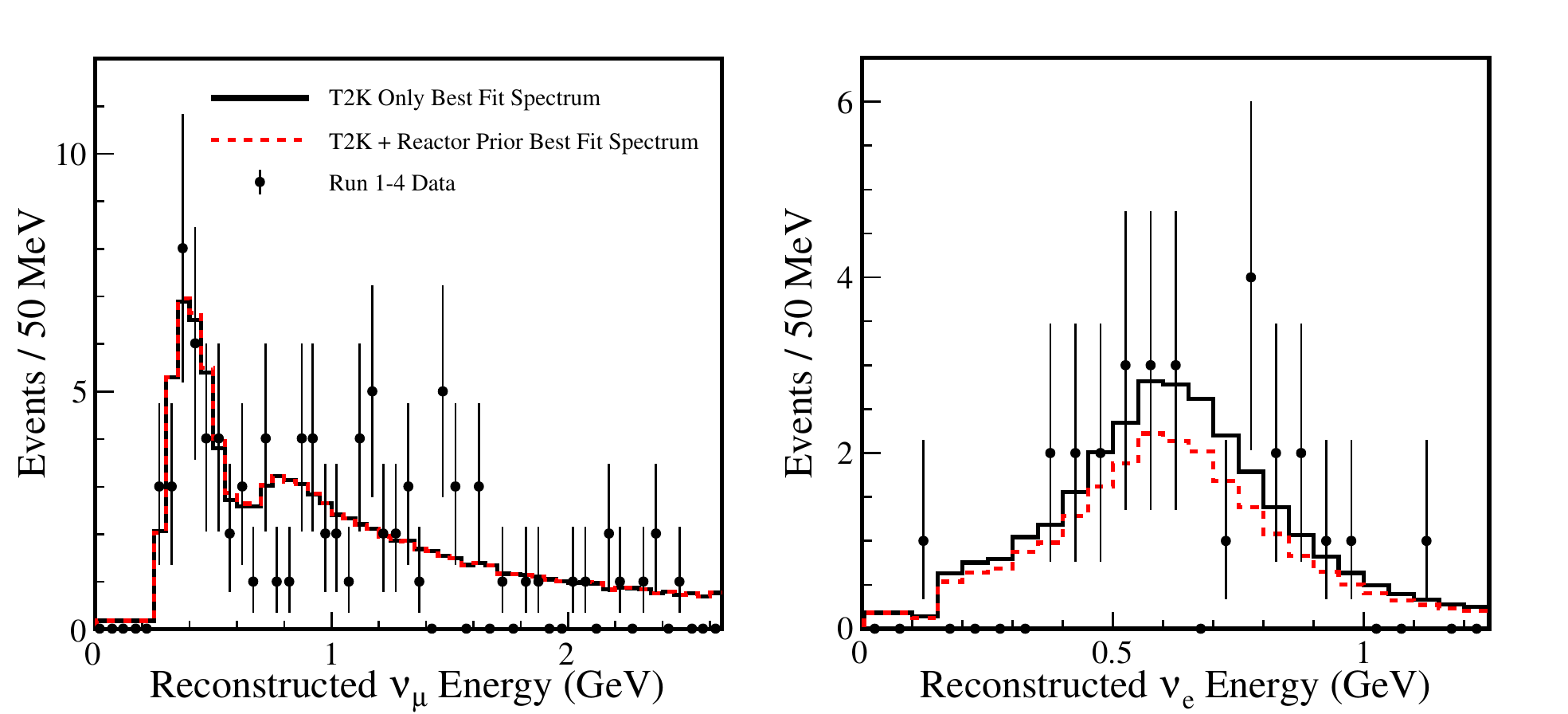}
\caption{T2K-only and T2K+reactor prior best-fit spectra overlaid with SK $\nu_\mu$ CC and $\nu_e$ CC candidate samples.}
\label{fig:bayes_overlay}
\end{figure*}

Figures~\ref{fig:th13_dcp_reactor} and~\ref{fig:th23_dm32_both} show the \dcp\ versus $\sin^2\theta_{13}$ and $\Delta m^2_{32}$ versus $\sin^2\theta_{23}$ credible regions for the T2K-only and T2K+reactor analyses. Note that the contours in Fig.~\ref{fig:th13_dcp_reactor} are marginalized over the mass hierarchy; in particular, the most probable value line appears to be offset from the center of the credible region. This is because the most probable value line is for the preferred inverted hierarchy, and the credible intervals are marginalized over hierarchy. Fig.~\ref{fig:dcp_marg_MCMC} shows the posterior probability for \dcp\ with 68\% and 90\% credible intervals for the T2K+reactor combined analysis. Figure~\ref{fig:bayes_overlay} shows comparisons of SK $\nu_\mu$ CC and $\nu_e$ CC candidate events with the best-fit spectra produced from the T2K-only and T2K+reactor combined analyses. 
Each best-fit spectrum is formed by calculating the most probable value for the predicted
number of events in each energy bin, using all of the MCMC points from the corresponding analysis.
The fit spectrum for $\nu_{\mu}$ CC events does not change appreciably when the reactor prior is included, but the $\nu_e$ CC fit spectrum shows a noticeable reduction in the number of events. 

\begin{figure*}
\centering
\includegraphics[width=0.8\textwidth]{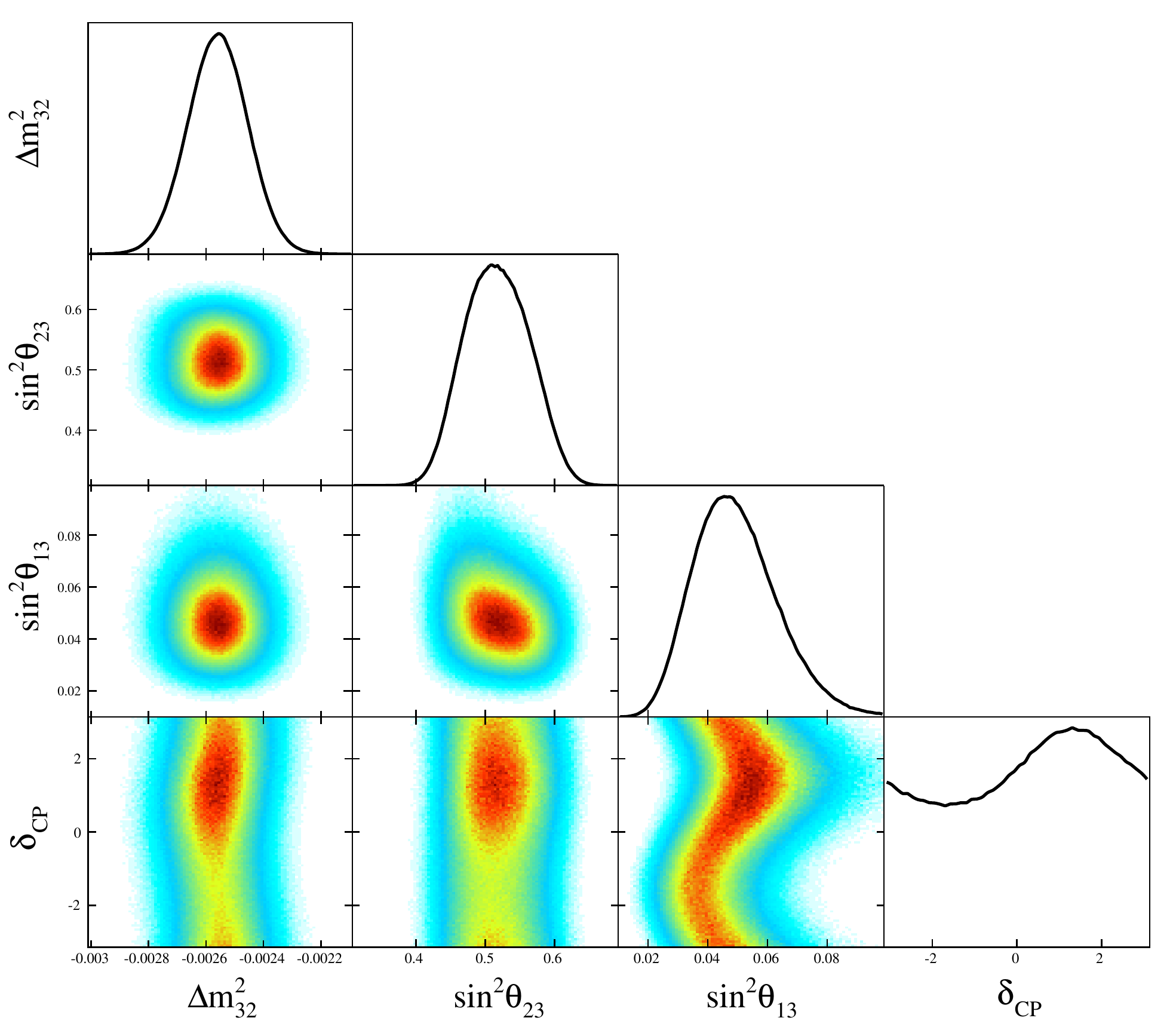}
\caption{Distributions of posterior probability between the oscillation parameters of interest for the T2K-only analysis. These posteriors use only MCMC points that are in the inverted hierarchy.}
\label{fig:osc_tri_t2konly}
\end{figure*}

\begin{figure*}
\centering
\includegraphics[width=0.8\textwidth]{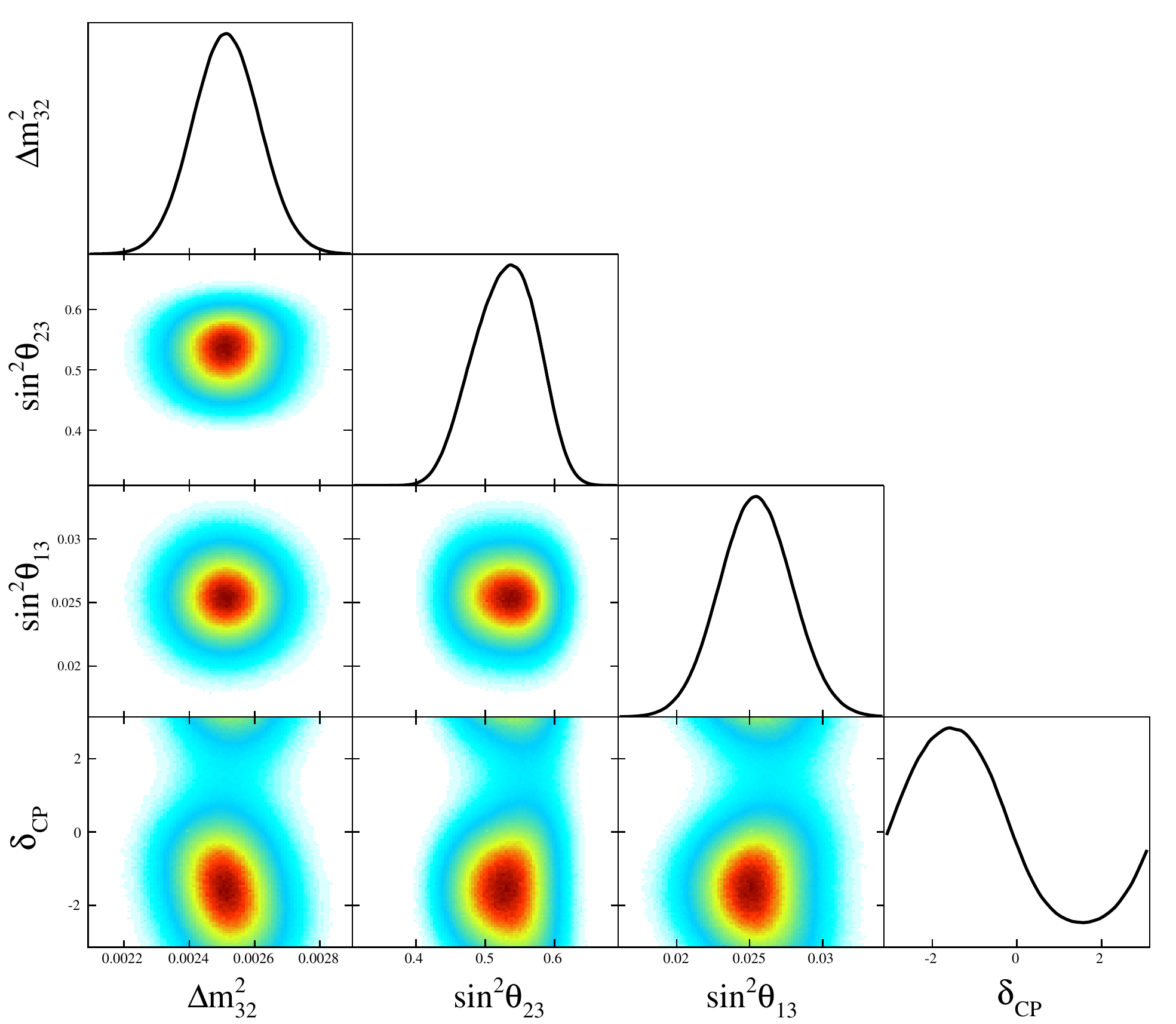}
\caption{Distributions of posterior probability between the oscillation parameters of interest for the T2K+reactor analysis. These posteriors use only MCMC points that are in the normal hierarchy.
In comparing to Fig.~\ref{fig:osc_tri_t2konly}, note the change in scales for some parameters.}
\label{fig:osc_tri_reactor}
\end{figure*}

Figures~\ref{fig:osc_tri_t2konly} and~\ref{fig:osc_tri_reactor} show the posterior PDFs for the oscillation parameters both singly and pairwise, using MCMC points from the inverted and normal hierarchy respectively, which reflect the most probable mass hierarchy 
for the T2K-only and T2K+reactor analysis respectively.
The plots along the diagonal show the posterior PDFs for each of the four oscillation parameters of interest, marginalized over all other parameters, except for the mass hierarchy. The off-diagonal elements show the pairwise posterior PDFs.

Another interesting feature of this analysis is that it provides a natural way to study the preference of the data for normal versus inverted hierarchy and lower versus upper octant in $\theta_{23}$.  This is done simply by comparing the total probability (that is, the number of MCMC steps) in the region of interest. Table~\ref{tab:dm_model_comp_t2konly} shows the probability for the various cases for the T2K-only analysis. Note that the inverted hierarchy is preferred in this analysis, but the posterior odds 
ratio\footnote{with the prior odds assumed to be 1, the posterior odds ratio is equivalent
to the Bayes Factor}
is only 1.2. Table~\ref{tab:dm_model_comp} shows the same for the T2K+reactor combined analysis. In this analysis, the normal hierarchy is preferred, but with a posterior odds ratio of 2.2, 
the inverted hierarchy is not significantly excluded with the present analysis.

To evaluate the dependency of this analysis on the form of the prior of the oscillation parameters, the analysis was repeated with a uniform prior in $\theta_{13}$ and $\theta_{23}$. The credible intervals and model comparison probabilities do not change appreciably with these alternative priors.

\begin{table}[tbp]
\centering
\caption{Model comparison probabilities for normal and inverted mass hierarchies, as well as upper and lower octants, without including reactor data.}
    \begin{tabular}{c c c c} 
\hline\hline
    & NH & IH & Sum\\	
    \hline
    $\sin^2\theta_{23}\leq0.5$\ \ &\ \ 0.165\ \ &\ \ 0.200\ \ &\ \ 0.365\ \ \\
    $\sin^2\theta_{23}>0.5$&0.288&0.347&0.635\\
    \hline
    Sum & 0.453& 0.547&1.0\\
\hline\hline
    \end{tabular}
\label{tab:dm_model_comp_t2konly}
\end{table}

\begin{table}[tbp]
\centering
\caption{Model comparison probabilities for normal and inverted mass hierarchies, as well as upper and lower octants, including reactor data.}
    \begin{tabular}{c c c c}
\hline\hline 
    & NH & IH & Sum\\	
    \hline
    $\sin^2\theta_{23}\leq0.5$\ \ &\ \ 0.179\ \ &\ \ 0.078\ \ &\ \ 0.257\ \ \\
    $\sin^2\theta_{23}>0.5$&0.505&0.238&0.743\\
    \hline
    Sum & 0.684& 0.316&1.0\\
\hline\hline
    \end{tabular}
\label{tab:dm_model_comp}
\end{table}

\subsection{Cross-check analysis}
\label{sec:crosscheck_analysis}
A second Bayesian joint analysis (JB2) is used to cross-check the results from the analysis
described above (JB1).
Like the frequentist analyses, JB2
uses the output from the ND280 analysis described in Sec.~\ref{sec:BANFF} to constrain some of the systematic uncertainties, by applying them as prior
probability densities.
Also, JB2 does not use (by default) the reconstructed energy spectrum for $\nue$ candidate events, but instead the 2D distribution of the momentum and angle with respect to beam direction $(p_{e},\theta_{e})$  of the particle reconstructed as an electron in those events. This is similar to what was used in the previously reported electron neutrino appearance observation~\cite{Abe:2013hdq}. JB2 can also use the shape of the reconstructed energy spectrum for $\nue$ candidate events, so that the results of the two analyses can be compared in both cases. On a technical level, MCMC is not used in this second analysis to marginalize over the nuisance parameters; the integration is done numerically by averaging the posterior probability over 10,000 throws of those parameters following their prior distribution. Finally, a second technical difference is that in JB2 the weighting is not done event by event but by $(p_{e},\theta_{e})$ bin.

\subsection{Comparison of analyses}
\subsubsection{Comparison of Bayesian joint analyses}
\begin{figure}
\includegraphics[width=0.6\textwidth]{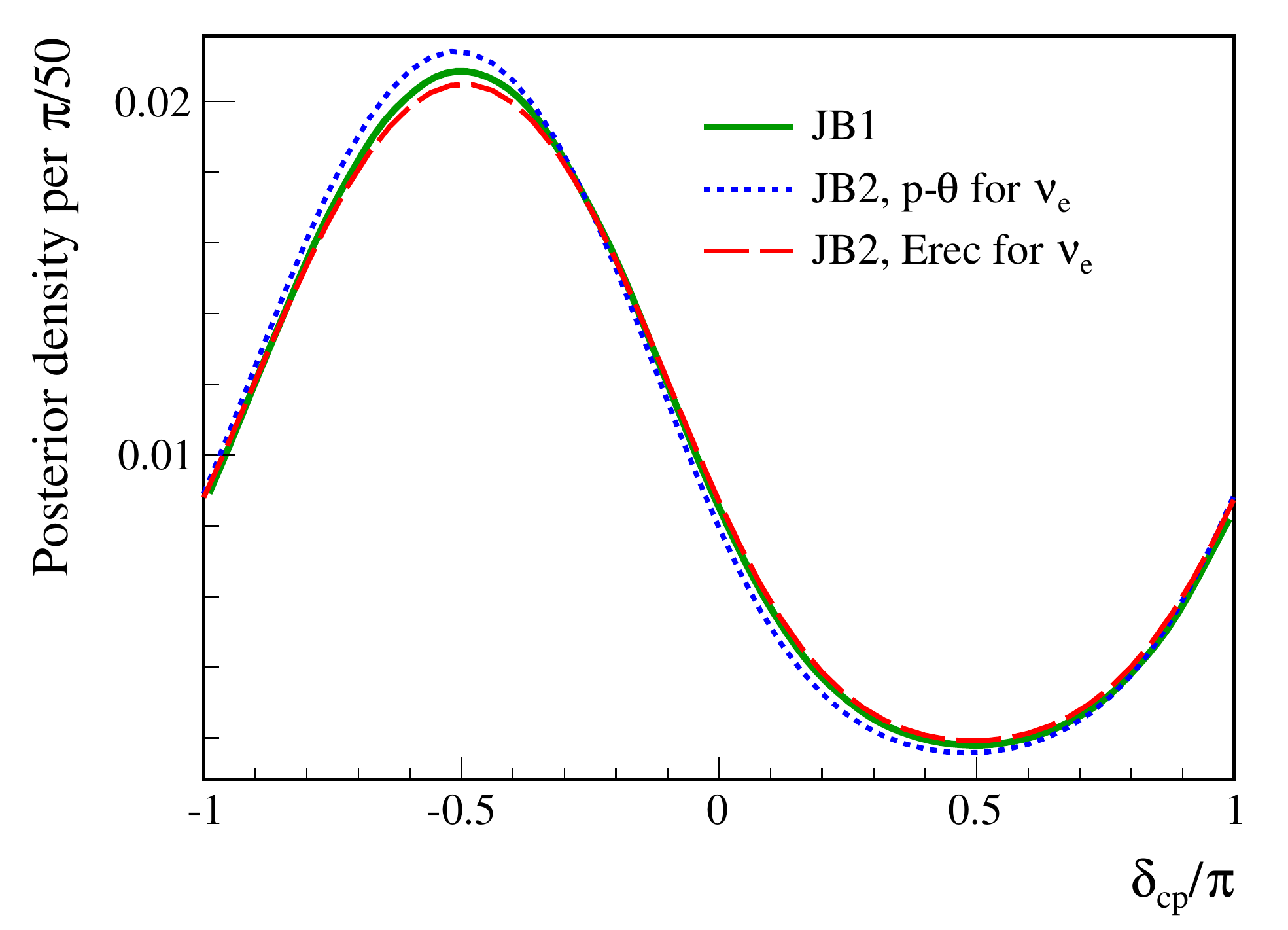}
\caption{Posterior probabilities for $\delta_{CP}$ obtained by the two joint Bayesian analyses using the reactor experiments prior for $\sin^{2}\theta_{13}$.}
\label{fig:CompPosterior}
\end{figure}

The results obtained with the two joint Bayesian analyses are very similar, both in terms of posterior probabilities for the different models and credible intervals for the oscillation parameters. The comparison in the case of the posterior probability for $\delta_{CP}$ is shown in Fig.~\ref{fig:CompPosterior}: the posterior probabilities obtained by the two analyses are similar, and most of the difference comes from JB2 using the $(p_{e},\theta_{e})$ spectrum shape for $\nue$ candidate events instead of the reconstructed energy spectrum shape as JB1 does. This also shows that at the current statistics, fitting the near and far detector samples at the same time and using the output of the near detector analysis described in Sec.~\ref{sec:BANFF} are equivalent.

\subsubsection{Treatment of the systematic uncertainties}
\begin{figure}
\includegraphics[width=0.6\textwidth]{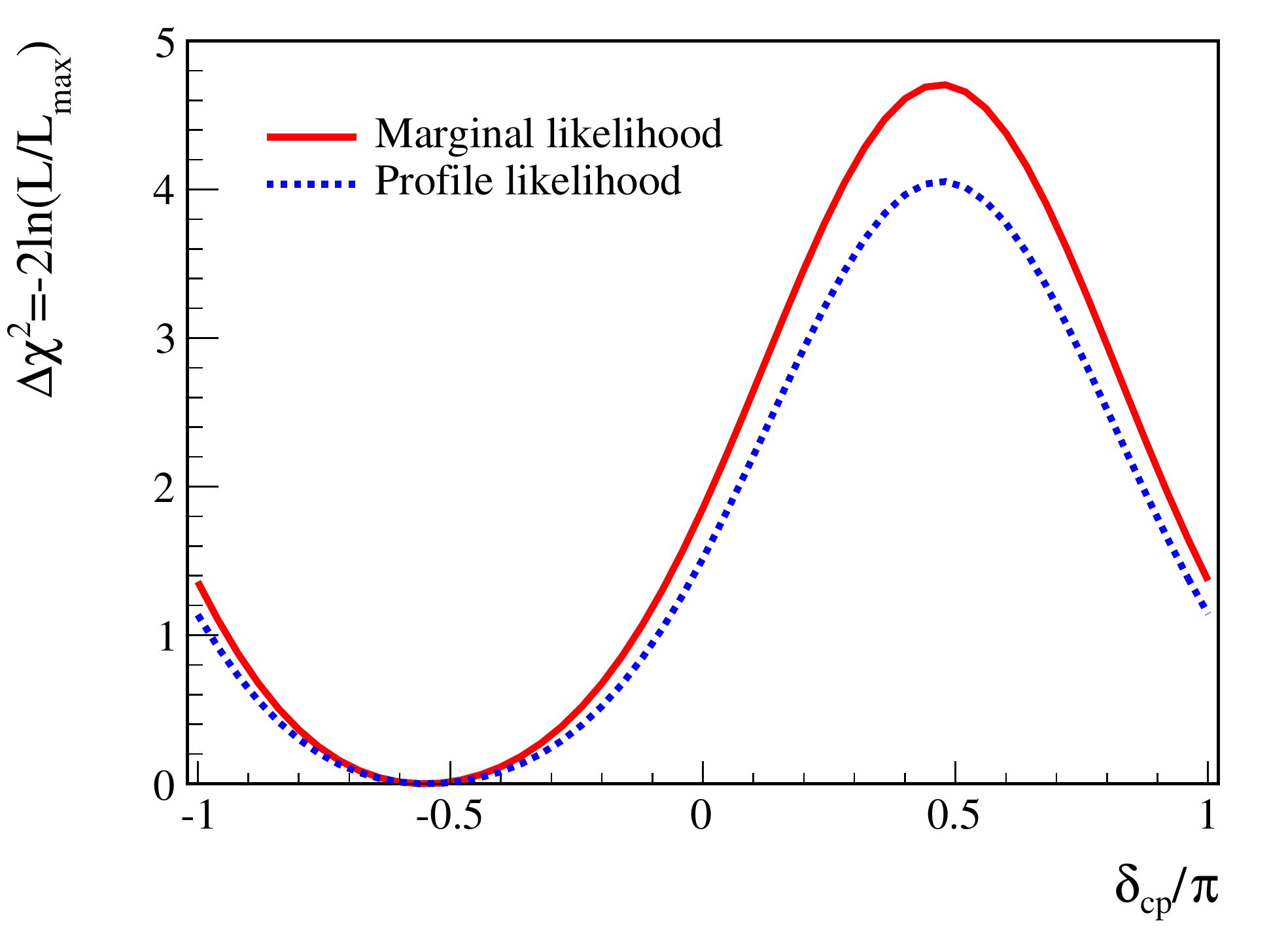}
\caption{Marginal and profile likelihoods of the T2K data with reactor constraint assuming normal hierarchy.}
\label{fig:CompTreatment}
\end{figure}

We also compare, using JB2, the marginalization and profiling approaches described in Sec.~\ref{sec:OA}C to reduce the dimensionality of the likelihood. 
In the case of $\delta_{CP}$, the marginal (obtained by integrating the product of the likelihood and priors over the nuisance parameters) and profile (obtained by maximizing the likelihood with respect to those parameters) likelihoods are visibly different, as can be seen on figure \ref{fig:CompTreatment}. Such differences are expected as some of the nuisance parameters appear in a non-Gaussian form and have a non-linear dependence. 
Within the Bayesian framework, only marginalization is well motivated.

\clearpage
\section{\label{sec:conclusions} Conclusions}
With the data collected between 2010 and 2013
we have analyzed the \num -disappearance to estimate the
two oscillation parameters, \Dmsq\ and \stt.
For the first time, we have used a combined analysis of \num -disappearance 
and \nue -appearance, to advance our knowledge of the
oscillation parameters \Dmsq, \stt, \sot, \dcp, 
and the mass hierarchy.

Uncertainty arising from systematic factors has been carefully assessed in
the analyses and its effect is small compared to statistical errors.
Our understanding of neutrino oscillation will continue to improve
as we collect more data in the coming years, in both
neutrino and anti-neutrino mode~\cite{Abe:2014tzr}.
The general approach followed in this paper that couples the
separate analysis of the beamline, neutrino interactions, near detectors,
and far detector, through sets of systematic parameters and their covariances,
will be extended to deal with additional information from
anti-neutrino data and from additional selections with the near detector data.

\begin{acknowledgments}
We thank the J-PARC staff for superb accelerator performance and the 
CERN NA61/SHINE collaboration for providing valuable particle production data.
 We acknowledge the support of MEXT, Japan; 
NSERC, NRC and CFI, Canada; 
CEA and CNRS/IN2P3, France; 
DFG, Germany; 
INFN, Italy; 
National Science Centre (NCN), Poland; 
RSF, RFBR and MES, Russia; 
MINECO and ERDF funds, Spain; 
SNSF and SER, Switzerland; 
STFC, UK; 
and DOE, USA. 
We also thank CERN for the UA1/NOMAD magnet, 
DESY for the HERA-B magnet mover system, 
NII for SINET4, 
the WestGrid and SciNet consortia in Compute Canada, 
GridPP, UK, 
and the Emerald High Performance Computing facility 
in the Centre for Innovation, UK. 
In addition participation of individual researchers 
and institutions has been further supported by funds from: ERC (FP7), EU; 
JSPS, Japan; 
Royal Society, UK; 
DOE Early Career program, USA.
\end{acknowledgments}

\clearpage
\raggedright

\bibliographystyle{apsrev4-1}
\bibliography{oa2014_prd}

\end{document}